\documentclass[11pt,notitlepage,a4paper]{article}
\usepackage[margin=1in]{geometry}
\usepackage{amsmath,amssymb}
\usepackage{graphicx}
\usepackage{booktabs}
\usepackage{longtable}
\usepackage{tabularx}
\usepackage{amsfonts}
\usepackage{algorithm}
\usepackage{algorithmic}
\usepackage{tikz}
\usetikzlibrary{arrows.meta,positioning,shapes.geometric,calc,backgrounds,fit,decorations.pathreplacing}
\usepackage{float}
\usepackage[round]{natbib}
\usepackage[singlespacing]{setspace}
\usepackage[dvipsnames]{xcolor}
\definecolor{linkcolor}{rgb}{0, 0, 0.35}
\usepackage{hyperref}
\hypersetup{
    colorlinks=true,
    linkcolor=linkcolor,
    citecolor=linkcolor,
    urlcolor=linkcolor,
    bookmarks=true,
}
\DeclareUnicodeCharacter{202F}{~} 

\newcommand{\ExpOneARevFactorOne}{Number of bidders}
\newcommand{\ExpOneARevFactorTOne}{17.8}
\newcommand{\ExpOneARevFactorTwo}{Discount factor ($\gamma$)}
\newcommand{\ExpOneARevFactorTTwo}{9.2}
\newcommand{\ExpOneARevFactorThree}{Update mode}
\newcommand{\ExpOneARevFactorTThree}{7.5}
\newcommand{\ExpOneARevTopPctEffect}{21.3}
\newcommand{\ExpOneARevTopVsAuctionRatio}{28.2}

\newcommand{\ExpOneARsqMin}{0.31}
\newcommand{\ExpOneARsqMax}{0.48}

\newcommand{\ExpOneARevAuctionT}{0.6}

\newcommand{\ExpOneARevAuctionPFmt}{= 0.529}
\newcommand{\ExpOneARevMeanFPA}{0.819}
\newcommand{\ExpOneARevMeanSPA}{0.813}
\newcommand{\ExpOneARevGapPct}{0.8}
\newcommand{\ExpOneAConvMedian}{55,000}

\newcommand{\ExpOneARevAuctionxNbidAbsT}{5.6}

\newcommand{\ExpOneARevAuctionxGammaAbsT}{4.4}

\newcommand{\ExpOneBRevFactorOne}{Number of bidders}
\newcommand{\ExpOneBRevFactorTOne}{6.5}
\newcommand{\ExpOneBRevFactorTwo}{State information}
\newcommand{\ExpOneBRevFactorTTwo}{5.9}
\newcommand{\ExpOneBRevFactorThree}{Auction format $\times$ Number of bidders}
\newcommand{\ExpOneBRevFactorTThree}{3.4}
\newcommand{\ExpOneBRevTopPctEffect}{24.0}

\newcommand{\ExpOneBRevAuctionT}{3.2}

\newcommand{\ExpOneBRevAuctionPFmt}{= 0.002}
\newcommand{\ExpOneBRevMeanFPA}{0.432}
\newcommand{\ExpOneBRevMeanSPA}{0.486}
\newcommand{\ExpOneBRevGapPct}{-11.0}
\newcommand{\ExpOneBConvMedian}{89,000}

\newcommand{\ExpOneBRevNbidAbsT}{6.5}

\newcommand{\ExpOneBAllRevAuctionAbsT}{1.4}
\newcommand{\ExpOneBAllRevAuctionPFmt}{= 0.152}

\newcommand{\ExpOneBAllRevNbidAbsT}{4.2}

\newcommand{\ExpOneBAllRevStateAbsT}{4.5}

\newcommand{\ExpOneBAllRevAuctionxNbidAbsT}{2.7}
\newcommand{\ExpOneBAllRevAuctionxNbidPFmt}{= 0.008}
\newcommand{\ExpOneBAllRevRsq}{0.254}
\newcommand{\ExpOneBAllMeanFPA}{0.546}
\newcommand{\ExpOneBAllMeanSPA}{0.521}
\newcommand{\ExpOneBEndMeanFPA}{0.432}
\newcommand{\ExpOneBEndMeanSPA}{0.486}
\newcommand{\ExpOneBFPAPremium}{0.114}
\newcommand{\ExpOneBSPAPremium}{0.035}
\newcommand{\ExpOneBAllFPATwoBid}{0.533}
\newcommand{\ExpOneBAllFPAFourBid}{0.559}
\newcommand{\ExpOneBAllSPATwoBid}{0.461}
\newcommand{\ExpOneBAllSPAFourBid}{0.581}

\newcommand{\ExpTwoANobs}{768}

\newcommand{\ExpTwoARevFactorOne}{Number of bidders}
\newcommand{\ExpTwoARevFactorTOne}{33.9}
\newcommand{\ExpTwoARevFactorTwo}{Auction format}
\newcommand{\ExpTwoARevFactorTTwo}{12.5}
\newcommand{\ExpTwoARevFactorThree}{Number of bidders $\times$ Reserve price}
\newcommand{\ExpTwoARevFactorTThree}{10.4}
\newcommand{\ExpTwoARevTopPctEffect}{57.0}

\newcommand{\ExpTwoARevAuctionT}{12.5}

\newcommand{\ExpTwoARevMeanFPA}{0.411}
\newcommand{\ExpTwoARevMeanSPA}{0.507}
\newcommand{\ExpTwoARevGapPct}{-19.0}
\newcommand{\ExpTwoAConvMedian}{16,000}

\newcommand{\ExpTwoBNobs}{192}

\newcommand{\ExpTwoBRevFactorOne}{Number of bidders}
\newcommand{\ExpTwoBRevFactorTOne}{8.7}
\newcommand{\ExpTwoBRevFactorTwo}{Reserve price}
\newcommand{\ExpTwoBRevFactorTTwo}{7.9}
\newcommand{\ExpTwoBRevFactorThree}{Number of bidders $\times$ Context richness}
\newcommand{\ExpTwoBRevFactorTThree}{4.7}
\newcommand{\ExpTwoBRevTopPctEffect}{20.6}

\newcommand{\ExpTwoBRevMeanFPA}{0.486}
\newcommand{\ExpTwoBRevMeanSPA}{0.532}
\newcommand{\ExpTwoBRevGapPct}{-8.7}
\newcommand{\ExpTwoBConvMedian}{10,000}

\newcommand{\ExpThreeANobs}{512}

\newcommand{\ExpThreeARevFactorOne}{Budget multiplier}
\newcommand{\ExpThreeARevFactorTOne}{39.4}
\newcommand{\ExpThreeARevFactorTwo}{Bidder objective}
\newcommand{\ExpThreeARevFactorTTwo}{28.1}
\newcommand{\ExpThreeARevFactorThree}{Bidder objective $\times$ Budget multiplier}
\newcommand{\ExpThreeARevFactorTThree}{27.3}
\newcommand{\ExpThreeARevTopPctEffect}{94.3}

\newcommand{\ExpThreeARsqMin}{0.12}
\newcommand{\ExpThreeARsqMax}{0.96}

\newcommand{\ExpThreeAPRESSGapMax}{0.081}

\newcommand{\ExpThreeABHPct}{84}

\newcommand{\ExpThreeARevAuctionT}{3.6}

\newcommand{\ExpThreeARevMeanFPA}{4440.570}
\newcommand{\ExpThreeARevMeanSPA}{4072.503}
\newcommand{\ExpThreeARevGapPct}{9.0}
\newcommand{\ExpThreeAConvMedian}{N/A}

\newcommand{\ExpThreeAAllocEffObjxNbidT}{10.2}

\newcommand{\ExpThreeAAllocEffObjxNbidPFmt}{< 10^{-21}}

\newcommand{\ExpThreeARevNbidAbsT}{22.8}

\newcommand{\ExpThreeAWarmObjT}{5.4}

\newcommand{\ExpThreeAIEVolObjT}{5.1}

\newcommand{\ExpThreeAWelfareRsqPct}{80}
\newcommand{\ExpThreeADriftRsq}{0.12}
\newcommand{\ExpThreeADriftRsqPct}{12}

\newcommand{\ExpThreeADriftFPFmt}{< 10^{-5}}
\newcommand{\ExpThreeAUtilEffLow}{0.43}
\newcommand{\ExpThreeAUtilEffHigh}{1.00}
\newcommand{\ExpThreeAValFourEffLow}{0.37}
\newcommand{\ExpThreeAValFourEffHigh}{0.75}

\newcommand{\ExpThreeBRevFactorOne}{Budget multiplier}
\newcommand{\ExpThreeBRevFactorTOne}{31.7}
\newcommand{\ExpThreeBRevFactorTwo}{Number of bidders}
\newcommand{\ExpThreeBRevFactorTTwo}{16.1}
\newcommand{\ExpThreeBRevFactorThree}{Auction format}
\newcommand{\ExpThreeBRevFactorTThree}{11.0}
\newcommand{\ExpThreeBRevTopPctEffect}{72.1}

\newcommand{\ExpThreeBRevMeanFPA}{4109.423}
\newcommand{\ExpThreeBRevMeanSPA}{3192.762}
\newcommand{\ExpThreeBRevGapPct}{28.7}
\newcommand{\ExpThreeBConvMedian}{N/A}

\newcommand{\ExpOneABNEMean}{1.079}

\newcommand{\ExpOneABNEStd}{0.403}
\newcommand{\ExpOneABNEPctTen}{32.4}

\newcommand{\ExpOneBBNEMedian}{0.990}

\newcommand{\ExpOneBBNEPctTen}{82.3}

\newcommand{\ExpTwoBBNEMedian}{1.105}
\newcommand{\ExpTwoBBNEStd}{0.311}

\title{Designing Auctions when Algorithms Learn to Bid}
\author{Pranjal Rawat\thanks{I thank John Rust, Nathan Miller, and Harry Paarsch for helpful comments and discussions. I am especially grateful to Harry Paarsch for the opportunity to present this paper at the WEAI Conference. I also thank participants at the Georgetown EGSO seminar and EconBrew for valuable feedback.}}
\date{\today}

\begin{document}

\maketitle

\begin{abstract}
Algorithms increasingly automate bidding in online auctions, raising concerns about tacit bid suppression and revenue shortfalls. Prior work identifies individual mechanisms behind algorithmic bid suppression, but it remains unclear which factors matter most and how they interact, and policy conclusions rest on algorithms unlike those deployed in practice. This paper develops a computational laboratory framework, based on factorial experimental designs and large-scale Monte Carlo simulation, that addresses bid suppression across multiple algorithm classes within a common methodology. Each simulation is treated as a black-box input-output observation; the framework varies inputs and ranks factors by association with outcomes, without explaining algorithms' internal mechanisms. Across six sub-experiments spanning Q-learning, contextual bandits, and budget-constrained pacing, the framework ranks the relative importance of auction format, competitive pressure, learning parameters, and budget constraints on seller revenue. The central finding is that structural market parameters dominate algorithmic design choices. In unconstrained settings, competitive pressure is the strongest predictor of revenue; under budget constraints, budget tightness takes over. The auction-format effect is context-dependent, favouring second-price under learning algorithms but reversing to favour first-price under budget-constrained pacing. Because the optimal format depends on the prevailing bidding technology, no single auction format is universally superior when bidders are algorithms, and applying format recommendations from one algorithm class to another leads to counterproductive design interventions.
\end{abstract}

\newpage
\tableofcontents
\newpage

\section{Introduction}

Intelligent algorithms are rapidly taking over real-time computerized auctions and markets in areas such as online marketplaces for pre-owned items, display advertising, sponsored search, financial trading, electricity, transportation, and public procurement. The prospect that these algorithms might autonomously learn to suppress bids or coordinate on supra-competitive outcomes has drawn attention from competition authorities worldwide \citep{OECD2017, EzrachiStucke2017}, yet the legal framework for addressing autonomous algorithmic convergence remains unsettled because existing antitrust doctrine requires evidence of agreement or concerted practice. Auctions originally designed for human participants may not remain efficient under algorithmic bidding. For instance, in large-scale advertising exchanges like Google AdSense, even minor inefficiencies or bid suppression can cause million-dollar losses. Recent work shows that reinforcement-learning agents can converge to sub-competitive bidding patterns through bid shading, failure to converge, or budget-mediated rationing \citep{Calvano2020, Klein2021, BanchioSkrzypacz2022}. 

The direction and magnitude of these revenue shortfalls depend on the bidding technology deployed by the participants. In particular, the effect of auction format on revenue reverses sign between unconstrained learning algorithms and budget-constrained pacing agents. Yet existing studies focus on narrow settings with few factors varied at a time (Section~\ref{sec:literature}). They employ algorithms (Q-learning and simple reinforcement-learning variants) that bear little resemblance to the budget-pacing systems deployed on major auction platforms, and do not systematically parse how algorithmic factors (learning rates, discount factors, synchronization modes, budget constraints) are associated with larger or smaller revenue shortfalls. Empirical data on real-world algorithmic auctions are scarce because most auctions are proprietary. Theoretical analysis faces a parallel barrier: modelling high-dimensional Q-learning systems is analytically intractable, and existing theory typically focuses on small state/action spaces and simple algorithms.

This paper develops a computational laboratory framework that jointly addresses algorithmic bid suppression and optimal auction design under autobidding. Factorial experimental designs, paired with large-scale Monte Carlo simulation, provide the controlled variation that theory cannot tractably derive and that proprietary market data do not make available. A parallel literature examines budget-constrained pacing in online advertising auctions \citep{Aggarwal2019, Balseiro2019Learning, Conitzer2022FPPE, Gaitonde2023}, where first-price auctions yield unique pacing equilibria and stronger welfare guarantees. This body of work reaches policy conclusions that directly contradict the Q-learning bid-suppression findings. Pacing theory favours first-price formats, whereas Q-learning studies find first-price auctions more vulnerable to bid suppression. The contradiction arises because the two literatures study fundamentally different bidding technologies; Section~\ref{sec:literature} reviews both in detail. The literature uses ``algorithmic collusion'' to denote any algorithmic revenue shortfall relative to competitive benchmarks \citep{Calvano2020, BanchioSkrzypacz2022}. That body of work supplies both theoretical predictions and experimentally identified factors, including exploration parameters, discount rates, synchronisation modes, and state representations, that directly inform the factorial designs in this paper. This paper measures those same shortfalls but treats them as black-box outcomes; it does not claim that they arise from strategic coordination or reward-punishment mechanisms. We use ``bid suppression'' throughout to describe the observed revenue gap without presupposing its cause.

Format recommendations derived from Q-learning experiments do not generalise to the pacing algorithms that dominate real advertising markets. This paper treats each simulation as an empirical input-output observation. We vary algorithmic and institutional inputs, measure auction outcomes, and report which factors are associated with larger or smaller revenue shortfalls, in what direction, at what magnitude, and how these associations differ across algorithm classes. We do not attempt to explain the internal mechanisms by which algorithms produce observed outcomes. In particular, we do not investigate whether agents develop reward-punishment strategies, whether observed bid suppression reflects genuine tacit coordination or merely suboptimal convergence, or whether budget-rational bid shading is involved. An algorithm that fails to explore sufficiently, one that rationally manages a finite budget, and one that learns to punish deviators can all produce identical revenue shortfalls in outcome data. The relevant question for auction design is not why revenue shortfalls occur, but which design parameters most effectively prevent them. The factorial framework provides a diagnostic. This paper addresses the gap by conducting a fully randomized factorial experiment comprising four experiments (six sub-experiments), each involving hundreds of independent trials and up to 100,000 simulated auctions per trial. Bidders apply reinforcement learning or bandit-based exploration under varied institutional elements (e.g., private vs.\ affiliated values, number of bidders, reserve prices) and algorithmic parameters (e.g., discount factors, Q-learning rates, synchronous vs.\ asynchronous updates, different exploration rules).

The experiments span Q-learning, contextual bandits (LinUCB and Thompson Sampling, analysed separately), and budget-constrained pacing algorithms, incorporate affiliated values to capture the continuum between private and common values, and measure revenue, price volatility, no-sale rates, and winner identity patterns. By varying auction format, the number of bidders, reserve prices, exploration schemes, and budget constraints in orthogonal factorial designs, the analysis cleanly separates main effects from interactions and ranks factors by statistical and economic significance. A further contribution is a false-positive warning. Budget-constrained pacing produces bid suppression that resembles algorithmic collusion in outcome data, yet responds to opposite design interventions, so misdiagnosis carries real costs for platform designers and regulators.

Two patterns emerge consistently. First, the degree of competition matters far more than algorithmic design choices for seller outcomes. In unconstrained settings the number of bidders is the strongest driver of revenue, consistent with classical auction theory \citep{BulowKlemperer1996}, while under budget constraints the budget multiplier takes over as the primary determinant. Second, and most consequential for policy, the effect of auction format on revenue depends on the bidding technology. Under Q-learning with constant valuations the two formats are equivalent; under contextual bandits first-price yields lower revenue; under budget-constrained pacing first-price yields higher revenue. The same bid-suppression pattern would suggest different causes and remedies depending on whether bidders use learning or pacing algorithms.

\subsection{Overview of Experimental Results}

We first establish a baseline using tabular Q-learning agents in unconstrained, constant-value environments (Experiment~1a). The number of bidders overwhelmingly dominates algorithmic hyperparameters in determining revenue, with an effect size of approximately 11\% of the grand mean. Auction format has a negligible main effect, establishing a boundary condition for recent claims regarding first-price auction vulnerability \citep{BanchioSkrzypacz2022}. Experiment~1b extends these agents to affiliated-value environments. The number of bidders remains the strongest factor at roughly 12\% of the grand mean, while auction format turns slightly negative, indicating a modest revenue advantage for second-price auctions that is small relative to competition.

Experiments~2a and~2b replace Q-learning with contextual bandits (LinUCB and Thompson Sampling). The number of bidders produces its largest effect in Experiment~2a, at roughly 29\% of the grand mean and a total-order Sobol index of 0.55; in Experiment~2b the effect is smaller but remains the dominant factor. Auction format becomes clearly negative under both algorithms, with second-price auctions generating higher revenue, a pattern directionally aligned with the \citet{MilgromWeber1982} revenue ranking for affiliated signals, though revenue equivalence holds in equilibrium because signals are independent (Appendix~\ref{sec:equilibria}). The emergence of a significant format effect under contextual bandits, but not under tabular Q-learning, demonstrates that the revenue consequences of auction design depend on the bidding technology.

\begin{figure}[H]
  \centering
  \includegraphics[width=\textwidth,keepaspectratio]{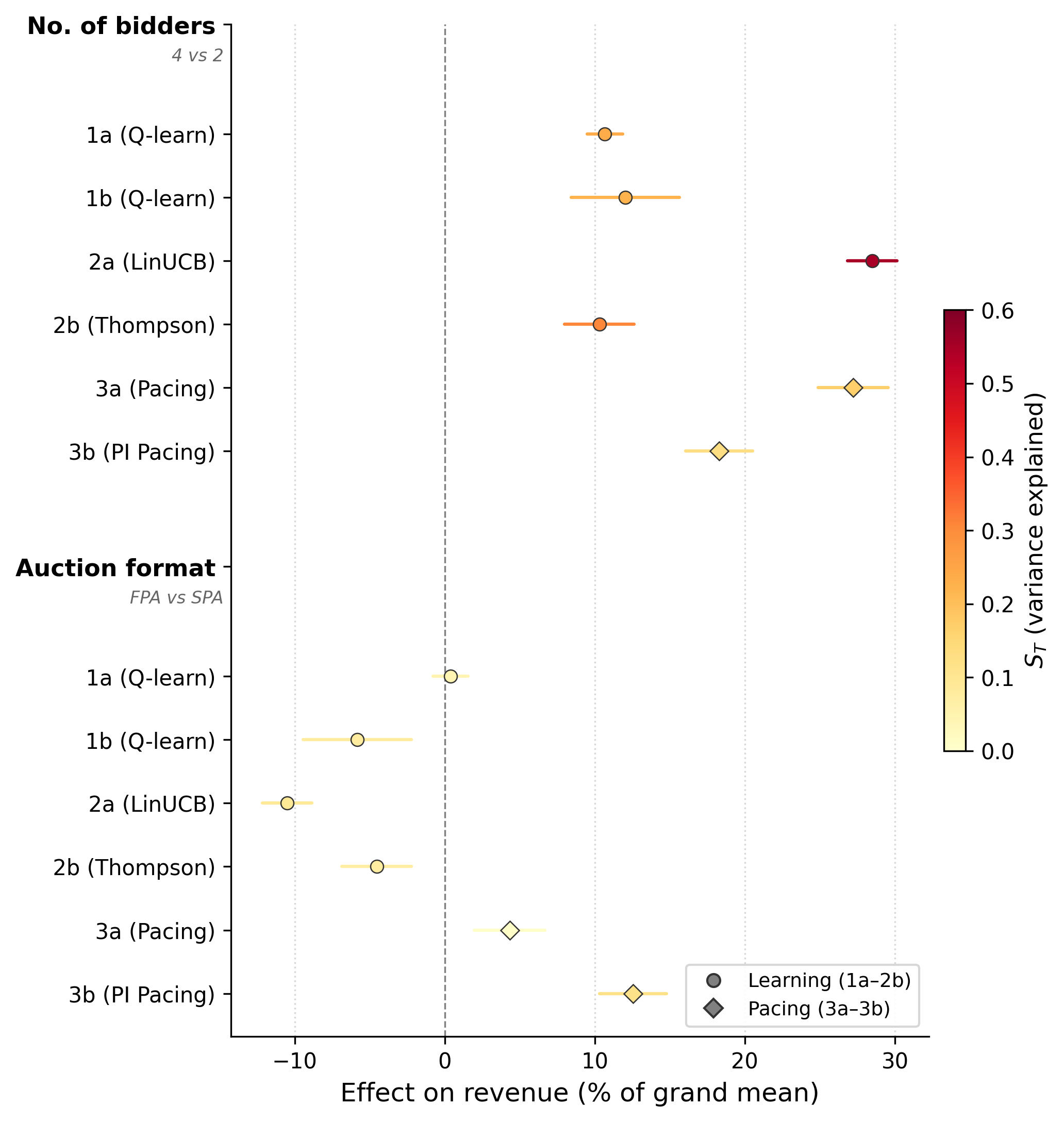}
  \caption{Standardised marginal effects of competition and auction format on platform revenue across all six experiments. Each point is an OLS coefficient (high vs.\ low factor level), normalised by experiment grand mean; bars show 95\% confidence intervals. Colour encodes total-order Sobol index ($S_T$); circles = learning experiments, diamonds = pacing experiments.}
  \label{fig:intro_forest}
\end{figure}

Experiments~3a and~3b replace unconstrained learners with budget-constrained pacing algorithms (dual-variable and PI controller). The number of bidders remains significant and positive, accounting for roughly 27\% and 18\% of the grand mean in Experiments~3a and~3b respectively, confirming that competition continues to matter under budget constraints. The auction format effect reverses sign; first-price auctions now generate strictly higher revenue than second-price in both experiments, directly contradicting the pattern observed under learning algorithms. Figure~\ref{fig:intro_forest} summarises these two cross-cutting patterns across all six experiments, with the sign reversal in auction format visible as the shift from negative coefficients (circles) to positive coefficients (diamonds).

\subsection{Contributions}

This paper makes four contributions. First, it resolves a major contradiction in the economic literature by bridging two previously isolated research programmes. Prior Q-learning studies conclude that first-price auctions are vulnerable to bid suppression, while the autobidding literature concludes that first-price auctions are highly efficient; by placing both within a single experimental framework, we show that both conclusions are correct in their respective contexts. Second, we demonstrate that the effect of auction format on revenue reverses sign across algorithm classes; first-price auctions reduce revenue under contextual bandits but strictly increase revenue under budget-constrained pacing, so no single auction format is universally superior. Third, across six sub-experiments comprising hundreds of thousands of simulated auctions, the degree of competition (number of bidders, budget tightness) overwhelmingly dominates algorithmic hyperparameters (learning rates, memory decay, exploration intensity) in determining platform revenue. Fourth, the paper introduces a rigorous computational laboratory methodology, combining fractional factorial designs, Sobol\' variance decomposition, and quantile regression, to algorithmic game theory and concludes with practical recommendations that give auction platforms and antitrust regulators a standardised pipeline for diagnosing algorithmic behaviour and testing market designs systematically. The remainder of the paper is organised as follows. Section~\ref{sec:literature} reviews the four bodies of literature that motivate this work. Section~\ref{sec:auctions_and_algorithms} presents the theoretical framework. Section~\ref{sec:inference} describes the experimental design and statistical methodology. Sections~\ref{sec:qlearning_results}--\ref{sec:pacing_results} report results by experiment, with each section self-contained. Section~\ref{sec:discussion} synthesises findings across all experiments and discusses policy implications.

\section{Related Work}
\label{sec:literature}

This paper sits at the intersection of auction theory, algorithmic collusion, autobidding, and competition policy. The theoretical predictions of rational auction models supply the benchmarks against which algorithmic behaviour is measured; the bid-suppression literature documents factors and conditions under which learning agents depart from those benchmarks; the autobidding literature asks how budget-constrained pacing reshapes welfare and efficiency; and legal scholarship clarifies why autonomous algorithmic convergence falls outside current enforcement frameworks. The remainder of this section reviews each body of work in turn and identifies the gaps that motivate this paper. Classical theories suggest hypotheses that we can test through Experiments~1a--2b. \citet{Vickrey1961} establishes revenue equivalence under independent private values; \citet{MilgromWeber1982} show that second-price auctions generate weakly higher revenue under affiliated signals; \citet{Myerson1981} derives the revenue-maximising mechanism with optimal reserves; and \citet{BulowKlemperer1996} demonstrate that an additional bidder raises revenue more than any optimal reserve. In repeated settings, the Folk Theorem implies that sufficiently patient bidders can sustain suppressed bids as equilibria, and \citet{Ivaldi2003} characterise the structural determinants of tacit collusion in a simple model, finding that fewer participants, symmetric conditions, high-frequency interaction, and greater transparency all facilitate coordination.

\subsection{Algorithmic Collusion}

A growing literature studies whether reinforcement-learning agents converge to supra-competitive outcomes, primarily in pricing games. \citet{Calvano2020} show that Q-learning agents in simultaneous Bertrand competition develop reward-punishment strategies resembling those predicted by the Folk Theorem, sustaining prices above the Nash equilibrium. \citet{Klein2021} extends this to sequential pricing, finding that the granularity of the action space matters, as finer grids facilitate collusion by enabling small, retaliatory price adjustments. \citet{Hansen2021} demonstrate that collusion can arise from misspecified demand prediction and correlated price experimentation, without any strategic intent. \citet{Calvano2023} introduce a distinction between genuine collusion, sustained by learned reward-punishment schemes (carrot-stick strategies like tit-for-tat), and spurious collusion, where agents converge to high prices simply because exploration is too limited to discover profitable deviations. \citet{Douglas2024} formalise this insight, proving analytically that deterministic bandit algorithms always converge to collusive outcomes in repeated games, while persistent stochastic exploration prevents such convergence even when agents observe only their own payoff history. \citet{Abada2022} and \citet{Dolgopolov2021} show that the choice of exploration strategy, whether epsilon-greedy, Boltzmann, or upper-confidence-bound, materially affects the likelihood of collusive convergence. \citet{Asker2022} highlight synchronous versus asynchronous updating as a further determinant. \citet{Hettich2021}, \citet{Han2022}, and \citet{Zhang2021} extend the analysis to deep reinforcement learning, comparing different network architectures and sampling strategies. A common criticism of this simulation-based literature is that the algorithms studied bear little resemblance to those deployed in real markets \citep{KuhnTadelis2017, Schwalbe2018}.

Fewer studies examine auctions specifically. \citet{BanchioSkrzypacz2022} compare first- and second-price auctions with Q-learning bidders and find that first-price auctions experience greater bid suppression. \citet{WaltmanKaymak2008} study reinforcement learning in Cournot-style and reverse auction settings, showing that mixed-strategy equilibria can be attained in simple cases. \citet{Bandyopadhyay2008} had reached similar conclusions for reverse auctions. \citet{Tellidou2007} find that tacit collusion is sustained even under competitive conditions in simulated electricity markets. Every paper cited above uses Q-learning, simple bandit algorithms, or shallow reinforcement-learning variants, yet the dominant bidding systems on major auction platforms employ budget pacing, dual-variable optimisation, and proportional-integral controllers that solve a fundamentally different problem: allocating a finite budget across auctions rather than learning value estimates from repeated play. The distinction matters for policy. The headline finding that first-price auctions are more vulnerable to algorithmic bid suppression than second-price auctions \citep{BanchioSkrzypacz2022} was derived entirely from Q-learning experiments. If that conclusion is an artefact of the algorithm class rather than an intrinsic property of the auction format, then platform design recommendations built on it may be misleading. This paper addresses both gaps by varying ten or more factors simultaneously within a factorial design and by embedding Q-learning, contextual bandits, and budget-constrained pacing within a common experimental framework, so that the robustness of each finding to the choice of algorithm can be assessed rather than assumed.

Contextual bandit algorithms offer a more sophisticated approach to exploration-exploitation than tabular Q-learning. \citet{Li2010LinUCB} introduce LinUCB, which models expected payoffs as linear functions of observable context features and selects actions using upper confidence bounds on these linear estimates. \citet{AgrawalGoyal2013} provide the first theoretical regret guarantees for Thompson Sampling in contextual bandits with linear payoffs, extending the Bayesian probability-matching idea introduced by \citet{Thompson1933}. Both algorithms build on the confidence set construction of \citet{AbbasiYadkori2011}, and \citet{Russo2018Thompson} provide a comprehensive survey of Thompson Sampling across diverse problem settings. Despite their growing deployment in recommendation and advertising systems, little prior work has studied the behaviour of contextual bandit algorithms in auction environments or assessed their potential for supra-competitive convergence. Experiments~2a and~2b of this paper address this gap by embedding LinUCB and Thompson Sampling within the same factorial framework as Q-learning, enabling a direct comparison of how algorithm class and bidding technology affects bidding outcomes.

Recent theoretical work has formalised the mechanisms underlying algorithmic collusion. \citet{BanchioMantegazza2022} identify ``spontaneous coupling'' as a key driver. When Q-learning agents explore infrequently, correlated estimation errors synchronise their play, creating stochastic cycles that sustain supra-competitive outcomes without deliberate reward-punishment schemes. \citet{Bertrand2025} strengthen this result by proving that Q-learners with one-step memory converge to the cooperative Pavlov (win-stay, lose-shift) policy in iterated prisoner's dilemmas, establishing the first formal convergence guarantee for collusion by standard stochastic Q-learning. \citet{Calzolari2021} extend the analysis to settings with imperfect monitoring of competitors' actions, showing that the degree of market observability interacts with algorithmic learning to determine whether collusive outcomes can be sustained. \citet{Zhang2025Noise} offers a distinct perspective by modelling collusion as a belief-averaging process mediated by shared market data, and demonstrates that calibrated noise injection can disrupt coordination without eliminating the informational benefits that pricing algorithms provide to individual firms. These mechanisms are not confined to tabular reinforcement learning. \citet{Fish2024} show that large language models autonomously reach supra-competitive prices in oligopoly experiments, and that seemingly innocuous prompt modifications substantially alter collusive outcomes. \citet{Arunachaleswaran2025} demonstrate that supra-competitive prices can arise even when algorithms explicitly lack the capacity for threats. If one firm deploys any no-regret learning algorithm and the other merely optimises its own revenue, near-monopoly prices emerge without either party encoding punishments or responding to competitor actions. These findings establish that supra-competitive convergence is a robust phenomenon driven by fundamental properties of adaptive algorithms, not idiosyncrasies of particular implementations.

Empirical evidence motivates the concern, but existing data cannot isolate which algorithmic or market features drive the observed outcomes. \citet{Chen2016} find that over a third of Amazon best-sellers use algorithmic pricing. \citet{Assad2020} document margin increases of 9--28\% following the adoption of algorithmic pricing in German gasoline markets. \citet{BrownMacKay2023} estimate 10\% price increases among large online retailers using pricing algorithms. These studies establish prevalence and magnitude but cannot isolate specific drivers.

\subsection{Autobidding, Pacing, and Platform Welfare}

A parallel literature in algorithmic game theory examines autobidding and pacing in online advertising auctions. Where the former asks whether learning agents converge to supra-competitive prices, the latter asks how much welfare is lost when budget-constrained advertisers delegate bidding to automated pacing agents. \citet{Aggarwal2019} show that uniform bid-scaling is the optimal single-agent strategy in truthful auctions and establish a worst-case Price of Anarchy of~2 for second-price auctions under budget constraints. \citet{Conitzer2022FPPE} prove that first-price pacing equilibria are unique and computable via the Eisenberg--Gale convex programme, connecting autobidding to classical competitive equilibrium theory. \citet{Deng2021TowardsEfficient} show that first-price auctions achieve a Price of Anarchy of~1 under return-on-spend constraints, though this result does not extend to budget-constrained settings where the applicable bound remains~2. \citet{Deng2022FPAEfficiency} extend this analysis to mixed populations of value-maximising autobidders and traditional utility-maximising bidders, finding that the first-price PoA degrades to approximately~$0.457$ in the mixed case. \citet{Mehta2022AuctionDesign} shows that randomised truthful auctions can achieve efficiency strictly better than VCG in the two-bidder autobidding setting, though the improvement vanishes as the number of bidders grows. \citet{Liaw2022EfficiencyNonTruthful} and \citet{Liaw2024EfficiencyBudget} establish tighter welfare bounds for non-truthful auctions and budget-constrained settings respectively. \citet{Aggarwal2024Survey} provide a recent survey of this rapidly growing field.

On the algorithmic side, \citet{Balseiro2019Learning} develop dual-based pacing algorithms in which each agent adjusts a shadow price on its budget constraint, suppressing bids when overspending and raising them when underspending, and prove $O(\sqrt{T})$ individual regret guarantees. \citet{Gaitonde2023} strengthen this by showing that gradient-based pacing achieves at least half the optimal liquid welfare without requiring convergence to equilibrium, a result that holds for any core auction format. \citet{PaesLeme2024} demonstrate that autobidding dynamics can exhibit bi-stability and periodic orbits even in simple market structures, raising the question of whether pacing systems reach equilibrium in practice. This literature characterises efficiency from the bidder's perspective but has not examined whether budget-mediated bid suppression produces systematic revenue shortfalls for the seller. Nor has it tested whether the format and thickness interactions observed in unconstrained learning settings persist under budget rationing.

Recent work has established format-specific welfare guarantees that hold even without convergence to equilibrium. \citet{FikiorisTardos2023} prove that in sequential first-price auctions with budget-constrained no-regret learners, liquid welfare is within a factor of approximately~2.41 of the optimum, whereas in second-price auctions the ratio can be arbitrarily bad even when all bidders satisfy no-regret guarantees. This sharp format comparison provides a welfare-theoretic rationale for the ongoing industry transition from second-price to first-price auction formats. \citet{Lucier2023PacingDynamics} extend the welfare analysis to the practically relevant setting with joint budget and return-on-investment constraints, proposing a gradient-based autobidding algorithm that achieves at least half the optimal liquid welfare for any core auction format without requiring convergence to equilibrium. These results strengthen the theoretical motivation for Experiments~3a and~3b, which test whether pacing algorithms realise these welfare guarantees in practice across varied market configurations. The theoretical predictions also create a direct tension with the Q-learning bid-suppression findings. \citet{FikiorisTardos2023} prove that first-price auctions are strictly superior for liquid welfare under budget-constrained pacing, and \citet{Conitzer2022FPPE} show that first-price pacing equilibria are unique and efficient, whereas \citet{BanchioSkrzypacz2022} find that first-price auctions are more vulnerable to bid suppression under Q-learning. This apparent contradiction may reflect genuine differences across algorithm classes, or merely the stylised settings in which each result was obtained. The present paper resolves this question empirically.

The two literatures study different algorithm classes operating under different constraints. Experiments~1a--2b of this paper study unconstrained learning agents; Experiments~3a and~3b study budget-constrained pacing agents. No prior work has studied both algorithm families within a common experimental framework. The consequence is that the two literatures reach opposite conclusions about auction format without any empirical basis for reconciliation. The present paper provides that basis. As Section~\ref{sec:discussion} reports, the auction-format effect reverses sign across algorithm classes, establishing that format recommendations derived from one bidding technology do not generalise to another.

\subsection{Legal and Regulatory Context}

\citet{EzrachiStucke2017} distinguish four modes of algorithmic participation in collusion, ranging from algorithms as messengers carrying out human-devised cartels to fully autonomous learning agents that converge to collusive outcomes without agreement or intent. \citet{Mehra2016} frames the fundamental challenge as an ``agreement gap,'' arguing that existing antitrust doctrine was built around human conspiracies and lacks the conceptual vocabulary to address autonomous algorithmic convergence. Both US antitrust law (Sherman Act \S1) and European competition law (Article~101 TFEU) require evidence of ``agreement'' or ``concerted practice'' to establish liability for horizontal coordination \citep{Harrington2018, Gal2019}. When algorithms converge to supra-competitive outcomes through independent learning rather than explicit communication, this evidentiary threshold is not met. The OECD recognised this gap in landmark competition policy roundtables, warning that algorithms amplify the conditions for tacit coordination by increasing market transparency, enabling rapid retaliation, and reducing the benefits of deviation \citep{OECD2017, OECD2023}. Despite growing regulatory attention, no jurisdiction has sanctioned purely autonomous tacit collusion by independent learning algorithms \citep{OECD2023, Harrington2018}. Hub-and-spoke collusion via shared third-party pricing software, exemplified by the Department of Justice's case against RealPage in rental housing markets, is a separate and active enforcement area \citep{EzrachiStucke2024} that falls outside the scope of this paper.

Several reform strategies have been proposed. \citet{Harrington2018} proposes per se prohibition of collusion-facilitating algorithm designs; \citet{Hartline2024, Hartline2025} propose calibrated regret auditing; and \citet{Gal2019} argues that algorithmic inspectability should expand rather than narrow the regulatory scope. Market design approaches use format choice and information disclosure as regulatory levers \citep{BanchioSkrzypacz2022, Zhang2025Noise}.\footnote{Legislative responses include the EU 2023 Horizontal Cooperation Guidelines addressing algorithmic coordination, the CJEU ruling in \emph{Eturas} (C-74/14) on algorithmic information exchanges, the UK CMA's investment in algorithmic screening tools, and the proposed US Preventing Algorithmic Collusion Act \citep{OECD2023}. \citet{Crane2024} argues that general-purpose AI may ultimately undermine the foundations of the antitrust order.} Against these proposals, \citet{Petit2017} and \citet{Schwalbe2018} caution that the theoretical conditions for algorithmic collusion are more restrictive than popular discussion suggests. Section~\ref{sec:discussion} synthesises our findings in light of these debates.

\section{Auctions and Algorithms}
\label{sec:auctions_and_algorithms}
\label{sec:auctions}


This section first describes the auction environment shared by all experiments, then specifies the learning algorithms that bidders employ. We consider a repeated-auction environment with $n$ bidders, each participating in multiple rounds. In each round, the bidders simultaneously submit a bid from a fixed discrete grid, and the auction outcome (who wins and how much is paid) depends on the chosen format (first- or second-price) as well as a reserve price. Ties for the top bid are resolved by uniformly random selection among all tying bidders. Each bidder's objective is to maximise her per-round payoff, given by her valuation minus her payment. The experiments differ in two dimensions: valuations range from a fixed value of 1 for all bidders to signals that are partially affiliated across bidders, and bidders may observe different informational states between rounds.

\subsection{Auction Environment}

In Experiment~1a, every bidder's valuation is fixed at 1, so the private-value assumption is trivially satisfied and there is no direct impact of signals. In Experiments~1b--2b, however, each bidder $i$ draws a signal $s_{i}\in [0,1]$ (using a finite set of possible signal realisations) and forms a valuation via a linear-affiliation function
\[
v_{i}(s_{i}, s_{-i}) \;=\; \bigl(1 - 0.5\,\eta\bigr)\,s_{i} \;+\; 0.5\,\eta\;\frac{1}{n-1}\sum_{j\neq i} s_{j},
\]
where $\eta\in[0,1]$ measures how strongly bidder\,$i$'s value depends on the others' signals. When $\eta=0$, each bidder's value depends only on her own signal; as $\eta$ increases, the environment moves closer to a ``common-value'' setting. In each round, every bidder draws a fresh signal independently across bidders and rounds.

Bids are constrained to lie in a finite grid, typically $\{0,0.1,0.2,\ldots,1.0\}$ or a similar set. This restriction is imposed in all three experiments to simplify the strategy space. Despite this simplification, each bidder remains free to adapt her bids over repeated rounds as she observes outcomes. We adopt a moderate granularity (e.g.\ 11 equally spaced points from 0 to 1) that is fine enough to allow distinct bidding behaviours and multiple points of convergence, yet small enough to allow full exploration. Robustness checks confirm that grid size does not materially affect results.

All experiments allow a reserve price $r\ge 0$. If every submitted bid is below $r$, then no sale occurs in that round; otherwise, any bid below $r$ is excluded from contention.

Both first- and second-price formats are studied. Under first-price, the winner pays her own highest valid bid; under second-price, the winner pays the second-highest valid bid (if any). In all cases, if multiple bidders tie for the highest valid bid, one among them is chosen at random to be the winner, and the payment is then computed according to the standard rule for that auction format. The resulting payoff for bidder~$i$ is
\[
u_{i} \;=\;
\begin{cases}
v_{i} - \text{(own bid)} & \text{(first-price),}\\
v_{i} - \text{(second-highest bid)} & \text{(second-price),}
\end{cases}
\]
and is zero for those bidders who do not win. Figure~\ref{fig:auction_timeline} summarises the sequence of events within each round.

\begin{figure}[H]
\centering
\resizebox{\textwidth}{!}{%
\begin{tikzpicture}[
  node distance=1.2cm,
  block/.style={rectangle, draw, rounded corners, minimum width=2.2cm,
                minimum height=0.9cm, text centered, font=\small},
  arrow/.style={-{Stealth[length=2mm]}, thick},
]
  \node[block] (val) {Valuations};
  \node[block, right=of val] (sig) {Signals};
  \node[block, right=of sig] (bid) {Bids};
  \node[block, right=of bid] (win) {Winner};
  \node[block, right=of win] (pay) {Payment};
  \node[block, right=of pay] (payoff) {Payoffs};

  \draw[arrow] (val) -- (sig);
  \draw[arrow] (sig) -- (bid);
  \draw[arrow] (bid) -- (win);
  \draw[arrow] (win) -- (pay);
  \draw[arrow] (pay) -- (payoff);

  \node[below=0.25cm of val, font=\scriptsize, align=center] {$v_i$ drawn or\\fixed};
  \node[below=0.25cm of sig, font=\scriptsize, align=center] {$s_i$ observed\\(if affiliated)};
  \node[below=0.25cm of bid, font=\scriptsize, align=center] {$b_i$ from\\strategy};
  \node[below=0.25cm of win, font=\scriptsize, align=center] {highest\\valid bid};
  \node[below=0.25cm of payoff, font=\scriptsize, align=center] {$v_i - p$\\(winner)};

  \node[below=1.1cm of pay, font=\scriptsize, align=center] (fpa) {FPA: $p = b_{\text{win}}$};
  \node[below=1.1cm of payoff, font=\scriptsize, align=center] (spa) {SPA: $p = b^{(2)}$};
  \draw[-{Stealth[length=1.5mm]}, thin, gray] (pay.south) -- (fpa.north);
  \draw[-{Stealth[length=1.5mm]}, thin, gray] (pay.south) -- (spa.north);
\end{tikzpicture}%
}
\caption{Sequence of events within a single auction round. Valuations are drawn (or fixed), bidders observe signals and submit bids, the highest valid bid wins, and payment depends on the auction format.}
\label{fig:auction_timeline}
\end{figure}

A defining feature of these repeated interactions is that bidders may condition their future bids on information from prior rounds. In Experiment~1a, the possible states include the previous round's winning bid. In Experiments~1b--2b, states can similarly incorporate bidder-specific signals $s_{i}$, as well as the previous winning bid. Each bidder knows her own signal in each round and can rely on these state variables to guide her subsequent bid choice. The affiliation parameter~$\eta$ thus captures how interdependent each bidder's underlying valuation is with the other signals, but each bidder's private signal still enters only her own valuation (albeit modulated by average signals).

\subsection{Budget Constraints and Welfare}

Experiment~3a introduces hard budget constraints and autobidding pacing agents. Each of $n \in \{2, 4\}$ advertisers participates in $D = 100$ episodes, each comprising $T = 1{,}000$ single-item auctions. Between episodes, budgets regenerate to their initial level while dual variables persist (warm-starting), mimicking the daily budget cycle in real ad exchanges. Valuations follow a log-normal model with bidder-specific asymmetry, where each bidder~$i$ draws a mean $\mu_i \sim \mathrm{Uniform}(0.5, 1.5)$ once per seed, and in each round $v_{it} \sim \mathrm{LogNormal}(\mu_i, \sigma)$. The budget per episode is $B_i = m \cdot \mathbb{E}[v_{it}] \cdot T$. All bidders use multiplicative dual pacing, computing bids as $v_t / \mu_t$ (value-maximizers) or $v_t / (1 + \mu_t)$ (utility-maximizers), subject to a hard budget cap that prevents spending from exceeding the remaining budget. The dual variable $\mu_t$ is updated via an exponential rule after each round (Section~\ref{sec:pacing}). This setup allows us to study how budget constraints interact with auction format and bidder objectives, bridging the bid suppression literature with the pacing efficiency literature.

All experiments repeat auctions for many rounds, allowing long-run behaviour to emerge across varied reserve prices, degrees of valuation interdependence, and budget regimes. The discrete setup (both for signals and for bids) is maintained for computational tractability, but the core elements of a standard first- or second-price auction, with or without a reserve, are preserved, and the ultimate allocation and payment follow the standard textbook rules.

\label{sec:welfare}

Experiments~3a and~3b evaluate allocative efficiency using liquid welfare and the Price of Anarchy. Following \citet{Gaitonde2023} and \citet{Deng2021TowardsEfficient}, the liquid welfare of an allocation $x$ for $N$ bidders over $T$ rounds is
\begin{equation}
  W(x) = \sum_{i=1}^{N} W_i(x), \qquad
  W_i(x) = \min\!\Big\{B_i,\; \sum_{t=1}^{T} x_{ti}\, v_{ti}\Big\},
  \label{eq:liquid_welfare_def}
\end{equation}
where $v_{ti}$ is bidder~$i$'s valuation in round~$t$, $B_i$ is bidder~$i$'s budget, and $x_{ti} \in [0,1]$ is the fraction of the item allocated to bidder~$i$ in round~$t$. In the integer case ($x_{ti} \in \{0,1\}$, at most one winner per round), liquid welfare reduces to $W(x) = \sum_{i=1}^{N} \min\{B_i,\; \sum_{t:\, \text{winner}_t = i} v_{ti}\}$, which is how it is computed in the experimental code.

The optimal fractional liquid welfare is the solution to
\begin{equation}
  W^* = \max_{x} \sum_{i=1}^{N} \min\!\Big\{B_i,\; \sum_{t=1}^{T} x_{ti}\, v_{ti}\Big\}
  \quad \text{s.t.} \quad
  \sum_{i=1}^{N} x_{ti} \leq 1 \;\; \forall t, \quad
  x_{ti} \in [0,1] \;\; \forall t,i.
  \label{eq:welfare_opt}
\end{equation}
The $\min$ operator makes this a non-linear program. We linearise it via the LP relaxation $\mathrm{LP}^* = \max_{x} \sum_{t,i} v_{ti}\, x_{ti}$ subject to $\sum_i x_{ti} \le 1$ per round and $\sum_t v_{ti}\, x_{ti} \le B_i$ per bidder. At any LP-feasible point the budget constraint ensures $\min\{B_i, \sum_t v_{ti}\, x_{ti}\} = \sum_t v_{ti}\, x_{ti}$, giving $\mathrm{LP}^* = W^*$. The LP relaxation allows fractional allocations, so $\mathrm{LP}^*$ is an upper bound on the optimal integer liquid welfare, making it a conservative benchmark.

The Price of Anarchy is defined as the ratio of optimal to realised liquid welfare:
\begin{equation}
  \text{PoA} = \frac{W^*}{W(x^{\text{obs}})},
  \label{eq:poa_def}
\end{equation}
where $x^{\text{obs}}$ is the allocation produced by the bidding algorithm. By construction, $\text{PoA} \geq 1$, with values closer to~1 indicating higher efficiency. The literature defines worst-case PoA as the supremum over all instances and equilibria. Our experiments instead report per-instance empirical PoA. The PoA~$= 1$ result of \citet{Deng2021TowardsEfficient} requires return-on-spend constraints, not budget constraints. Experiments~3a and~3b use cumulative budget constraints, for which the applicable theoretical bound is PoA~$\leq 2$ regardless of auction format.

\begin{table}[H]
\centering
\small
\caption{Theoretical Price of Anarchy bounds for autobidding.}
\label{tab:poa_bounds}
\begin{tabular}{lll}
\toprule
\textbf{Setting} & \textbf{PoA bound} & \textbf{Source} \\
\midrule
SPA + budget constraints & $\leq 2$ (tight) & \citet{Aggarwal2019} \\
FPA + RoS constraints (no budgets) & $= 1$ & \citet{Deng2021TowardsEfficient} \\
FPA + budget constraints & $\leq 2$ (if $v_{ij} \leq B_i$) & \citet{Aggarwal2024Survey} \\
Any core auction + learning dynamics & $\leq 2$ & \citet{Gaitonde2023} \\
\bottomrule
\end{tabular}
\end{table}


\label{sec:algorithms}

Reinforcement learning (RL) typically involves an agent interacting with an environment through states, actions, and rewards. In \emph{Q-learning}, the agent learns to approximate an optimal action-value function $Q(s,a)$, which represents the expected discounted reward for taking action $a$ in state $s$. By comparison, \emph{bandit} algorithms (including multi-armed and contextual bandits) assume no or minimal state transitions and learn which action yields the highest expected payoff.

\subsection{Q-Learning}
\label{sec:qlearning}

An \emph{asynchronous} Q-learning agent updates its $Q$-values only for the action actually taken at each step. Let $s_t$ be the current state, $a_t$ the chosen action, and $r_t$ the immediate reward upon transitioning to $s_{t+1}$. The Q-update is:
\begin{equation}
Q(s_t,a_t)
\;\leftarrow\;
Q(s_t,a_t)
\;+\;
\alpha \Bigl[
  r_t
  \;+\;
  \gamma \,\max_{a'} Q(s_{t+1},a')
  \;-\;
  Q(s_t,a_t)
\Bigr],
\label{eq:qlearning_async}
\end{equation}
where $\alpha$ is the learning rate and $\gamma$ the discount factor. Only $Q(s_t,a_t)$ changes, reflecting the experience from action $a_t$.

A \emph{synchronous} variant updates the Q-values for \emph{all} actions from the same state in a single step, using counterfactual rewards. Let $A$ be the set of possible actions. If the agent took action $a_t$ but also computes hypothetical rewards $r_t(a)$ for each $a \in A$, the synchronous update is:
\begin{equation}
Q\bigl(s_t,a\bigr)
\;\leftarrow\;
\bigl(1-\alpha\bigr)\,Q\bigl(s_t,a\bigr)
\;+\;
\alpha \Bigl[
  r_t(a)
  \;+\;
  \gamma \,\max_{a'} Q\bigl(s_{t+1},a'\bigr)
\Bigr]
\quad
\forall a \in A.
\label{eq:qlearning_sync}
\end{equation}
Thus every $Q(s_t,a)$ is updated in a single step, including actions the agent did not actually choose. This accelerates convergence at the cost of additional computation.

Agents select actions using one of two exploration strategies. \emph{Boltzmann} (or \emph{softmax}) exploration draws an action $a$ from a distribution favouring higher estimated $Q$-values:
\begin{equation}
P\bigl(a \mid s\bigr)
\;=\;
\frac{\exp\!\bigl(\beta\,Q(s,a)\bigr)}{\sum_{b}\,\exp\!\bigl(\beta\,Q(s,b)\bigr)},
\label{eq:boltzmann}
\end{equation}
where $\beta = 1/\tau$ is the inverse temperature parameter. Larger $\beta$ (lower temperature $\tau$) makes the distribution more peaked, favouring higher-valued actions; smaller $\beta$ (higher $\tau$) flattens the distribution, promoting more exploration. Alternatively, \emph{$\varepsilon$-greedy} exploration selects the action that maximises $Q(s,a)$ with probability $1-\varepsilon$ and selects a random action with probability $\varepsilon$:
\begin{equation}
P\bigl(a \mid s\bigr)
\;=\;
\begin{cases}
1-\varepsilon, & \text{if }a = \displaystyle\arg\max_{b}\,Q(s,b),\\
\displaystyle\frac{\varepsilon}{|A|}, & \text{otherwise}.
\end{cases}
\label{eq:egreedy}
\end{equation}
As $\varepsilon$ diminishes, the policy exploits more aggressively based on current $Q$-estimates. Both exploration methods use a decaying schedule that transitions from full exploration to near-greedy exploitation.\footnote{$\varepsilon$ starts at 1.0 and decays over the first 90\% of episodes to a floor of 0.01; agents then switch to pure exploitation ($\varepsilon = 0$) for the final 10\%. For Boltzmann exploration, the temperature $\tau$ follows an analogous schedule (1.0 to 0.01). Both schedules admit linear and exponential variants (Experiment~1a). Q-tables may be initialised to all zeros or to small random values, influencing early exploration. In any given state, if multiple actions share the same $Q$-value, ties are broken uniformly at random.}

\subsection{Contextual Bandits}
\label{sec:contextual_bandits}

A \emph{multi-armed bandit} scenario omits state transitions, focusing instead on learning which of several actions (arms) maximises expected reward. Let $\hat{\mu}_a$ be the estimated mean reward of arm $a$, and $u_a$ be an uncertainty term. A typical approach is Upper Confidence Bound (UCB), which selects
\begin{equation}
a_t
\;=\;
\arg\max_{a}\Bigl[\hat{\mu}_a + u_a\Bigr],
\end{equation}
then updates $\hat{\mu}_a$ and $u_a$ using the reward observed after pulling arm $a$.

\emph{Contextual bandits} generalise this problem by providing a context vector $\mathbf{x}\in\mathbb{R}^d$ prior to choosing an action. The \emph{LinUCB} algorithm assumes a linear payoff model, so each action $a$ has parameters $\boldsymbol{\theta}_a$, and the reward is approximately $\boldsymbol{\theta}_a^\top \mathbf{x}$. For each action $a$, the algorithm maintains a matrix $\mathbf{A}_a$ and vector $\mathbf{b}_a$, initialised as $\mathbf{A}_a = \lambda \mathbf{I}$ and $\mathbf{b}_a = \mathbf{0}$, where the regularisation parameter $\lambda$ ensures that $\mathbf{A}_a$ is invertible from the start and provides numerical stability. The algorithm selects $a$ via:
\begin{equation}
a_t
\;=\;
\arg\max_{a}\Bigl[\hat{\boldsymbol{\theta}}_{a}^\top \mathbf{x}
\;+\;
c\sqrt{\mathbf{x}^\top \mathbf{A}_a^{-1}\mathbf{x}}\Bigr],
\quad
\text{where}
\quad
\hat{\boldsymbol{\theta}}_{a} \;=\; \mathbf{A}_a^{-1}\mathbf{b}_a,
\end{equation}
and $c>0$ controls exploration. After observing the reward, $\mathbf{A}_a$ and $\mathbf{b}_a$ are updated, refining the estimate of $\boldsymbol{\theta}_a$.

As an alternative to the optimism-based exploration of LinUCB, \emph{Contextual Thompson Sampling} (CTS) uses posterior sampling from a Bayesian linear model to balance exploration and exploitation. Each action $a$ maintains a Bayesian linear regression model with posterior $\boldsymbol{\theta}_a \mid \text{data} \sim \mathcal{N}(\hat{\boldsymbol{\theta}}_a,\, \sigma^2 \mathbf{A}_a^{-1})$, where $\hat{\boldsymbol{\theta}}_a = \mathbf{A}_a^{-1}\mathbf{b}_a$ and $\sigma^2$ is a noise variance parameter. At each step, the algorithm draws a sample $\tilde{\boldsymbol{\theta}}_a$ from the posterior for every action and selects greedily with respect to the samples:
\begin{equation}
a_t \;=\; \arg\max_{a}\; \tilde{\boldsymbol{\theta}}_a^\top \mathbf{x}, \quad \tilde{\boldsymbol{\theta}}_a \sim \mathcal{N}\!\bigl(\hat{\boldsymbol{\theta}}_a,\, \sigma^2 \mathbf{A}_a^{-1}\bigr).
\end{equation}
The parameter $\sigma^2$ plays an analogous role to $c$ in LinUCB. Larger $\sigma^2$ produces wider posterior draws, encouraging more exploration; smaller $\sigma^2$ concentrates the posterior, promoting exploitation. The two methods are compared in Experiments~2a and~2b.

In standard implementations, the precision matrices $\mathbf{A}_a$ accumulate all past observations with equal weight, progressively reducing uncertainty and concentrating the policy around the estimated optimum. Experiment~2a introduces an optional \emph{memory decay} parameter $\delta \in (0,1]$ that applies exponential discounting to historical observations, preventing the algorithm from over-weighting early experiences.\footnote{At each update, $\mathbf{A}_a \leftarrow \delta\,\mathbf{A}_a + \mathbf{x}\mathbf{x}^\top$ and $\mathbf{b}_a \leftarrow \delta\,\mathbf{b}_a + r\,\mathbf{x}$. Setting $\delta = 1$ recovers the standard algorithm; $\delta < 1$ produces an effective observation window of approximately $1/(1-\delta)$ rounds.} This mechanism tests whether the rigidity of full-memory bandit learners contributes to bid suppression, following \citet{Douglas2024}, who show that deterministic convergence is a key driver of what the literature terms ``naive algorithmic collusion''. The discounted variant draws on \citet{Russac2019}, who analyse weighted linear bandits in non-stationary environments.

\subsection{Budget-Constrained Pacing}
\label{sec:pacing}

Experiments~3a and~3b replace the unconstrained bidding of prior experiments with \emph{budget-constrained pacing agents}. Each agent~$i$ has a total budget $B^i$ over a horizon of $T$ rounds. At each round~$t$, the agent observes its private valuation $v_t^i$ and computes a bid using a Lagrangian dual variable $\mu_t^i$ that controls bid shading. Following the dual decomposition framework of \citet{Balseiro2019Learning}, the bid depends on the agent's objective:
\begin{align}
b_t^i &\;=\; \min\!\bigl(v_t^i / \mu_t^i,\; B^i - S_t^i\bigr) & &\text{(value-maximizer),} \label{eq:bid_vmax_algo}\\
b_t^i &\;=\; \min\!\bigl(v_t^i / (1 + \mu_t^i),\; B^i - S_t^i\bigr) & &\text{(utility-maximizer),} \label{eq:bid_umax_algo}
\end{align}
where $S_t^i$ is the cumulative spend at round~$t$. The hard budget cap ensures spending never exceeds the remaining budget.

Following \citet{Balseiro2019Learning}, the dual variable is updated via an exponential multiplicative rule:
\begin{equation}
\mu_{t+1}^i \;=\; \operatorname{clip}\!\bigl(\mu_t^i \cdot \exp\!\bigl(\alpha_p\,(c_t^i - B^i/T)\bigr),\; \mu_{\min},\; \mu_{\max}\bigr),
\label{eq:dual_update}
\end{equation}
where $\alpha_p = 1/\!\sqrt{T}$ is the dual step size and $c_t^i$ is the payment made by agent~$i$ in round~$t$.\footnote{The dual variable is clipped to $[\mu_{\min}, \mu_{\max}] = [10^{-4}, 100]$ for numerical stability.} When the per-round cost exceeds the target spend rate $B^i/T$, the dual variable rises, reducing bids; underspending reverses this adjustment.

As an alternative to the multiplicative dual update, Experiment~3b implements a proportional--integral (PI) controller that operates directly on the cumulative spending error:
\begin{align}
e_t^i &\;=\; \tfrac{t}{T}\,B^i \;-\; {\textstyle\sum_{\tau \le t}} c_\tau^i, \label{eq:pid_error}\\
\lambda_{t+1}^i &\;=\; \operatorname{clip}\!\bigl(\lambda_t^i + K_P\,e_t^i + K_I\!\textstyle\sum_{\tau} e_\tau^i,\; 0.01,\; 1.5\bigr), \label{eq:pid_update}
\end{align}
where $K_P = 0.30\times\text{aggressiveness}$ and $K_I = 0.05\times\text{aggressiveness}$. A positive error (underspending) raises $\lambda$, encouraging higher bids; overspending drives $\lambda$ down. The derivative term is omitted ($K_D = 0$) following the literature consensus that it is counterproductive in stochastic auction environments.

Both algorithms implement the same closed-loop pacing structure illustrated in Figure~\ref{fig:pacing_loop}. The multiplier governs bids, the auction determines payments, and the control law updates the multiplier in response. They differ only in how the error signal is defined and propagated.

\begin{figure}[H]
\centering
\begin{tikzpicture}[
  node distance=1.8cm and 2.5cm,
  block/.style={rectangle, draw, rounded corners, minimum width=2.6cm,
                minimum height=0.9cm, text centered, font=\small},
  arrow/.style={-{Stealth[length=2mm]}, thick},
]
  \node[block] (mult)   {Dual $\mu_t$};
  \node[block, right=of mult] (bid)    {Bid $b_t = \min(v_t/\mu_t,\,\text{rem})$};
  \node[block, right=of bid]  (auction){Auction};
  \node[block, below=1.2cm of auction] (cost)  {Cost $c_t$};
  \node[block, below=1.2cm of mult]  (ctrl)  {Control law};

  \draw[arrow] (mult)    -- (bid);
  \draw[arrow] (bid)     -- (auction);
  \draw[arrow] (auction) -- (cost);
  \draw[arrow] (cost)    -- (ctrl);
  \draw[arrow] (ctrl)    -- node[left,font=\scriptsize]{$\mu_{t+1}$} (mult);

  \node[font=\scriptsize, align=center, below=0.2cm of ctrl]
    {Multiplicative exp.\ update};
\end{tikzpicture}
\caption{Pacing feedback loop shared by both algorithms. The dual variable scales bids; auction outcomes feed back through the control law to update $\mu$.}
\label{fig:pacing_loop}
\end{figure}

The budget per episode is $B^i = m \cdot \mathbb{E}[v_t^i] \cdot T$, where $\mathbb{E}[v_t^i] = \exp(\mu_i + \sigma^2/2)$ under the log-normal valuation model. Note that $\eta$ throughout this paper denotes the affiliation parameter (Experiments~1b--2b); the dual step size above uses $\alpha_p$ to avoid notation collision.

\label{sec:appendix_algorithms}

In summary, Experiments~1a and~1b employ Q-learning under varying valuation models; Experiments~2a and~2b replace Q-learning with contextual bandits (LinUCB and Thompson Sampling, respectively); and Experiments~3a and~3b shift to budget-constrained pacing agents (multiplicative dual pacing and PI control, respectively). Figure~\ref{fig:algo_taxonomy} summarises the three decision loops.

\begin{figure}[H]
\centering
\begin{tikzpicture}[
  block/.style={rectangle, draw, rounded corners, minimum width=3.4cm,
                minimum height=0.8cm, text centered, font=\small},
  arrow/.style={-{Stealth[length=2mm]}, thick},
  title/.style={font=\small\bfseries, text centered},
]
  \def\colA{0}
  \def\colB{5.0}
  \def\colC{10.0}
  \def\rowA{0}
  \def\rowB{-1.3}
  \def\rowC{-2.6}
  \def\rowD{-3.9}

  \node[title] at (\colA, 1.0) {Q-Learning};
  \node[title] at (\colB, 1.0) {Contextual Bandits};
  \node[title] at (\colC, 1.0) {Budget Pacing};

  \node[block] (q1) at (\colA, \rowA) {Observe state $s_t$};
  \node[block] (q2) at (\colA, \rowB) {$\varepsilon$-greedy select};
  \node[block] (q3) at (\colA, \rowC) {Submit bid $b_t$};
  \node[block] (q4) at (\colA, \rowD) {Update $Q(s,a)$};

  \node[block] (b1) at (\colB, \rowA) {Observe context $\mathbf{x}_t$};
  \node[block] (b2) at (\colB, \rowB) {UCB / Thompson};
  \node[block] (b3) at (\colB, \rowC) {Submit bid $b_t$};
  \node[block] (b4) at (\colB, \rowD) {Update $\boldsymbol{\theta}$};

  \node[block] (p1) at (\colC, \rowA) {Observe $v_t$, budget};
  \node[block] (p2) at (\colC, \rowB) {Shade bid $v_t/\mu_t$};
  \node[block] (p3) at (\colC, \rowC) {Submit bid $b_t$};
  \node[block] (p4) at (\colC, \rowD) {Update dual $\mu_t$};

  \foreach \col in {q, b, p} {
    \draw[arrow] (\col 1) -- (\col 2);
    \draw[arrow] (\col 2) -- (\col 3);
    \draw[arrow] (\col 3) -- (\col 4);
  }

  \draw[arrow, rounded corners=8pt] (q4.west) -- ++(-1.2, 0) |- (q1.west);
  \draw[arrow, rounded corners=8pt] (b4.west) -- ++(-1.2, 0) |- (b1.west);
  \draw[arrow, rounded corners=8pt] (p4.east) -- ++(1.2, 0) |- (p1.east);

  \begin{pgfonlayer}{background}
    \node[fill=RoyalBlue!8, rounded corners=6pt, fit=(q1)(q4),
          inner xsep=10pt, inner ysep=10pt] {};
    \node[fill=ForestGreen!8, rounded corners=6pt, fit=(b1)(b4),
          inner xsep=10pt, inner ysep=10pt] {};
    \node[fill=RedOrange!8, rounded corners=6pt, fit=(p1)(p4),
          inner xsep=10pt, inner ysep=10pt] {};
  \end{pgfonlayer}
\end{tikzpicture}
\caption{Decision loops for the three algorithm families. Each column shows the per-round cycle of observation, action selection, bid submission, and parameter update. Loopback arrows indicate that the cycle repeats each round.}
\label{fig:algo_taxonomy}
\end{figure}

\section{Statistical Inference}
\label{sec:inference}

This section presents the experimental design and statistical methods shared across all six sub-experiments. It covers experiment descriptions, the factorial design framework, estimation and inference procedures, model adequacy diagnostics, and the global sensitivity analysis methodology.

\subsection{Experimental Overview}
\label{sec:experiments}

We design three experiment families, each comprising two sub-experiments, to study repeated sealed-bid auctions under varying valuations and learning strategies. The auction environment is described in Section~\ref{sec:auctions_and_algorithms}; all factors are coded as $-1$ (low level) and $+1$ (high level) for statistical analysis, ensuring orthogonal estimation of main effects and interactions.

Experiment~1a deploys Q-learning agents with constant valuations ($v_i = 1$) in a $2^{10-1}$ Resolution~V half-fraction spanning 10 factors (learning rates, discount factors, exploration modes, update synchronisation, number of bidders, reserve prices) and 1{,}024 observations. Under independent private values, revenue equivalence \citep{Vickrey1961} predicts that first- and second-price auctions yield identical expected revenue. The experiment tests whether Q-learning preserves this equivalence or breaks it.

Experiment~1b generalises Q-learning to affiliated valuations via $v_i = (1 - 0.5\eta)s_i + 0.5\eta \frac{1}{n-1}\sum_{j \neq i} s_j$, where $\eta \in \{0, 0.5, 1\}$ ranges from private to near-common values. Revenue equivalence holds for all $\eta$ in this model because signals are drawn independently (Appendix~\ref{sec:equilibria}). The $3 \times 2^3$ mixed-level design crosses auction format, affiliation, number of bidders, and state information, with Q-learning hyperparameters fixed at levels identified in Experiment~1a.

Experiment~2a replaces Q-learning with LinUCB contextual bandits under the same affiliated valuation model. LinUCB uses optimism-based exploration and Thompson Sampling uses posterior sampling. Comparing the two tests how exploration style affects outcomes. Bandit algorithms should exploit contextual structure better than tabular Q-learning; the experiment tests whether this improves seller outcomes.

The $3 \times 2^7$ mixed-level design for 2a has 8 factors (7 binary plus $\eta$) and 384 cells, replicated twice for 768 observations. It includes regularisation ($\lambda$) and memory decay as factors, both of which strongly affect LinUCB performance. Experiment~2b deploys Contextual Thompson Sampling with a smaller $3 \times 2^5$ design (6 factors, 96 cells, 192 observations). It drops regularisation and memory decay because Thompson Sampling's posterior dominates the prior after approximately 20 rounds, making these factors irrelevant.\footnote{Thompson Sampling explores via posterior sampling, which is inherently stochastic even as the posterior tightens.} Separating the two algorithms avoids wasting design cells on factors that are structurally irrelevant to one algorithm class.

Experiment~3a introduces budget-constrained autobidding agents using multiplicative dual pacing, implementing the framework of \citet{Balseiro2019Learning}. Each of $n \in \{2, 4\}$ advertisers participates in 100 episodes of 1{,}000 auctions each, with budgets regenerating between episodes while dual variables persist. The theoretical worst-case Price of Anarchy is bounded by~2 for both auction formats \citep{Aggarwal2019, Gaitonde2023}. The $2^6$ full factorial crosses auction format, bidder objective (value- vs.\ utility-maximizer), number of bidders, budget multiplier (tight vs.\ loose), reserve price, and value dispersion, replicated across 8 seeds for 512 observations.

Experiment~3b retains the budget-constrained environment of Experiment~3a but replaces multiplicative dual pacing with a proportional-integral (PI) controller. The PI agent bids $\lambda \cdot v$ and updates $\lambda$ additively based on cumulative spending error. The experiment tests whether the control law matters or whether budget structure is the primary driver of outcomes. The $2^6$ full factorial crosses auction format, controller aggressiveness (conservative vs.\ aggressive PI gains), number of bidders, budget multiplier, reserve price, and value dispersion, replicated across 8 seeds for 512 observations. Detailed parameter tables for each experiment appear in their respective results sections. Table~\ref{tab:params} summarises the parameters used across all experiments.

All experiments use two-level factorial designs (with a three-level $\eta$ factor in Experiments~1b--2b). This screening strategy identifies which factors matter, rather than mapping precise functional forms \citep{Box2005, Montgomery2017}. Experiment~1a uses a Resolution~V half-fraction that aliases only four-factor and higher interactions. The programme follows \citeauthor{Box2005}'s sequential experimentation philosophy; Experiments~1a and~2a screen many factors with large designs, and their findings inform the reduced factor sets in Experiments~1b and~2b.

\begin{table}[H]
\centering
\caption{Parameter Ranges and Their Usage Across Experiments}
\label{tab:params}
\small
\begin{tabular}{l l c c c c c c}
\toprule
\textbf{Name} & \textbf{Description} & \textbf{1a} & \textbf{1b} & \textbf{2a} & \textbf{2b} & \textbf{3a} & \textbf{3b}\\
\midrule
$\alpha$ & Q-learning rate & \checkmark & & & & & \\
$\gamma$ & Discount factor & \checkmark & & & & & \\
$\varepsilon$ & E-greedy exploration prob & \checkmark & & & & & \\
Boltzmann & Softmax exploration & \checkmark & & & & & \\
$c$ / $\sigma^2$ & Bandit exploration param &  &  & \checkmark & \checkmark & & \\
$\lambda$ & Regularisation &  &  & \checkmark & & & \\
\textit{Memory decay} & Observation weighting &  &  & \checkmark & & & \\
$r$ & Reserve price & \checkmark & & \checkmark & \checkmark & \checkmark & \checkmark \\
$n$ & Number of bidders & \checkmark & \checkmark & \checkmark & \checkmark & \checkmark & \checkmark \\
$\eta$ & Affiliation parameter &  & \checkmark & \checkmark & \checkmark & & \\
\textit{Episodes} & Total training rounds & \checkmark & \checkmark & \checkmark & \checkmark & \checkmark & \checkmark \\
\textit{Sync/Async} & Q-learning modes & \checkmark & & & & & \\
\textit{State info} & State features & \checkmark & \checkmark & & & & \\
\textit{Context richness} & Context features &  &  & \checkmark & \checkmark & & \\
\textit{Decay type}  & Epsilon decay schedule         & \checkmark & & & & & \\
\textit{Objective}   & Value- vs.\ utility-maximizer  & & & & & \checkmark & \\
\textit{Aggressiveness} & PI controller gain scaling & & & & & & \checkmark \\
$m$ & Budget multiplier & & & & & \checkmark & \checkmark \\
$\sigma$ & Value dispersion & & & & & \checkmark & \checkmark \\
\bottomrule
\end{tabular}
\end{table}

Our core outcome metrics are consistent across all experiments. We measure the \emph{average revenue in later rounds}, which reflects how well the auction performs after strategies have stabilised.\footnote{Specifically, we average revenue over the final 1{,}000 episodes for Experiments~1a--2b and over all post-burn-in episodes for Experiments~3a--3b.} We measure \emph{time to converge} as the earliest round after which rolling-average revenue stays within $\pm 5\%$ of its final mean. We compute a \emph{no-sale rate} by noting the fraction of rounds in which no valid bid exceeds the reserve. We track \emph{price volatility} by taking the sample standard deviation of winning bids in later rounds. Finally, we measure \emph{winner entropy} to assess whether outcomes concentrate among few bidders, calculating the Shannon entropy of the empirical distribution of winners. Table~\ref{tab:outcomes} provides a concise summary of these common metrics that allow systematic comparisons of convergence speed, bidding stability, and revenue performance across all experiments.

\begin{table}[h]
\centering
\caption{Outcome metrics used in all experiments.}
\label{tab:outcomes}
\begin{tabular}{l l}
\toprule
\textbf{Metric} & \textbf{Description}\\
\midrule
Average revenue (later rounds) & Mean revenue in the final 1{,}000 rounds \\
Time to converge & Round at which revenue stays in a $\pm 5\%$ band \\
No-sale rate & Fraction of rounds with all bids below $r$\\
Price volatility & Standard deviation of winning bids in later rounds \\
Winner entropy & Shannon entropy of bidder identity distribution \\
\bottomrule
\end{tabular}
\end{table}

Experiments~1b, 2a, 2b, 3a, and~3b introduce additional response variables specific to their designs; these are defined in the respective results sections (Sections~\ref{sec:qlearning_results}--\ref{sec:pacing_results}). Figure~\ref{fig:factorial_cube} illustrates the geometry of a $2^3$ factorial for three generic factors; our experiments extend this structure to 6--10 factors.

\begin{figure}[H]
\centering
\begin{tikzpicture}[
  arrow/.style={-{Stealth[length=2mm]}, thick},
]
  \def\dx{3.0}   
  \def\dy{3.0}   
  \def\ox{1.4}   
  \def\oy{0.8}   

  \coordinate (A) at (0, 0);        
  \coordinate (B) at (\dx, 0);      
  \coordinate (C) at (0, \dy);      
  \coordinate (D) at (\dx, \dy);    

  \coordinate (E) at (\ox, \oy);           
  \coordinate (F) at (\dx+\ox, \oy);       
  \coordinate (G) at (\ox, \dy+\oy);       
  \coordinate (H) at (\dx+\ox, \dy+\oy);   

  \draw[thick, dashed] (A) -- (E);
  \draw[thick, dashed] (E) -- (F);
  \draw[thick, dashed] (E) -- (G);

  \draw[thick] (A) -- (B);
  \draw[thick] (A) -- (C);
  \draw[thick] (B) -- (D);
  \draw[thick] (C) -- (D);

  \draw[thick] (B) -- (F);
  \draw[thick] (C) -- (G);
  \draw[thick] (D) -- (H);

  \draw[thick] (G) -- (H);
  \draw[thick] (F) -- (H);

  \filldraw (A) circle (2pt) node[below left, font=\scriptsize] {$(-,-,-)$};
  \filldraw (B) circle (2pt) node[below right, font=\scriptsize] {$(+,-,-)$};
  \filldraw (C) circle (2pt) node[above left, font=\scriptsize] {$(-,+,-)$};
  \filldraw (D) circle (2pt) node[above right=0.05cm and -0.2cm, font=\scriptsize] {$(+,+,-)$};
  \filldraw (E) circle (2pt) node[below left=-0.05cm and 0.1cm, font=\scriptsize] {$(-,-,+)$};
  \filldraw (F) circle (2pt) node[below right, font=\scriptsize] {$(+,-,+)$};
  \filldraw (G) circle (2pt) node[above left, font=\scriptsize] {$(-,+,+)$};
  \filldraw (H) circle (2pt) node[above right, font=\scriptsize] {$(+,+,+)$};

  \draw[arrow, gray] (A) -- ++(-0.7, 0) node[left, font=\scriptsize] {Factor A};
  \draw[arrow, gray] (A) -- ++(0, -0.7) node[below, font=\scriptsize] {Factor B};
  \draw[arrow, gray] (A) -- ++(-0.5, -0.3) node[below left, font=\scriptsize] {Factor C};

  \draw[decorate, decoration={brace, amplitude=5pt, raise=3pt}]
    ($(F)+(0.15,0)$) -- ($(H)+(0.15,0)$)
    node[midway, right=10pt, font=\scriptsize, align=left] {Main effect of B\\$= \bar{Y}_{B=+} - \bar{Y}_{B=-}$};
\end{tikzpicture}
\caption{Geometry of a $2^3$ factorial design. Each vertex represents one experimental cell defined by the sign combination of three factors. The brace illustrates how a main effect is computed as the contrast between high- and low-level marginal means. Our experiments extend this structure to 6--10 factors.}
\label{fig:factorial_cube}
\end{figure}

\subsection{Design and Estimation}
\label{sec:appendix_inference}

The factorial designs described above are analysed using ordinary least squares with effects coding ($x_i \in \{-1, +1\}$). This coding ensures orthogonality, meaning each main effect and two-way interaction can be estimated independently of all others, enabling clean attribution of outcome variation to specific factors. Experiment~1a's $2^{10-1}$ half-fraction has Resolution~V, ensuring that all main effects and two-way interactions are estimable without bias from three-way terms, which are assumed negligible under effect sparsity.\footnote{In the smaller mixed-level designs (Experiments~1b--2b), the degrees of freedom for error scale directly with replication count. Experiment~1b's $3 \times 2^3$ design has 24 cells and 16 model parameters; with eight replicates the residual degrees of freedom are adequate. HC3 robust standard errors and wild bootstrap $p$-values (Appendix~\ref{sec:appendix_robustness}) confirm that findings are stable across all sub-experiments.} Replication in all experiments enables pure error estimation and lack-of-fit testing.

The stochastic variation in our experiments arises from seed-induced randomness in exploration, valuation draws, and tie-breaking. Inference is over realizations of the stochastic simulation, not over a population of real markets. This is methodologically standard in computational economics and agent-based modelling \citep{Tesfatsion2006}. Factorial designs applied to stochastic simulations yield unbiased effect estimates under the same assumptions as physical experiments, with seed variation playing the role of experimental noise.

Experiment~1a uses a fractional factorial design to reduce experimental burden. The key trade-off is aliasing, whereby some higher-order interactions become confounded with lower-order terms. Experiments~1b--2b use mixed-level designs (combining the three-level $\eta$ factor with binary factors), and Experiments~3a and~3b each use a full $2^6$ factorial, so aliasing is not a concern for these experiments.

In a Resolution~V\footnote{In a Resolution~$R$ design, no $p$-factor interaction is aliased with any interaction of fewer than $R - p$ factors. Resolution~V therefore guarantees that main effects are free of two-way interaction bias and two-way interactions are free of other two-way interaction bias.} design, no main effect is aliased with any interaction of fewer than four factors, and no two-way interaction is aliased with any interaction of fewer than three factors. This ensures that all main effects and two-way interactions are estimable without bias from three-way terms, assumed negligible under effect sparsity. We validate this assumption using LASSO variable selection and nonparametric model comparison.

For each response variable $Y$ (e.g., average revenue, convergence time, price volatility), we fit a linear model with main effects and all two-way interactions via ordinary least squares (OLS).
\begin{align}
  Y_i = \beta_0 + \sum_{j=1}^{k} \beta_j\, x_{ji} + \sum_{1 \le j < l \le k} \beta_{jl}\, x_{ji}\, x_{li} + \varepsilon_i,
  \label{eq:ols}
\end{align}
where $x_{ji} \in \{-1, +1\}$ is the coded level of factor $j$ in observation $i$, $\beta_j$ represents the main effect of factor $j$, and $\beta_{jl}$ captures the two-way interaction between factors $j$ and $l$. Under effects coding, $\beta_j$ equals the mean response at the high level minus the mean response at the low level, divided by two, holding all other factors at their centre values. We use Type~III ANOVA decomposition to assess the marginal contribution of each term,\footnote{Under the $\pm 1$ effects coding in a balanced factorial, the design matrix is orthogonal ($\mathbf{X}^\top\mathbf{X}$ is diagonal), so the OLS $t$-test for each coefficient is algebraically equivalent to the Type~III ANOVA $F$-test for the same term ($F = t^2$), yielding identical $p$-values. This equivalence depends on orthogonality; under $0/1$ dummy coding the columns would be correlated, main effect estimates would be conditional on the reference level rather than marginal, and the two tests would diverge.} testing each coefficient against the null $H_0\!: \beta = 0$ via the $t$-statistic $\hat{\beta}/\text{SE}(\hat{\beta})$.\footnote{The ANOVA $F$-tests are unadjusted for multiplicity. In the factorial design tradition \citep{Box2005}, the primary screening tools are the half-normal probability plot and Lenth's pseudo-standard-error method \citep{Lenth1989}, both of which identify active effects without requiring an independent error estimate. Formal multiplicity corrections (Holm--Bonferroni, Benjamini--Hochberg) are applied in the robustness analysis described below.} Model adequacy is validated by comparing OLS $R^2$ to a gradient-boosted machine (LightGBM) $R^2$ trained with five-fold cross-validation. Gaps below 0.05 support adequate linear approximation; gaps above 0.05 reveal detectable nonlinearity that warrants checking whether effect rankings change under the nonparametric model.

\subsection{Inference and Diagnostics}

We compute HC3 robust standard errors \citep{MacKinnonWhite1985} to account for non-constant error variance across the design space. Comparing OLS standard errors to HC3 robust standard errors, we verify that significance claims hold under heteroscedasticity. The fraction of effects changing significance status under HC3 ranges from 0\% (Experiment~3a) to 6.7\% (Experiment~1a); per-experiment details appear in Appendix~\ref{sec:appendix_robustness}.

With $k$ main effects, $\binom{k}{2}$ two-way interactions, and $m$ response variables, each experiment involves many simultaneous hypothesis tests. We apply Holm--Bonferroni sequential correction \citep{Holm1979} to control the family-wise error rate at $\alpha = 0.05$. Key findings, including the auction type main effect and the auction type $\times$ exploration interaction, survive this stringent correction across all responses.

We also apply Benjamini--Hochberg (BH) correction \citep{BenjaminiHochberg1995} to control the false discovery rate, a less conservative criterion than FWER, providing greater power when many effects are truly active.

For the top~10 effects by absolute $t$-statistic in each response, we compute Rademacher wild bootstrap $p$-values \citep{DavidsonFlachaire2008} (1{,}000 iterations) to validate inference under minimal distributional assumptions. Bootstrap $p$-values align closely with HC3 asymptotic $p$-values (differences below 0.01), confirming robustness.

We fit quantile regressions \citep{KoenkerBassett1978} at the 10th, 25th, 50th, 75th, and 90th percentiles of each response to assess whether factor effects are uniform across the outcome distribution or concentrated in the tails. This complements the OLS analysis, which estimates effects at the conditional mean.

With two replicates per cell, we decompose the residual sum of squares into pure error (within-cell variation) and lack of fit (deviation of cell means from model predictions). The $F$-test for lack of fit assesses whether the linear model with two-way interactions adequately fits the data. Per-experiment lack-of-fit results are reported in Appendix~\ref{sec:appendix_robustness}.

We compute leave-one-out cross-validated $R^2$ using the PRESS statistic. A gap between $R^2$ and Pred-$R^2$ smaller than 0.10 indicates minimal overfitting. Predicted $R^2$ and PRESS gaps vary across experiments and are reported in each experiment's model adequacy table (Tables~\ref{tab:exp1a_adequacy}--\ref{tab:exp3b_adequacy}).

Gradient-boosted machines (LightGBM, 200 trees, maximum depth~4) are fit using five-fold cross-validation to establish a nonparametric upper bound on achievable $R^2$. If OLS $R^2$ is within 0.05 of LightGBM $R^2$, we conclude that the linear model with two-way interactions captures most of the signal and that higher-order or nonlinear terms contribute negligibly. In all experiments, OLS $R^2$ meets or exceeds LightGBM cross-validated $R^2$, confirming that the parametric model is well suited to the balanced factorial structure (see per-experiment diagnostics in Tables~\ref{tab:exp1a_adequacy}--\ref{tab:exp3b_adequacy}).

We fit five-fold cross-validated LASSO models with regularisation parameter $\lambda$ chosen to minimise mean squared error. Surviving variables are compared to OLS significance. In all experiments, LASSO retains auction type and number of bidders, corroborating OLS effect selection. Heredity is verified by checking that no interaction survives whose parent main effects were dropped.

The LASSO estimator \citep{Tibshirani1996} performs simultaneous estimation and variable selection via $L_1$ penalisation, driving small coefficients exactly to zero. The regularisation parameter $\lambda$ is chosen by five-fold cross-validation to minimise mean squared error. Heredity requires that an interaction $\beta_{jl}$ can be nonzero only if both parent main effects $\beta_j$ and $\beta_l$ are also nonzero.

Four types of diagnostic plots accompany each response variable. Pareto charts rank effects by absolute $t$-statistic. Main effects plots display the mean response at the low and high levels of each factor. Interaction plots show the mean response across factor-level combinations for the top six interactions. Half-normal probability plots \citep{Daniel1959} separate active effects (large departures from the reference line) from inert effects following the half-normal distribution under effect sparsity. Residual diagnostics include quantile--quantile plots and residuals-versus-fitted plots.

For each response variable, we compute the minimum detectable effect (MDE) at 80\% power and $\alpha = 0.05$, using the coefficient standard error $\hat{\sigma} / \sqrt{\sum_i x_{ij}^2}$. For binary $\pm 1$ factors in a balanced design this reduces to $\hat{\sigma}/\sqrt{n}$; for orthogonal polynomial contrasts with three-level factors, the sum of squares differs by column.\footnote{The linear contrast $\{-1, 0, +1\}$ has $\sum x^2 = 2n/3$ (lower power), and the quadratic contrast $\{+1, -2, +1\}$ has $\sum x^2 = 2n$ (higher power).} All effects reported as statistically significant in the results sections survive Benjamini--Hochberg correction.

\subsection{Robustness and Sensitivity}

We validate all findings through thirteen robustness checks applied to each experiment. These include HC3 robust standard errors, Holm--Bonferroni and Benjamini--Hochberg corrections, Rademacher wild bootstrap $p$-values, quantile regressions at five percentiles, lack-of-fit $F$-tests, leave-one-out cross-validated $R^2$ via the PRESS statistic, LightGBM nonparametric benchmarking, and LASSO variable selection. No multiplicity correction is applied across the six sub-experiments; the probability that the number of bidders ranks first among $k$ factors in all six independent sub-experiments by chance is bounded above by $(1/k)^6 < 0.0014$ for $k \geq 3$.\footnote{$t$-statistic magnitudes are not comparable across experiments due to differences in sample size, model specification, and error variance; the comparison is strictly ordinal.} MDEs range from 2--5\% of the mean response, indicating sufficient power to detect practically meaningful effects. Per-experiment diagnostics are reported in Appendix~\ref{sec:appendix_robustness}.

\label{sec:appendix_sensitivity}

As a complement to the factorial ANOVA, we decompose output variance through the Sobol--Hoeffding framework \citep{Sobol2001}. The first-order Sobol' index $S_i$ measures the fraction of output variance attributable to factor $X_i$ alone. The total-order index $S_{T_i}$ captures both the direct and all interaction contributions involving $X_i$. The gap $S_{T_i} - S_i$ quantifies how much of factor $i$'s influence operates through interactions with other factors rather than through its own main effect.

For balanced factorial designs with effects coding ($\pm 1$), the columns of the design matrix are mutually orthogonal. Under orthogonality, the Type~III ANOVA sum of squares decomposition coincides with the Sobol--Hoeffding decomposition \citep{ArcherSaltelliSobol1997, Saltelli2008}. The first-order index for factor $i$ is
\begin{equation}
S_i = \frac{\mathrm{SS}_i}{\mathrm{SS}_{\mathrm{total}}},
\end{equation}
where $\mathrm{SS}_i$ is the Type~III sum of squares for factor $i$ and $\mathrm{SS}_{\mathrm{total}}$ is the total (corrected) sum of squares. The second-order index $S_{ij} = \mathrm{SS}_{ij} / \mathrm{SS}_{\mathrm{total}}$ is computed analogously from the interaction sum of squares. This equivalence provides exact, closed-form Sobol' indices without Monte Carlo sampling.\footnote{For three-level factors such as affiliation ($\eta$), which are represented by two orthogonal contrast columns (linear and quadratic), the first-order index is the sum of the two contrast contributions, $S_\eta = (\mathrm{SS}_{\eta,\mathrm{lin}} + \mathrm{SS}_{\eta,\mathrm{quad}}) / \mathrm{SS}_{\mathrm{total}}$.} For orthogonal designs, this decomposition is not only exact but A-optimal for first-order Sobol' index estimation \citep{MorrisMooreMcKay2008}. The three-level factors in Experiments~1b--2b further improve estimation of higher-order indices relative to two-level designs \citep{WangTangZhang2012}.

\label{sec:power_analysis}

To assess robustness of the factor rankings, seven independent sensitivity methods are applied to each response variable in each experiment. In addition to analytical Sobol' indices, we compute random forest permutation importance, SHAP values via TreeSHAP on a LightGBM surrogate, Morris elementary effects \citep{Morris1991}, Monte Carlo Sobol' indices via three surrogate models (neural network, LightGBM, Kriging), Fourier amplitude sensitivity testing (FAST), and Borgonovo's $\delta$ moment-independent measure \citep{Borgonovo2007}. Cross-method concordance is assessed by computing pairwise Spearman rank correlations across all seven methods; the mean Spearman correlation provides a scalar measure of agreement on which factors matter most. Per-experiment Sobol' tables appear in Appendix~\ref{sec:appendix_sensitivity_tables}, with concordance metrics and cross-experiment synthesis in Sections~\ref{sec:sens_concordance}--\ref{sec:sens_synthesis}.

\section{Q-Learning}
\label{sec:qlearning_results}

\subsection{Design}
\label{sec:appendix_design}

\subsubsection{Constant Valuations}

Each bidder has a constant valuation $v_i=1.0$. Table~\ref{tab:exp1a_params} details the 10 factorial factors and fixed parameters. The $2^{10-1}$ Resolution~V half-fraction yields 512 cells, each replicated twice for 1{,}024 observations. The bid grid resolution is fixed at 11 discrete actions; a discretisation sensitivity analysis across grid sizes of 6, 11, and 21 is reported in Section~\ref{sec:appendix_robustness}.

\begin{table}[h]
\centering
\caption{Parameter settings for Experiment~1a (Constant Valuations).}
\label{tab:exp1a_params}
\begin{tabular}{l l l}
\toprule
\textbf{Factor} & \textbf{Low ($-1$)} & \textbf{High ($+1$)}\\
\midrule
Auction format & Second-price & First-price \\
Learning rate ($\alpha$) & 0.01 & 0.10 \\
Discount factor ($\gamma$) & 0.0 & 0.95 \\
Reserve price ($r$) & 0.0 & 0.5 \\
Initialisation & Zeros & Optimistic \\
Exploration & $\varepsilon$-greedy & Boltzmann \\
Update mode & Synchronous & Asynchronous \\
Number of bidders ($n$) & 2 & 4 \\
Information feedback & None (stateless) & Previous winning bid \\
Decay type & Linear & Exponential \\
\midrule
\multicolumn{3}{l}{\textbf{Fixed parameters}} \\
\midrule
Bid grid resolution & \multicolumn{2}{l}{11 discrete actions\footnotemark} \\
Number of episodes & \multicolumn{2}{l}{100{,}000} \\
\bottomrule
\end{tabular}
\footnotetext{Bid grid resolution is held constant at 11 levels. A discretisation sensitivity analysis varying the grid size is reported in Section~\ref{sec:appendix_robustness}.}
\end{table}

Experiment~1a compares $\varepsilon$-greedy and Boltzmann exploration using identical parameter schedules. Both decay from $1.0$ to $0.01$ over the first 90\% of training. To ensure comparability, Boltzmann exploration uses Q-range normalised temperatures, with the effective temperature $\tau_{\text{eff}} = \tau \cdot \max(\Delta Q, \epsilon)$ where $\Delta Q = \max_a Q(s,a) - \min_a Q(s,a)$ and $\epsilon = 10^{-8}$.\footnote{This normalisation ensures the softmax probabilities $\pi(a|s) \propto \exp(Q(s,a)/\tau_{\text{eff}})$ depend on the relative spacing of Q-values rather than their absolute scale, making the Boltzmann temperature comparable in effect to $\varepsilon$-greedy across different discount factors and training stages.} The decay type factor contrasts linear decay ($\varepsilon_t = 1 - t/t_{\max}$) with exponential decay ($\varepsilon_t = \varepsilon_0 \cdot (\varepsilon_{\min}/\varepsilon_0)^{t/t_{\max}}$); at the midpoint of training, exponential decay retains a higher exploration rate than linear decay (ratio approximately 1.6:1).\footnote{This asymmetry is by design. The factor tests whether front-loaded exploration (linear) or sustained exploration (exponential) better supports learning. The small estimated effect sizes across response variables confirm that decay shape is a secondary consideration relative to factors such as auction format and discount factor.}

\subsubsection{Affiliated Valuations}

Bidders receive private signals $s_i\in[0,1]$, and their valuations become interdependent through an affiliation parameter $\eta\in[0,1]$ via $v_i=(1-0.5\,\eta)\,s_i + 0.5\,\eta\,\bigl(\tfrac{1}{n-1}\sum_{j\neq i}s_j\bigr)$. Thus $\eta=0$ corresponds to purely private values and $\eta=1$ to strong common-value elements. Table~\ref{tab:exp1b_params} highlights the key parameters.

\begin{table}[h]
\centering
\caption{Parameter settings for Experiment~1b (Affiliated Valuations + Q-learning).}
\label{tab:exp1b_params}
\begin{tabular}{l l l}
\toprule
\textbf{Factor} & \textbf{Low ($-1$)} & \textbf{High ($+1$)}\\
\midrule
Auction format & Second-price & First-price \\
Number of bidders ($n$) & 2 & 4 \\
State information & Signal only & Signal + winner \\
\midrule
\textbf{Factor} & \multicolumn{2}{l}{\textbf{Levels}} \\
\midrule
Affiliation ($\eta$) & \multicolumn{2}{l}{0, 0.5, 1 (linear + quadratic contrasts)} \\
\midrule
\multicolumn{3}{l}{\textbf{Fixed parameters (from Experiment~1a results)}} \\
\midrule
Learning rate ($\alpha$) & \multicolumn{2}{l}{0.1} \\
Discount factor ($\gamma$) & \multicolumn{2}{l}{0.95} \\
Exploration & \multicolumn{2}{l}{$\varepsilon$-greedy} \\
Update mode & \multicolumn{2}{l}{Asynchronous} \\
Decay type & \multicolumn{2}{l}{Linear} \\
Reserve price ($r$) & \multicolumn{2}{l}{0.0} \\
Initialisation & \multicolumn{2}{l}{Zeros} \\
Bid grid resolution & \multicolumn{2}{l}{11 discrete actions} \\
Number of episodes & \multicolumn{2}{l}{100{,}000} \\
\bottomrule
\end{tabular}
\end{table}

\subsection{Constant Valuations (Experiment~1a)}
\label{sec:exp1a_results}

Experiment~1a deploys Q-learning agents with constant valuations ($v=1$) in a $2^{10-1}$ Resolution~V half-fraction spanning 10 factors and 1{,}024 observations. Under independent private values with constant valuations, revenue equivalence \citep{Vickrey1961} predicts identical expected revenue across auction formats. Despite this theoretical equivalence, average revenues in the final 1{,}000 episodes range from approximately 0.3 to 1.0, revealing substantial variability across configurations of learning parameters and market structure. The factorial analysis decomposes this variability into systematic effects attributable to individual factors and their interactions.

\begin{figure}[H]
  \centering
  \includegraphics[width=\textwidth]{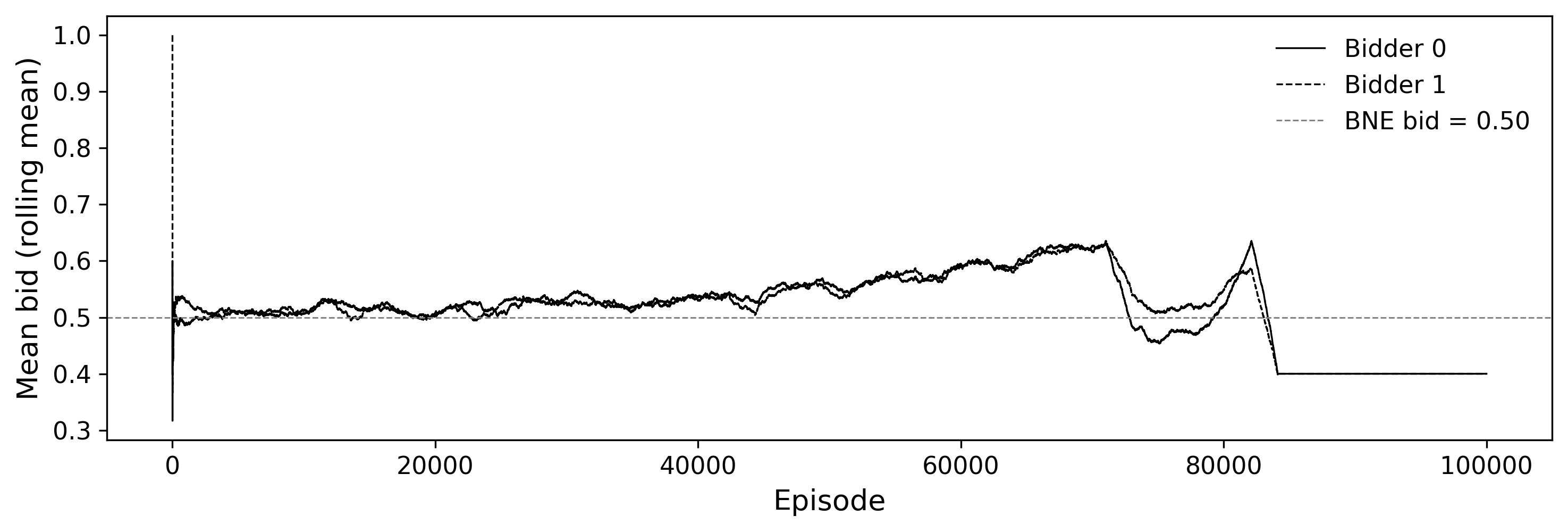}
  \caption{Representative learning trajectory for a single trial of Experiment~1a (first-price auction, 2~bidders, Boltzmann exploration, 10{,}000 episodes). Per-bidder mean bids (200-episode rolling mean) converge toward competitive levels over the course of training.}
  \label{fig:e1a_trace}
\end{figure}

\subsubsection{Revenue}

\begin{figure}[H]
  \centering
  \includegraphics[width=0.7\textwidth]{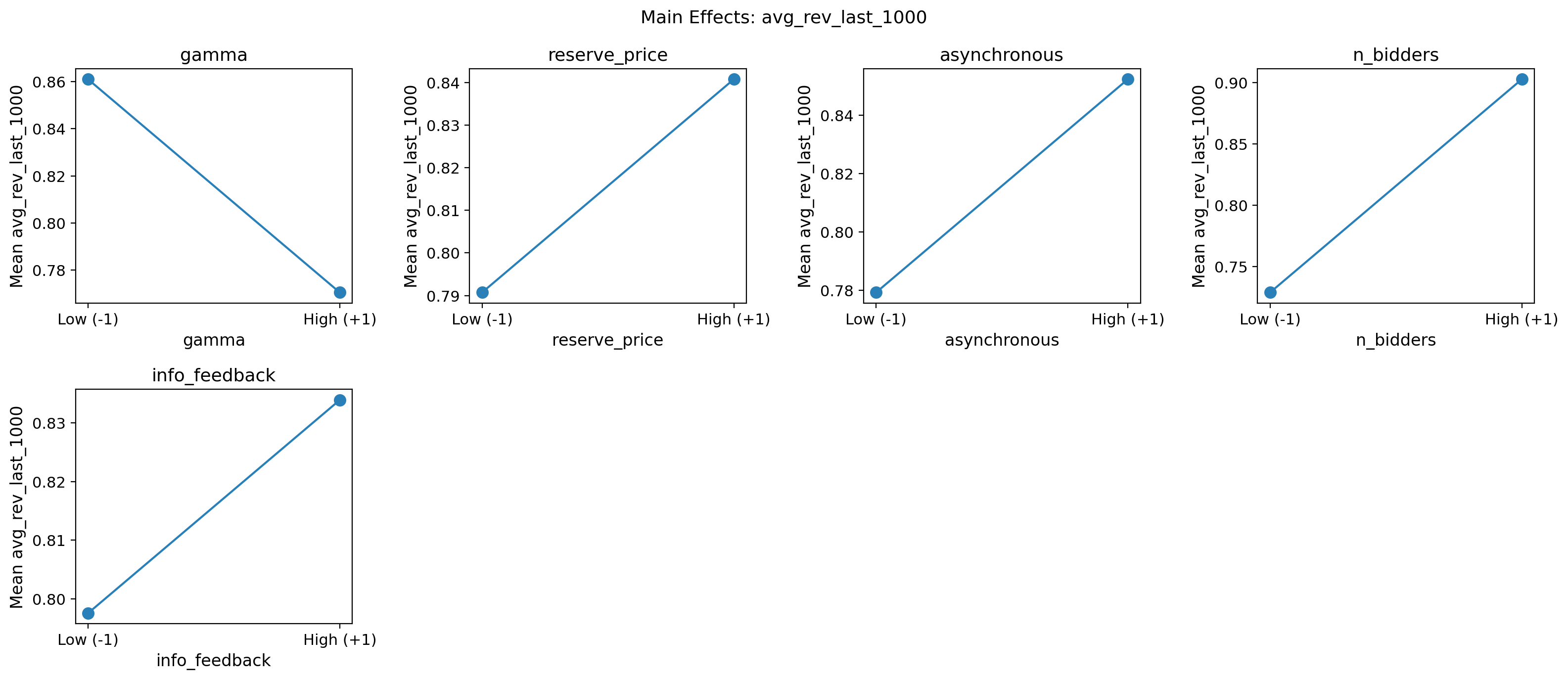}
  \caption{Experiment~1a: Main effects plot for average revenue. Number of bidders and discount factor dominate; auction format has a negligible main effect.}
  \label{fig:e1a_main_rev}
\end{figure}

Number of bidders is the dominant factor for revenue ($|t| = \ExpOneARevFactorTOne$), followed by the discount factor $\gamma$ ($|t| = \ExpOneARevFactorTTwo$) and update mode ($|t| = \ExpOneARevFactorTThree$). Table~\ref{tab:exp1a_ranked_rev} ranks significant effects by absolute $t$-statistic. Moving from two to four bidders increases average revenue by \ExpOneARevTopPctEffect\% of the grand mean, an effect \ExpOneARevTopVsAuctionRatio{} times larger than that of auction format. Auction format, by contrast, does not exert a statistically significant main effect on revenue ($|t| = \ExpOneARevAuctionT$, $p \ExpOneARevAuctionPFmt$). First-price and second-price auctions produce nearly identical average revenues in the final 1{,}000 episodes (FPA mean \ExpOneARevMeanFPA{} vs.\ SPA mean \ExpOneARevMeanSPA). The main effects plot (Figure~\ref{fig:e1a_main_rev}) confirms these findings.

\begin{table}[H]
\centering
\caption{Experiment 1a: Significant effects for average revenue ($p < 0.05$), ranked by $|t|$.}
\label{tab:exp1a_ranked_rev}
\begin{tabular}{lrrl}
\toprule
\textbf{Effect} & \textbf{Coeff.} & \textbf{$|t|$} & \textbf{Direction} \\
\midrule
Number of bidders & 0.0871 & 17.75 & + \\
Discount factor ($\gamma$) & -0.0453 & 9.23 & $-$ \\
Update mode & 0.0366 & 7.45 & + \\
Discount factor ($\gamma$) $\times$ Number of bidders & 0.0298 & 6.09 & + \\
Auction format $\times$ Number of bidders & -0.0276 & 5.63 & $-$ \\
Reserve price & 0.0250 & 5.10 & + \\
Reserve price $\times$ Information feedback & -0.0219 & 4.47 & $-$ \\
Auction format $\times$ Discount factor ($\gamma$) & -0.0215 & 4.38 & $-$ \\
Auction format $\times$ Update mode & -0.0193 & 3.94 & $-$ \\
Information feedback & 0.0182 & 3.71 & + \\
\bottomrule
\end{tabular}
\end{table}

\begin{figure}[H]
  \centering
  \includegraphics[width=0.7\textwidth]{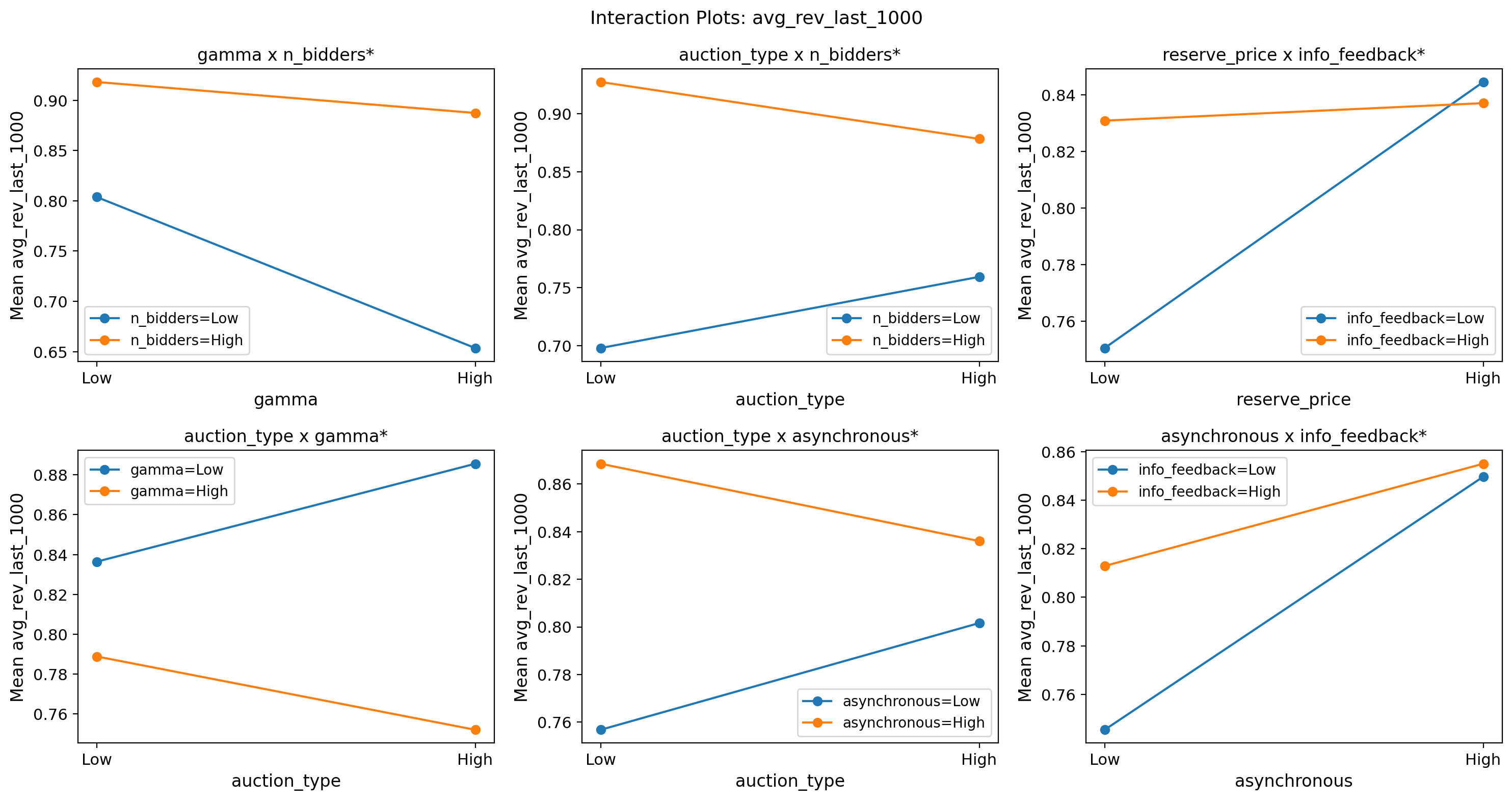}
  \caption{Experiment~1a: Interaction plot for average revenue, showing the top two-way interactions. Non-parallel lines indicate that the effect of one factor depends on the level of another.}
  \label{fig:e1a_int_rev}
\end{figure}

Although auction format has no significant main effect on final-episode revenue, it does interact significantly with other factors. The auction format $\times$ number of bidders interaction ($|t| = \ExpOneARevAuctionxNbidAbsT$) indicates that the format gap is larger in thin markets than in thick markets. The auction format $\times$ discount factor interaction ($|t| = \ExpOneARevAuctionxGammaAbsT$) shows that the auction format effect varies with the discount factor. Figure~\ref{fig:e1a_int_rev} displays these top interaction effects.

\subsubsection{Price Volatility}

\begin{figure}[H]
  \centering
  \includegraphics[width=0.7\textwidth]{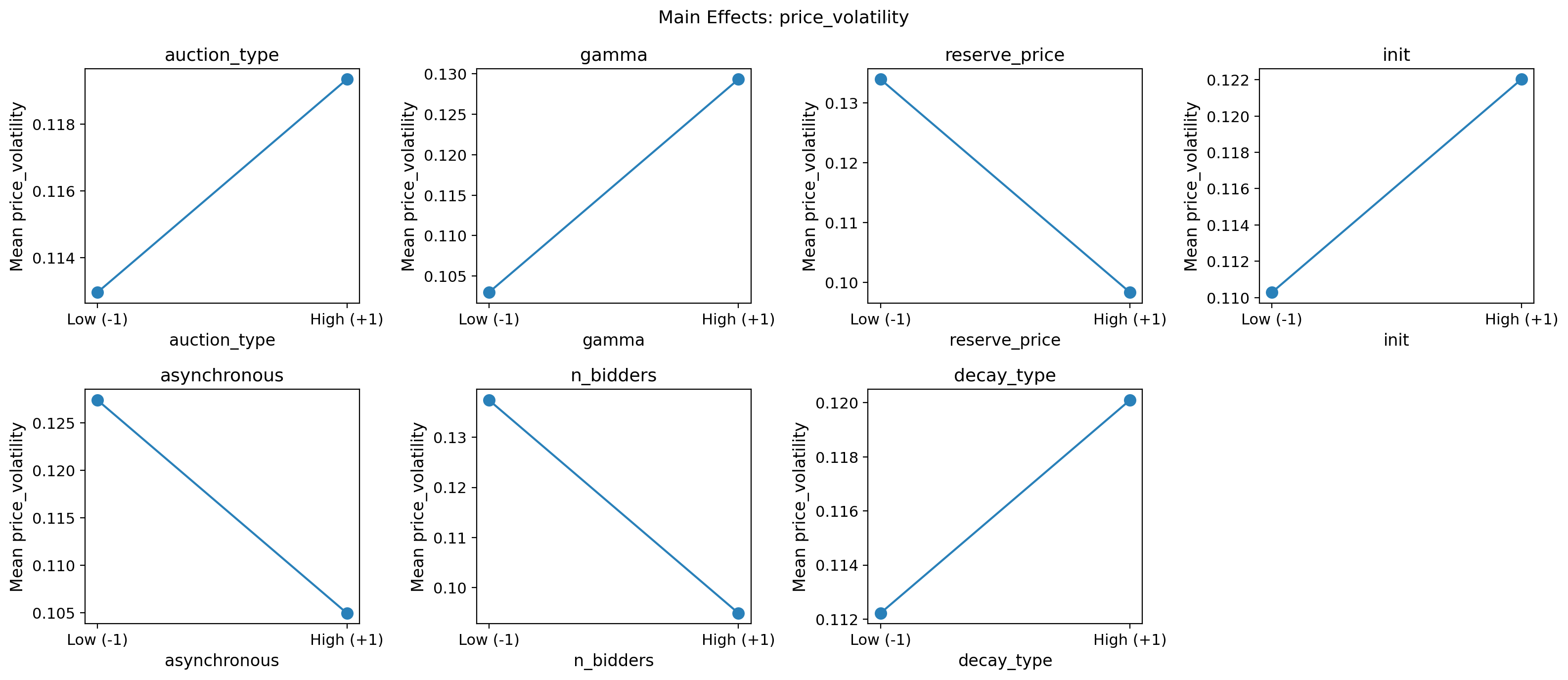}
  \caption{Experiment~1a: Main effects plot for price volatility. Number of bidders, reserve price, and discount factor are the primary drivers.}
  \label{fig:e1a_main_vol}
\end{figure}

Table~\ref{tab:exp1a_ranked_vol} ranks the significant effects for price volatility. Market thickness, reserve price, discount factor, and update mode are the primary drivers (Figure~\ref{fig:e1a_main_vol}). Neither auction format nor exploration strategy has a significant direct effect. Model adequacy diagnostics confirm these findings (Appendix~\ref{sec:appendix_robustness}).

\begin{table}[H]
\centering
\caption{Experiment 1a: Significant effects for price volatility ($p < 0.05$), ranked by $|t|$.}
\label{tab:exp1a_ranked_vol}
\begin{tabular}{lrrl}
\toprule
\textbf{Effect} & \textbf{Coeff.} & \textbf{$|t|$} & \textbf{Direction} \\
\midrule
Number of bidders & -0.0213 & 15.60 & $-$ \\
Reserve price & -0.0178 & 13.07 & $-$ \\
Discount factor ($\gamma$) & 0.0132 & 9.66 & + \\
Update mode & -0.0112 & 8.23 & $-$ \\
Reserve price $\times$ Update mode & 0.0069 & 5.08 & + \\
Exploration strategy $\times$ Information feedback & -0.0059 & 4.30 & $-$ \\
Initialisation & 0.0059 & 4.30 & + \\
Reserve price $\times$ Number of bidders & 0.0053 & 3.89 & + \\
Number of bidders $\times$ Information feedback & -0.0051 & 3.71 & $-$ \\
Discount factor ($\gamma$) $\times$ Reserve price & -0.0042 & 3.11 & $-$ \\
\bottomrule
\end{tabular}
\end{table}

\subsection{Affiliated Valuations (Experiment~1b)}
\label{sec:exp1b_results}

Experiment~1b generalises the framework to affiliated valuations with $\eta \in \{0, 0.5, 1\}$, introducing stochastic signals and value interdependence. Revenue equivalence holds for all $\eta$ in this model because signals are drawn independently (Appendix~\ref{sec:equilibria}). The $3 \times 2^3 = 24$ mixed-level factorial crosses auction format, affiliation strength, number of bidders, and state information, with hyperparameters fixed at levels identified in Experiment~1a.

\begin{figure}[H]
  \centering
  \includegraphics[width=\textwidth]{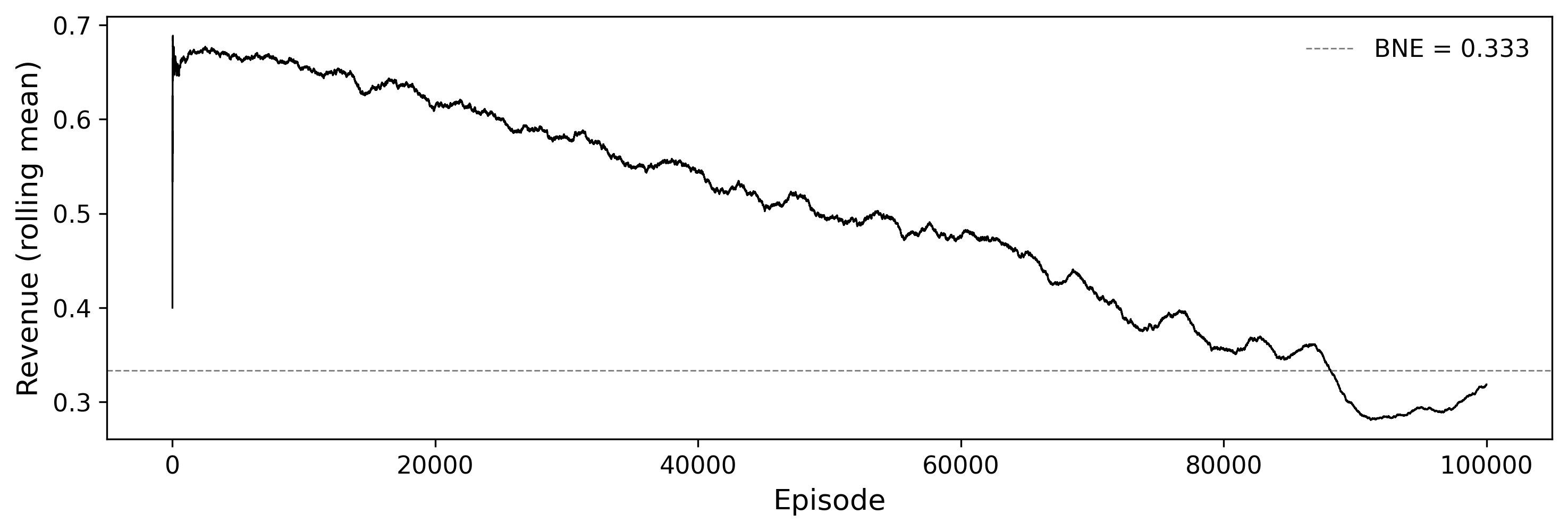}
  \caption{Representative learning trajectory for a single trial of Experiment~1b (first-price auction, $\eta = 0.5$, 2~bidders, 10{,}000 episodes). Rolling-mean revenue converges toward competitive levels.}
  \label{fig:e1b_trace}
\end{figure}

\subsubsection{Revenue}

\begin{figure}[H]
  \centering
  \includegraphics[width=0.7\textwidth]{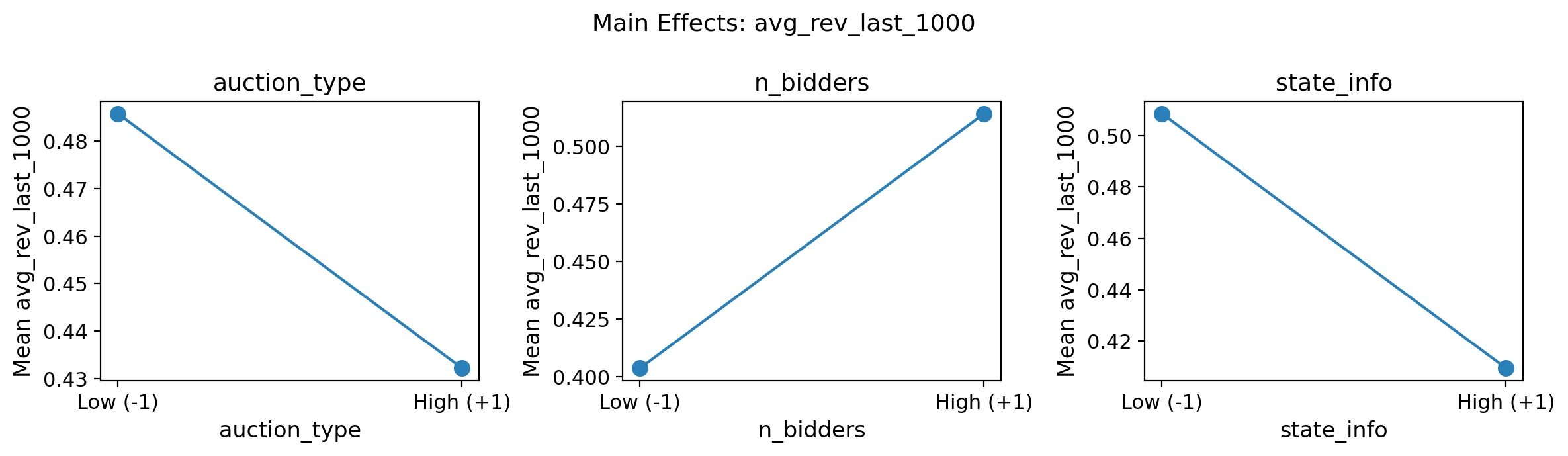}
  \caption{Experiment~1b: Main effects plot for average revenue. First-price auctions reduce revenue across all factor configurations.}
  \label{fig:e1b_main_rev}
\end{figure}

Number of bidders is the dominant factor for revenue ($|t| = \ExpOneBRevNbidAbsT$), with the low-to-high shift increasing revenue by \ExpOneBRevTopPctEffect\% of the grand mean. Auction format is statistically significant but economically modest ($|t| = \ExpOneBRevAuctionT$, $p \ExpOneBRevAuctionPFmt$). Table~\ref{tab:exp1b_ranked_rev} reports the full effect hierarchy; Figure~\ref{fig:e1b_main_rev} displays the directional patterns.

\begin{table}[H]
\centering
\caption{Experiment 1b: Significant effects for average revenue ($p < 0.05$), ranked by $|t|$.}
\label{tab:exp1b_ranked_rev}
\begin{tabular}{lrrl}
\toprule
\textbf{Effect} & \textbf{Coeff.} & \textbf{$|t|$} & \textbf{Direction} \\
\midrule
Number of bidders & 0.0552 & 6.53 & + \\
State information & -0.0495 & 5.85 & $-$ \\
Auction format $\times$ Number of bidders & -0.0285 & 3.38 & $-$ \\
Auction format & -0.0268 & 3.17 & $-$ \\
Number of bidders $\times$ State information & -0.0263 & 3.11 & $-$ \\
\bottomrule
\end{tabular}
\end{table}

\begin{figure}[H]
  \centering
  \includegraphics[width=0.7\textwidth]{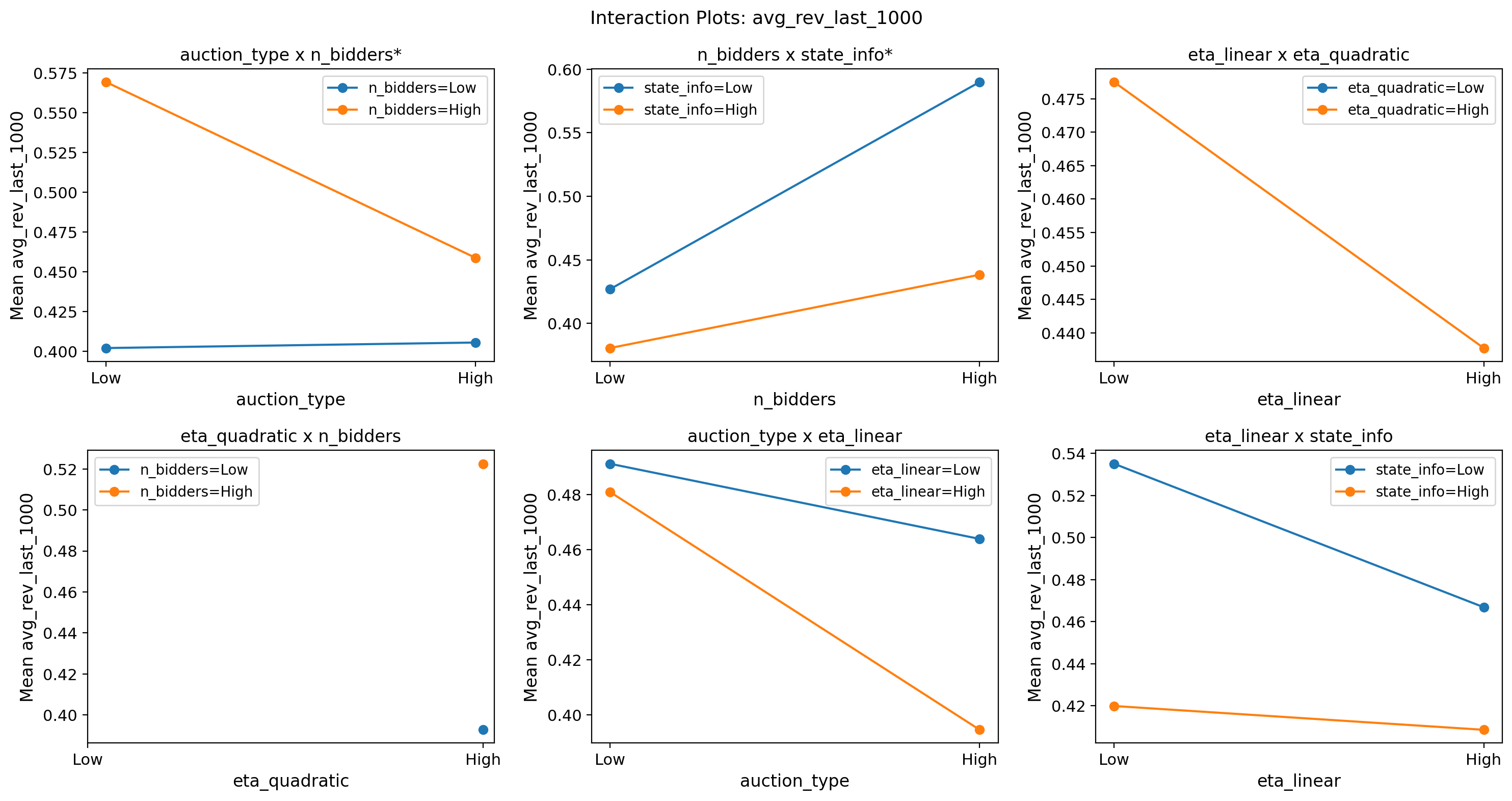}
  \caption{Experiment~1b: Interaction plot for average revenue. The auction type $\times$ number of bidders and auction type $\times$ state information interactions are among the strongest two-way effects.}
  \label{fig:e1b_int_rev}
\end{figure}

The affiliation parameter $\eta$ has no significant effect on any primary outcome, despite spanning the full range from independent private values to near-common values, consistent with the revenue equivalence that holds for all $\eta$ (Appendix~\ref{sec:equilibria}). The number of bidders moderates the auction type effect, with the first-price revenue penalty smaller under four bidders than under two. State information also influences revenue outcomes. The interaction plot (Figure~\ref{fig:e1b_int_rev}) displays these moderating relationships.

End-state revenue (final 1{,}000 episodes) and lifetime revenue (all episodes) reveal different patterns across auction formats. Both formats generate higher revenue during the learning phase than at convergence. First-price auctions show a learning-phase premium roughly three times larger than second-price, which offsets the end-state revenue gap.\footnote{First-price auctions show a learning-phase premium of \ExpOneBFPAPremium{} (mean lifetime revenue \ExpOneBAllMeanFPA{} vs.\ end-state \ExpOneBEndMeanFPA), while second-price auctions show a premium of \ExpOneBSPAPremium{} (\ExpOneBAllMeanSPA{} vs.\ \ExpOneBEndMeanSPA).} In the factorial model for lifetime revenue, the number of bidders and state information remain significant, but auction format is no longer significant, rendering the two formats statistically indistinguishable over the full trajectory.\footnote{Lifetime revenue model: $R^2 = \ExpOneBAllRevRsq$; number of bidders $|t| = \ExpOneBAllRevNbidAbsT$, state information $|t| = \ExpOneBAllRevStateAbsT$, auction format $|t| = \ExpOneBAllRevAuctionAbsT$ ($p \ExpOneBAllRevAuctionPFmt$).} The auction format $\times$ number of bidders interaction persists for lifetime revenue. With two bidders, first-price auctions generate higher lifetime revenue, while with four bidders, second-price auctions lead.\footnote{Interaction: $|t| = \ExpOneBAllRevAuctionxNbidAbsT$, $p \ExpOneBAllRevAuctionxNbidPFmt$. Two bidders: FPA \ExpOneBAllFPATwoBid{} vs.\ SPA \ExpOneBAllSPATwoBid; four bidders: SPA \ExpOneBAllSPAFourBid{} vs.\ FPA \ExpOneBAllFPAFourBid.}

\subsubsection{Price Volatility}

\begin{figure}[H]
  \centering
  \includegraphics[width=0.7\textwidth]{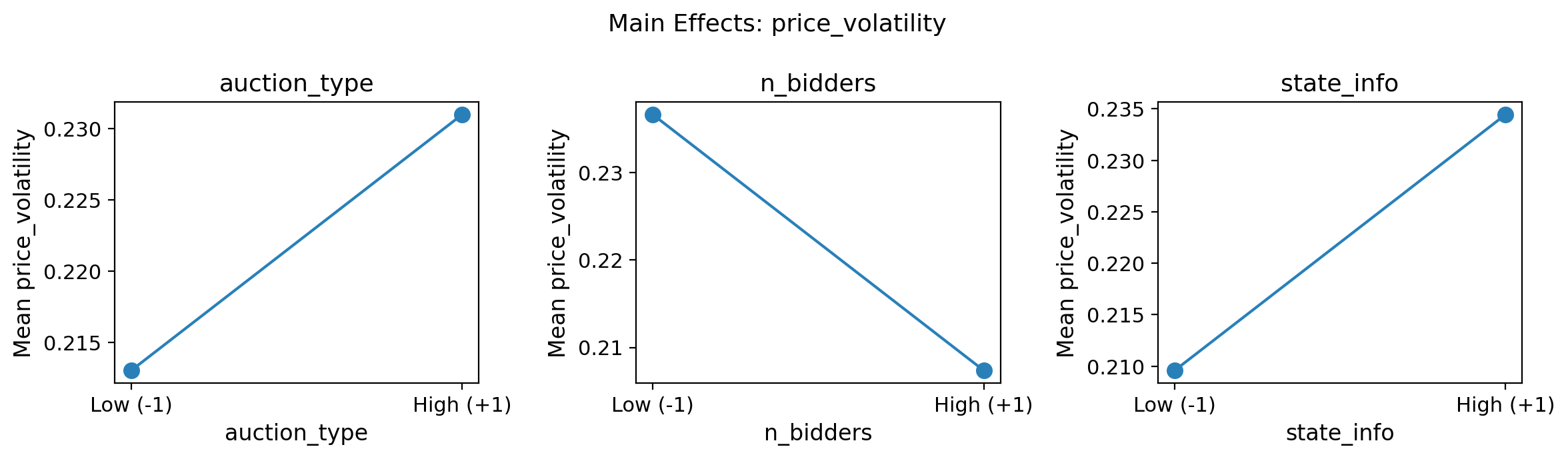}
  \caption{Experiment~1b: Main effects plot for price volatility.}
  \label{fig:e1b_main_vol}
\end{figure}

Unlike Experiment~1a, where auction type had minimal effect on volatility, first-price auctions now significantly raise price volatility under affiliated valuations. Table~\ref{tab:exp1b_ranked_vol} shows auction type among the significant effects for volatility. The main effects plot (Figure~\ref{fig:e1b_main_vol}) confirms the directional increase. The number of bidders and affiliation strength also contribute to volatility differences across configurations. Model adequacy diagnostics confirm these findings (Appendix~\ref{sec:appendix_robustness}).

\begin{table}[H]
\centering
\caption{Experiment 1b: Significant effects for price volatility ($p < 0.05$), ranked by $|t|$.}
\label{tab:exp1b_ranked_vol}
\begin{tabular}{lrrl}
\toprule
\textbf{Effect} & \textbf{Coeff.} & \textbf{$|t|$} & \textbf{Direction} \\
\midrule
Number of bidders & -0.0147 & 6.15 & $-$ \\
State information & 0.0124 & 5.22 & + \\
Auction format & 0.0090 & 3.77 & + \\
Auction format $\times$ Affiliation (linear) & 0.0074 & 2.54 & + \\
Auction format $\times$ Number of bidders & 0.0056 & 2.34 & + \\
\bottomrule
\end{tabular}
\end{table}

\subsection{Sensitivity Analysis}
\label{sec:qlearning_sensitivity}

Global sensitivity analysis independently confirms the factorial rankings. In Experiment~1a, the number of bidders achieves the highest total-order Sobol' index for revenue ($S_T = 0.24$) and dominates four of five responses, while reserve price dominates the no-sale rate ($S_T = 0.27$). Learning rate and initialisation are negligible across all responses. Cross-method concordance is high (mean Spearman $\rho = 0.87$), confirming stable factor rankings across all seven methods. In Experiment~1b, the number of bidders again dominates revenue ($S_T = 0.22$), while auction format is the most important factor for winner's curse frequency ($S_T = 0.36$). The affiliation parameter $\eta$ overwhelmingly drives signal responsiveness ($S_T = 0.40$), confirming that the experimental manipulation successfully modulates bidding behaviour. Cross-method concordance is low ($\rho = 0.33$), reflecting the small design (24 cells). Detailed per-factor Sobol' indices appear in Appendix~\ref{sec:appendix_sensitivity_tables} (Tables~\ref{tab:sens_exp1a}--\ref{tab:sens_exp1b_b}).

\subsection{Summary}

The number of bidders is the strongest determinant of seller outcomes across both Q-learning experiments. In Experiment~1a, it is followed by the discount factor and update mode, while auction format does not significantly affect end-state revenue but interacts with competitive pressure and temporal discounting. This null result for auction format provides a baseline; subsequent experiments show that the importance of format is context-dependent, emerging under affiliated valuations and budget-constrained pacing. Price volatility in Experiment~1a is governed by market thickness, reserve prices, and the discount factor.

In Experiment~1b, the number of bidders dominates all other factors ($|t| = \ExpOneBRevNbidAbsT$). Auction format reduces end-state revenue ($|t| = \ExpOneBRevAuctionT$) but not lifetime revenue, as the end-state gap is offset by a larger learning-phase premium under first-price auctions. The affiliation parameter $\eta$ has no significant impact on any outcome. The absence of any affiliation effect raises the question of whether more sophisticated algorithms can exploit the informational structure that Q-learning ignores. Global sensitivity analysis independently confirms these rankings (Section~\ref{sec:qlearning_sensitivity}). Robustness checks confirm these findings (Appendix~\ref{sec:appendix_robustness}).

\section{Contextual Bandits}
\label{sec:bandits_results}

\subsection{Design}
\label{sec:bandits_design}

This experiment preserves the affiliated valuation model from Experiment~1b but replaces Q-learning with contextual bandit methods. Table~\ref{tab:exp2_params} summarises the parameters. Seven binary factors and the three-level affiliation parameter $\eta$ define the experimental space. LinUCB includes two additional factors (regularisation $\lambda$ and memory decay) that preliminary screening found negligible for Thompson Sampling. These are excluded from the Thompson Sampling design to concentrate cells on active factors.\footnote{The exploration intensity factor maps to algorithm-specific parameters. For LinUCB, the confidence bound width $c \in \{0.5, 2.0\}$ (a 4$\times$ ratio), and for Contextual Thompson Sampling, the posterior variance $\sigma^2 \in \{0.1, 1.0\}$ (a 10$\times$ ratio). These ratios reflect the different units and scales of the two exploration mechanisms. LinUCB explores by inflating the estimated reward with $c\sqrt{\mathbf{x}^\top A^{-1}\mathbf{x}}$, where the uncertainty term shrinks with data; a 4$\times$ increase in $c$ roughly quadruples the exploration bonus. Thompson Sampling explores through posterior sampling variance $\sigma^2 A^{-1}$; a 10$\times$ increase broadens the sampling distribution proportionally. The factorial model captures this asymmetry through the algorithm $\times$ exploration intensity interaction term.}\footnote{The memory decay factor $\delta$ controls how much weight historical observations receive relative to recent ones. At $\delta = 1.0$ (no decay), both LinUCB and Thompson Sampling accumulate all past data, and after $T$ rounds the effective sample size is $T$. At $\delta < 1$, the effective sample size plateaus at approximately $1/(1-\delta)$. This factor is motivated by \citet{Douglas2024}, who show that deterministic convergence in bandit learners drives supra-competitive outcomes, and by \citet{Russac2019}, who provide regret guarantees for discounted linear bandits in non-stationary environments.}

\begin{table}[H]
\centering
\caption{Parameter settings for Experiments~2a and~2b (Affiliated Valuations + Bandits).}
\label{tab:exp2_params}
\begin{tabular}{l l l}
\toprule
\textbf{Factor} & \textbf{Low ($-1$)} & \textbf{High ($+1$)}\\
\midrule
Algorithm & LinUCB & CTS \\
Auction format & Second-price & First-price \\
Number of bidders ($n$) & 2 & 4 \\
Reserve price ($r$) & 0.0 & 0.3 \\
Exploration intensity & Low & High \\
Context richness & Signal only & Signal + winner \\
Regularisation ($\lambda$) & 0.1 & 5.0 \\
Memory decay ($\delta$) & 1.0 (no decay) & 0.999 (active) \\
\midrule
\textbf{Factor} & \multicolumn{2}{l}{\textbf{Levels}} \\
\midrule
Affiliation ($\eta$) & \multicolumn{2}{l}{0, 0.5, 1 (linear + quadratic contrasts)} \\
\midrule
\multicolumn{3}{l}{\textbf{Fixed parameters}} \\
\midrule
Bid grid resolution & \multicolumn{2}{l}{11 discrete actions} \\
Number of rounds & \multicolumn{2}{l}{100{,}000} \\
\bottomrule
\end{tabular}
\end{table}

\subsection{LinUCB (Experiment~2a)}
\label{sec:exp2a_results}

Experiment~2a deploys LinUCB agents under the affiliated valuation model. The design is a $3 \times 2^7 = 384$ mixed-level factorial with 8 factors (7 binary plus the three-level affiliation parameter $\eta$), including the regularisation parameter $\lambda$ and memory decay, both of which strongly affect LinUCB performance. The design is replicated twice for \ExpTwoANobs{} observations.

\begin{figure}[H]
  \centering
  \includegraphics[width=\textwidth]{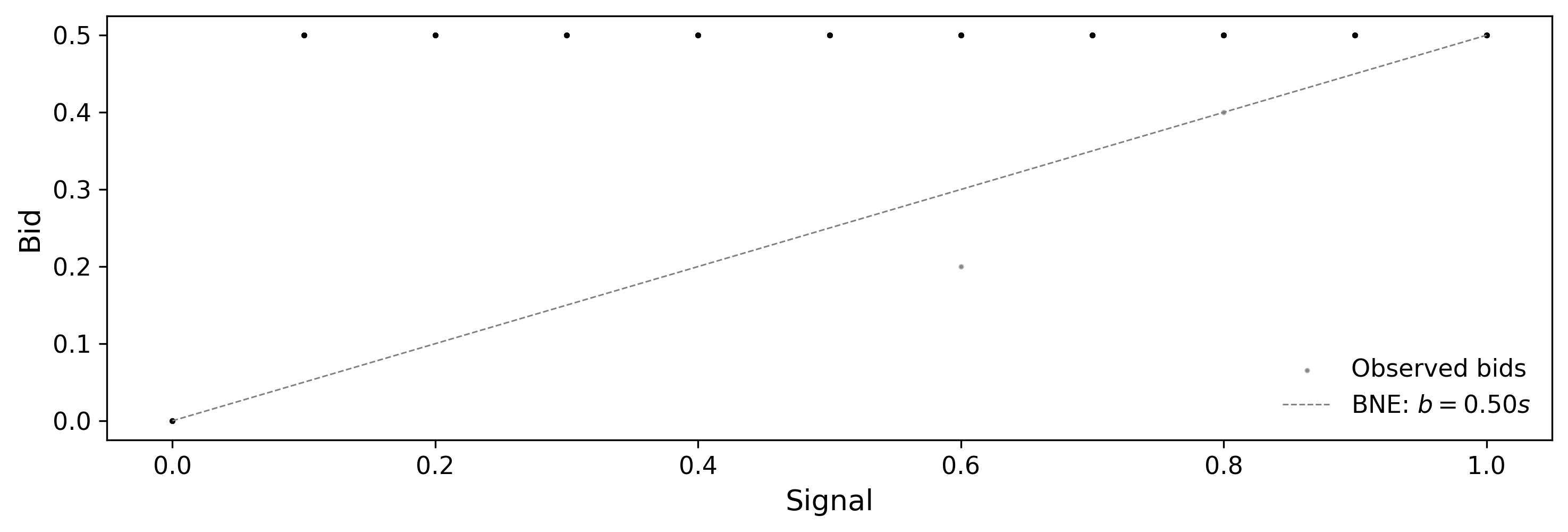}
  \caption{Representative bid function for a single trial of Experiment~2a (LinUCB, first-price auction, $\eta = 0.5$, 2~bidders, 5{,}000 rounds). Bids versus signals in the final 1{,}000 rounds, overlaid with the theoretical bid function.}
  \label{fig:e2a_trace}
\end{figure}

\subsubsection{Revenue}

\begin{figure}[H]
  \centering
  \includegraphics[width=0.7\textwidth]{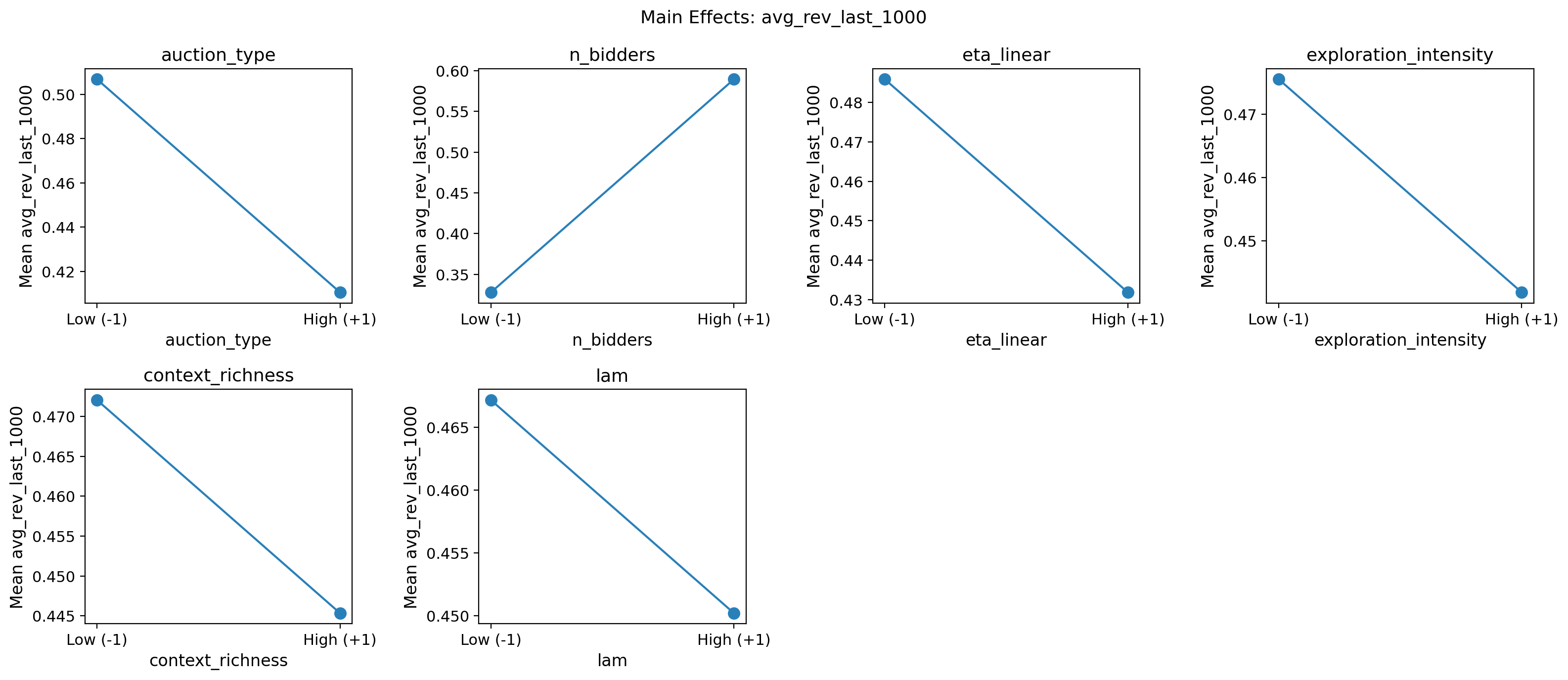}
  \caption{Experiment~2a: Main effects plot for average revenue under LinUCB.}
  \label{fig:e2a_main_rev}
\end{figure}

\begin{figure}[H]
  \centering
  \includegraphics[width=0.7\textwidth]{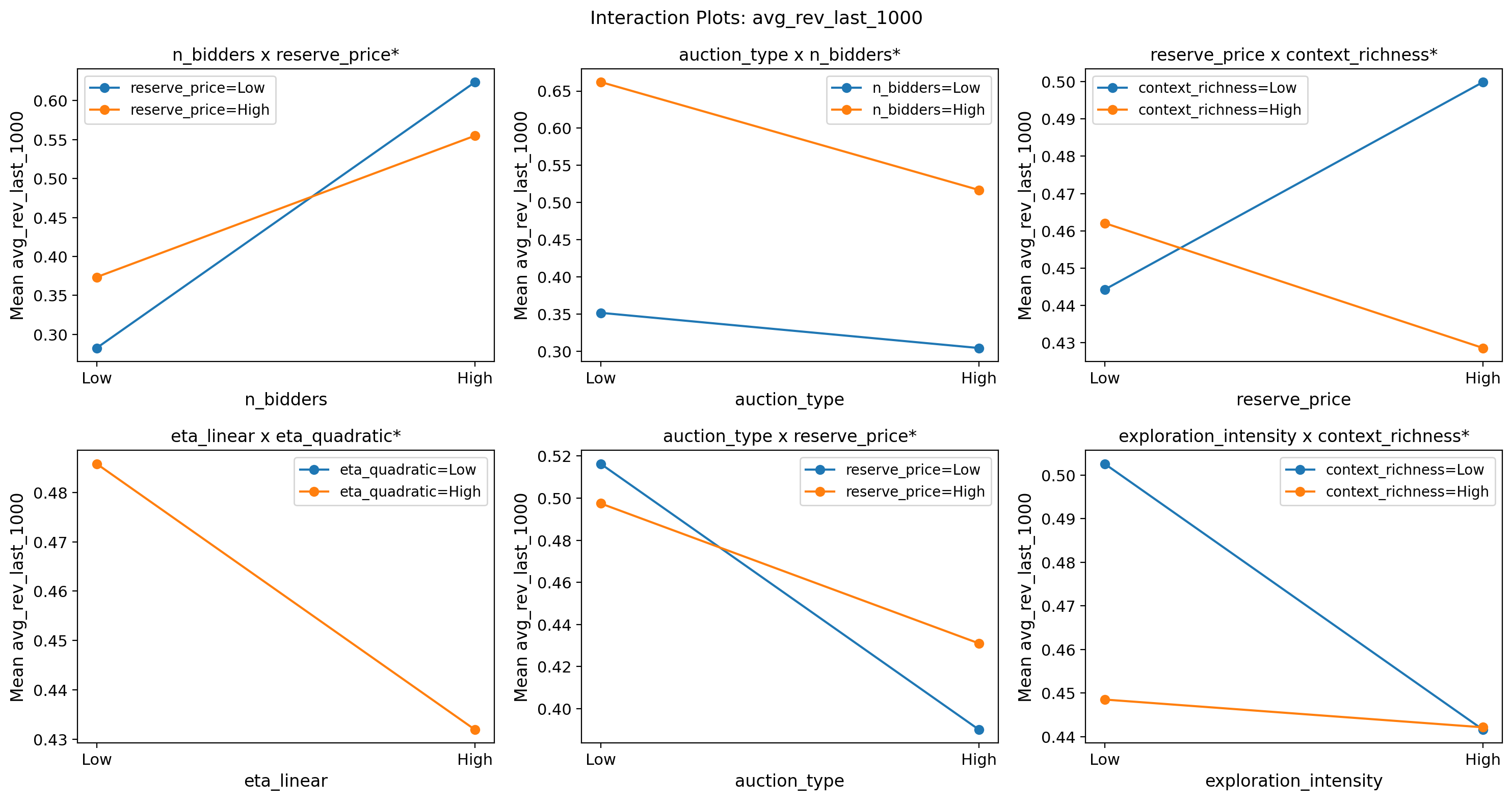}
  \caption{Experiment~2a: Interaction plot for average revenue under LinUCB. Non-parallel lines indicate factor interdependencies.}
  \label{fig:e2a_int_rev}
\end{figure}

Number of bidders dominates ($|t| = \ExpTwoARevFactorTOne$), with the shift from two to four bidders increasing revenue by \ExpTwoARevTopPctEffect\% of the grand mean. The next largest effects are \ExpTwoARevFactorTwo{} ($|t| = \ExpTwoARevFactorTTwo$) and \ExpTwoARevFactorThree{} ($|t| = \ExpTwoARevFactorTThree$). Table~\ref{tab:exp2a_ranked_rev} ranks all significant effects by absolute $t$-statistic. First-price auctions yield lower final revenue ($|t| = \ExpTwoARevAuctionT$), as the main effects plot (Figure~\ref{fig:e2a_main_rev}) confirms. The top interaction effects appear in Figure~\ref{fig:e2a_int_rev}.

\begin{table}[H]
\centering
\caption{Experiment 2a: Significant effects for average revenue ($p < 0.05$), ranked by $|t|$.}
\label{tab:exp2a_ranked_rev}
\begin{tabular}{lrrl}
\toprule
\textbf{Effect} & \textbf{Coeff.} & \textbf{$|t|$} & \textbf{Direction} \\
\midrule
Number of bidders & 0.1308 & 33.90 & + \\
Auction format & -0.0482 & 12.49 & $-$ \\
Number of bidders $\times$ Reserve price & -0.0400 & 10.36 & $-$ \\
Auction format $\times$ Number of bidders & -0.0245 & 6.35 & $-$ \\
Reserve price $\times$ Context richness & -0.0223 & 5.78 & $-$ \\
Affiliation (linear) & -0.0135 & 5.71 & $-$ \\
Affiliation (linear) $\times$ Affiliation (quadratic) & -0.0135 & 5.71 & $-$ \\
Exploration intensity & -0.0168 & 4.37 & $-$ \\
Auction format $\times$ Reserve price & 0.0150 & 3.88 & + \\
Exploration intensity $\times$ Context richness & 0.0137 & 3.54 & + \\
\bottomrule
\end{tabular}
\end{table}

\subsubsection{Price Volatility}

\begin{figure}[H]
  \centering
  \includegraphics[width=0.7\textwidth]{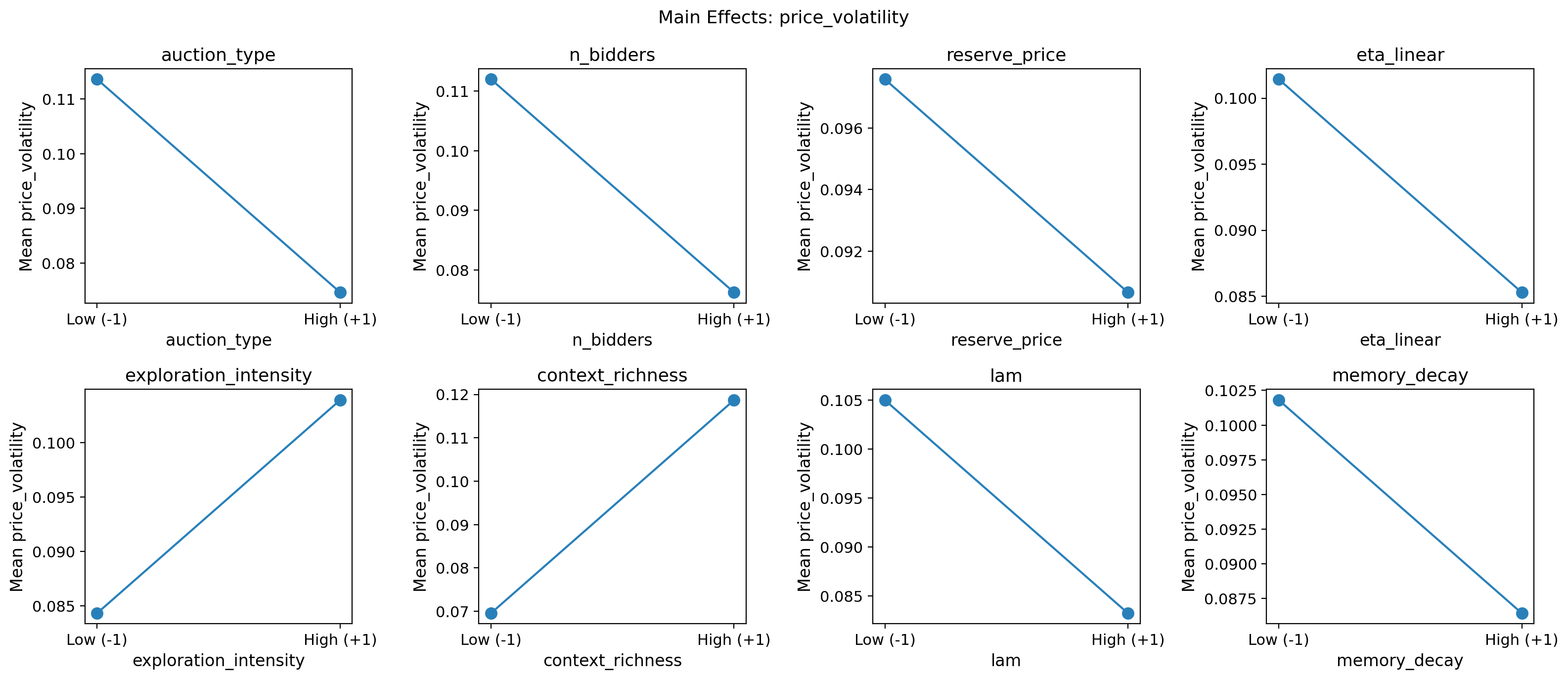}
  \caption{Experiment~2a: Main effects plot for price volatility under LinUCB.}
  \label{fig:e2a_main_vol}
\end{figure}

Table~\ref{tab:exp2a_ranked_vol} and the main effects plot (Figure~\ref{fig:e2a_main_vol}) show the factor rankings for price volatility under LinUCB. Model adequacy diagnostics confirm these findings (Appendix~\ref{sec:appendix_robustness}).

\begin{table}[H]
\centering
\caption{Experiment 2a: Significant effects for price volatility ($p < 0.05$), ranked by $|t|$.}
\label{tab:exp2a_ranked_vol}
\begin{tabular}{lrrl}
\toprule
\textbf{Effect} & \textbf{Coeff.} & \textbf{$|t|$} & \textbf{Direction} \\
\midrule
Context richness & 0.0246 & 16.91 & + \\
Auction format $\times$ Number of bidders & 0.0227 & 15.61 & + \\
Auction format & -0.0194 & 13.38 & $-$ \\
Number of bidders & -0.0179 & 12.31 & $-$ \\
Regularisation ($\lambda$) & -0.0109 & 7.48 & $-$ \\
Number of bidders $\times$ Reserve price & 0.0100 & 6.91 & + \\
Exploration intensity & 0.0098 & 6.75 & + \\
Memory decay ($\gamma_m$) & -0.0077 & 5.29 & $-$ \\
Auction format $\times$ Reserve price & -0.0070 & 4.81 & $-$ \\
Affiliation (linear) $\times$ Affiliation (quadratic) & -0.0040 & 4.53 & $-$ \\
\bottomrule
\end{tabular}
\end{table}

\subsection{Contextual Thompson Sampling (Experiment~2b)}

Experiment~2b deploys Contextual Thompson Sampling agents under the same affiliated valuation model. The design is a $3 \times 2^5 = 96$ mixed-level factorial with 6 factors (5 binary plus $\eta$). Preliminary screening confirmed that $\lambda$ and memory decay have no detectable effect on Thompson Sampling outcomes, so both are excluded to avoid allocating design cells to factors with no detectable effect. The design is replicated twice for \ExpTwoBNobs{} observations.

\begin{figure}[H]
  \centering
  \includegraphics[width=\textwidth]{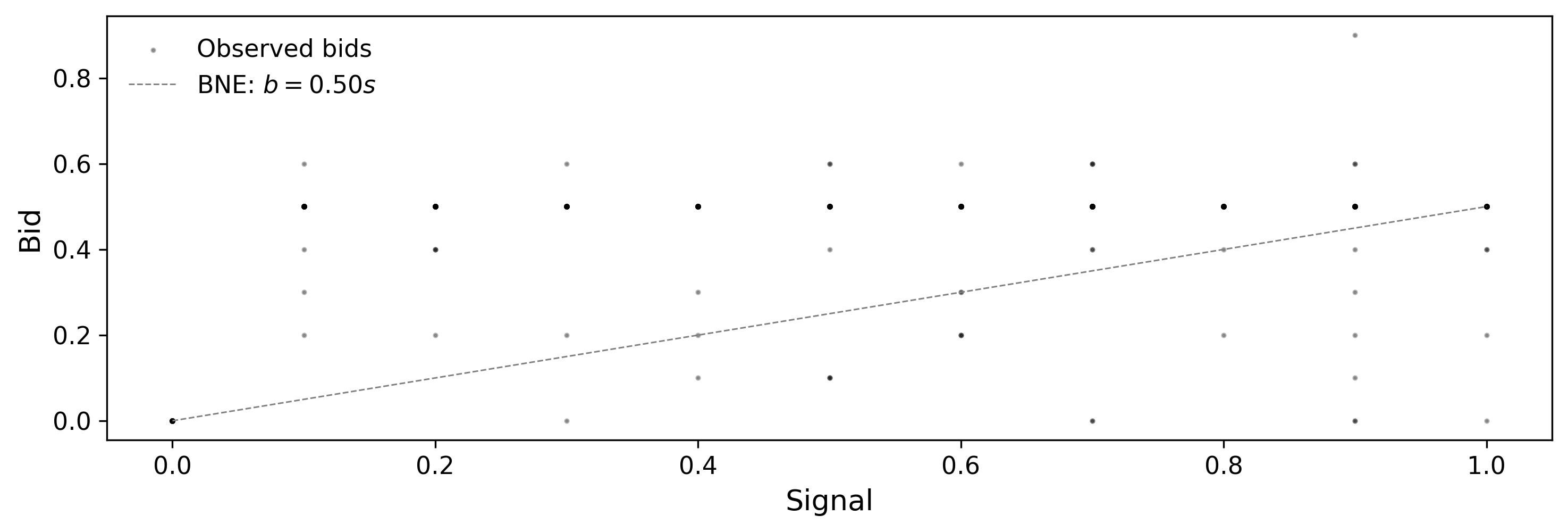}
  \caption{Representative bid function for a single trial of Experiment~2b (Thompson Sampling, first-price auction, $\eta = 0.5$, 2~bidders, 5{,}000 rounds).}
  \label{fig:e2b_trace}
\end{figure}

\subsubsection{Revenue}

Under Thompson Sampling, \ExpTwoBRevFactorOne{} dominates ($|t| = \ExpTwoBRevFactorTOne$), with the effect amounting to \ExpTwoBRevTopPctEffect\% of the grand mean, followed by \ExpTwoBRevFactorTwo{} ($|t| = \ExpTwoBRevFactorTTwo$). Table~\ref{tab:exp2b_ranked_rev} ranks all significant effects.

\begin{table}[H]
\centering
\caption{Experiment 2b: Significant effects for average revenue ($p < 0.05$), ranked by $|t|$.}
\label{tab:exp2b_ranked_rev}
\begin{tabular}{lrrl}
\toprule
\textbf{Effect} & \textbf{Coeff.} & \textbf{$|t|$} & \textbf{Direction} \\
\midrule
Number of bidders & 0.0525 & 8.69 & + \\
Reserve price & -0.0480 & 7.95 & $-$ \\
Number of bidders $\times$ Context richness & 0.0284 & 4.71 & + \\
Auction format $\times$ Number of bidders & -0.0245 & 4.07 & $-$ \\
Auction format & -0.0231 & 3.83 & $-$ \\
Exploration intensity & -0.0204 & 3.38 & $-$ \\
Affiliation (linear) $\times$ Affiliation (quadratic) & -0.0117 & 3.18 & $-$ \\
Affiliation (linear) & -0.0117 & 3.18 & $-$ \\
Context richness & -0.0175 & 2.89 & $-$ \\
Number of bidders $\times$ Reserve price & -0.0174 & 2.88 & $-$ \\
\bottomrule
\end{tabular}
\end{table}

\begin{figure}[H]
  \centering
  \includegraphics[width=0.7\textwidth]{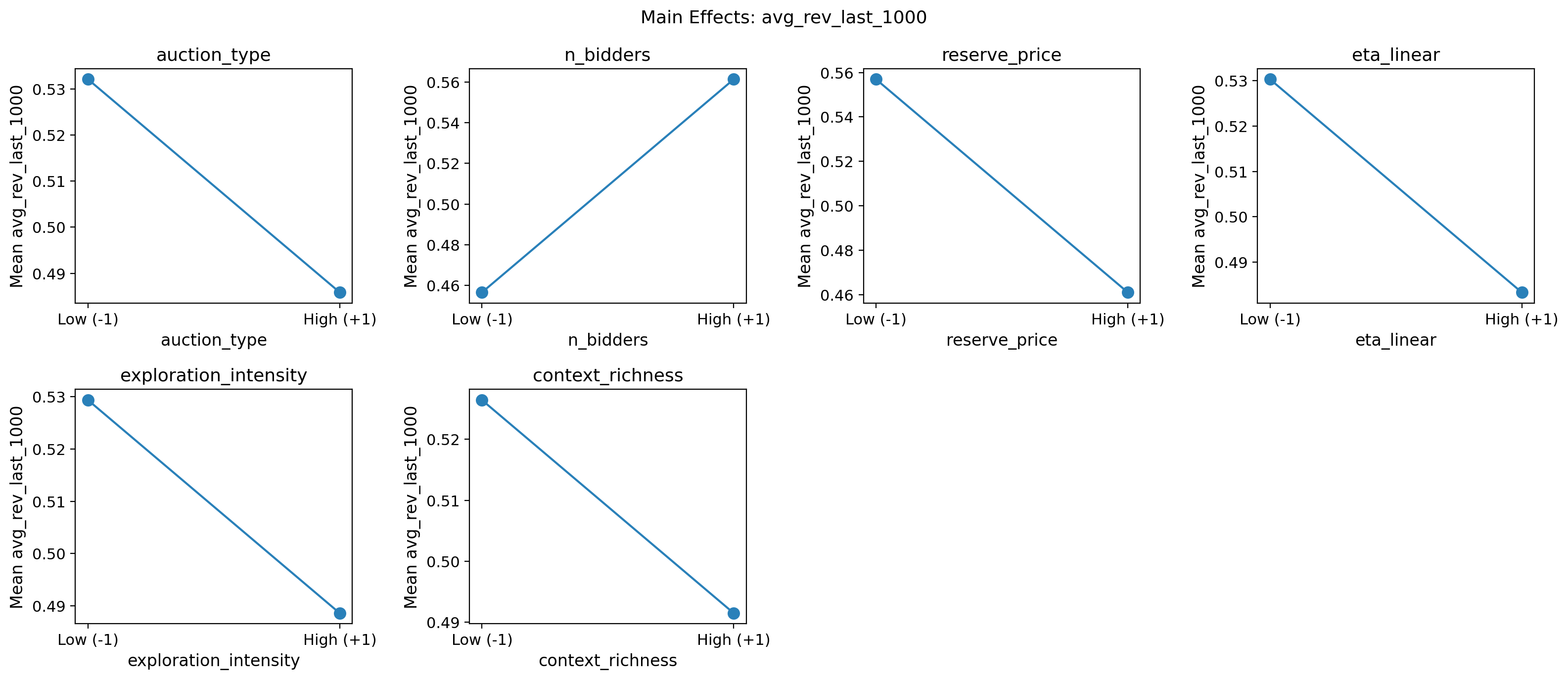}
  \caption{Experiment~2b: Main effects plot for average revenue under Thompson Sampling.}
  \label{fig:e2b_main_rev}
\end{figure}

\begin{figure}[H]
  \centering
  \includegraphics[width=0.7\textwidth]{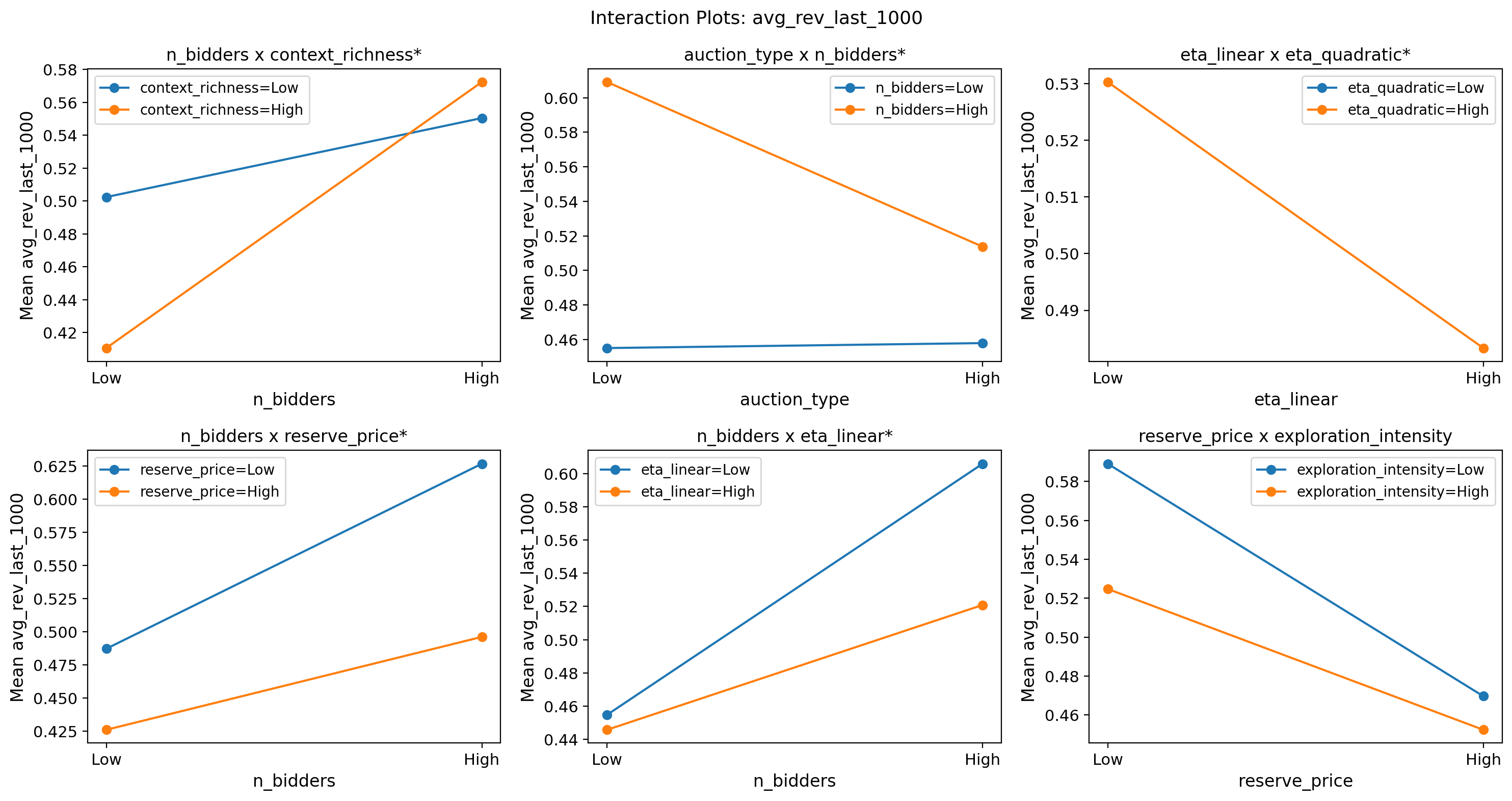}
  \caption{Experiment~2b: Interaction plot for average revenue under Thompson Sampling.}
  \label{fig:e2b_int_rev}
\end{figure}

\subsubsection{Price Volatility}

\begin{figure}[H]
  \centering
  \includegraphics[width=0.7\textwidth]{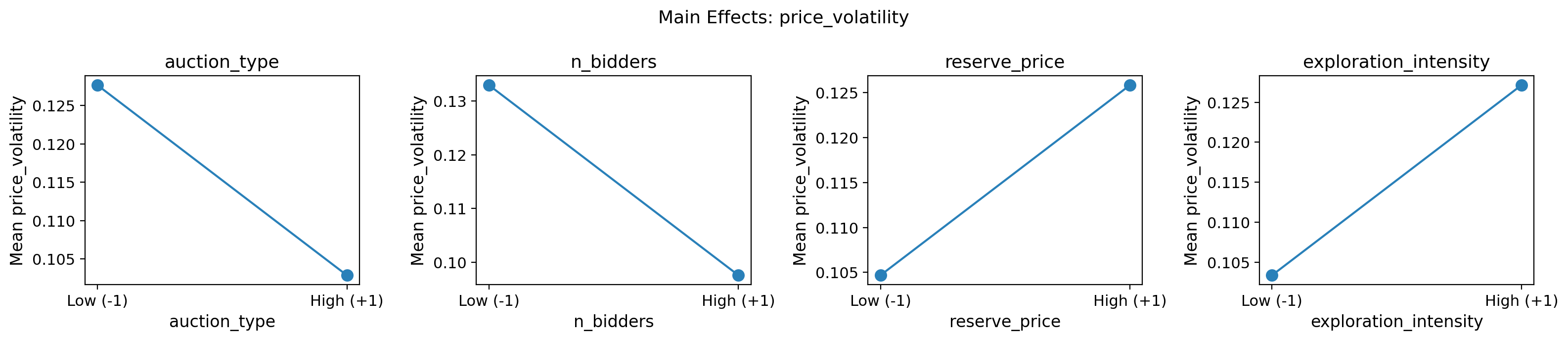}
  \caption{Experiment~2b: Main effects plot for price volatility under Thompson Sampling.}
  \label{fig:e2b_main_vol}
\end{figure}

Table~\ref{tab:exp2b_ranked_vol} and the main effects plot (Figure~\ref{fig:e2b_main_vol}) show the factor rankings for price volatility under Thompson Sampling. Model adequacy diagnostics confirm these findings (Appendix~\ref{sec:appendix_robustness}).

\begin{table}[H]
\centering
\caption{Experiment 2b: Significant effects for price volatility ($p < 0.05$), ranked by $|t|$.}
\label{tab:exp2b_ranked_vol}
\begin{tabular}{lrrl}
\toprule
\textbf{Effect} & \textbf{Coeff.} & \textbf{$|t|$} & \textbf{Direction} \\
\midrule
Auction format $\times$ Number of bidders & 0.0192 & 6.57 & + \\
Number of bidders $\times$ Reserve price & 0.0180 & 6.16 & + \\
Number of bidders & -0.0177 & 6.04 & $-$ \\
Number of bidders $\times$ Context richness & -0.0156 & 5.33 & $-$ \\
Auction format & -0.0124 & 4.24 & $-$ \\
Exploration intensity & 0.0119 & 4.06 & + \\
Reserve price & 0.0106 & 3.62 & + \\
Auction format $\times$ Context richness & -0.0077 & 2.63 & $-$ \\
Reserve price $\times$ Context richness & -0.0072 & 2.45 & $-$ \\
Affiliation (linear) $\times$ Context richness & 0.0081 & 2.27 & + \\
\bottomrule
\end{tabular}
\end{table}

\subsection{Sensitivity Analysis}

Global sensitivity analysis confirms the factorial rankings. Under LinUCB, the number of bidders dominates four of five responses, explaining approximately half of revenue variance ($S_T = 0.55$) and nearly all winner entropy variance ($S_T = 0.75$). Algorithmic parameters (regularisation, memory decay, exploration intensity) are consistently negligible ($S_T < 0.04$). Cross-method concordance is moderate ($\rho = 0.33$). Under Thompson Sampling, factor importance is more diffuse, with no single factor exceeding $S_T = 0.31$ for revenue. Convergence time is the most diffusely driven response, with five of six factors exceeding $S_T = 0.10$. Cross-method concordance is notably low ($\rho = 0.04$), reflecting disagreement when many factors have similar importance. Detailed per-factor Sobol' indices appear in Appendix~\ref{sec:appendix_sensitivity_tables} (Tables~\ref{tab:sens_exp2a}--\ref{tab:sens_exp2b}).

\subsection{Summary}

Replacing Q-learning with contextual bandits does not improve seller welfare. Richer algorithm design does not translate into higher revenue for the auctioneer. LinUCB and Thompson Sampling respond to different design levers, validating the decision to analyse them separately. Higher exploration intensity reduces revenue under both algorithms, but the magnitudes and moderating factors differ. Global sensitivity analysis confirms these hierarchies (Appendix~\ref{sec:appendix_sensitivity_tables}). Across both algorithms, tuning parameters are consistently negligible ($S_T < 0.04$ for revenue), reinforcing the conclusion that market structure dominates algorithm configuration. Robustness checks confirm these findings (Appendix~\ref{sec:appendix_robustness}). Experiments~1a--2b share a common feature, namely that agents bid directly from value estimates without external constraints. The next section tests whether the patterns documented above persist when agents face budget constraints, the dominant paradigm in modern advertising exchanges.

\section{Budget-Constrained Pacing}
\label{sec:pacing_results}
\label{sec:exp3_results}

This section studies budget-constrained pacing agents, in contrast to the unconstrained learning algorithms of earlier experiments. Experiment~3a uses multiplicative dual pacing \citep{Balseiro2019Learning}; Experiment~3b uses proportional-integral control.

\subsection{Design}
\label{sec:shared_env_pacing}
\label{sec:exp3_appendix}

Experiments~3a and~3b share a common auction environment. Each of $n \in \{2, 4\}$ advertisers participates in a sequence of $D = 100$ episodes, each comprising $T = 1{,}000$ single-item auctions. Between episodes, budgets regenerate to their initial level while control variables persist (warm-starting).

Valuations follow a log-normal model with bidder-specific asymmetry. Each bidder $i$ draws a mean $\mu_i \sim \mathrm{Uniform}(0.5, 1.5)$ once per seed. In each round $t$, the valuation $v_{it} \sim \mathrm{LogNormal}(\mu_i, \sigma)$ with $\sigma \in \{0.1, 0.5\}$. The budget per bidder per episode is $B_i = m \cdot \mathbb{E}[v_{it}] \cdot T$, where $\mathbb{E}[v_{it}] = \exp(\mu_i + \sigma^2/2)$ and $m \in \{0.25, 1.0\}$ is the budget multiplier. The auction may impose a reserve price $r \in \{0.0, 0.3\}$. Bids below $r$ are excluded, and in second-price auctions the payment is the maximum of the second-highest bid and $r$.

\label{sec:exp3a_design}
\label{sec:exp3a_appendix}

All bidders use a multiplicative dual pacing algorithm. In each round the agent computes a bid as a function of its current dual variable $\mu_t$:
\begin{align}
  b_{it} &= \min\!\Big(\frac{v_{it}}{\mu_t},\; B_i - S_{it}\Big) & &\text{(value-maximizer),} \label{eq:bid_vmax} \\
  b_{it} &= \min\!\Big(\frac{v_{it}}{1 + \mu_t},\; B_i - S_{it}\Big) & &\text{(utility-maximizer),} \label{eq:bid_umax}
\end{align}
where $S_{it}$ is the cumulative spend at round $t$. The dual update follows
\begin{equation}
  \mu_{t+1} = \mathrm{clip}\!\Big(\mu_t \cdot \exp\!\big(\alpha_p\,(p_t - \rho)\big),\; 10^{-4},\; 100\Big),
  \label{eq:dual_update_exp3a}
\end{equation}
with step size $\alpha_p = 1/\sqrt{T}$ and target spend rate $\rho = B_i / T$.

\subsubsection{Factorial Design}

The experiment uses a $2^6 = 64$ cell full factorial with six factors (Table~\ref{tab:exp3a_params}). Each cell is replicated across independent seeds, yielding \ExpThreeANobs{} total runs. This design estimates all main effects and two-way interactions without aliasing.

\begin{table}[H]
\centering
\caption{Parameter settings for Experiment~3a (Budget-Constrained Pacing).}
\label{tab:exp3a_params}
\begin{tabular}{l l l}
\toprule
\textbf{Factor} & \textbf{Low ($-1$)} & \textbf{High ($+1$)}\\
\midrule
Auction format & Second-price & First-price \\
Bidder objective & Value-maximizer & Utility-maximizer \\
Number of bidders ($n$) & 2 & 4 \\
Budget multiplier ($m$) & 0.25 (tight) & 1.0 (loose) \\
Reserve price ($r$) & 0.0 & 0.3 \\
Value dispersion ($\sigma$) & 0.1 (low) & 0.5 (high) \\
\midrule
\multicolumn{3}{l}{\textbf{Fixed parameters}} \\
\midrule
Episodes ($D$) & \multicolumn{2}{l}{100} \\
Rounds per episode ($T$) & \multicolumn{2}{l}{1{,}000} \\
Log-normal $\mu$ range & \multicolumn{2}{l}{$[0.5, 1.5]$} \\
Initial $\mu_0$ & \multicolumn{2}{l}{1.0} \\
Burn-in episodes & \multicolumn{2}{l}{10} \\
\bottomrule
\end{tabular}
\end{table}

\label{sec:exp3b_design}
\label{sec:exp3b_appendix}

Experiment~3b replaces the multiplicative dual pacing with a proportional-integral (PI) controller. The PI controller computes a spending error $e_t = (t/T)\,B^i - S_{it}$ at each round, where $S_{it}$ is cumulative spend, and updates the bid multiplier:
\begin{align}
  e_t^i &= \tfrac{t}{T}\,B^i - {\textstyle\sum_{\tau \le t}} c_\tau^i, \label{eq:pid_error_exp3b}\\
  \lambda_{t+1}^i &= \operatorname{clip}\!\bigl(\lambda_t^i + K_P\,e_t^i + K_I\!\textstyle\sum_{\tau} e_\tau^i,\; 0.01,\; 1.5\bigr), \label{eq:pid_update_exp3b}
\end{align}
with $K_P = 0.30 \times a$ and $K_I = 0.05 \times a$, where $a$ is the aggressiveness parameter. The agent bids $b_t = \min(\lambda_t \cdot v_t,\; B^i - S_t^i)$. The competitive benchmark is $\lambda = 1$ (full-value bidding) for both auction formats.

\subsubsection{Factorial Design}

The experiment uses a $2^6 = 64$ cell full factorial with six factors (Table~\ref{tab:exp3b_params}). The aggressiveness factor replaces the bidder objective factor from Experiment~3a, testing how controller responsiveness affects bidding outcomes. Each cell is replicated across 8 independent seeds for 512 total runs.

\begin{table}[H]
\centering
\caption{Parameter settings for Experiment~3b (PI Controller Pacing).}
\label{tab:exp3b_params}
\begin{tabular}{l l l}
\toprule
\textbf{Factor} & \textbf{Low ($-1$)} & \textbf{High ($+1$)}\\
\midrule
Auction format & Second-price & First-price \\
Aggressiveness $a$ & 0.3 (conservative) & 3.0 (aggressive) \\
Number of bidders ($n$) & 2 & 4 \\
Budget multiplier ($m$) & 0.25 (tight) & 1.0 (loose) \\
Reserve price ($r$) & 0.0 & 0.3 \\
Value dispersion ($\sigma$) & 0.1 (low) & 0.5 (high) \\
\midrule
\multicolumn{3}{l}{\textbf{Fixed parameters}} \\
\midrule
Episodes ($D$) & \multicolumn{2}{l}{100} \\
Rounds per episode ($T$) & \multicolumn{2}{l}{1{,}000} \\
Log-normal $\mu$ range & \multicolumn{2}{l}{$[0.5, 1.5]$} \\
Initial $\lambda_0$ & \multicolumn{2}{l}{1.0} \\
Burn-in episodes & \multicolumn{2}{l}{10} \\
$K_P$ & \multicolumn{2}{l}{$0.30 \times a$} \\
$K_I$ & \multicolumn{2}{l}{$0.05 \times a$} \\
\bottomrule
\end{tabular}
\end{table}

The aggressiveness factor scales the PI controller gains by a factor of $10\times$ (from $a = 0.3$ to $a = 3.0$). The proportional gain $K_P$ ranges from $0.09$ to $0.90$ and the integral gain $K_I$ from $0.015$ to $0.15$. Higher aggressiveness produces faster budget consumption and more reactive pacing, while lower aggressiveness yields conservative, slow-adjusting behaviour. The $10\times$ ratio ensures meaningful separation in controller dynamics without pushing either level into degeneracy (budget exhaustion in the first few rounds or near-zero spending).\footnote{In Experiment~3a, the bidder objective factor contrasts value-maximisation ($b = v/\mu$) with utility-maximisation ($b = v/(1+\mu)$), two structurally different bidding rules drawn from the price of anarchy literature. The asymmetry is definitional, not parametric.}

The following metrics are computed per run by averaging over the 90 post-burn-in episodes ($d \geq 10$):

\begin{table}[H]
\centering
\small
\caption{Experiment~3a response variables.}
\label{tab:exp3a_responses}
\begin{tabular}{ll}
\toprule
\textbf{Metric} & \textbf{Definition} \\
\midrule
Platform revenue       & Total payments per episode \\
Liquid welfare         & $\sum_i \min(B_i,\; \text{total value won by } i)$ per episode \\
Effective PoA          & LP offline optimum / liquid welfare \\
Budget utilisation     & Mean spend/budget across bidders \\
Bid-to-value ratio     & Mean $b/v$ across all bids \\
Allocative efficiency  & Fraction of rounds won by highest-value bidder \\
Dual variable CV       & CV of dual in last 200 rounds \\
No-sale rate           & Fraction of rounds with no valid bids \\
Winner entropy         & Shannon entropy of winner distribution \\
Warm-start benefit     & Revenue improvement episode~2 vs.~episode~1 \\
Inter-episode volatility & CV of revenue across post-burn-in episodes \\
Bid suppression ratio  & Observed btv / competitive btv \\
Cross-episode drift    & Slope of btv across episodes \\
\bottomrule
\end{tabular}
\end{table}

The effective PoA metric compares realised liquid welfare against offline optima computed with full hindsight. Section~\ref{sec:welfare} establishes the LP relaxation as an upper bound on achievable liquid welfare. Experiment~3b uses the same response variables (Table~\ref{tab:exp3a_responses}), with the dual variable CV computed from $\lambda$ history rather than $\mu$ history and the bid suppression ratio using a competitive benchmark of $\lambda = 1.0$ (full-value bidding). Analysis applies the same factorial ANOVA engine used in Experiments~1a through~2b to the run-level data, treating each seed as an independent replicate within its factorial cell. Figure~\ref{fig:pacing_comparison} contrasts the two pacing mechanisms.

\begin{figure}[H]
\centering
\begin{tikzpicture}[
  block/.style={rectangle, draw, rounded corners, minimum width=6.0cm,
                minimum height=0.9cm, text centered, font=\small},
  arrow/.style={-{Stealth[length=2mm]}, thick},
  title/.style={font=\small\bfseries, text centered},
]
  \def\colA{0}
  \def\colB{7.5}
  \def\rowA{0}
  \def\rowB{-1.4}
  \def\rowC{-2.8}
  \def\rowD{-4.2}

  \node[title] at (\colA, 1.0) {Multiplicative Dual (Exp~3a)};
  \node[title] at (\colB, 1.0) {PI Controller (Exp~3b)};

  \node[block] (d1) at (\colA, \rowA) {Observe $v_t$, remaining budget};
  \node[block] (d2) at (\colA, \rowB) {Bid $\min(v_t / \mu_t,\;\text{rem})$};
  \node[block] (d3) at (\colA, \rowC) {Auction $\to$ cost $c_t$};
  \node[block, text width=5.4cm] (d4) at (\colA, \rowD) {$\mu \leftarrow \mu \cdot \exp\!\big(\alpha_p(c_t - \rho)\big)$};

  \node[block] (p1) at (\colB, \rowA) {Observe $v_t$, remaining budget};
  \node[block] (p2) at (\colB, \rowB) {Bid $\min(\lambda_t \cdot v_t,\;\text{rem})$};
  \node[block] (p3) at (\colB, \rowC) {Auction $\to$ cost $c_t$};
  \node[block, text width=5.4cm] (p4) at (\colB, \rowD) {$e_t = \text{target} - \text{spent}$;\; $\lambda \leftarrow \lambda + K_P e + K_I \Sigma e$};

  \foreach \col in {d, p} {
    \draw[arrow] (\col 1) -- (\col 2);
    \draw[arrow] (\col 2) -- (\col 3);
    \draw[arrow] (\col 3) -- (\col 4);
  }

  \draw[arrow, rounded corners=8pt] (d4.west) -- ++(-1.2, 0) |- (d1.west);
  \draw[arrow, rounded corners=8pt] (p4.east) -- ++(1.2, 0) |- (p1.east);

  \node[font=\scriptsize, text=RedOrange!70!black, rotate=90, anchor=south] at (-1.2, -2.1) {multiplicative update};
  \node[font=\scriptsize, text=Violet!70!black, rotate=-90, anchor=south] at (8.7, -2.1) {additive error-based update};

  \begin{pgfonlayer}{background}
    \node[fill=RedOrange!8, rounded corners=6pt, fit=(d1)(d4),
          inner xsep=10pt, inner ysep=10pt] {};
    \node[fill=Violet!8, rounded corners=6pt, fit=(p1)(p4),
          inner xsep=10pt, inner ysep=10pt] {};
  \end{pgfonlayer}
\end{tikzpicture}
\caption{Per-round decision loops for the two pacing algorithms. Multiplicative dual pacing updates the dual variable $\mu$ via an exponential step proportional to the spending error. The PI controller updates the bid multiplier $\lambda$ via proportional and integral error terms. Both cap bids at the remaining budget.}
\label{fig:pacing_comparison}
\end{figure}

\subsection{Dual Pacing Results (Experiment~3a)}
\label{sec:exp3a_results}

\begin{figure}[H]
  \centering
  \includegraphics[width=\textwidth]{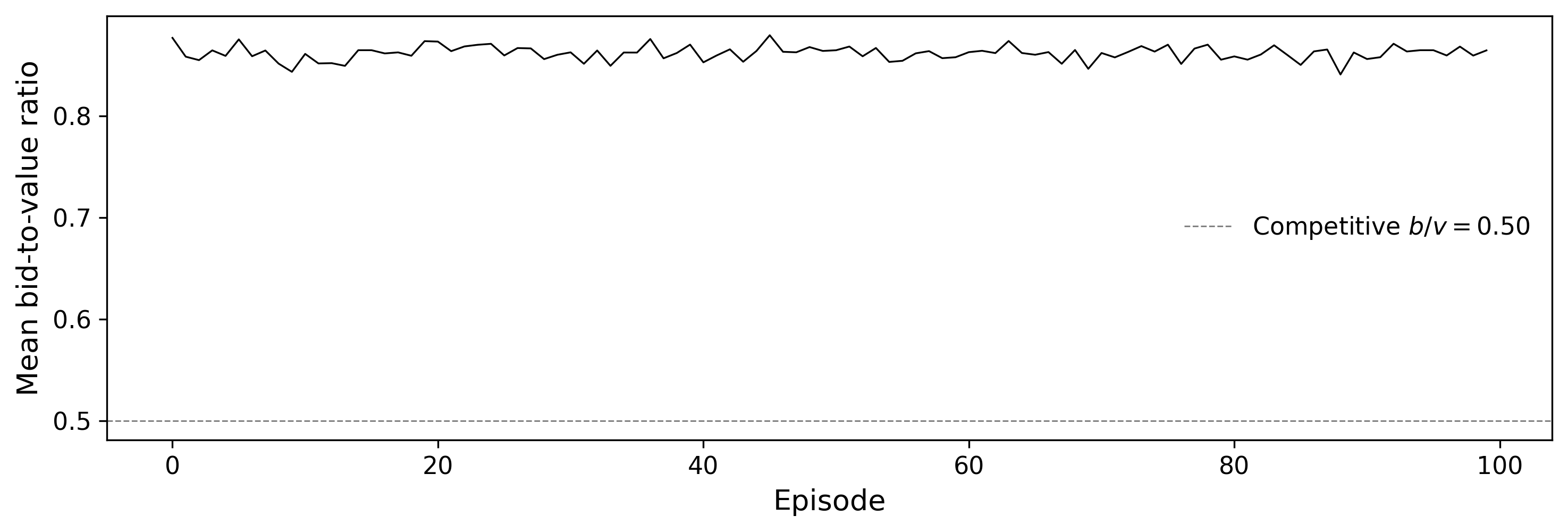}
  \caption{Representative learning trajectory for a single trial of Experiment~3a (first-price auction, value-maximizer objective, 2~bidders, 100 episodes $\times$ 1{,}000 rounds). Mean bid-to-value ratio per episode with the competitive benchmark (unconstrained Nash equilibrium $b/v = 0.5$).}
  \label{fig:e3a_trace}
\end{figure}

\subsubsection{Efficiency and Welfare}

The design and factor levels are given in Table~\ref{tab:exp3a_params} (\ExpThreeANobs{} runs total). Tables~\ref{tab:exp3a_ranked_rev} and~\ref{tab:exp3a_ranked_poa} report the ranked significant effects for platform revenue and effective Price of Anarchy, respectively. Response variable definitions appear in Table~\ref{tab:exp3a_responses}.

\begin{table}[H]
\centering
\caption{Experiment 3a: Significant effects for average revenue ($p < 0.05$), ranked by $|t|$.}
\label{tab:exp3a_ranked_rev}
\begin{tabular}{lrrl}
\toprule
\textbf{Effect} & \textbf{Coeff.} & \textbf{$|t|$} & \textbf{Direction} \\
\midrule
Budget multiplier & 2006.6019 & 39.44 & + \\
Bidder objective & -1428.0108 & 28.07 & $-$ \\
Bidder objective $\times$ Budget multiplier & -1387.4815 & 27.27 & $-$ \\
Number of bidders & 1158.2655 & 22.76 & + \\
Bidder objective $\times$ Number of bidders & -629.6958 & 12.38 & $-$ \\
Number of bidders $\times$ Budget multiplier & 448.9495 & 8.82 & + \\
Value dispersion ($\sigma$) & 356.7849 & 7.01 & + \\
Auction format & 184.0336 & 3.62 & + \\
Auction format $\times$ Bidder objective & 182.7252 & 3.59 & + \\
Auction format $\times$ Budget multiplier & 140.8645 & 2.77 & + \\
\bottomrule
\end{tabular}
\end{table}

\begin{table}[H]
\centering
\caption{Experiment 3a: Significant effects for effective Price of Anarchy ($p < 0.05$), ranked by $|t|$.}
\label{tab:exp3a_ranked_poa}
\begin{tabular}{lrrl}
\toprule
\textbf{Effect} & \textbf{Coeff.} & \textbf{$|t|$} & \textbf{Direction} \\
\midrule
Bidder objective $\times$ Budget multiplier & -0.0301 & 17.59 & $-$ \\
Bidder objective & -0.0205 & 12.01 & $-$ \\
Number of bidders & 0.0201 & 11.76 & + \\
Value dispersion ($\sigma$) & -0.0182 & 10.63 & $-$ \\
Budget multiplier & 0.0149 & 8.69 & + \\
Number of bidders $\times$ Value dispersion ($\sigma$) & -0.0142 & 8.28 & $-$ \\
Bidder objective $\times$ Number of bidders & -0.0042 & 2.48 & $-$ \\
Bidder objective $\times$ Value dispersion ($\sigma$) & 0.0039 & 2.29 & + \\
\bottomrule
\end{tabular}
\end{table}

All observed PoA values are well within the theoretical bound of~2 \citep{Aggarwal2019, Gaitonde2023}. The absence of a significant auction format effect on PoA contrasts with the revenue results. First-price bid suppression reduces payments to the auctioneer without proportionally distorting allocative outcomes. Allocative efficiency, the fraction of rounds in which the highest-value bidder wins, provides a complementary view. Under utility-maximizing objectives, agents achieve near-perfect efficiency. Under value-maximizing objectives with four bidders, efficiency drops substantially, with bid-to-value ratios exceeding 1.5.\footnote{Utility-maximizing efficiency ranges from \ExpThreeAUtilEffLow{} to \ExpThreeAUtilEffHigh{} across cells; value-maximizing four-bidder efficiency drops to \ExpThreeAValFourEffLow--\ExpThreeAValFourEffHigh, with bid-to-value ratios exceeding 1.5.} The objective$\times$n\_bidders interaction ($t = \ExpThreeAAllocEffObjxNbidT$, $p \ExpThreeAAllocEffObjxNbidPFmt$) captures this pattern, with the efficiency cost of value-maximizing objectives larger under four bidders than under two.

\subsubsection{Revenue and Budget Dynamics}

\begin{figure}[H]
  \centering
  \includegraphics[width=0.7\textwidth]{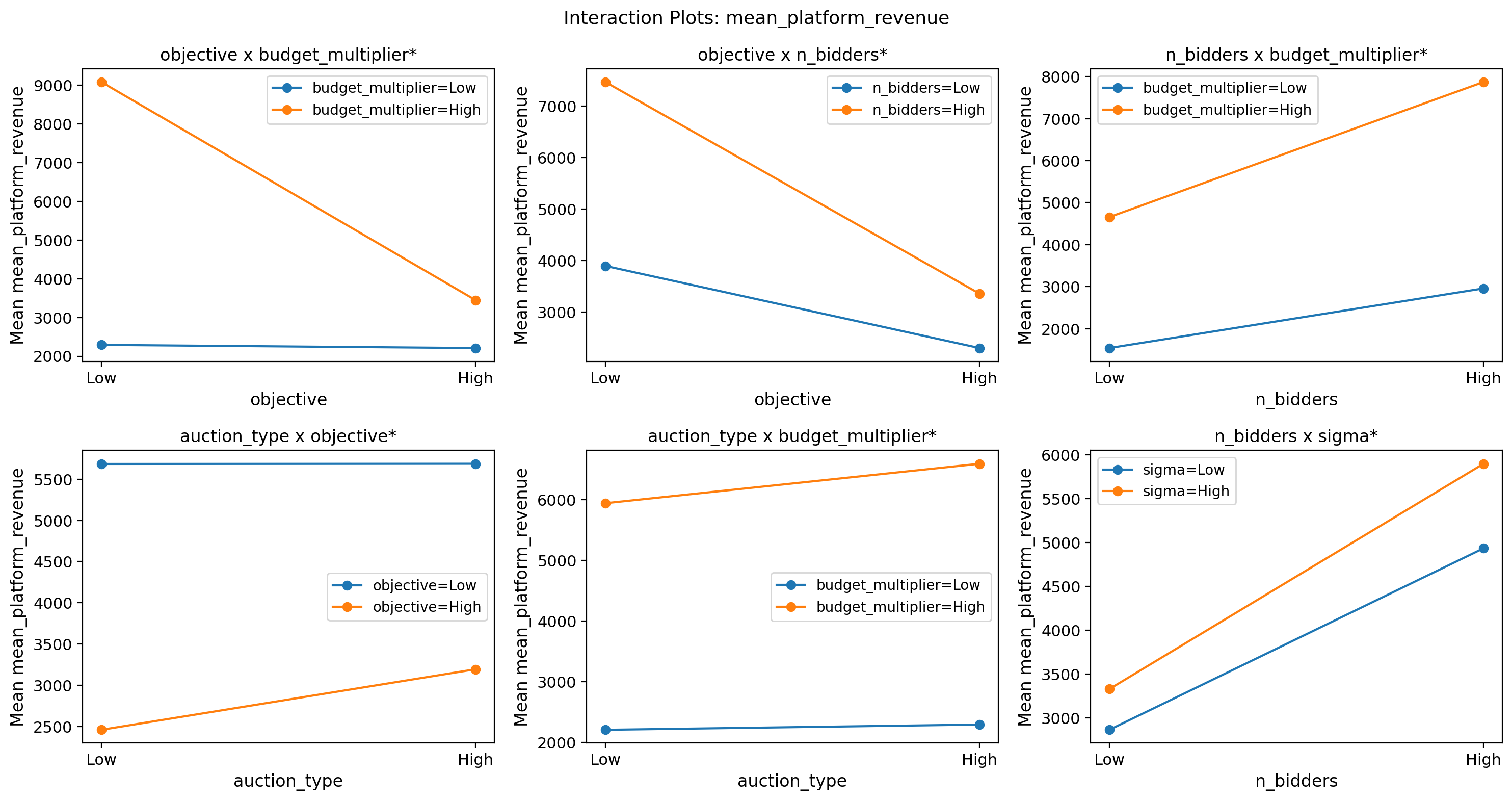}
  \caption{Experiment~3a: Interaction plot for platform revenue.}
  \label{fig:e3a_int_rev}
\end{figure}

The budget multiplier dominates platform revenue ($|t| = \ExpThreeARevFactorTOne$), shifting revenue by \ExpThreeARevTopPctEffect\% of the grand mean. Tight budgets produce lower bids regardless of other design choices, while loose budgets produce higher bids and greater competitive intensity. The number of bidders ranks fourth ($|t| = \ExpThreeARevNbidAbsT$) and auction format fifth ($|t| = \ExpThreeARevAuctionT$). Table~\ref{tab:exp3a_ranked_rev} reports the full hierarchy. The interaction plot (Figure~\ref{fig:e3a_int_rev}) reveals how factor combinations jointly shape revenue outcomes. Market thickness interacts with auction format, with the revenue gap between formats widening as the number of bidders increases.

\subsubsection{Bidding Behaviour and Bid Suppression Indicators}

\begin{table}[H]
\centering
\caption{Experiment 3a: Significant effects for price volatility ($p < 0.05$), ranked by $|t|$.}
\label{tab:exp3a_ranked_vol}
\begin{tabular}{lrrl}
\toprule
\textbf{Effect} & \textbf{Coeff.} & \textbf{$|t|$} & \textbf{Direction} \\
\midrule
Budget multiplier & 1.2878 & 3.26 & + \\
Bidder objective & -1.1490 & 2.91 & $-$ \\
Bidder objective $\times$ Budget multiplier & -1.1266 & 2.85 & $-$ \\
Auction format & -0.8413 & 2.13 & $-$ \\
Auction format $\times$ Bidder objective & 0.7996 & 2.02 & + \\
Auction format $\times$ Value dispersion ($\sigma$) & -0.7853 & 1.99 & $-$ \\
\bottomrule
\end{tabular}
\end{table}

The factorial model explains \ExpThreeAWelfareRsqPct\% of the variance in liquid welfare but only \ExpThreeADriftRsqPct\% in cross-episode drift, reflecting substantial seed-level noise in the latter metric. Cross-episode drift, measured as the slope of the bid-to-value ratio across post-burn-in episodes, assesses the \citet{PaesLeme2024} prediction that bidders progressively suppress bids across campaign cycles. Although the factorial ANOVA for drift is statistically significant ($F$-test $p \ExpThreeADriftFPFmt$), the model explains only \ExpThreeADriftRsqPct\% of the variance ($R^2 = \ExpThreeADriftRsq$), and all cell-level drift estimates remain near zero.\footnote{The large sample size (\ExpThreeANobs{} runs) allows detection of statistically significant but economically negligible effects.} This near-zero drift does not support the prediction of \citet{PaesLeme2024} that dual-based pacing produces progressive bid suppression, suggesting that the theoretical result may not generalise to noisy, discrete-time pacing environments with finite horizons.

\subsubsection{Learning Dynamics}

Warm-start benefits are dominated by bidder objective ($t = \ExpThreeAWarmObjT$). Under utility-maximizing objectives, warm-start produces measurable gains, while value-maximizing agents show negligible warm-start gains. Inter-episode volatility is similarly dominated by objective ($t = \ExpThreeAIEVolObjT$), with utility-maximizing agents exhibiting higher volatility. Model adequacy diagnostics (Appendix~\ref{sec:appendix_robustness}) confirm clean inference.\footnote{$R^2$ ranges from \ExpThreeARsqMin{} to \ExpThreeARsqMax, PRESS gaps are below \ExpThreeAPRESSGapMax, and \ExpThreeABHPct\% of significant effects survive Benjamini--Hochberg correction.}

\subsubsection{Summary}

The budget multiplier dominates revenue (\ExpThreeARevFactorOne{}, $|t| = \ExpThreeARevFactorTOne$), with bidder objective ranking second and their interaction third. The number of bidders ranks fourth ($|t| = \ExpThreeARevNbidAbsT$), and auction format is a comparatively minor factor ($|t| = \ExpThreeARevAuctionT$). Global sensitivity analysis confirms budget multiplier dominance across all response variables (Section~\ref{sec:sens_exp3a}).

\subsection{PI Controller Results (Experiment~3b)}
\label{sec:exp3b_results}

Experiment~3b uses the same $2^6$ factorial structure as Experiment~3a (Table~\ref{tab:exp3b_params}), replacing the bidder objective factor with aggressiveness. The 512 runs are analysed with the same factorial ANOVA engine.

\begin{figure}[H]
  \centering
  \includegraphics[width=\textwidth]{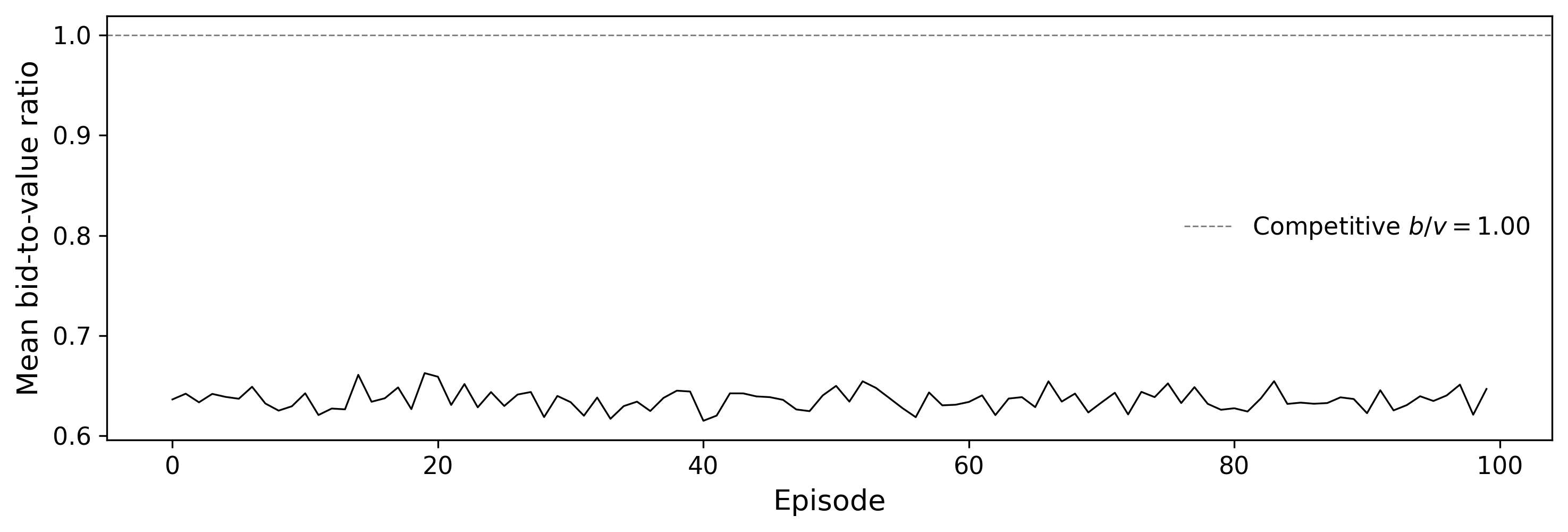}
  \caption{Representative learning trajectory for a single trial of Experiment~3b (first-price auction, default aggressiveness, 2~bidders, 100 episodes $\times$ 1{,}000 rounds). Mean bid-to-value ratio per episode with the competitive benchmark ($b/v = 1.0$).}
  \label{fig:e3b_trace}
\end{figure}

\subsubsection{Efficiency and Welfare}

The PI controller produces welfare outcomes that can be compared directly with the dual pacing results of Experiment~3a. Platform revenue and liquid welfare are computed identically, and the effective Price of Anarchy uses the same LP-based offline optimum as denominator. As in Experiment~3a, the budget multiplier dominates revenue (Table~\ref{tab:exp3b_ranked_rev}), with the effect amounting to \ExpThreeBRevTopPctEffect\% of the grand mean.

\begin{table}[H]
\centering
\caption{Experiment 3b: Significant effects for average revenue ($p < 0.05$), ranked by $|t|$.}
\label{tab:exp3b_ranked_rev}
\begin{tabular}{lrrl}
\toprule
\textbf{Effect} & \textbf{Coeff.} & \textbf{$|t|$} & \textbf{Direction} \\
\midrule
Budget multiplier & 1316.1821 & 31.68 & + \\
Number of bidders & 667.7779 & 16.07 & + \\
Auction format & 458.3301 & 11.03 & + \\
Auction format $\times$ Budget multiplier & 393.6340 & 9.47 & + \\
Value dispersion ($\sigma$) & 360.6121 & 8.68 & + \\
Auction format $\times$ Value dispersion ($\sigma$) & 233.6535 & 5.62 & + \\
Budget multiplier $\times$ Value dispersion ($\sigma$) & 175.0143 & 4.21 & + \\
Number of bidders $\times$ Value dispersion ($\sigma$) & 146.9008 & 3.54 & + \\
\bottomrule
\end{tabular}
\end{table}

\begin{table}[H]
\centering
\caption{Experiment 3b: Significant effects for effective Price of Anarchy ($p < 0.05$), ranked by $|t|$.}
\label{tab:exp3b_ranked_poa}
\begin{tabular}{lrrl}
\toprule
\textbf{Effect} & \textbf{Coeff.} & \textbf{$|t|$} & \textbf{Direction} \\
\midrule
Auction format & 0.0141 & 10.69 & + \\
Number of bidders $\times$ Budget multiplier & -0.0130 & 9.87 & $-$ \\
Auction format $\times$ Budget multiplier & 0.0073 & 5.56 & + \\
Value dispersion ($\sigma$) & -0.0064 & 4.84 & $-$ \\
Number of bidders & 0.0039 & 2.93 & + \\
Auction format $\times$ Reserve price & 0.0036 & 2.73 & + \\
Number of bidders $\times$ Reserve price & 0.0031 & 2.35 & + \\
Number of bidders $\times$ Value dispersion ($\sigma$) & -0.0029 & 2.21 & $-$ \\
Budget multiplier $\times$ Reserve price & -0.0029 & 2.18 & $-$ \\
Reserve price $\times$ Value dispersion ($\sigma$) & 0.0029 & 2.16 & + \\
\bottomrule
\end{tabular}
\end{table}

For the effective Price of Anarchy, auction format is the dominant factor ($|t| = 10.7$), with first-price auctions producing higher PoA values. The number of bidders $\times$ budget multiplier interaction ranks second ($|t| = 9.9$). Thin markets with tight budgets produce the highest effective PoA values, indicating the largest gap between achieved and optimal liquid welfare. Value dispersion reduces PoA ($|t| = 4.8$). As in Experiment~3a, all observed PoA values remain within the theoretical bound. Allocative efficiency follows a similar pattern. The budget multiplier dominates ($|t| = 44.7$), with loose budgets yielding near-efficient allocations.

\subsubsection{Revenue and Budget Dynamics}

\begin{figure}[H]
  \centering
  \includegraphics[width=0.7\textwidth]{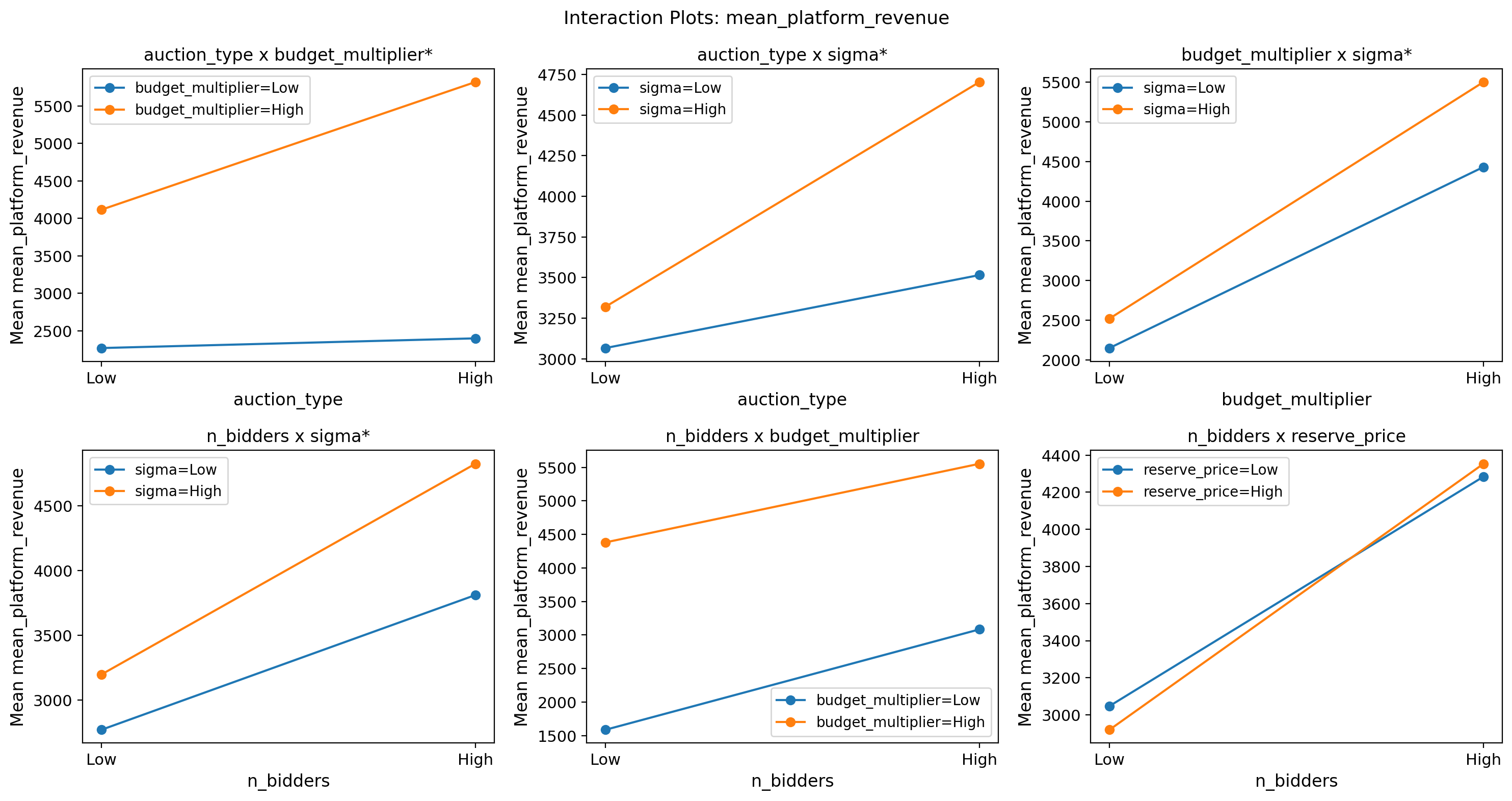}
  \caption{Experiment~3b: Interaction plot for platform revenue.}
  \label{fig:e3b_int_rev}
\end{figure}

The auction type $\times$ budget multiplier interaction is the strongest two-way effect on revenue ($|t| = 9.5$), indicating that the revenue advantage of first-price auctions is concentrated in loose-budget settings where agents can bid aggressively. Under tight budgets, both formats produce similar revenue as pacing constraints dominate. The auction type $\times$ value dispersion interaction ($|t| = 5.6$) shows that the first-price revenue advantage grows with value dispersion. The budget multiplier $\times$ value dispersion ($|t| = 4.2$) and number of bidders $\times$ value dispersion ($|t| = 3.5$) interactions round out the significant two-way effects. Figure~\ref{fig:e3b_int_rev} displays these interaction patterns.

\subsubsection{Bidding Behaviour}

The bid-to-value ratio is well-explained by the factorial model ($R^2 = 0.89$). The budget multiplier is the dominant factor ($|t| = 54.4$). Loose budgets permit higher bids relative to value, while tight budgets force aggressive shading. Auction format ranks second ($|t| = 22.3$), with first-price auctions producing substantially lower bid-to-value ratios than second-price, consistent with the theoretical prediction that first-price bidders shade below value. The auction type $\times$ budget multiplier interaction ($|t| = 17.3$) indicates that bid shading under first-price rules is amplified when budgets are tight. The auction type $\times$ number of bidders interaction ($|t| = 5.8$) reflects that additional bidders narrow the bid-to-value gap between formats. Compared with the multiplicative pacing of Experiment~3a, the PI controller produces qualitatively similar bid-to-value patterns but with less extreme shading under tight budgets. Table~\ref{tab:exp3b_ranked_vol} reports the full effect hierarchy.

\begin{table}[H]
\centering
\caption{Experiment 3b: Significant effects for price volatility ($p < 0.05$), ranked by $|t|$.}
\label{tab:exp3b_ranked_vol}
\begin{tabular}{lrrl}
\toprule
\textbf{Effect} & \textbf{Coeff.} & \textbf{$|t|$} & \textbf{Direction} \\
\midrule
Budget multiplier & 0.3873 & 54.37 & + \\
Auction format & -0.1585 & 22.25 & $-$ \\
Auction format $\times$ Budget multiplier & 0.1233 & 17.31 & + \\
Auction format $\times$ Number of bidders & 0.0410 & 5.76 & + \\
\bottomrule
\end{tabular}
\end{table}

\subsection{Sensitivity Analysis}
\label{sec:sens_pacing}

\subsubsection{Experiment~3a: Dual Pacing Autobidders}
\label{sec:sens_exp3a}

Budget multiplier is the dominant factor for seven of fourteen responses, including revenue ($S_T = 0.54$), liquid welfare ($S_T = 0.52$), and pacing stability ($S_T = 0.43$). The bidder objective dominates budget utilisation ($S_T = 0.66$), allocative efficiency ($S_T = 0.48$), and effective price of anarchy ($S_T = 0.35$). Auction format and reserve price are negligible across primary responses ($S_T < 0.04$). Cross-method concordance is strong ($\rho = 0.72$). Detailed Sobol' indices appear in Appendix~\ref{sec:appendix_sensitivity_tables} (Tables~\ref{tab:sens_exp3a_primary}--\ref{tab:sens_exp3a_secondary}).

\subsubsection{Experiment~3b: PI Controller Pacing}
\label{sec:sens_exp3b}

Budget multiplier again dominates the majority of responses, including revenue ($S_T = 0.53$) and budget utilisation ($S_T = 0.70$). A notable difference from Experiment~3a emerges for the price of anarchy and inter-episode volatility, where auction format becomes the most important factor ($S_T = 0.19$ and $0.33$ respectively), whereas it was negligible under dual pacing. Aggressiveness is negligible across all responses ($S_T < 0.04$). Cross-method concordance is strong ($\rho = 0.77$). Detailed Sobol' indices appear in Appendix~\ref{sec:appendix_sensitivity_tables} (Tables~\ref{tab:sens_exp3b_primary}--\ref{tab:sens_exp3b_secondary}).

\subsubsection{Cross-Method Concordance}
\label{sec:sens_concordance}

Table~\ref{tab:sens_concordance} summarises the agreement among the seven sensitivity methods across all experiments. The mean pairwise Spearman rank correlation provides a scalar measure of whether different methods agree on which factors matter most. Concordance is highest for Experiment~1a ($\rho = 0.87$) and Experiment~3b ($\rho = 0.77$), both of which have clear dominant factors and large sample sizes. Concordance is lowest for Experiment~2b ($\rho = 0.04$), where many factors have similar total-order indices and the methods disagree on their relative ranking. In all experiments, the identity of the top one or two factors is consistent across methods even when the full ranking shows disagreement; the low concordance values primarily reflect instability in the ordering of minor factors.

\begin{table}[H]
\centering
\caption{Cross-method concordance: mean pairwise Spearman $\rho$ across seven sensitivity methods, averaged over all response variables within each experiment.}
\label{tab:sens_concordance}
\small
\begin{tabular}{lcc}
\toprule
Experiment & Mean Spearman $\rho$ & Number of responses \\
\midrule
1a: Q-Learning (constant)       & 0.87 & 5 \\
1b: Q-Learning (affiliated)     & 0.33 & 8 \\
2a: LinUCB                     & 0.33 & 5 \\
2b: Thompson Sampling          & 0.04 & 5 \\
3a: Dual Pacing                & 0.72 & 16 \\
3b: PI Pacing                  & 0.77 & 16 \\
\bottomrule
\end{tabular}
\end{table}

\subsection{Summary}

The comparison between multiplicative dual pacing and PI control isolates the role of the control mechanism from the budget constraint itself. In both experiments, the budget multiplier is the dominant revenue driver ($|t| = \ExpThreeARevFactorTOne$ in Experiment~3a; $|t| = \ExpThreeBRevFactorTOne$ in Experiment~3b), confirming that budget tightness, not the pacing algorithm, is the primary determinant of revenue outcomes. The number of bidders ranks among the top factors in both cases ($|t| = \ExpThreeARevNbidAbsT$ in 3a; $|t| = 16.1$ in 3b), while auction format plays a stronger role under PI control ($|t| = 11.0$) than under dual pacing ($|t| = \ExpThreeARevAuctionT$). PoA values remain within the theoretical bound in both experiments (Section~\ref{sec:exp3a_results}). The two pacing mechanisms produce different bid-to-value ratio levels, but the factor hierarchy is consistent across both approaches, indicating that the structural features of the budget-constrained environment dominate the choice of control algorithm. This consistency suggests that the factor hierarchy is more stable within the pacing algorithm family than across unconstrained learning algorithms, where switching from Q-learning to contextual bandits changed the direction of the auction-format effect. Global sensitivity analysis independently confirms these rankings (Sections~\ref{sec:sens_exp3a}--\ref{sec:sens_exp3b}). Cross-method concordance is strong in both cases ($\rho = 0.72$ in 3a, $\rho = 0.77$ in 3b), indicating robust agreement across all seven sensitivity methods (Section~\ref{sec:sens_concordance}).

\label{sec:sens_synthesis}

Table~\ref{tab:sens_dominant} summarises the dominant factor for each response variable that is comparable across multiple experiments. The entries report the factor with the highest total-order index and its $S_T$ value.

\begin{table}[H]
\centering
\caption{Dominant factor (highest $S_T$) for comparable responses across experiments. Factor abbreviations: N = number of bidders, R = reserve price, A = auction format, B = budget multiplier, O = bidder objective.}
\label{tab:sens_dominant}
\small
\setlength{\tabcolsep}{4pt}
\begin{tabular}{l cccccc}
\toprule
Response & Exp~1a & Exp~1b & Exp~2a & Exp~2b & Exp~3a & Exp~3b \\
\midrule
Revenue        & N (0.24) & N (0.22) & N (0.55) & N (0.31) & B (0.54) & B (0.53) \\
Conv.\ time    & N (0.11) & A (0.11) & N (0.21) & N (0.16) & -- & -- \\
No-sale rate   & R (0.27) & -- & R (0.45) & N (0.41) & B (0.40) & B (0.33) \\
Volatility     & N (0.17) & N (0.17) & A (0.24) & N (0.37) & -- & -- \\
Winner entropy & N (0.31) & N (0.16) & N (0.75) & N (0.32) & N (0.63) & N (0.58) \\
\midrule
Eff.\ PoA      & -- & -- & -- & -- & O (0.35) & A (0.19) \\
Liq.\ welfare  & -- & -- & -- & -- & B (0.52) & B (0.50) \\
Budget util.   & -- & -- & -- & -- & O (0.66) & B (0.70) \\
Alloc.\ eff.   & -- & -- & -- & -- & O (0.48) & B (0.74) \\
\bottomrule
\end{tabular}
\end{table}

Market thickness is the primary determinant of algorithmic auction outcomes across all six experiments. In the four learning experiments (1a, 1b, 2a, 2b), the number of bidders is the dominant factor for revenue in every case, achieving total-order indices from 0.22 to 0.55 (Table~\ref{tab:sens_dominant}). It also dominates winner entropy in all six experiments ($S_T = 0.16$--$0.75$), reflecting the mechanical relationship between market size and competitive balance. In the two autobidding experiments (3a, 3b), budget multiplier replaces number of bidders as the dominant revenue factor ($S_T = 0.53$--$0.54$) and drives most welfare and efficiency metrics. This transition is consistent with the different constraints faced by the two algorithm families: in learning experiments the number of competitors determines equilibrium revenue, while in autobidding the budget constraint is the binding force that shapes market outcomes.

Reserve price and auction format occupy distinct niches in the factor importance hierarchy. Reserve price is the dominant factor for no-sale rate in Experiments~1a and~2a ($S_T = 0.27$ and $0.45$), confirming its direct role as a bid-acceptance threshold. In Experiment~2b, reserve price ranks second to the number of bidders for no-sale rate ($S_T = 0.41$ vs $0.41$), essentially tied. Auction format is rarely the top-ranked factor; across 51 well-fitted response variables, it is dominant in only five cases.\footnote{Well-fitted responses are those with $R^2 > 0.15$, excluding three Experiment~3a responses and two Experiment~3b responses where the model explains negligible variance.} Its strongest contributions are to winner's curse frequency in Experiment~1b ($S_T = 0.36$), inter-episode volatility in Experiment~3b ($S_T = 0.33$), price volatility in Experiment~2a ($S_T = 0.24$), effective price of anarchy in Experiment~3b ($S_T = 0.19$), and convergence time in Experiment~1b ($S_T = 0.11$). The choice between first-price and second-price formats thus matters most for stability and strategic behaviour rather than for aggregate revenue.

Algorithmic tuning parameters are consistently negligible across all four algorithm families tested. In Experiment~1a, learning rate ($\alpha$), decay type, and initialisation each have $S_T < 0.03$ across all five responses. In Experiments~2a and 3b, regularisation ($\lambda$), memory decay, and exploration intensity fall below $S_T = 0.04$ for both revenue and no-sale rate. In Experiment~3b, the aggressiveness parameter has $S_T < 0.04$ for every response. This pattern suggests that market outcomes are robust to moderate variation in algorithmic hyperparameters, and that the market environment matters far more than the algorithm's internal configuration.

Interaction effects are pervasive across all experiments. The gap between total-order and first-order indices ($S_T - S_1$) frequently exceeds 0.10, indicating that a factor's influence on the response depends substantially on the levels of other factors. This pattern is most pronounced for the no-sale rate in Experiment~2b, where three factors each have $S_T > 0.33$ despite $S_1 < 0.20$. Even dominant factors operate partly through interactions: the number of bidders in Experiment~2b has $S_1 = 0.18$ but $S_T = 0.31$ for revenue, and budget multiplier in Experiment~3a has $S_1 = 0.35$ but $S_T = 0.54$. The residual fraction (variance not explained by any first- or second-order term) ranges from 0.04 (Experiment~3a budget utilisation) to 0.97 (Experiment~3b warm-start benefit), reflecting both replication noise and higher-order interactions beyond the two-way terms in the model. These interaction effects mean that isolated factor-by-factor analysis can substantially understate a factor's true importance, reinforcing the value of total-order indices over first-order indices alone.

\section{Discussion}
\label{sec:discussion}

\subsection{Primary Findings}

The experiments reported in the preceding sections treat each simulation as a black-box observation: we varied inputs, measured outcomes, and ranked factors by statistical importance. The patterns below summarise these input-output associations and compare them to prior theoretical predictions; they do not claim to explain the mechanisms by which algorithms produce observed outcomes.

The most robust empirical finding across all experiments is that the degree of competition matters far more than algorithmic design choices. In unconstrained settings (Experiments~1a--2b), the number of bidders yields the largest effect on revenue in every case. Table~\ref{tab:cross_exp_summary} ranks the top factors by absolute $t$-statistic for revenue across all six sub-experiments. The dominance of market thickness confirms the prediction of \citet{BulowKlemperer1996} that an additional bidder matters more than an optimal reserve, and extends it from static equilibria to settings where bidders are learning algorithms. Under budget constraints (Experiments~3a--3b), the budget multiplier takes over as the primary revenue driver. The structural shift from market thickness to budget tightness is the defining feature of budget-constrained pacing environments.

\begin{table}[H]
\centering
\caption{Cross-experiment summary of revenue drivers, format effects, and convergence. Factor rankings by absolute $t$-statistic for revenue; FPA and SPA columns show mean revenue in final evaluation period; Gap is $(\text{FPA} - \text{SPA})/\text{SPA}$; convergence in rounds (episodes for Experiments~3a--3b).}
\label{tab:cross_exp_summary}
\resizebox{\textwidth}{!}{%
\begin{tabular}{l l lll ccc c}
\toprule
& & \multicolumn{3}{c}{\textbf{Top Revenue Drivers ($|t|$)}}
& \multicolumn{3}{c}{\textbf{Revenue by Format}}
& \textbf{Conv.} \\
\cmidrule(lr){3-5} \cmidrule(lr){6-8} \cmidrule(lr){9-9}
\textbf{Exp} & \textbf{Algorithm}
& \textbf{Rank 1} & \textbf{Rank 2} & \textbf{Rank 3}
& \textbf{FPA} & \textbf{SPA} & \textbf{Gap (\%)}
& \textbf{Median} \\
\midrule
1a & Q-learn (const.) & \ExpOneARevFactorOne{} (\ExpOneARevFactorTOne) & \ExpOneARevFactorTwo{} (\ExpOneARevFactorTTwo) & \ExpOneARevFactorThree{} (\ExpOneARevFactorTThree) & \ExpOneARevMeanFPA & \ExpOneARevMeanSPA & $\ExpOneARevGapPct$ & \ExpOneAConvMedian \\
1b & Q-learn (affil.) & \ExpOneBRevFactorOne{} (\ExpOneBRevFactorTOne) & \ExpOneBRevFactorTwo{} (\ExpOneBRevFactorTTwo) & \ExpOneBRevFactorThree{} (\ExpOneBRevFactorTThree) & \ExpOneBRevMeanFPA & \ExpOneBRevMeanSPA & $\ExpOneBRevGapPct$ & \ExpOneBConvMedian \\
2a & LinUCB & \ExpTwoARevFactorOne{} (\ExpTwoARevFactorTOne) & \ExpTwoARevFactorTwo{} (\ExpTwoARevFactorTTwo) & \ExpTwoARevFactorThree{} (\ExpTwoARevFactorTThree) & \ExpTwoARevMeanFPA & \ExpTwoARevMeanSPA & $\ExpTwoARevGapPct$ & \ExpTwoAConvMedian \\
2b & Thompson & \ExpTwoBRevFactorOne{} (\ExpTwoBRevFactorTOne) & \ExpTwoBRevFactorTwo{} (\ExpTwoBRevFactorTTwo) & \ExpTwoBRevFactorThree{} (\ExpTwoBRevFactorTThree) & \ExpTwoBRevMeanFPA & \ExpTwoBRevMeanSPA & $\ExpTwoBRevGapPct$ & \ExpTwoBConvMedian \\
3a & Pacing (Mult.) & \ExpThreeARevFactorOne{} (\ExpThreeARevFactorTOne) & \ExpThreeARevFactorTwo{} (\ExpThreeARevFactorTTwo) & \ExpThreeARevFactorThree{} (\ExpThreeARevFactorTThree) & \ExpThreeARevMeanFPA & \ExpThreeARevMeanSPA & $\ExpThreeARevGapPct$ & \ExpThreeAConvMedian \\
3b & Pacing (PI) & \ExpThreeBRevFactorOne{} (\ExpThreeBRevFactorTOne) & \ExpThreeBRevFactorTwo{} (\ExpThreeBRevFactorTTwo) & \ExpThreeBRevFactorThree{} (\ExpThreeBRevFactorTThree) & \ExpThreeBRevMeanFPA & \ExpThreeBRevMeanSPA & $\ExpThreeBRevGapPct$ & \ExpThreeBConvMedian \\
\bottomrule
\end{tabular}}
\end{table}

The effect of auction format on revenue reverses across algorithm classes. Under unconstrained learning (Experiments~1a--2b), second-price auctions weakly dominate; under budget-constrained pacing (Experiments~3a--3b), first-price auctions dominate (Table~\ref{tab:cross_exp_summary}). This reversal qualifies the universal first-price penalty found by \citet{BanchioSkrzypacz2022}, who studied a single algorithm class. A platform choosing between formats must first assess which bidding technology its agents deploy.

The auction format $\times$ number of bidders interaction is significant in Experiments~1a and~2a but with opposite signs. Under Q-learning (Experiment~1a), additional bidders reduce the format effect; under LinUCB (Experiment~2a), additional bidders widen it. The measurement horizon also matters; when lifetime revenue is used instead of end-state revenue, the two formats become statistically indistinguishable in Experiment~1b.\footnote{First-price learning-phase premium: \ExpOneBFPAPremium; second-price: \ExpOneBSPAPremium. Lifetime revenue model: $|t| = \ExpOneBAllRevAuctionAbsT$, $p \ExpOneBAllRevAuctionPFmt$.} All primary findings survive comprehensive robustness checks (Appendix~\ref{sec:appendix_robustness}).

Quantile regression reveals that these mean effects mask distributional heterogeneity. In Experiment~1a, the auction format coefficient reverses sign across quantiles, indicating that first-price auctions compress the revenue distribution even though the average effect is negligible. In Experiments~2b, 3a, and~3b, the format effect grows toward the upper tail, so that the best-performing configurations are most sensitive to format choice (Appendix~\ref{sec:appendix_robustness}).

\subsection{Algorithm-Class Dependence of Secondary Effects}

Beyond the headline findings, a pervasive meta-pattern emerges across experiments. The direction and significance of nearly every secondary effect depends on the algorithm class. The affiliation parameter $\eta$ (spanning independent private to near-common values) shows no significant effect on revenue in Experiment~1b ($p = 0.056$). It becomes significant in both contextual bandit experiments (Experiment~2a, $t = -5.7$, $p < 0.001$; Experiment~2b, $t = -3.2$, $p = 0.002$). Revenue equivalence holds for all $\eta$ because signals are independent (Appendix~\ref{sec:equilibria}), yet under contextual bandits the second-price advantage widens with affiliation strength, directionally aligned with the ranking \citet{MilgromWeber1982} derive for affiliated signals. Reserve prices show similarly algorithm-dependent effects, significant and positive under Q-learning with constant valuations (Experiment~1a, $t = 5.1$) but significant and negative under Thompson Sampling (Experiment~2b, $t = -7.9$). Classical theory predictions are mediated by the learning algorithm.

Efficiency outcomes under budget-constrained pacing illustrate a further dimension of algorithm-class dependence. Under utility-maximizing objectives, dual pacing agents achieve near-perfect allocative efficiency; under value-maximizing objectives with four bidders, efficiency drops substantially, with bid-to-value ratios exceeding 1.5.\footnote{Utility-maximizing efficiency ranges from \ExpThreeAUtilEffLow{} to \ExpThreeAUtilEffHigh{} across cells; value-maximizing four-bidder efficiency drops to \ExpThreeAValFourEffLow--\ExpThreeAValFourEffHigh.} All observed Price of Anarchy values fall well within the theoretical worst-case bound of~2 \citep{Aggarwal2019, Gaitonde2023} for both auction formats. However, the effect of auction format on the Price of Anarchy differs sharply between the two pacing mechanisms, with negligible contribution under dual pacing ($S_T = 0.005$) and dominant contribution under PI control ($S_T = 0.19$). Even within the same algorithm family, the mechanism choice matters for welfare.

Convergence speed varies substantially across algorithm classes. Table~\ref{tab:cross_exp_summary} compares convergence times across all experiments. In unconstrained settings, contextual bandits converge substantially faster than Q-learning, yet faster convergence does not translate into higher revenue. Experiment~2a converges far faster than Experiment~1a (\ExpTwoAConvMedian{} vs.\ \ExpOneAConvMedian{} rounds) but produces lower revenues.\footnote{Pacing experiments use a different convergence concept (within-episode budget depletion rather than cross-episode strategy settling), so direct comparison of convergence times between pacing and learning experiments is not meaningful.} First-price auctions produce higher winner entropy than second-price auctions under LinUCB (Experiment~2a), but this pattern does not replicate under Thompson Sampling (Experiment~2b) or Q-learning (Experiment~1b), where the auction format effect on winner entropy is not statistically significant. Higher exploration intensity reduces revenue under both bandit algorithms.

One finding is consistent across all algorithm classes: algorithmic tuning parameters contribute negligibly to revenue. Learning rate, discount factor, decay type, regularisation, memory decay, and aggressiveness each achieve $S_T < 0.04$ in every experiment where they are varied. Market thickness dominates algorithm configuration universally. This dominance holds within the professionally relevant ranges tested here; extreme misconfiguration of any algorithm (for example, a near-zero learning rate or an implausibly large budget) would naturally cause systemic failure regardless of market structure.

\subsection{Implications}

The most effective way to improve seller revenue is ensuring sufficient competition. In unconstrained settings, increasing the number of bidders improves revenue more than any other intervention tested, including switching auction formats. In budget-constrained environments, relaxing budget tightness has the largest impact. The effect of auction format is context-dependent; sellers should choose format based on the prevailing bidding technology rather than adopting a blanket preference.

The dominance of market thickness and budget tightness over algorithmic design choices carries a direct implication for competition policy. Interventions that increase competition, such as lowering barriers to entry or preventing excessive market concentration among bidders, are likely to be more effective than algorithm-specific regulation. Across all unconstrained experiments, adding bidders produces a larger revenue improvement than any algorithmic intervention, and under budget constraints, relaxing budget tightness dominates all other factors. The context-dependence of auction format effects complicates prescriptive regulation. A blanket mandate requiring a particular auction format ignores the finding that the welfare consequences of format choice depend on the learning algorithm, the value environment, and the presence of budget constraints. Under Q-learning with constant valuations, the two formats produce statistically indistinguishable revenue. Under contextual bandits with affiliated values, second-price auctions yield higher revenue. Under budget-constrained pacing, first-price auctions perform better. This heterogeneity supports outcome-based monitoring over rule-based mandates \citep{OECD2023, Johnson2023}. The ranked-effects tables produced by the factorial analysis quantify the marginal contribution of each factor, offering a transparent and reproducible evidentiary basis that could support such outcome-based regulatory approaches \citep{Harrington2018, Gal2019}.

\subsection{Practical Recommendations}
\label{sec:practitioners_guide}

The factorial laboratory is a general-purpose diagnostic for algorithmic markets, not a method tied to the specific algorithms or auction formats studied here. Any researcher who suspects that autonomous bidding agents may suppress prices, inflate costs, or distort allocations can apply the same workflow. The first step is to characterise the algorithm-mechanism pair under study. Nearly every secondary finding depends on the algorithm class, so the pair must be identified before any design decisions are made. The researcher should determine whether agents are learning values from scratch (Q-learning, contextual bandits), managing budget constraints across repeated auctions (multiplicative pacing, PI controllers), or following a different approach altogether (deep reinforcement learning, transformer-based bidders, LLM agents). The market rules matter equally. Sealed-bid versus open formats, first-price versus second-price payment rules, reserve price policies, and single-item versus combinatorial allocation all shape which theoretical predictions serve as benchmarks and which factors are worth screening.

The second step is to survey the relevant theory and empirical literature for candidate factors, plausible ranges, and theoretical predictions. Classical auction theory supplies equilibrium benchmarks against which simulated outcomes can be compared, including revenue equivalence under independent private values, revenue rankings under affiliation, and the marginal value of an additional bidder. Prior work on learning algorithms identifies mechanism-specific factors such as exploration parameters, discount rates, synchronisation modes, and state representations. The autobidding literature highlights budget multipliers, pacing objectives, and the distinction between value-maximising and utility-maximising agents. Each candidate factor should have a theoretical or empirical basis for inclusion and a range spanning practically relevant levels. Section~\ref{sec:literature} discusses the four bodies of literature this paper draws on.

The third step is to screen the candidate factors down to a manageable set. No single experiment can test everything, so the screening stage selects the factors most likely to matter. A sequential approach begins with a large screening design that spans many factors at two levels each, then follows up on active factors with focused designs that add levels or new response variables. The key discipline is to separate structural market parameters (number of bidders, reserve prices, value environments) from algorithmic tuning parameters (learning rates, discount factors, exploration intensity), because the two categories may operate at different effect scales. Factor levels should span the range encountered in practice rather than being chosen for theoretical convenience. Table~\ref{tab:params} lists the factor definitions and ranges used in this paper.

The fourth step is to select an experimental design that provides orthogonal or near-orthogonal estimation of main effects and interactions. Two-level factorial designs, whether full or fractional, are the standard choice for screening. When some factors have more than two natural levels, as with the affiliation parameter spanning independent, moderate, and common values, mixed-level designs accommodate the extra levels while preserving balance. Space-filling designs such as Latin hypercubes or Sobol sequences are an alternative when factors are continuous and the researcher wants to map the response surface rather than screen for active factors. The choice depends on the research question; factor identification favours factorials, while functional-form mapping favours space-filling designs. Replication should scale inversely with design size to supply pure-error degrees of freedom for lack-of-fit testing. Section~\ref{sec:inference} details the designs used in this paper.

Estimation combines ANOVA with a surrogate model. An OLS model with main effects and all two-way interactions, fit using effects coding ($-1$/$+1$), guarantees orthogonal estimation of all estimable effects (Section~\ref{sec:appendix_inference}). The linear model should always be accompanied by a nonparametric surrogate, such as gradient-boosted trees with cross-validation, to benchmark model adequacy; a meaningful $R^2$ gap signals detectable nonlinearity that the linear model misses. Pareto charts of absolute $t$-statistics and half-normal probability plots are the primary tools for separating active effects from noise, providing a visual complement to the numerical significance tests.

The final step is to report results in a structured way. Variance decomposition through Sobol indices, both first-order $S_i$ and total-order $S_{T_i}$, quantifies each factor's contribution to outcome variability; for balanced factorials these coincide with the ANOVA sum-of-squares decomposition (Section~\ref{sec:appendix_sensitivity}). Effect direction and magnitude come from the OLS coefficients. Multiple-testing corrections are essential when many effects are tested simultaneously; family-wise error rate control via Holm--Bonferroni and false discovery rate control via Benjamini--Hochberg guard against false positives. Quantile regressions at multiple percentiles reveal whether effects are uniform across the outcome distribution or concentrated in the tails. Power analysis establishes what the design can and cannot detect. Together, these diagnostics provide a transparent, reproducible evidentiary basis for ranking the factors that matter most. Appendix~\ref{sec:appendix_robustness} presents the full battery applied in this paper. Figure~\ref{fig:lab_pipeline} summarises this workflow.

The same framework lends itself to regulatory auditing. A competition authority could either build its own factorial laboratory using the methodology described above, or mandate that auction platforms run standardised factorial experiments on their bidding systems and report the results. The Sobol variance decomposition and ranked-effects tables provide a transparent, reproducible diagnostic that reduces a high-dimensional algorithmic system to a short list of factors ordered by their contribution to outcome variability. Because the designs are pre-registered and the analysis is mechanical, auditors can verify compliance without requiring access to proprietary source code; the platform submits its design matrix, raw outcomes, and estimated effects, and the regulator checks whether the reported decomposition is internally consistent and whether any flagged risk factors, such as bid suppression under specific format--algorithm combinations, exceed acceptable thresholds.

\begin{figure}[H]
\centering
\begin{tikzpicture}[
  node distance=0.45cm,
  block/.style={rectangle, draw, rounded corners, minimum width=2.0cm,
                minimum height=0.9cm, text centered, font=\small},
  arrow/.style={-{Stealth[length=2mm]}, thick},
]
  \node[block] (spec) {Specify};
  \node[block, right=of spec] (design) {Design};
  \node[block, right=of design] (sim) {Simulate};
  \node[block, right=of sim] (est) {Estimate};
  \node[block, right=of est] (val) {Validate};
  \node[block, right=of val] (dec) {Decide};

  \draw[arrow] (spec) -- (design);
  \draw[arrow] (design) -- (sim);
  \draw[arrow] (sim) -- (est);
  \draw[arrow] (est) -- (val);
  \draw[arrow] (val) -- (dec);

  \node[below=0.15cm of spec, font=\scriptsize, align=center] {algorithm,\\mechanism, factors};
  \node[below=0.15cm of design, font=\scriptsize, align=center] {factorial or\\space-filling};
  \node[below=0.15cm of sim, font=\scriptsize, align=center] {run auction\\experiments};
  \node[below=0.15cm of est, font=\scriptsize, align=center] {OLS + surrogate\\benchmarking};
  \node[below=0.15cm of val, font=\scriptsize, align=center] {robustness\\checks};
  \node[below=0.15cm of dec, font=\scriptsize, align=center] {Sobol indices,\\policy guidance};

  \draw[arrow, rounded corners=10pt, gray]
    (val.north) -- ++(0, 0.8) -| (spec.north)
    node[pos=0.25, above, font=\scriptsize] {iterate};
\end{tikzpicture}
\caption{Laboratory pipeline for diagnosing algorithmic markets. The six stages correspond to the workflow described in Practical Recommendations. A feedback loop from validation to specification supports iterative refinement of factor sets and design choices.}
\label{fig:lab_pipeline}
\end{figure}

\subsection{Limitations and Future Directions}

The qualitative patterns documented here are likely to hold in settings characterised by repeated interactions, autonomous learning, and sealed-bid formats, conditions that closely approximate programmatic advertising markets. External validity to specific real-world markets requires empirical validation with proprietary auction data. The algorithms studied here are stylised; the experiments do not include deep reinforcement learning, LLM-based bidders, or transformer architectures. Agents have no explicit communication channels, episode lengths are fixed, bidder populations are symmetric, and no combinatorial formats are tested. Settings involving few-shot auctions or human bidders may produce qualitatively different outcomes.

This work can be extended in three directions. Response surface designs can map the precise functional forms of the factors identified as important by the screening factorials, moving from factor identification to functional-form estimation. Multi-agent architectures beyond tabular methods and PI controllers, including transformers, LLM-based bidders, and deep reinforcement learning, would test whether the dominance of market thickness persists when algorithmic complexity increases. Empirical validation with proprietary auction data would test whether the laboratory patterns replicate in real advertising exchanges.

\bibliographystyle{plainnat}
\bibliography{references}

\begin{thebibliography}{80}
\providecommand{\natexlab}[1]{#1}
\providecommand{\url}[1]{\texttt{#1}}
\expandafter\ifx\csname urlstyle\endcsname\relax
  \providecommand{\doi}[1]{doi: #1}\else
  \providecommand{\doi}{doi: \begingroup \urlstyle{rm}\Url}\fi

\bibitem[Abada et~al.(2022)Abada, Lambin, and Tchakarov]{Abada2022}
Ibrahim Abada, Xavier Lambin, and Nikolay Tchakarov.
\newblock Collusion by mistake: Does algorithmic sophistication drive
  supra-competitive profits?
\newblock \emph{SSRN Electronic Journal}, 2022.
\newblock \doi{10.2139/ssrn.4099361}.

\bibitem[Abbasi-Yadkori et~al.(2011)Abbasi-Yadkori, P\'{a}l, and
  Szepesv\'{a}ri]{AbbasiYadkori2011}
Yasin Abbasi-Yadkori, D\'{a}vid P\'{a}l, and Csaba Szepesv\'{a}ri.
\newblock Improved algorithms for linear stochastic bandits.
\newblock In \emph{Advances in Neural Information Processing Systems 24
  (NeurIPS)}, pages 2312--2320, 2011.

\bibitem[Aggarwal et~al.(2019)Aggarwal, Badanidiyuru, and Mehta]{Aggarwal2019}
Gagan Aggarwal, Ashwinkumar Badanidiyuru, and Aranyak Mehta.
\newblock Autobidding with constraints.
\newblock In \emph{Web and Internet Economics: 15th International Conference,
  WINE 2019, New York, NY, USA, December 10--12, 2019, Proceedings 15}, pages
  17--30. Springer, 2019.

\bibitem[Aggarwal et~al.(2024)Aggarwal, Badanidiyuru, Balseiro, Bhawalkar,
  Deng, Feng, Goel, Liaw, Lu, Mahdian, et~al.]{Aggarwal2024Survey}
Gagan Aggarwal, Ashwinkumar Badanidiyuru, Santiago~R Balseiro, Kshipra
  Bhawalkar, Yuan Deng, Zhe Feng, Gagan Goel, Christopher Liaw, Haihao Lu,
  Mohammad Mahdian, et~al.
\newblock Auto-bidding and auctions in online advertising: A survey.
\newblock \emph{ACM SIGecom Exchanges}, 22\penalty0 (1):\penalty0 159--183,
  2024.

\bibitem[Agrawal and Goyal(2013)]{AgrawalGoyal2013}
Shipra Agrawal and Navin Goyal.
\newblock Thompson sampling for contextual bandits with linear payoffs.
\newblock In \emph{Proceedings of the 30th International Conference on Machine
  Learning (ICML)}, pages 127--135, 2013.

\bibitem[Archer et~al.(1997)Archer, Saltelli, and
  Sobol']{ArcherSaltelliSobol1997}
G.~E.~B. Archer, Andrea Saltelli, and I.~M. Sobol'.
\newblock Sensitivity measures, {ANOVA}-like techniques and the use of
  bootstrap.
\newblock \emph{Journal of Statistical Computation and Simulation}, 58\penalty0
  (2):\penalty0 99--120, 1997.

\bibitem[Arunachaleswaran et~al.(2025)Arunachaleswaran, Collina, Kannan, Roth,
  and Ziani]{Arunachaleswaran2025}
Eshwar~Ram Arunachaleswaran, Natalie Collina, Sampath Kannan, Aaron Roth, and
  Juba Ziani.
\newblock Algorithmic collusion without threats.
\newblock 2025.
\newblock University of Pennsylvania working paper.

\bibitem[Asker et~al.(2022)Asker, Fershtman, and Pakes]{Asker2022}
John Asker, Chaim Fershtman, and Ariel Pakes.
\newblock Artificial intelligence, algorithm design, and pricing.
\newblock \emph{AEA Papers and Proceedings}, 112:\penalty0 452--456, 2022.
\newblock \doi{10.1257/pandp.20221059}.

\bibitem[Assad et~al.(2020)Assad, Clark, Ershov, and Xu]{Assad2020}
Stephanie Assad, Robert Clark, Daniel Ershov, and Lei Xu.
\newblock Algorithmic pricing and competition: Empirical evidence from the
  german retail gasoline market.
\newblock \emph{SSRN Electronic Journal}, 2020.
\newblock \doi{10.2139/ssrn.3682021}.

\bibitem[Balseiro and Gur(2019)]{Balseiro2019Learning}
Santiago~R Balseiro and Yonatan Gur.
\newblock Learning in repeated auctions with budgets: Regret minimization and
  equilibrium.
\newblock \emph{Management Science}, 65\penalty0 (9):\penalty0 3952--3968,
  2019.

\bibitem[Banchio and Mantegazza(2022)]{BanchioMantegazza2022}
E~H Banchio and F~Mantegazza.
\newblock Spontaneous coupling of learning agents in two-player games.
\newblock \emph{Physical Review E}, 105\penalty0 (3):\penalty0 034304, 2022.

\bibitem[Banchio and Skrzypacz(2022)]{BanchioSkrzypacz2022}
Martino Banchio and Andrzej Skrzypacz.
\newblock Artificial intelligence and auction design.
\newblock In \emph{Proceedings of the 23rd ACM Conference on Economics and
  Computation}, pages 30--31. ACM, 2022.
\newblock \doi{10.1145/3490486.3538244}.

\bibitem[Bandyopadhyay et~al.(2008)Bandyopadhyay, Rees, and
  Barron]{Bandyopadhyay2008}
Subhajyoti Bandyopadhyay, Jackie Rees, and John~M. Barron.
\newblock Reverse auctions with multiple reinforcement learning agents.
\newblock \emph{Decision Sciences}, 39\penalty0 (1):\penalty0 33--63, 2008.
\newblock \doi{10.1111/j.1540-5915.2008.00181.x}.

\bibitem[Benjamini and Hochberg(1995)]{BenjaminiHochberg1995}
Yoav Benjamini and Yosef Hochberg.
\newblock Controlling the false discovery rate: A practical and powerful
  approach to multiple testing.
\newblock \emph{Journal of the Royal Statistical Society: Series B
  (Methodological)}, 57\penalty0 (1):\penalty0 289--300, 1995.
\newblock \doi{10.1111/j.2517-6161.1995.tb02031.x}.

\bibitem[Bertrand et~al.(2025)Bertrand, Duque, Calvano, and
  Gidel]{Bertrand2025}
Quentin Bertrand, Juan~Agustin Duque, Emilio Calvano, and Gauthier Gidel.
\newblock Self-play q-learners can provably collude in the iterated prisoner's
  dilemma.
\newblock In \emph{Proceedings of the 42nd International Conference on Machine
  Learning (ICML)}, 2025.

\bibitem[Borgonovo(2007)]{Borgonovo2007}
Emanuele Borgonovo.
\newblock A new uncertainty importance measure.
\newblock \emph{Reliability Engineering \& System Safety}, 92\penalty0
  (6):\penalty0 771--784, 2007.

\bibitem[Box et~al.(2005)Box, Hunter, and Hunter]{Box2005}
George E.~P. Box, J.~Stuart Hunter, and William~G. Hunter.
\newblock \emph{Statistics for Experimenters: Design, Innovation, and
  Discovery}.
\newblock Wiley, Hoboken, NJ, 2nd edition, 2005.

\bibitem[Brown and MacKay(2023)]{BrownMacKay2023}
Zach~Y. Brown and Alexander MacKay.
\newblock Competition in pricing algorithms.
\newblock \emph{American Economic Journal: Microeconomics}, 15\penalty0
  (2):\penalty0 109--156, 2023.

\bibitem[Bulow and Klemperer(1996)]{BulowKlemperer1996}
Jeremy Bulow and Paul Klemperer.
\newblock Auctions versus negotiations.
\newblock \emph{American Economic Review}, 86\penalty0 (1):\penalty0 180--194,
  1996.

\bibitem[Calvano et~al.(2020)Calvano, Calzolari, Denicol\`{o}, and
  Pastorello]{Calvano2020}
Emilio Calvano, Giacomo Calzolari, Vincenzo Denicol\`{o}, and Sergio
  Pastorello.
\newblock Artificial intelligence, algorithmic pricing, and collusion.
\newblock \emph{American Economic Review}, 110\penalty0 (10):\penalty0
  3267--3297, 2020.
\newblock \doi{10.1257/aer.20190623}.

\bibitem[Calvano et~al.(2023)Calvano, Calzolari, Denicol\`{o}, and
  Pastorello]{Calvano2023}
Emilio Calvano, Giacomo Calzolari, Vincenzo Denicol\`{o}, and Sergio
  Pastorello.
\newblock Algorithmic collusion: Genuine or spurious?
\newblock \emph{International Journal of Industrial Organization}, 90:\penalty0
  102973, 2023.
\newblock \doi{10.1016/j.ijindorg.2023.102973}.

\bibitem[Calzolari and Denicol\`{o}(2021)]{Calzolari2021}
Giacomo Calzolari and Vincenzo Denicol\`{o}.
\newblock Algorithmic collusion with imperfect monitoring.
\newblock \emph{International Journal of Industrial Organization}, 79:\penalty0
  102717, 2021.
\newblock \doi{10.1016/j.ijindorg.2021.102717}.

\bibitem[Chen et~al.(2016)Chen, Mislove, and Wilson]{Chen2016}
Le~Chen, Alan Mislove, and Christo Wilson.
\newblock An empirical analysis of algorithmic pricing on amazon marketplace.
\newblock In \emph{Proceedings of the 25th International Conference on World
  Wide Web}, pages 1339--1349, 2016.
\newblock \doi{10.1145/2872427.2883089}.

\bibitem[Conitzer et~al.(2022)Conitzer, Kroer, Panigrahi, Schrijvers,
  Stier-Moses, Sodomka, and Wilkens]{Conitzer2022FPPE}
Vincent Conitzer, Christian Kroer, Debmalya Panigrahi, Okke Schrijvers,
  Nicolas~E Stier-Moses, Eric Sodomka, and Christopher~A Wilkens.
\newblock Pacing equilibrium in first price auction markets.
\newblock \emph{Management Science}, 2022.
\newblock Forthcoming, previously arXiv:1811.07166v4.

\bibitem[Crane(2024)]{Crane2024}
Daniel~A. Crane.
\newblock Antitrust after the coming wave.
\newblock \emph{New York University Law Review}, 99\penalty0 (4), 2024.

\bibitem[Daniel(1959)]{Daniel1959}
Cuthbert Daniel.
\newblock Use of half-normal plots in interpreting factorial two-level
  experiments.
\newblock \emph{Technometrics}, 1\penalty0 (4):\penalty0 311--341, 1959.
\newblock \doi{10.1080/00401706.1959.10489866}.

\bibitem[Davidson and Flachaire(2008)]{DavidsonFlachaire2008}
Russell Davidson and Emmanuel Flachaire.
\newblock The wild bootstrap, tamed at last.
\newblock \emph{Journal of Econometrics}, 146\penalty0 (1):\penalty0 162--169,
  2008.
\newblock \doi{10.1016/j.jeconom.2008.08.003}.

\bibitem[Deng et~al.(2021)Deng, Mao, Mirrokni, and
  Zuo]{Deng2021TowardsEfficient}
Yuan Deng, Jieming Mao, Vahab Mirrokni, and Song Zuo.
\newblock Towards efficient auctions in an auto-bidding world.
\newblock In \emph{Proceedings of the Web Conference 2021}, pages 3965--3973,
  2021.

\bibitem[Deng et~al.(2022)Deng, Mao, Mirrokni, Zhang, and
  Zuo]{Deng2022FPAEfficiency}
Yuan Deng, Jieming Mao, Vahab Mirrokni, Hanrui Zhang, and Song Zuo.
\newblock Efficiency of the first-price auction in the autobidding world.
\newblock In \emph{Advances in Neural Information Processing Systems 35
  (NeurIPS 2022)}, 2022.
\newblock Also arXiv:2208.10650.

\bibitem[Dolgopolov(2021)]{Dolgopolov2021}
Arthur Dolgopolov.
\newblock Algorithmic pricing and liquidity in securities markets.
\newblock 2021.
\newblock Working paper; verify bibliographic details.

\bibitem[Douglas et~al.(2024)Douglas, Provost, and Sundararajan]{Douglas2024}
Connor Douglas, Foster~J. Provost, and Arun Sundararajan.
\newblock Naive algorithmic collusion: When do bandit learners cooperate and
  when do they compete?
\newblock In \emph{Proceedings of the International Conference on Information
  Systems (ICIS)}, 2024.
\newblock Also arXiv:2411.16574.

\bibitem[Ezrachi and Stucke(2017)]{EzrachiStucke2017}
Ariel Ezrachi and Maurice~E. Stucke.
\newblock Artificial intelligence \& collusion: When computers inhibit
  competition.
\newblock \emph{University of Illinois Law Review}, 2017:\penalty0 1775--1810,
  2017.

\bibitem[Ezrachi and Stucke(2024)]{EzrachiStucke2024}
Ariel Ezrachi and Maurice~E. Stucke.
\newblock Sustainable and algorithmic collusion.
\newblock \emph{Vanderbilt Journal of Entertainment \& Technology Law}, 27,
  2024.

\bibitem[Fikioris and Tardos(2023)]{FikiorisTardos2023}
Giannis Fikioris and \'{E}va Tardos.
\newblock Liquid welfare guarantees for no-regret learning in sequential
  budgeted auctions.
\newblock In \emph{Proceedings of the 24th ACM Conference on Economics and
  Computation (EC '23)}, 2023.
\newblock Also in Mathematics of Operations Research.

\bibitem[Fish et~al.(2024)Fish, Gonczarowski, and Shorrer]{Fish2024}
Sara Fish, Yannai~A. Gonczarowski, and Ran Shorrer.
\newblock Algorithmic collusion by large language models.
\newblock \emph{arXiv preprint arXiv:2404.00806}, 2024.

\bibitem[Gaitonde et~al.(2023)Gaitonde, Li, Light, Lucier, and
  Slivkins]{Gaitonde2023}
Jason Gaitonde, Yingkai Li, Bar Light, Brendan Lucier, and Aleksandrs Slivkins.
\newblock Budget pacing in repeated auctions: Regret and efficiency without
  convergence.
\newblock In \emph{Proceedings of the 14th Innovations in Theoretical Computer
  Science Conference (ITCS)}, 2023.
\newblock \doi{10.4230/LIPIcs.ITCS.2023.52}.

\bibitem[Gal(2019)]{Gal2019}
Michal~S. Gal.
\newblock Algorithms as illegal agreements.
\newblock \emph{Berkeley Technology Law Journal}, 34:\penalty0 67--118, 2019.

\bibitem[Han(2022)]{Han2022}
Zhijun Han.
\newblock Deep reinforcement learning for algorithmic trading and collusion.
\newblock 2022.
\newblock Working paper; verify bibliographic details.

\bibitem[Hansen et~al.(2021)Hansen, Misra, and Pai]{Hansen2021}
Karsten~T. Hansen, Kanishka Misra, and Mallesh~M. Pai.
\newblock Frontiers: Algorithmic collusion: Supra-competitive prices via
  independent algorithms.
\newblock \emph{Marketing Science}, 40\penalty0 (1):\penalty0 1--12, 2021.

\bibitem[Harrington(2018)]{Harrington2018}
Joseph~E. Harrington, Jr.
\newblock Developing competition law for collusion by autonomous price-setting
  agents.
\newblock \emph{Journal of Competition Law \& Economics}, 14\penalty0
  (4):\penalty0 495--559, 2018.
\newblock \doi{10.1093/joclec/nhy016}.

\bibitem[Hartline et~al.(2024)Hartline, Johnsen, and Li]{Hartline2024}
Jason~D. Hartline, Aleck Johnsen, and Ao~Li.
\newblock Regulation of algorithmic collusion.
\newblock 2024.
\newblock Northwestern University working paper.

\bibitem[Hartline et~al.(2025)Hartline, Wang, and Zhang]{Hartline2025}
Jason~D. Hartline, Chang Wang, and Chenhao Zhang.
\newblock Regulation of algorithmic collusion, refined: Testing pessimistic
  calibrated regret.
\newblock \emph{arXiv preprint arXiv:2501.09740}, 2025.

\bibitem[Hettich(2021)]{Hettich2021}
Matthias Hettich.
\newblock Algorithmic collusion: Insights from deep learning.
\newblock \emph{SSRN Electronic Journal}, 2021.

\bibitem[Holm(1979)]{Holm1979}
Sture Holm.
\newblock A simple sequentially rejective multiple test procedure.
\newblock \emph{Scandinavian Journal of Statistics}, 6\penalty0 (2):\penalty0
  65--70, 1979.

\bibitem[Ivaldi et~al.(2003)Ivaldi, Jullien, Rey, Seabright, and
  Tirole]{Ivaldi2003}
Marc Ivaldi, Bruno Jullien, Patrick Rey, Paul Seabright, and Jean Tirole.
\newblock The economics of tacit collusion.
\newblock In \emph{Final Report for DG Competition, European Commission}. 2003.
\newblock IDEI Working Paper.

\bibitem[Johnson et~al.(2023)Johnson, Rhodes, and Wildenbeest]{Johnson2023}
Justin~P. Johnson, Andrew Rhodes, and Matthijs Wildenbeest.
\newblock Platform design when sellers use pricing algorithms.
\newblock \emph{Econometrica}, 91\penalty0 (5):\penalty0 1841--1879, 2023.
\newblock \doi{10.3982/ECTA19978}.

\bibitem[Klein(2021)]{Klein2021}
Timo Klein.
\newblock Autonomous algorithmic collusion: Q-learning under sequential
  pricing.
\newblock \emph{The RAND Journal of Economics}, 52\penalty0 (3):\penalty0
  538--558, 2021.
\newblock \doi{10.1111/1756-2171.12383}.

\bibitem[Koenker and Bassett(1978)]{KoenkerBassett1978}
Roger Koenker and Gilbert Bassett, Jr.
\newblock Regression quantiles.
\newblock \emph{Econometrica}, 46\penalty0 (1):\penalty0 33--50, 1978.
\newblock \doi{10.2307/1913643}.

\bibitem[K\"{u}hn and Tadelis(2018)]{KuhnTadelis2017}
Kai-Uwe K\"{u}hn and Steve Tadelis.
\newblock The economics of algorithmic pricing: Is collusion really inevitable?
\newblock 2018.
\newblock Working paper.

\bibitem[Lenth(1989)]{Lenth1989}
Russell~V. Lenth.
\newblock Quick and easy analysis of unreplicated factorials.
\newblock \emph{Technometrics}, 31\penalty0 (4):\penalty0 469--473, 1989.

\bibitem[Li et~al.(2010)Li, Chu, Langford, and Schapire]{Li2010LinUCB}
Lihong Li, Wei Chu, John Langford, and Robert~E. Schapire.
\newblock A contextual-bandit approach to personalized news article
  recommendation.
\newblock In \emph{Proceedings of the 19th International Conference on World
  Wide Web (WWW 2010)}, pages 661--670, 2010.
\newblock \doi{10.1145/1772690.1772758}.

\bibitem[Liaw et~al.(2022)Liaw, Mehta, and
  Perlroth]{Liaw2022EfficiencyNonTruthful}
Christopher Liaw, Aranyak Mehta, and Andres Perlroth.
\newblock Efficiency of non-truthful auctions under auto-bidding.
\newblock In \emph{Proceedings of the ACM Web Conference 2023}, pages
  3561--3571, 2022.
\newblock Also arXiv:2207.03630.

\bibitem[Liaw et~al.(2024)Liaw, Mehta, and Zhu]{Liaw2024EfficiencyBudget}
Christopher Liaw, Aranyak Mehta, and Wennan Zhu.
\newblock Efficiency of non-truthful auctions in auto-bidding with budget
  constraints.
\newblock In \emph{Proceedings of the ACM Web Conference 2024}, pages 223--234,
  2024.
\newblock Also arXiv:2310.09271.

\bibitem[Lucier et~al.(2023)Lucier, Pattathil, Slivkins, and
  Zhang]{Lucier2023PacingDynamics}
Brendan Lucier, Sarath Pattathil, Aleksandrs Slivkins, and Mengxiao Zhang.
\newblock Autobidders with budget and roi constraints: Efficiency, regret, and
  pacing dynamics.
\newblock In \emph{Proceedings of the 24th ACM Conference on Economics and
  Computation}, pages 678--698, 2023.

\bibitem[MacKinnon and White(1985)]{MacKinnonWhite1985}
James~G. MacKinnon and Halbert White.
\newblock Some heteroskedasticity-consistent covariance matrix estimators with
  improved finite sample properties.
\newblock \emph{Journal of Econometrics}, 29\penalty0 (3):\penalty0 305--325,
  1985.
\newblock \doi{10.1016/0304-4076(85)90158-7}.

\bibitem[Mehra(2016)]{Mehra2016}
Salil~K. Mehra.
\newblock Antitrust and the robo-seller: Competition in the time of algorithms.
\newblock \emph{Minnesota Law Review}, 100:\penalty0 1323--1375, 2016.

\bibitem[Mehta(2022)]{Mehta2022AuctionDesign}
Aranyak Mehta.
\newblock Auction design in an auto-bidding setting: Randomization improves
  efficiency beyond vcg.
\newblock In \emph{Proceedings of the ACM Web Conference 2022}, pages 173--181,
  2022.

\bibitem[Milgrom and Weber(1982)]{MilgromWeber1982}
Paul~R. Milgrom and Robert~J. Weber.
\newblock A theory of auctions and competitive bidding.
\newblock \emph{Econometrica}, 50\penalty0 (5):\penalty0 1089--1122, 1982.
\newblock \doi{10.2307/1911865}.

\bibitem[Montgomery(2017)]{Montgomery2017}
Douglas~C. Montgomery.
\newblock \emph{Design and Analysis of Experiments}.
\newblock Wiley, Hoboken, NJ, 9th edition, 2017.

\bibitem[Morris(1991)]{Morris1991}
Max~D Morris.
\newblock Factorial sampling plans for preliminary computational experiments.
\newblock \emph{Technometrics}, 33\penalty0 (2):\penalty0 161--174, 1991.

\bibitem[Morris et~al.(2008)Morris, Moore, and McKay]{MorrisMooreMcKay2008}
Max~D. Morris, Leslie~M. Moore, and Michael~D. McKay.
\newblock Using orthogonal arrays in the sensitivity analysis of computer
  models.
\newblock \emph{Technometrics}, 50\penalty0 (2):\penalty0 205--215, 2008.

\bibitem[Myerson(1981)]{Myerson1981}
Roger~B Myerson.
\newblock Optimal auction design.
\newblock \emph{Mathematics of operations research}, 6\penalty0 (1):\penalty0
  58--73, 1981.

\bibitem[{OECD}(2017)]{OECD2017}
{OECD}.
\newblock Algorithms and collusion: Competition policy in the digital age.
\newblock Technical report, Organisation for Economic Co-operation and
  Development, Paris, 2017.

\bibitem[{OECD}(2023)]{OECD2023}
{OECD}.
\newblock Algorithmic competition.
\newblock Technical report, Organisation for Economic Co-operation and
  Development, Paris, 2023.
\newblock OECD Competition Policy Roundtable Background Note.

\bibitem[Paes~Leme et~al.(2024)Paes~Leme, Piliouras, Schneider, Spendlove, and
  Zuo]{PaesLeme2024}
Renato Paes~Leme, Georgios Piliouras, Jon Schneider, Kelly Spendlove, and Song
  Zuo.
\newblock Complex dynamics in autobidding systems.
\newblock In \emph{Proceedings of the 25th ACM Conference on Economics and
  Computation (EC '24)}, 2024.
\newblock To appear, also arXiv:2406.19350v2.

\bibitem[Petit(2017)]{Petit2017}
Nicolas Petit.
\newblock Antitrust and artificial intelligence: A research agenda.
\newblock \emph{Journal of European Competition Law \& Practice}, 8\penalty0
  (6):\penalty0 361--362, 2017.
\newblock \doi{10.1093/jeclap/lpx033}.

\bibitem[Russac et~al.(2019)Russac, Vernade, and Capp\'{e}]{Russac2019}
Yoan Russac, Claire Vernade, and Olivier Capp\'{e}.
\newblock Weighted linear bandits for non-stationary environments.
\newblock In \emph{Advances in Neural Information Processing Systems 32
  (NeurIPS)}, pages 12017--12026, 2019.

\bibitem[Russo et~al.(2018)Russo, Van~Roy, Kazerouni, Osband, and
  Wen]{Russo2018Thompson}
Daniel~J. Russo, Benjamin Van~Roy, Abbas Kazerouni, Ian Osband, and Zheng Wen.
\newblock A tutorial on {Thompson Sampling}.
\newblock \emph{Foundations and Trends in Machine Learning}, 11\penalty0
  (1):\penalty0 1--96, 2018.
\newblock \doi{10.1561/2200000070}.

\bibitem[Saltelli et~al.(2008)Saltelli, Ratto, Andres, Campolongo, Cariboni,
  Gatelli, Saisana, and Tarantola]{Saltelli2008}
Andrea Saltelli, Marco Ratto, Terry Andres, Francesca Campolongo, Jessica
  Cariboni, Daniela Gatelli, Michaela Saisana, and Stefano Tarantola.
\newblock \emph{Global Sensitivity Analysis: The Primer}.
\newblock John Wiley \& Sons, 2008.

\bibitem[Schwalbe(2018)]{Schwalbe2018}
Ulrich Schwalbe.
\newblock Algorithms, machine learning, and collusion.
\newblock \emph{Journal of Competition Law \& Economics}, 14\penalty0
  (4):\penalty0 568--607, 2018.

\bibitem[Sobol'(2001)]{Sobol2001}
Ilya~M Sobol'.
\newblock Global sensitivity indices for nonlinear mathematical models and
  their {Monte Carlo} estimates.
\newblock \emph{Mathematics and Computers in Simulation}, 55\penalty0
  (1--3):\penalty0 271--280, 2001.

\bibitem[Tellidou and Bakirtzis(2007)]{Tellidou2007}
Athina~C. Tellidou and Anastasios~G. Bakirtzis.
\newblock Agent-based analysis of capacity withholding and tacit collusion in
  electricity markets.
\newblock \emph{IEEE Transactions on Power Systems}, 22\penalty0 (4):\penalty0
  1735--1742, 2007.

\bibitem[Tesfatsion(2006)]{Tesfatsion2006}
Leigh Tesfatsion.
\newblock Agent-based computational economics: A constructive approach to
  economic theory.
\newblock In Leigh Tesfatsion and Kenneth~L. Judd, editors, \emph{Handbook of
  Computational Economics}, volume~2, pages 831--880. North-Holland, 2006.

\bibitem[Thompson(1933)]{Thompson1933}
William~R. Thompson.
\newblock On the likelihood that one unknown probability exceeds another in
  view of the evidence of two samples.
\newblock \emph{Biometrika}, 25\penalty0 (3--4):\penalty0 285--294, 1933.
\newblock \doi{10.1093/biomet/25.3-4.285}.

\bibitem[Tibshirani(1996)]{Tibshirani1996}
Robert Tibshirani.
\newblock Regression shrinkage and selection via the lasso.
\newblock \emph{Journal of the Royal Statistical Society: Series B
  (Methodological)}, 58\penalty0 (1):\penalty0 267--288, 1996.
\newblock \doi{10.1111/j.2517-6161.1996.tb02080.x}.

\bibitem[Vickrey(1961)]{Vickrey1961}
William Vickrey.
\newblock Counterspeculation, auctions, and competitive sealed tenders.
\newblock \emph{The Journal of finance}, 16\penalty0 (1):\penalty0 8--37, 1961.

\bibitem[Waltman and Kaymak(2008)]{WaltmanKaymak2008}
Ludo Waltman and Uzay Kaymak.
\newblock Q-learning agents in a cournot oligopoly model.
\newblock \emph{Journal of Economic Dynamics and Control}, 32\penalty0
  (10):\penalty0 3275--3293, 2008.

\bibitem[Wang et~al.(2012)Wang, Tang, and Zhang]{WangTangZhang2012}
Xiaodi Wang, Boxin Tang, and Runchu Zhang.
\newblock Orthogonal arrays for estimating global sensitivity indices of
  non-parametric models based on {ANOVA} high-dimensional model representation.
\newblock \emph{Journal of Statistical Planning and Inference}, 142\penalty0
  (7):\penalty0 1801--1810, 2012.

\bibitem[Zhang(2025)]{Zhang2025Noise}
Niuniu Zhang.
\newblock Too noisy to collude? algorithmic collusion under laplacian noise.
\newblock \emph{UCLA Anderson School of Management Working Paper}, 2025.

\bibitem[Zhang(2021)]{Zhang2021}
Xu~Zhang.
\newblock Deep reinforcement learning and algorithmic collusion.
\newblock 2021.
\newblock Working paper; verify bibliographic details.

\end{thebibliography}

\appendix

\section{Appendix: Equilibria under Discretized Bidding}
\label{sec:equilibria}

This section derives the theoretical equilibrium benchmarks against which the learning experiments are evaluated.

\subsection{Constant Valuations}

In a first-price auction where each bidder's valuation is 1 and allowable bids are in \(\{0,0.1,\dots,1\}\), a symmetric profile \((b,b,\dots,b)\) yields an expected payoff of \(\frac{1}{n}(1-b)\). A unilateral upward deviation to \(b+0.1\) (provided \(b+0.1 \le 1\)) yields a payoff of \(1-(b+0.1)\). No deviation is profitable if 
\[
1 - (b + 0.1) \;\le\; \frac{1}{n}(1 - b)
\;\;\Longrightarrow\;\;
b \;\ge\; \frac{0.9\,n - 1}{n - 1}.
\]
Thus, for each \(n\), any \(b\) at or above this threshold (rounded to 0.1) is a pure-strategy Nash equilibrium. For example, when \(n=2\), \(b\ge 0.8\), and when \(n=3\), \(b\ge 0.85\). In a second-price auction with identical valuations all equal to 1, bidding 1 is weakly dominant. When valuations are known to be identical, any profile in which at least one bidder bids 1 is a Nash equilibrium.

\subsection{Affiliated Valuations}

In Experiments~1b--2b, each bidder~$i$ draws a signal $s_i \sim \text{Uniform}[0,1]$ independently and forms a valuation
\[
v_i \;=\; \alpha\, s_i \;+\; \beta \sum_{j \neq i} s_j,
\qquad
\alpha = 1 - \tfrac{\eta}{2},
\quad
\beta = \frac{\eta}{2(n-1)},
\]
where $\eta \in [0,1]$ controls affiliation. The symmetric Bayesian Nash Equilibrium features linear bidding strategies in both auction formats \citep{MilgromWeber1982}.

\subsubsection{Model and Efficiency}

The valuation can be written as $(\alpha-\beta) s_i + \beta S$ where $S$ is the sum of all signals. For $\eta$ in $[0,1]$ and $n\ge2$ the coefficient $(\alpha-\beta)$ is nonnegative, so the highest signal bidder has the highest valuation. The efficient allocation assigns the object to the highest signal.

\subsubsection{BNE Bid Functions}

The symmetric bid in the second price auction equals the expected value conditional on the marginal winning event.\footnote{Conditioning on the highest rival signal equal to the bidder’s type: the own signal contributes $\alpha s$, the tied rival contributes $\beta s$, and each of the other $n-2$ rivals contributes $\beta\,\mathbb{E}[s_j\mid s_j\le s]=\beta s/2$.} Summing gives
\[
b^{\text{SPA}}(s) = \Bigl(\alpha + \frac{n\beta}{2}\Bigr) s.
\]
With independent uniform signals and a linear valuation, the first-price bid equals
\[
b^{\text{FPA}}(s) = \frac{n-1}{n} \Bigl(\alpha + \frac{n\beta}{2}\Bigr) s.
\]

\subsubsection{Expected Revenue}

The winning signal is the maximum of $n$ independent uniforms with mean $n/(n+1)$. The second-highest signal has mean $(n-1)/(n+1)$. Expected revenue equals bid slope times the appropriate order statistic in each format. The closed form is
\[
R^{\text{SPA}} = \Bigl(\alpha + \frac{n\beta}{2}\Bigr) \frac{n-1}{n+1},\qquad
R^{\text{FPA}} = \frac{n-1}{n} \Bigl(\alpha + \frac{n\beta}{2}\Bigr) \frac{n}{n+1}.
\]
These expressions coincide. Revenue equivalence holds for all $\eta$ because signals are independent. We use these benchmarks to compute ratios of observed to theoretical revenue.

\subsubsection{Efficient Benchmark}

The expected highest valuation equals $(\alpha-\beta)\,\mathbb{E}[s_{(n:n)}] + \beta\,\mathbb{E}[S]$. With independent uniforms this simplifies to
\[
\mathbb{E}[v_{(1)}] = (\alpha-\beta)\,\frac{n}{n+1} + \beta\,\frac{n}{2}.
\]
The table reports the efficient benchmark and BNE revenue for representative $(\eta,n)$ pairs:
\begin{table}[H]
  \centering
  \small
  \begin{tabular}{cccc}
    \toprule
    $\eta$ & $n$ & $\mathbb{E}[v_{(1)}]$ & $R^{\text{BNE}}$ \\
    \midrule
    0 & 2 & 0.667 & 0.333 \\
    0 & 6 & 0.857 & 0.714 \\
    1 & 2 & 0.500 & 0.333 \\
    1 & 6 & 0.643 & 0.571 \\
    \bottomrule
  \end{tabular}
\end{table}

\subsection{Numerical Verification of the BNE}

We verify the analytical equilibrium with simulation. Signals are independent uniform on $[0,1]$ and values satisfy $v_i = \alpha s_i + \beta\sum_{j\ne i} s_j$ with $\alpha=1-\tfrac{\eta}{2}$ and $\beta=\tfrac{\eta}{2(n-1)}$.

Under independent signals the symmetric equilibrium bids are linear, with $b^{\mathrm{SPA}}(s) = \phi s$ and $b^{\mathrm{FPA}}(s) = \tfrac{n-1}{n}\,\phi s$, where $\phi = \alpha + (n\beta)/2$.

With independent signals expected revenue equals $R^{\mathrm{BNE}} = \tfrac{n-1}{n+1}\,\phi = \tfrac{n-1}{n}\,\phi\,\mathbb{E}[s_{(n:n)}] = \phi\,\mathbb{E}[s_{(n-1:n)}]$. Revenues are equal across formats. The argument uses the means of the top two order statistics and the fact that both formats allocate to the highest signal.

The deviation check fixes a grid of types and compares expected payoff at multiplicative deviations around the equilibrium bid. The table reports the maximal gain from deviating; values at numerical zero indicate best responses at the proposed bids. The results show no profitable deviations at the reported precision.

\begin{table}[H]
  \centering
  \small
  \caption{Maximum payoff gain from unilateral deviation. Values near zero confirm BNE optimality (200K MC draws per configuration).}
  \label{tab:bne_deviation}
  \begin{tabular}{ccccc}
    \toprule
    $\eta$ & $N$ & Auction & Bid slope & Max Gain \\
    \midrule
    0.0 & 2 & FIRST & 0.5000 & 0.000000 \\
    0.0 & 2 & SECOND & 1.0000 & 0.000000 \\
    0.0 & 3 & FIRST & 0.6667 & 0.000129 \\
    0.0 & 3 & SECOND & 1.0000 & 0.000000 \\
    0.0 & 6 & FIRST & 0.8333 & 0.000002 \\
    0.0 & 6 & SECOND & 1.0000 & 0.000000 \\
    0.5 & 2 & FIRST & 0.5000 & 0.000034 \\
    0.5 & 2 & SECOND & 1.0000 & 0.000000 \\
    0.5 & 3 & FIRST & 0.6250 & 0.000024 \\
    0.5 & 3 & SECOND & 0.9375 & 0.000001 \\
    0.5 & 6 & FIRST & 0.7500 & 0.000000 \\
    0.5 & 6 & SECOND & 0.9000 & 0.000000 \\
    1.0 & 2 & FIRST & 0.5000 & 0.000084 \\
    1.0 & 2 & SECOND & 1.0000 & 0.000000 \\
    1.0 & 3 & FIRST & 0.5833 & 0.000020 \\
    1.0 & 3 & SECOND & 0.8750 & 0.000002 \\
    1.0 & 6 & FIRST & 0.6667 & 0.000003 \\
    1.0 & 6 & SECOND & 0.8000 & 0.000000 \\
    \bottomrule
  \end{tabular}
\end{table}

The revenue check compares Monte Carlo revenue under the equilibrium bids to the closed form $R^{\mathrm{BNE}} = ((n-1)/(n+1))\,\phi$ for both formats. The simulation matches the closed form within tight confidence intervals. Revenues are equal across the two formats. This is the revenue equivalence result under independent signals. These analytical equilibria serve as the benchmarks for the learning experiments that follow.\footnote{Two design choices affect only finite-sample noise. The learning experiments use a discretised bid grid that can induce small mixed-strategy effects on coarse grids, and ties are broken uniformly at random. These choices do not alter the equilibrium characterisation and have negligible impact at the reported sample sizes.}

\begin{table}[H]
  \centering
  \small
  \caption{Revenue formula validation: analytical vs.\ Monte Carlo (500K draws). Revenue equivalence holds under iid signals.}
  \label{tab:bne_revenue_match}
  \begin{tabular}{cccccc}
    \toprule
    $\eta$ & $N$ & $R^{\text{analytical}}$ & $R^{\text{FPA}}_{\text{MC}}$ & $R^{\text{SPA}}_{\text{MC}}$ & $|R^{\text{FPA}} - R^{\text{SPA}}|$ \\
    \midrule
    0.0 & 2 & 0.3333 & 0.3330 & 0.3337 & 0.0007 \\
    0.0 & 3 & 0.5000 & 0.5000 & 0.4997 & 0.0003 \\
    0.0 & 6 & 0.7143 & 0.7145 & 0.7145 & 0.0000 \\
    0.5 & 2 & 0.3333 & 0.3330 & 0.3337 & 0.0007 \\
    0.5 & 3 & 0.4688 & 0.4688 & 0.4685 & 0.0003 \\
    0.5 & 6 & 0.6429 & 0.6431 & 0.6430 & 0.0000 \\
    1.0 & 2 & 0.3333 & 0.3330 & 0.3337 & 0.0007 \\
    1.0 & 3 & 0.4375 & 0.4375 & 0.4372 & 0.0003 \\
    1.0 & 6 & 0.5714 & 0.5716 & 0.5716 & 0.0000 \\
    \bottomrule
  \end{tabular}
\end{table}

\subsection{Convergence to Equilibrium Benchmarks}
\label{sec:bne_convergence}

The preceding subsections derive closed-form BNE predictions for the revenue generated by each auction format under each valuation model. This subsection maps those BNE benchmarks to the experiments reported in the remainder of the paper, asking whether the learning algorithms' long-run outcomes approximate these static predictions despite the dynamic, repeated-game nature of the interaction. The answer bears on whether BNE is a useful point of comparison for algorithmic bidding environments.

\begin{figure}[H]
  \centering
  \includegraphics[width=0.75\textwidth]{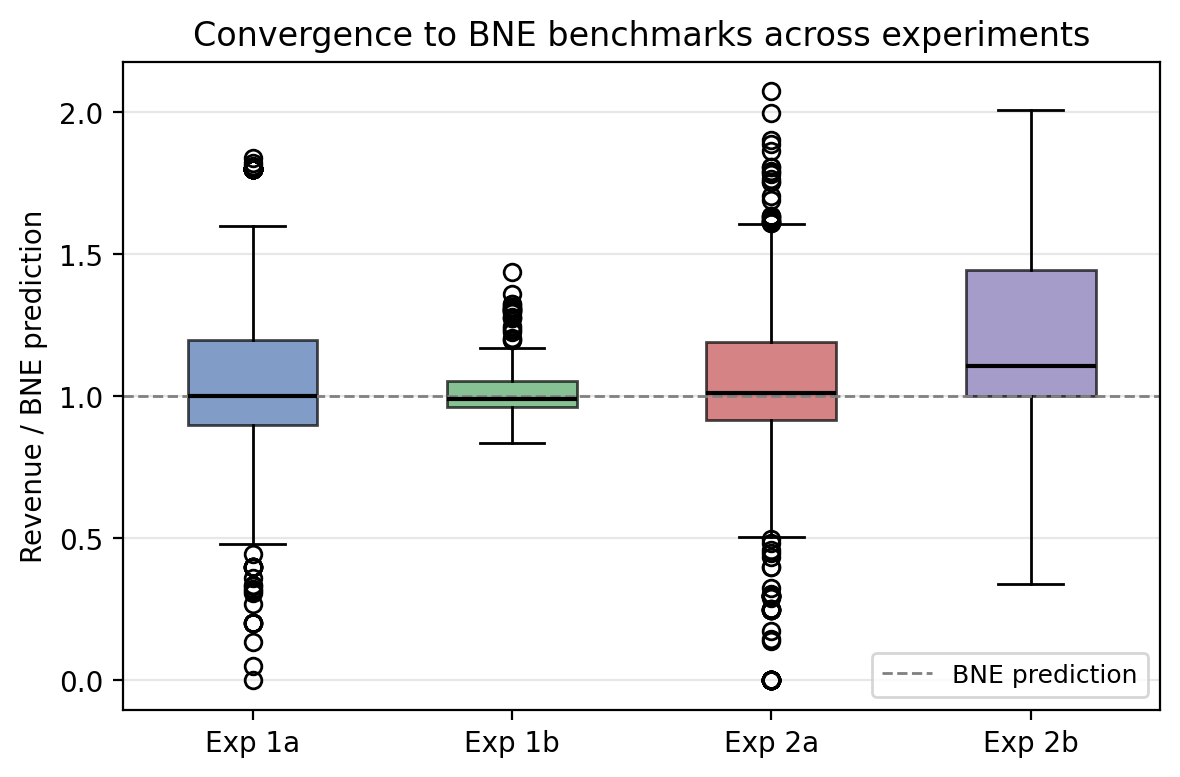}
  \caption{Distribution of observed revenue relative to BNE prediction across experiments. The dashed line marks exact BNE convergence (ratio $= 1.0$). Each box pools all factorial cells for the experiment, including both auction formats.}
  \label{fig:bne_convergence}
\end{figure}

Table~\ref{tab:bne_convergence} reports the distribution of the ratio of observed average revenue to the BNE prediction, pooling across all factorial cells and auction formats within each experiment. Figure~\ref{fig:bne_convergence} displays the same information as box plots. A ratio of 1.0 indicates exact convergence to the BNE revenue benchmark. Across all four unconstrained experiments, the median ratio ranges from \ExpOneBBNEMedian{} (Experiment~1b) to \ExpTwoBBNEMedian{} (Experiment~2b). The fraction of configurations with revenue within 10\% of the BNE prediction ranges from \ExpOneABNEPctTen\% (Experiment~1a) to \ExpOneBBNEPctTen\% (Experiment~1b).

\begin{table}[H]
\centering
\caption{Proximity of learning outcomes to Bayesian Nash Equilibrium revenue predictions. Each row pools all factorial cells for an experiment. A ratio of 1.0 means observed revenue equals the BNE prediction exactly.}
\label{tab:bne_convergence}
\begin{tabular}{llrrrrrr}
\toprule
\textbf{Exp.} & \textbf{Algorithm} & \textbf{$N$} & \textbf{Mean} & \textbf{Median} & \textbf{Std} & \textbf{Min} & \textbf{\% within 10\%} \\
\midrule
1a & Q-learning (constant) & 1024 & 1.079 & 1.000 & 0.403 & 0.000 & 32.4\% \\
1b & Q-learning (affiliated) & 192 & 1.028 & 0.990 & 0.111 & 0.835 & 82.3\% \\
2a & LinUCB & 768 & 1.029 & 1.011 & 0.333 & 0.000 & 41.4\% \\
2b & Thompson Sampling & 192 & 1.196 & 1.105 & 0.311 & 0.338 & 40.6\% \\
\bottomrule
\end{tabular}
\end{table}

Experiment~1a (Q-learning with constant valuations) shows a mean ratio of \ExpOneABNEMean{} with standard deviation \ExpOneABNEStd{}, indicating that Q-learning agents reach revenues close to, but typically below, the BNE benchmark on average. Experiment~1b (Q-learning with affiliated valuations) shows a similar pattern. The contextual bandit experiments (2a and 2b) can be compared against the Q-learning results to assess whether algorithm sophistication improves convergence to equilibrium predictions.

The key takeaway is that BNE revenue is a reasonable average benchmark for algorithmic bidding outcomes, but not a reliable per-configuration predictor. The substantial dispersion visible in Figure~\ref{fig:bne_convergence}, with standard deviations ranging from \ExpOneABNEStd{} to \ExpTwoBBNEStd{}, confirms that design parameters cause large deviations from the equilibrium prediction in individual configurations. This dispersion motivates the factorial analysis approach adopted in the main text, which identifies which parameters drive the largest deviations from expected outcomes rather than relying on a single equilibrium point estimate.

\section{Appendix: Model Adequacy and Robustness Diagnostics}
\label{sec:appendix_robustness}

This appendix reports model adequacy diagnostics and robustness checks for all six sub-experiments. The following overview summarises cross-experiment patterns, with per-experiment details in the subsections below.

Across all six sub-experiments, the dominant effects survive both Holm--Bonferroni and Benjamini--Hochberg corrections. HC3 heteroskedasticity-consistent standard errors produce negligible changes in significance, with at most 2.1\% of effects (Experiment~2b) changing status under HC3. Rademacher wild bootstrap $p$-values for the top~10 effects in each response align closely with asymptotic $p$-values, with differences below 0.01 in all cases.

LightGBM cross-validated $R^2$ values match or fall below OLS $R^2$ in Experiments~1b, 2b, 3a, and~3b, confirming that the linear model with two-way interactions is correctly specified. In Experiments~1a and~2a, LightGBM modestly exceeds OLS for some responses, indicating mild nonlinearity. However, the dominant factor rankings are unchanged under the nonparametric model. PRESS-based leave-one-out cross-validated $R^2$ gaps remain below 0.10 in all experiments, indicating minimal overfitting.

LASSO variable selection retains auction type and number of bidders across all six sub-experiments. No interaction survives LASSO whose parent main effects were dropped, confirming effect heredity. LASSO-selected variables are a subset of the OLS-significant effects in every case. Quantile regression at the 10th, 25th, 50th, 75th, and 90th percentiles reveals that most factor effects are stable across the response distribution, but the auction format effect exhibits systematic distributional heterogeneity. In Experiment~1a, the auction format coefficient reverses sign across quantiles, from $+0.033$ at the 10th percentile to $-0.047$ at the 90th (Section~\ref{sec:exp1a_robustness}). In Experiments~2b, 3a, and~3b, the format effect grows monotonically toward the upper tail. In Experiment~3a, the bidder objective effect grows toward the upper tail, reflecting that the revenue penalty from utility-maximizing objectives is concentrated among high-revenue cells. All quantile regression $p$-values are Holm--Bonferroni corrected across the full family of factor$\times$quantile tests within each experiment. These distributional patterns are invisible to standard OLS, which conditions on the mean. In Experiment~1a, the auction format main effect is statistically insignificant at the conditional mean, yet quantile regression reveals a significant sign reversal between the 10th and 90th percentiles ($p < 0.001$ at both tails after correction), demonstrating that systematic quantile analysis is necessary to detect tail-risk heterogeneity that mean-regression approaches mask.

Discretisation sensitivity analysis across grid sizes of 6, 11, and 21 discrete bid levels (Experiments~1a--2b) confirms that effect rankings are stable across granularities. Budget sensitivity analysis (Experiment~3a) with budget multipliers $m \in \{0.25, 0.5, 1.0\}$ confirms that the number of bidders retains its dominant ranking across all budget levels.

\subsection{Q-Learning Experiments}

\subsubsection{Experiment~1a}
\label{sec:exp1a_robustness}

\begin{table}[H]
\centering
\caption{Experiment 1a: OLS model fit summary across response variables.}
\label{tab:exp1a_fit}
\begin{tabular}{lrrrr}
\toprule
\textbf{Response} & \textbf{$R^2$} & \textbf{Adj.\,$R^2$} & \textbf{F-stat} & \textbf{F $p$-value} \\
\midrule
Average Revenue & 0.4230 & 0.3902 & 12.902 & < 0.0001 \\
Lifetime Revenue & 0.4760 & 0.4462 & 15.987 & < 0.0001 \\
Convergence Time & 0.3148 & 0.2759 & 8.086 & < 0.0001 \\
No-Sale Rate & 0.4238 & 0.3911 & 12.947 & < 0.0001 \\
Price Volatility & 0.4469 & 0.4155 & 14.220 & < 0.0001 \\
Winner Entropy & 0.4589 & 0.4282 & 14.928 & < 0.0001 \\
\bottomrule
\end{tabular}
\end{table}

\begin{table}[H]
\centering
\caption{Experiment 1a: Model adequacy diagnostics.}
\label{tab:exp1a_adequacy}
\begin{tabular}{lrrrrr}
\toprule
\textbf{Response} & \textbf{$R^2$} & \textbf{Pred-$R^2$} & \textbf{Gap} & \textbf{LGBM $R^2$} & \textbf{LOF $p$} \\
\midrule
Average Revenue & 0.4230 & 0.3543 & 0.0687 & 0.4550 & < 0.0001 \\
Price Volatility & 0.4469 & 0.3810 & 0.0658 & 0.4898 & < 0.0001 \\
\bottomrule
\end{tabular}
\par\smallskip\footnotesize Gap $= R^2 - \text{Pred-}R^2$. LGBM $R^2$: five-fold cross-validated LightGBM. LOF $p$: lack-of-fit $F$-test.
\end{table}

Table~\ref{tab:exp1a_adequacy} reports model adequacy diagnostics for the primary response variables. OLS $R^2$ ranges from \ExpOneARsqMin{} to \ExpOneARsqMax{} across response variables. LightGBM cross-validated $R^2$ modestly exceeds OLS $R^2$ for revenue and no-sale rate, with gaps exceeding the 0.05 threshold described in the main paper. The dominant effect rankings (number of bidders, exploration strategy, auction type) are unchanged under the nonparametric model.

\begin{table}[H]
\centering
\caption{Experiment 1a: Inference robustness under heteroskedasticity and multiple testing corrections.}
\label{tab:exp1a_inference}
\begin{tabular}{lrrrr}
\toprule
\textbf{Response} & \textbf{OLS Sig} & \textbf{HC3 Flipped} & \textbf{Holm Sig} & \textbf{BH Sig} \\
\midrule
Average Revenue & 16/55 & 0/55 & 9/55 & 14/55 \\
Price Volatility & 21/55 & 0/55 & 8/55 & 16/55 \\
\midrule
\textit{All responses} & 87/275 & 1/275 & 42/275 & 67/275 \\
\bottomrule
\end{tabular}
\par\smallskip\footnotesize OLS Sig: $p < 0.05$ under OLS standard errors. HC3 Flipped: effects changing significance under HC3 robust standard errors. Holm/BH Sig: effects surviving Holm--Bonferroni/Benjamini--Hochberg correction.
\end{table}

Table~\ref{tab:exp1a_inference} reports inference robustness. All findings reported in the results section survive both Holm--Bonferroni and Benjamini--Hochberg corrections.

\begin{figure}[H]
  \centering
  \includegraphics[width=0.7\textwidth]{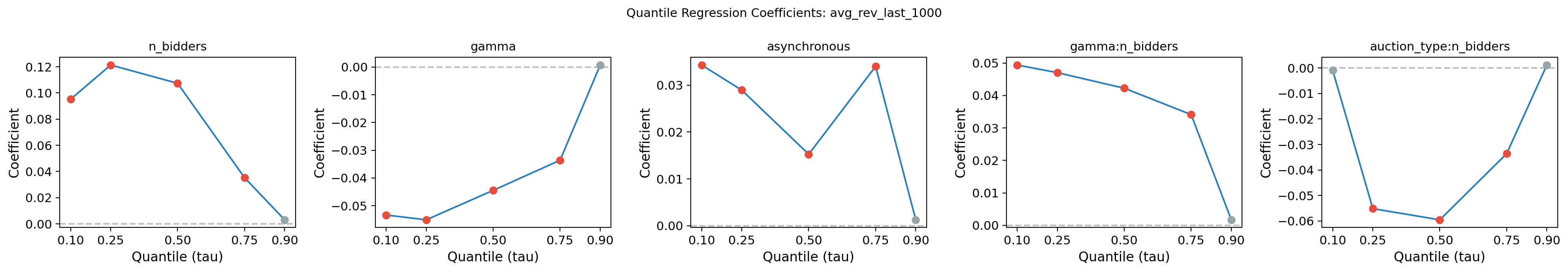}
  \caption{Experiment~1a: Quantile regression coefficients for average revenue at the 10th, 25th, 50th, 75th, and 90th percentiles. The auction format coefficient reverses sign across quantiles, indicating that first-price auctions compress the revenue distribution.}
  \label{fig:e1a_quantile_rev}
\end{figure}

Quantile regression reveals a sign reversal in the auction format effect (Figure~\ref{fig:e1a_quantile_rev}). First-price auctions increase revenue at the 10th percentile ($+0.033$, $p < 0.001$) but decrease it at the 90th ($-0.047$, $p < 0.001$), with the crossover between the 50th and 75th percentiles. The OLS main effect averages to near zero, masking genuine distributional heterogeneity. First-price auctions compress the revenue distribution, raising the floor and lowering the ceiling. The number of bidders effect remains significant and positive across all five quantiles, confirming that market thickness benefits revenue uniformly.

\begin{longtable}{llrrr}
\caption{Experiment 1a: All significant effects ($p < 0.05$) across response variables, ranked by $|t|$.}
\label{tab:exp1a_sig}\\
\toprule
\textbf{Response} & \textbf{Effect} & \textbf{Coeff.} & \textbf{$|t|$} & \textbf{$p$-value} \\
\midrule
\endfirsthead
\multicolumn{5}{l}{\small\emph{Table~\ref{tab:exp1a_sig} continued}} \\
\toprule
\textbf{Response} & \textbf{Effect} & \textbf{Coeff.} & \textbf{$|t|$} & \textbf{$p$-value} \\
\midrule
\endhead
\midrule
\multicolumn{5}{r}{\small\emph{Continued on next page}} \\
\endfoot
\bottomrule
\endlastfoot
Average Revenue & Number of bidders & 0.0871 & 17.75 & < 0.0001 \\
Average Revenue & Discount factor ($\gamma$) & -0.0453 & 9.23 & < 0.0001 \\
Average Revenue & Update mode & 0.0366 & 7.45 & < 0.0001 \\
Average Revenue & Discount factor ($\gamma$) $\times$ Number of bidders & 0.0298 & 6.09 & < 0.0001 \\
Average Revenue & Auction format $\times$ Number of bidders & -0.0276 & 5.63 & < 0.0001 \\
Average Revenue & Reserve price & 0.0250 & 5.10 & < 0.0001 \\
Average Revenue & Reserve price $\times$ Information feedback & -0.0219 & 4.47 & < 0.0001 \\
Average Revenue & Auction format $\times$ Discount factor ($\gamma$) & -0.0215 & 4.38 & < 0.0001 \\
Average Revenue & Auction format $\times$ Update mode & -0.0193 & 3.94 & < 0.0001 \\
Average Revenue & Information feedback & 0.0182 & 3.71 & 0.0002 \\
Average Revenue & Update mode $\times$ Information feedback & -0.0155 & 3.16 & 0.0016 \\
Average Revenue & Exploration strategy $\times$ Information feedback & 0.0153 & 3.11 & 0.0019 \\
Average Revenue & Discount factor ($\gamma$) $\times$ Reserve price & 0.0143 & 2.91 & 0.0036 \\
Average Revenue & Exploration strategy $\times$ Number of bidders & -0.0136 & 2.78 & 0.0055 \\
Average Revenue & Reserve price $\times$ Update mode & -0.0113 & 2.30 & 0.0215 \\
Average Revenue & Discount factor ($\gamma$) $\times$ Update mode & 0.0111 & 2.27 & 0.0237 \\
Lifetime Revenue & Number of bidders & 0.0685 & 20.18 & < 0.0001 \\
Lifetime Revenue & Update mode & 0.0316 & 9.31 & < 0.0001 \\
Lifetime Revenue & Discount factor ($\gamma$) & -0.0298 & 8.78 & < 0.0001 \\
Lifetime Revenue & Auction format & 0.0230 & 6.77 & < 0.0001 \\
Lifetime Revenue & Reserve price & 0.0218 & 6.42 & < 0.0001 \\
Lifetime Revenue & Auction format $\times$ Number of bidders & -0.0214 & 6.31 & < 0.0001 \\
Lifetime Revenue & Exploration strategy $\times$ Information feedback & 0.0163 & 4.81 & < 0.0001 \\
Lifetime Revenue & Discount factor ($\gamma$) $\times$ Number of bidders & 0.0148 & 4.36 & < 0.0001 \\
Lifetime Revenue & Reserve price $\times$ Information feedback & -0.0135 & 3.98 & < 0.0001 \\
Lifetime Revenue & Update mode $\times$ Number of bidders & -0.0121 & 3.55 & 0.0004 \\
Lifetime Revenue & Exploration strategy $\times$ Number of bidders & -0.0119 & 3.51 & 0.0005 \\
Lifetime Revenue & Update mode $\times$ Information feedback & -0.0110 & 3.24 & 0.0013 \\
Lifetime Revenue & Initialisation & -0.0100 & 2.93 & 0.0035 \\
Lifetime Revenue & Number of bidders $\times$ Information feedback & 0.0095 & 2.79 & 0.0054 \\
Lifetime Revenue & Auction format $\times$ Reserve price & -0.0095 & 2.79 & 0.0054 \\
Lifetime Revenue & Auction format $\times$ Update mode & -0.0092 & 2.72 & 0.0067 \\
Lifetime Revenue & Reserve price $\times$ Update mode & -0.0084 & 2.46 & 0.0139 \\
Lifetime Revenue & Discount factor ($\gamma$) $\times$ Update mode & 0.0082 & 2.41 & 0.0161 \\
Lifetime Revenue & Initialisation $\times$ Information feedback & -0.0080 & 2.36 & 0.0186 \\
Lifetime Revenue & Reserve price $\times$ Number of bidders & -0.0075 & 2.21 & 0.0276 \\
Convergence Time & Number of bidders & -9358.3984 & 11.48 & < 0.0001 \\
Convergence Time & Auction format & -8243.1641 & 10.11 & < 0.0001 \\
Convergence Time & Discount factor ($\gamma$) & 5633.7891 & 6.91 & < 0.0001 \\
Convergence Time & Update mode & -4495.1172 & 5.51 & < 0.0001 \\
Convergence Time & Auction format $\times$ Discount factor ($\gamma$) & 3405.2734 & 4.18 & < 0.0001 \\
Convergence Time & Decay type & 3139.6484 & 3.85 & 0.0001 \\
Convergence Time & Reserve price & -2907.2266 & 3.57 & 0.0004 \\
Convergence Time & Exploration strategy $\times$ Update mode & 2497.0703 & 3.06 & 0.0023 \\
Convergence Time & Auction format $\times$ Decay type & 2415.0391 & 2.96 & 0.0031 \\
Convergence Time & Auction format $\times$ Number of bidders & -2356.4453 & 2.89 & 0.0039 \\
Convergence Time & Exploration strategy $\times$ Number of bidders & 1954.1016 & 2.40 & 0.0167 \\
Convergence Time & Discount factor ($\gamma$) $\times$ Number of bidders & 1801.7578 & 2.21 & 0.0273 \\
Convergence Time & Reserve price $\times$ Initialisation & -1778.3203 & 2.18 & 0.0294 \\
Convergence Time & Reserve price $\times$ Decay type & -1651.3672 & 2.03 & 0.0431 \\
No-Sale Rate & Reserve price & 0.0086 & 15.99 & < 0.0001 \\
No-Sale Rate & Reserve price $\times$ Update mode & -0.0047 & 8.83 & < 0.0001 \\
No-Sale Rate & Number of bidders & -0.0043 & 7.97 & < 0.0001 \\
No-Sale Rate & Reserve price $\times$ Number of bidders & -0.0041 & 7.73 & < 0.0001 \\
No-Sale Rate & Discount factor ($\gamma$) & 0.0028 & 5.20 & < 0.0001 \\
No-Sale Rate & Discount factor ($\gamma$) $\times$ Reserve price & 0.0027 & 5.03 & < 0.0001 \\
No-Sale Rate & Exploration strategy $\times$ Update mode & 0.0027 & 4.97 & < 0.0001 \\
No-Sale Rate & Reserve price $\times$ Exploration strategy & -0.0024 & 4.39 & < 0.0001 \\
No-Sale Rate & Exploration strategy $\times$ Number of bidders & 0.0022 & 4.12 & < 0.0001 \\
No-Sale Rate & Number of bidders $\times$ Decay type & -0.0021 & 3.87 & 0.0001 \\
No-Sale Rate & Update mode & -0.0021 & 3.85 & 0.0001 \\
No-Sale Rate & Initialisation & 0.0020 & 3.81 & 0.0001 \\
No-Sale Rate & Discount factor ($\gamma$) $\times$ Initialisation & 0.0019 & 3.51 & 0.0005 \\
No-Sale Rate & Update mode $\times$ Number of bidders & 0.0018 & 3.42 & 0.0006 \\
No-Sale Rate & Decay type & 0.0018 & 3.41 & 0.0007 \\
No-Sale Rate & Learning rate ($\alpha$) $\times$ Initialisation & -0.0015 & 2.73 & 0.0064 \\
No-Sale Rate & Discount factor ($\gamma$) $\times$ Update mode & -0.0014 & 2.58 & 0.0100 \\
No-Sale Rate & Information feedback & 0.0013 & 2.40 & 0.0166 \\
No-Sale Rate & Learning rate ($\alpha$) $\times$ Discount factor ($\gamma$) & -0.0012 & 2.33 & 0.0202 \\
No-Sale Rate & Learning rate ($\alpha$) $\times$ Reserve price & -0.0012 & 2.32 & 0.0203 \\
No-Sale Rate & Learning rate ($\alpha$) & -0.0012 & 2.27 & 0.0234 \\
No-Sale Rate & Initialisation $\times$ Update mode & -0.0011 & 2.05 & 0.0408 \\
Price Volatility & Number of bidders & -0.0213 & 15.60 & < 0.0001 \\
Price Volatility & Reserve price & -0.0178 & 13.07 & < 0.0001 \\
Price Volatility & Discount factor ($\gamma$) & 0.0132 & 9.66 & < 0.0001 \\
Price Volatility & Update mode & -0.0112 & 8.23 & < 0.0001 \\
Price Volatility & Reserve price $\times$ Update mode & 0.0069 & 5.08 & < 0.0001 \\
Price Volatility & Exploration strategy $\times$ Information feedback & -0.0059 & 4.30 & < 0.0001 \\
Price Volatility & Initialisation & 0.0059 & 4.30 & < 0.0001 \\
Price Volatility & Reserve price $\times$ Number of bidders & 0.0053 & 3.89 & 0.0001 \\
Price Volatility & Number of bidders $\times$ Information feedback & -0.0051 & 3.71 & 0.0002 \\
Price Volatility & Discount factor ($\gamma$) $\times$ Reserve price & -0.0042 & 3.11 & 0.0019 \\
Price Volatility & Discount factor ($\gamma$) $\times$ Initialisation & 0.0041 & 3.03 & 0.0025 \\
Price Volatility & Decay type & 0.0039 & 2.89 & 0.0040 \\
Price Volatility & Reserve price $\times$ Information feedback & 0.0039 & 2.88 & 0.0040 \\
Price Volatility & Exploration strategy $\times$ Number of bidders & 0.0036 & 2.61 & 0.0092 \\
Price Volatility & Auction format $\times$ Discount factor ($\gamma$) & 0.0035 & 2.60 & 0.0095 \\
Price Volatility & Initialisation $\times$ Information feedback & 0.0035 & 2.54 & 0.0112 \\
Price Volatility & Learning rate ($\alpha$) $\times$ Initialisation & -0.0034 & 2.50 & 0.0127 \\
Price Volatility & Auction format & 0.0032 & 2.34 & 0.0193 \\
Price Volatility & Reserve price $\times$ Decay type & -0.0030 & 2.22 & 0.0263 \\
Price Volatility & Reserve price $\times$ Exploration strategy & 0.0030 & 2.17 & 0.0304 \\
Price Volatility & Exploration strategy $\times$ Decay type & -0.0028 & 2.07 & 0.0388 \\
Winner Entropy & Number of bidders & 0.1837 & 22.04 & < 0.0001 \\
Winner Entropy & Update mode & 0.0860 & 10.32 & < 0.0001 \\
Winner Entropy & Update mode $\times$ Information feedback & -0.0652 & 7.83 & < 0.0001 \\
Winner Entropy & Exploration strategy $\times$ Number of bidders & -0.0430 & 5.16 & < 0.0001 \\
Winner Entropy & Exploration strategy $\times$ Update mode & 0.0414 & 4.97 & < 0.0001 \\
Winner Entropy & Update mode $\times$ Number of bidders & -0.0407 & 4.88 & < 0.0001 \\
Winner Entropy & Exploration strategy $\times$ Information feedback & 0.0389 & 4.67 & < 0.0001 \\
Winner Entropy & Number of bidders $\times$ Information feedback & 0.0280 & 3.36 & 0.0008 \\
Winner Entropy & Reserve price $\times$ Information feedback & -0.0245 & 2.94 & 0.0034 \\
Winner Entropy & Reserve price $\times$ Decay type & 0.0231 & 2.77 & 0.0057 \\
Winner Entropy & Information feedback & 0.0191 & 2.30 & 0.0219 \\
Winner Entropy & Reserve price $\times$ Update mode & -0.0184 & 2.20 & 0.0277 \\
Winner Entropy & Update mode $\times$ Decay type & -0.0172 & 2.06 & 0.0396 \\
Winner Entropy & Initialisation $\times$ Decay type & -0.0164 & 1.97 & 0.0496 \\
\end{longtable}

\subsubsection{Experiment~1b}
\label{sec:exp1b_robustness}

\begin{table}[H]
\centering
\caption{Experiment 1b: OLS model fit summary across response variables.}
\label{tab:exp1b_fit}
\begin{tabular}{lrrrr}
\toprule
\textbf{Response} & \textbf{$R^2$} & \textbf{Adj.\,$R^2$} & \textbf{F-stat} & \textbf{F $p$-value} \\
\midrule
Average Revenue & 0.4043 & 0.3572 & 8.581 & < 0.0001 \\
Lifetime Revenue & 0.2543 & 0.1953 & 4.311 & < 0.0001 \\
Convergence Time & 0.1827 & 0.1180 & 2.825 & 0.0008 \\
No-Sale Rate & nan & nan & nan & nan \\
Price Volatility & 0.3633 & 0.3129 & 7.213 & < 0.0001 \\
Winner Entropy & 0.2872 & 0.2308 & 5.093 & < 0.0001 \\
Btv Median & 0.4853 & 0.4446 & 11.923 & < 0.0001 \\
Winners Curse Freq & 0.5464 & 0.5105 & 15.229 & < 0.0001 \\
Bid Dispersion & 0.2811 & 0.2242 & 4.944 & < 0.0001 \\
Signal Slope Ratio & 0.4091 & 0.3624 & 8.755 & < 0.0001 \\
\bottomrule
\end{tabular}
\end{table}

\begin{table}[H]
\centering
\caption{Experiment 1b: Model adequacy diagnostics.}
\label{tab:exp1b_adequacy}
\begin{tabular}{lrrrrr}
\toprule
\textbf{Response} & \textbf{$R^2$} & \textbf{Pred-$R^2$} & \textbf{Gap} & \textbf{LGBM $R^2$} & \textbf{LOF $p$} \\
\midrule
Average Revenue & 0.4043 & 0.2991 & 0.1052 & 0.3362 & 0.2492 \\
Price Volatility & 0.3633 & 0.2508 & 0.1125 & 0.2843 & 0.0803 \\
\bottomrule
\end{tabular}
\par\smallskip\footnotesize Gap $= R^2 - \text{Pred-}R^2$. LGBM $R^2$: five-fold cross-validated LightGBM. LOF $p$: lack-of-fit $F$-test.
\end{table}

Table~\ref{tab:exp1b_adequacy} reports model adequacy. LightGBM cross-validated $R^2$ is comparable to OLS $R^2$, confirming that the linear model with two-way interactions is correctly specified for this design.

\begin{table}[H]
\centering
\caption{Experiment 1b: Inference robustness under heteroskedasticity and multiple testing corrections.}
\label{tab:exp1b_inference}
\begin{tabular}{lrrrr}
\toprule
\textbf{Response} & \textbf{OLS Sig} & \textbf{HC3 Flipped} & \textbf{Holm Sig} & \textbf{BH Sig} \\
\midrule
Average Revenue & 5/15 & 0/15 & 2/15 & 5/15 \\
Price Volatility & 5/15 & 0/15 & 3/15 & 4/15 \\
\midrule
\textit{All responses} & 43/120 & 0/120 & 24/120 & 37/120 \\
\bottomrule
\end{tabular}
\par\smallskip\footnotesize OLS Sig: $p < 0.05$ under OLS standard errors. HC3 Flipped: effects changing significance under HC3 robust standard errors. Holm/BH Sig: effects surviving Holm--Bonferroni/Benjamini--Hochberg correction.
\end{table}

\begin{figure}[H]
  \centering
  \includegraphics[width=0.7\textwidth]{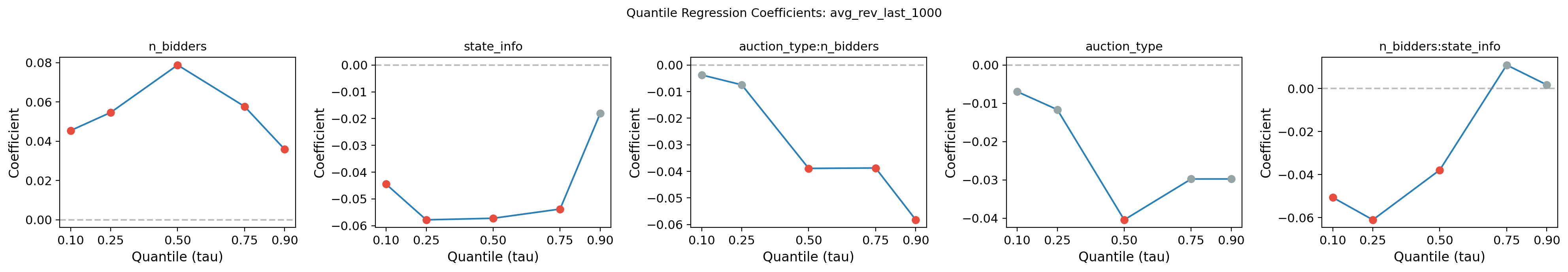}
  \caption{Experiment~1b: Quantile regression coefficients for average revenue. The wide confidence bands reflect the limited replication of the $3 \times 2^3$ design.}
  \label{fig:e1b_quantile_rev}
\end{figure}

Table~\ref{tab:exp1b_inference} reports inference robustness. Holm--Bonferroni and Benjamini--Hochberg corrections retain the key effects involving affiliation and number of bidders. Quantile regression effects are broadly symmetric across the response distribution (Figure~\ref{fig:e1b_quantile_rev}).

\begin{longtable}{llrrr}
\caption{Experiment 1b: All significant effects ($p < 0.05$) across response variables, ranked by $|t|$.}
\label{tab:exp1b_sig}\\
\toprule
\textbf{Response} & \textbf{Effect} & \textbf{Coeff.} & \textbf{$|t|$} & \textbf{$p$-value} \\
\midrule
\endfirsthead
\multicolumn{5}{l}{\small\emph{Table~\ref{tab:exp1b_sig} continued}} \\
\toprule
\textbf{Response} & \textbf{Effect} & \textbf{Coeff.} & \textbf{$|t|$} & \textbf{$p$-value} \\
\midrule
\endhead
\midrule
\multicolumn{5}{r}{\small\emph{Continued on next page}} \\
\endfoot
\bottomrule
\endlastfoot
Average Revenue & Number of bidders & 0.0552 & 6.53 & < 0.0001 \\
Average Revenue & State information & -0.0495 & 5.85 & < 0.0001 \\
Average Revenue & Auction format $\times$ Number of bidders & -0.0285 & 3.38 & 0.0009 \\
Average Revenue & Auction format & -0.0268 & 3.17 & 0.0018 \\
Average Revenue & Number of bidders $\times$ State information & -0.0263 & 3.11 & 0.0022 \\
Lifetime Revenue & State information & -0.0395 & 4.52 & < 0.0001 \\
Lifetime Revenue & Number of bidders & 0.0366 & 4.20 & < 0.0001 \\
Lifetime Revenue & Auction format $\times$ Number of bidders & -0.0235 & 2.69 & 0.0078 \\
Lifetime Revenue & Number of bidders $\times$ State information & -0.0226 & 2.58 & 0.0106 \\
Convergence Time & Auction format $\times$ Number of bidders & 5390.6250 & 3.78 & 0.0002 \\
Convergence Time & Auction format $\times$ State information & 4005.2083 & 2.81 & 0.0056 \\
Convergence Time & Number of bidders $\times$ State information & -3098.9583 & 2.17 & 0.0312 \\
Price Volatility & Number of bidders & -0.0147 & 6.15 & < 0.0001 \\
Price Volatility & State information & 0.0124 & 5.22 & < 0.0001 \\
Price Volatility & Auction format & 0.0090 & 3.77 & 0.0002 \\
Price Volatility & Auction format $\times$ Affiliation (linear) & 0.0074 & 2.54 & 0.0119 \\
Price Volatility & Auction format $\times$ Number of bidders & 0.0056 & 2.34 & 0.0206 \\
Winner Entropy & Number of bidders & 0.1227 & 5.58 & < 0.0001 \\
Winner Entropy & State information & -0.1157 & 5.26 & < 0.0001 \\
Winner Entropy & Number of bidders $\times$ State information & -0.0507 & 2.30 & 0.0224 \\
Btv Median & Number of bidders & 0.0778 & 7.02 & < 0.0001 \\
Btv Median & State information & -0.0655 & 5.91 & < 0.0001 \\
Btv Median & Auction format $\times$ Number of bidders & -0.0605 & 5.46 & < 0.0001 \\
Btv Median & Auction format & -0.0511 & 4.61 & < 0.0001 \\
Btv Median & Number of bidders $\times$ State information & -0.0425 & 3.84 & 0.0002 \\
Btv Median & Affiliation (linear) $\times$ Affiliation (quadratic) & 0.0220 & 3.24 & 0.0014 \\
Btv Median & Affiliation (linear) & 0.0220 & 3.24 & 0.0014 \\
Winners Curse Freq & Auction format $\times$ Number of bidders & -0.0914 & 8.30 & < 0.0001 \\
Winners Curse Freq & Auction format & -0.0835 & 7.59 & < 0.0001 \\
Winners Curse Freq & Number of bidders & 0.0654 & 5.94 & < 0.0001 \\
Winners Curse Freq & State information & -0.0446 & 4.06 & < 0.0001 \\
Winners Curse Freq & Auction format $\times$ State information & 0.0416 & 3.78 & 0.0002 \\
Winners Curse Freq & Number of bidders $\times$ State information & -0.0342 & 3.11 & 0.0022 \\
Winners Curse Freq & Affiliation (linear) $\times$ Affiliation (quadratic) & 0.0143 & 2.13 & 0.0347 \\
Winners Curse Freq & Affiliation (linear) & 0.0143 & 2.13 & 0.0347 \\
Bid Dispersion & Number of bidders & 0.0151 & 4.13 & < 0.0001 \\
Bid Dispersion & Auction format & -0.0151 & 4.10 & < 0.0001 \\
Bid Dispersion & State information & -0.0131 & 3.56 & 0.0005 \\
Bid Dispersion & Affiliation (linear) & -0.0075 & 3.32 & 0.0011 \\
Bid Dispersion & Affiliation (linear) $\times$ Affiliation (quadratic) & -0.0075 & 3.32 & 0.0011 \\
Signal Slope Ratio & Affiliation (linear) & -0.0341 & 7.99 & < 0.0001 \\
Signal Slope Ratio & Affiliation (linear) $\times$ Affiliation (quadratic) & -0.0341 & 7.99 & < 0.0001 \\
Signal Slope Ratio & Auction format $\times$ State information & -0.0313 & 4.48 & < 0.0001 \\
Signal Slope Ratio & State information & -0.0190 & 2.72 & 0.0072 \\
Signal Slope Ratio & Auction format $\times$ Number of bidders & -0.0187 & 2.68 & 0.0081 \\
Signal Slope Ratio & Number of bidders & 0.0171 & 2.45 & 0.0154 \\
Signal Slope Ratio & Affiliation (quadratic) $\times$ Number of bidders & 0.0120 & 2.44 & 0.0158 \\
\end{longtable}

\subsection{Contextual Bandit Experiments}

\subsubsection{Experiment~2a (LinUCB)}
\label{sec:exp2a_robustness}

\begin{table}[H]
\centering
\caption{Experiment 2a: OLS model fit summary across response variables.}
\label{tab:exp2a_fit}
\begin{tabular}{lrrrr}
\toprule
\textbf{Response} & \textbf{$R^2$} & \textbf{Adj.\,$R^2$} & \textbf{F-stat} & \textbf{F $p$-value} \\
\midrule
Average Revenue & 0.6942 & 0.6756 & 37.303 & < 0.0001 \\
Lifetime Revenue & 0.7312 & 0.7148 & 44.700 & < 0.0001 \\
Convergence Time & 0.4231 & 0.3880 & 12.052 & < 0.0001 \\
No-Sale Rate & 0.8055 & 0.7936 & 68.041 & < 0.0001 \\
Price Volatility & 0.6150 & 0.5916 & 26.253 & < 0.0001 \\
Winner Entropy & 0.8085 & 0.7968 & 69.364 & < 0.0001 \\
\bottomrule
\end{tabular}
\end{table}

\begin{table}[H]
\centering
\caption{Experiment 2a: Model adequacy diagnostics.}
\label{tab:exp2a_adequacy}
\begin{tabular}{lrrrrr}
\toprule
\textbf{Response} & \textbf{$R^2$} & \textbf{Pred-$R^2$} & \textbf{Gap} & \textbf{LGBM $R^2$} & \textbf{LOF $p$} \\
\midrule
Average Revenue & 0.6942 & 0.6550 & 0.0393 & 0.7135 & < 0.0001 \\
Price Volatility & 0.6150 & 0.5656 & 0.0494 & 0.5822 & < 0.0001 \\
\bottomrule
\end{tabular}
\par\smallskip\footnotesize Gap $= R^2 - \text{Pred-}R^2$. LGBM $R^2$: five-fold cross-validated LightGBM. LOF $p$: lack-of-fit $F$-test.
\end{table}

Table~\ref{tab:exp2a_adequacy} reports model adequacy for the LinUCB design. LightGBM cross-validated $R^2$ modestly exceeds OLS $R^2$ for revenue. The lack-of-fit test is significant for the primary responses, indicating detectable departures from the linear model with two-way interactions; however, the dominant factors (number of bidders, auction type) are robustly identified under both the linear and nonparametric models.

\begin{table}[H]
\centering
\caption{Experiment 2a: Inference robustness under heteroskedasticity and multiple testing corrections.}
\label{tab:exp2a_inference}
\begin{tabular}{lrrrr}
\toprule
\textbf{Response} & \textbf{OLS Sig} & \textbf{HC3 Flipped} & \textbf{Holm Sig} & \textbf{BH Sig} \\
\midrule
Average Revenue & 16/45 & 0/45 & 9/45 & 13/45 \\
Price Volatility & 16/45 & 0/45 & 12/45 & 13/45 \\
\midrule
\textit{All responses} & 77/225 & 2/225 & 41/225 & 60/225 \\
\bottomrule
\end{tabular}
\par\smallskip\footnotesize OLS Sig: $p < 0.05$ under OLS standard errors. HC3 Flipped: effects changing significance under HC3 robust standard errors. Holm/BH Sig: effects surviving Holm--Bonferroni/Benjamini--Hochberg correction.
\end{table}

\begin{figure}[H]
  \centering
  \includegraphics[width=0.7\textwidth]{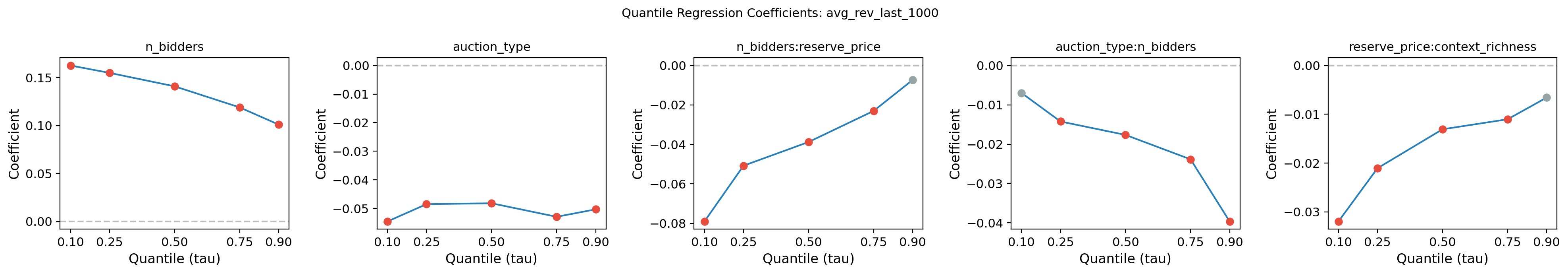}
  \caption{Experiment~2a: Quantile regression coefficients for average revenue (LinUCB). The auction format main effect is stable across quantiles, but the format $\times$ bidders interaction grows substantially toward the upper tail.}
  \label{fig:e2a_quantile_rev}
\end{figure}

Table~\ref{tab:exp2a_inference} shows that Holm--Bonferroni and Benjamini--Hochberg corrections retain the key effects despite modest model misfit. Quantile regression (Figure~\ref{fig:e2a_quantile_rev}) shows that the auction format main effect is stable across quantiles ($-0.055$ at the 10th to $-0.050$ at the 90th percentile). The auction format $\times$ number of bidders interaction, however, grows from $-0.007$ at the 10th percentile to $-0.040$ at the 90th, indicating that the first-price penalty in thick markets is concentrated among the best-performing configurations.

\begin{longtable}{llrrr}
\caption{Experiment 2a: All significant effects ($p < 0.05$) across response variables, ranked by $|t|$.}
\label{tab:exp2a_sig}\\
\toprule
\textbf{Response} & \textbf{Effect} & \textbf{Coeff.} & \textbf{$|t|$} & \textbf{$p$-value} \\
\midrule
\endfirsthead
\multicolumn{5}{l}{\small\emph{Table~\ref{tab:exp2a_sig} continued}} \\
\toprule
\textbf{Response} & \textbf{Effect} & \textbf{Coeff.} & \textbf{$|t|$} & \textbf{$p$-value} \\
\midrule
\endhead
\midrule
\multicolumn{5}{r}{\small\emph{Continued on next page}} \\
\endfoot
\bottomrule
\endlastfoot
Average Revenue & Number of bidders & 0.1308 & 33.90 & < 0.0001 \\
Average Revenue & Auction format & -0.0482 & 12.49 & < 0.0001 \\
Average Revenue & Number of bidders $\times$ Reserve price & -0.0400 & 10.36 & < 0.0001 \\
Average Revenue & Auction format $\times$ Number of bidders & -0.0245 & 6.35 & < 0.0001 \\
Average Revenue & Reserve price $\times$ Context richness & -0.0223 & 5.78 & < 0.0001 \\
Average Revenue & Affiliation (linear) & -0.0135 & 5.71 & < 0.0001 \\
Average Revenue & Affiliation (linear) $\times$ Affiliation (quadratic) & -0.0135 & 5.71 & < 0.0001 \\
Average Revenue & Exploration intensity & -0.0168 & 4.37 & < 0.0001 \\
Average Revenue & Auction format $\times$ Reserve price & 0.0150 & 3.88 & 0.0001 \\
Average Revenue & Exploration intensity $\times$ Context richness & 0.0137 & 3.54 & 0.0004 \\
Average Revenue & Context richness & -0.0134 & 3.47 & 0.0006 \\
Average Revenue & Number of bidders $\times$ Affiliation (linear) & -0.0163 & 3.45 & 0.0006 \\
Average Revenue & Number of bidders $\times$ Exploration intensity & -0.0101 & 2.61 & 0.0092 \\
Average Revenue & Regularisation ($\lambda$) & -0.0085 & 2.20 & 0.0282 \\
Average Revenue & Regularisation ($\lambda$) $\times$ Memory decay ($\gamma_m$) & -0.0080 & 2.08 & 0.0381 \\
Average Revenue & Auction format $\times$ Exploration intensity & -0.0078 & 2.02 & 0.0436 \\
Lifetime Revenue & Number of bidders & 0.1280 & 37.70 & < 0.0001 \\
Lifetime Revenue & Auction format & -0.0458 & 13.49 & < 0.0001 \\
Lifetime Revenue & Number of bidders $\times$ Reserve price & -0.0367 & 10.82 & < 0.0001 \\
Lifetime Revenue & Affiliation (linear) & -0.0135 & 6.48 & < 0.0001 \\
Lifetime Revenue & Affiliation (linear) $\times$ Affiliation (quadratic) & -0.0135 & 6.48 & < 0.0001 \\
Lifetime Revenue & Reserve price $\times$ Context richness & -0.0211 & 6.21 & < 0.0001 \\
Lifetime Revenue & Auction format $\times$ Number of bidders & -0.0208 & 6.14 & < 0.0001 \\
Lifetime Revenue & Exploration intensity & -0.0157 & 4.63 & < 0.0001 \\
Lifetime Revenue & Context richness & -0.0154 & 4.54 & < 0.0001 \\
Lifetime Revenue & Auction format $\times$ Reserve price & 0.0137 & 4.03 & < 0.0001 \\
Lifetime Revenue & Number of bidders $\times$ Affiliation (linear) & -0.0153 & 3.68 & 0.0002 \\
Lifetime Revenue & Exploration intensity $\times$ Context richness & 0.0099 & 2.92 & 0.0036 \\
Lifetime Revenue & Number of bidders $\times$ Exploration intensity & -0.0096 & 2.81 & 0.0050 \\
Lifetime Revenue & Regularisation ($\lambda$) & -0.0084 & 2.47 & 0.0137 \\
Lifetime Revenue & Affiliation (quadratic) $\times$ Exploration intensity & 0.0051 & 2.12 & 0.0343 \\
Convergence Time & Number of bidders & -15992.1875 & 14.07 & < 0.0001 \\
Convergence Time & Auction format & -12445.3125 & 10.95 & < 0.0001 \\
Convergence Time & Number of bidders $\times$ Reserve price & 6804.6875 & 5.99 & < 0.0001 \\
Convergence Time & Exploration intensity & 5747.3958 & 5.06 & < 0.0001 \\
Convergence Time & Reserve price $\times$ Context richness & -5661.4583 & 4.98 & < 0.0001 \\
Convergence Time & Auction format $\times$ Number of bidders & 4755.2083 & 4.18 & < 0.0001 \\
Convergence Time & Memory decay ($\gamma_m$) & 4567.7083 & 4.02 & < 0.0001 \\
Convergence Time & Auction format $\times$ Reserve price & -4434.8958 & 3.90 & 0.0001 \\
Convergence Time & Context richness $\times$ Memory decay ($\gamma_m$) & 3317.7083 & 2.92 & 0.0036 \\
Convergence Time & Regularisation ($\lambda$) $\times$ Memory decay ($\gamma_m$) & -2885.4167 & 2.54 & 0.0113 \\
Convergence Time & Auction format $\times$ Memory decay ($\gamma_m$) & -2872.3958 & 2.53 & 0.0117 \\
Convergence Time & Auction format $\times$ Affiliation (quadratic) & -1699.2188 & 2.11 & 0.0348 \\
Convergence Time & Number of bidders $\times$ Exploration intensity & 2375.0000 & 2.09 & 0.0370 \\
Convergence Time & Exploration intensity $\times$ Context richness & -2315.1042 & 2.04 & 0.0420 \\
Convergence Time & Regularisation ($\lambda$) & -2291.6667 & 2.02 & 0.0441 \\
Convergence Time & Reserve price $\times$ Exploration intensity & -2263.0208 & 1.99 & 0.0468 \\
No-Sale Rate & Reserve price & 0.0174 & 27.61 & < 0.0001 \\
No-Sale Rate & Number of bidders & -0.0139 & 22.11 & < 0.0001 \\
No-Sale Rate & Reserve price $\times$ Context richness & 0.0138 & 21.89 & < 0.0001 \\
No-Sale Rate & Context richness & 0.0134 & 21.24 & < 0.0001 \\
No-Sale Rate & Number of bidders $\times$ Reserve price & -0.0129 & 20.56 & < 0.0001 \\
No-Sale Rate & Number of bidders $\times$ Context richness & -0.0110 & 17.53 & < 0.0001 \\
No-Sale Rate & Auction format $\times$ Number of bidders & 0.0018 & 2.89 & 0.0039 \\
No-Sale Rate & Auction format & -0.0018 & 2.79 & 0.0053 \\
No-Sale Rate & Reserve price $\times$ Regularisation ($\lambda$) & -0.0017 & 2.76 & 0.0058 \\
No-Sale Rate & Auction format $\times$ Affiliation (linear) & -0.0021 & 2.70 & 0.0071 \\
No-Sale Rate & Auction format $\times$ Reserve price & -0.0017 & 2.65 & 0.0083 \\
No-Sale Rate & Regularisation ($\lambda$) & -0.0016 & 2.55 & 0.0110 \\
No-Sale Rate & Affiliation (linear) $\times$ Affiliation (quadratic) & 0.0009 & 2.37 & 0.0179 \\
No-Sale Rate & Affiliation (linear) & 0.0009 & 2.37 & 0.0179 \\
No-Sale Rate & Exploration intensity $\times$ Context richness & -0.0015 & 2.36 & 0.0187 \\
No-Sale Rate & Auction format $\times$ Memory decay ($\gamma_m$) & -0.0013 & 2.08 & 0.0377 \\
Price Volatility & Context richness & 0.0246 & 16.91 & < 0.0001 \\
Price Volatility & Auction format $\times$ Number of bidders & 0.0227 & 15.61 & < 0.0001 \\
Price Volatility & Auction format & -0.0194 & 13.38 & < 0.0001 \\
Price Volatility & Number of bidders & -0.0179 & 12.31 & < 0.0001 \\
Price Volatility & Regularisation ($\lambda$) & -0.0109 & 7.48 & < 0.0001 \\
Price Volatility & Number of bidders $\times$ Reserve price & 0.0100 & 6.91 & < 0.0001 \\
Price Volatility & Exploration intensity & 0.0098 & 6.75 & < 0.0001 \\
Price Volatility & Memory decay ($\gamma_m$) & -0.0077 & 5.29 & < 0.0001 \\
Price Volatility & Auction format $\times$ Reserve price & -0.0070 & 4.81 & < 0.0001 \\
Price Volatility & Affiliation (linear) $\times$ Affiliation (quadratic) & -0.0040 & 4.53 & < 0.0001 \\
Price Volatility & Affiliation (linear) & -0.0040 & 4.53 & < 0.0001 \\
Price Volatility & Exploration intensity $\times$ Context richness & -0.0062 & 4.29 & < 0.0001 \\
Price Volatility & Auction format $\times$ Context richness & -0.0043 & 2.97 & 0.0031 \\
Price Volatility & Reserve price $\times$ Exploration intensity & -0.0035 & 2.43 & 0.0155 \\
Price Volatility & Reserve price & -0.0035 & 2.39 & 0.0172 \\
Price Volatility & Context richness $\times$ Regularisation ($\lambda$) & -0.0033 & 2.27 & 0.0233 \\
Winner Entropy & Number of bidders & 0.3231 & 53.06 & < 0.0001 \\
Winner Entropy & Reserve price & -0.0457 & 7.50 & < 0.0001 \\
Winner Entropy & Exploration intensity & -0.0381 & 6.25 & < 0.0001 \\
Winner Entropy & Context richness & -0.0284 & 4.66 & < 0.0001 \\
Winner Entropy & Auction format & 0.0257 & 4.23 & < 0.0001 \\
Winner Entropy & Context richness $\times$ Regularisation ($\lambda$) & -0.0224 & 3.68 & 0.0002 \\
Winner Entropy & Regularisation ($\lambda$) & -0.0214 & 3.51 & 0.0005 \\
Winner Entropy & Reserve price $\times$ Regularisation ($\lambda$) & -0.0214 & 3.51 & 0.0005 \\
Winner Entropy & Exploration intensity $\times$ Regularisation ($\lambda$) & -0.0211 & 3.47 & 0.0005 \\
Winner Entropy & Auction format $\times$ Number of bidders & -0.0201 & 3.30 & 0.0010 \\
Winner Entropy & Number of bidders $\times$ Regularisation ($\lambda$) & -0.0174 & 2.86 & 0.0043 \\
Winner Entropy & Auction format $\times$ Context richness & -0.0146 & 2.40 & 0.0167 \\
Winner Entropy & Auction format $\times$ Memory decay ($\gamma_m$) & 0.0124 & 2.03 & 0.0425 \\
\end{longtable}

\subsubsection{Experiment~2b (Thompson Sampling)}
\label{sec:exp2b_robustness}

\begin{table}[H]
\centering
\caption{Experiment 2b: OLS model fit summary across response variables.}
\label{tab:exp2b_fit}
\begin{tabular}{lrrrr}
\toprule
\textbf{Response} & \textbf{$R^2$} & \textbf{Adj.\,$R^2$} & \textbf{F-stat} & \textbf{F $p$-value} \\
\midrule
Average Revenue & 0.6086 & 0.5442 & 9.447 & < 0.0001 \\
Lifetime Revenue & 0.6264 & 0.5649 & 10.186 & < 0.0001 \\
Convergence Time & 0.3409 & 0.2323 & 3.141 & < 0.0001 \\
No-Sale Rate & 0.8241 & 0.7952 & 28.461 & < 0.0001 \\
Price Volatility & 0.5809 & 0.5119 & 8.420 & < 0.0001 \\
Winner Entropy & 0.6316 & 0.5710 & 10.416 & < 0.0001 \\
\bottomrule
\end{tabular}
\end{table}

\begin{table}[H]
\centering
\caption{Experiment 2b: Model adequacy diagnostics.}
\label{tab:exp2b_adequacy}
\begin{tabular}{lrrrrr}
\toprule
\textbf{Response} & \textbf{$R^2$} & \textbf{Pred-$R^2$} & \textbf{Gap} & \textbf{LGBM $R^2$} & \textbf{LOF $p$} \\
\midrule
Average Revenue & 0.6086 & 0.4636 & 0.1450 & 0.5567 & < 0.0001 \\
Price Volatility & 0.5809 & 0.4256 & 0.1553 & 0.5140 & 0.0018 \\
\bottomrule
\end{tabular}
\par\smallskip\footnotesize Gap $= R^2 - \text{Pred-}R^2$. LGBM $R^2$: five-fold cross-validated LightGBM. LOF $p$: lack-of-fit $F$-test.
\end{table}

Table~\ref{tab:exp2b_adequacy} reports model adequacy for the Thompson Sampling design. LightGBM cross-validated $R^2$ and OLS $R^2$ are comparable, confirming that the linear model with two-way interactions is correctly specified.

\begin{table}[H]
\centering
\caption{Experiment 2b: Inference robustness under heteroskedasticity and multiple testing corrections.}
\label{tab:exp2b_inference}
\begin{tabular}{lrrrr}
\toprule
\textbf{Response} & \textbf{OLS Sig} & \textbf{HC3 Flipped} & \textbf{Holm Sig} & \textbf{BH Sig} \\
\midrule
Average Revenue & 11/28 & 0/28 & 5/28 & 11/28 \\
Price Volatility & 11/28 & 1/28 & 6/28 & 8/28 \\
\midrule
\textit{All responses} & 56/168 & 3/168 & 26/168 & 48/168 \\
\bottomrule
\end{tabular}
\par\smallskip\footnotesize OLS Sig: $p < 0.05$ under OLS standard errors. HC3 Flipped: effects changing significance under HC3 robust standard errors. Holm/BH Sig: effects surviving Holm--Bonferroni/Benjamini--Hochberg correction.
\end{table}

\begin{figure}[H]
  \centering
  \includegraphics[width=0.7\textwidth]{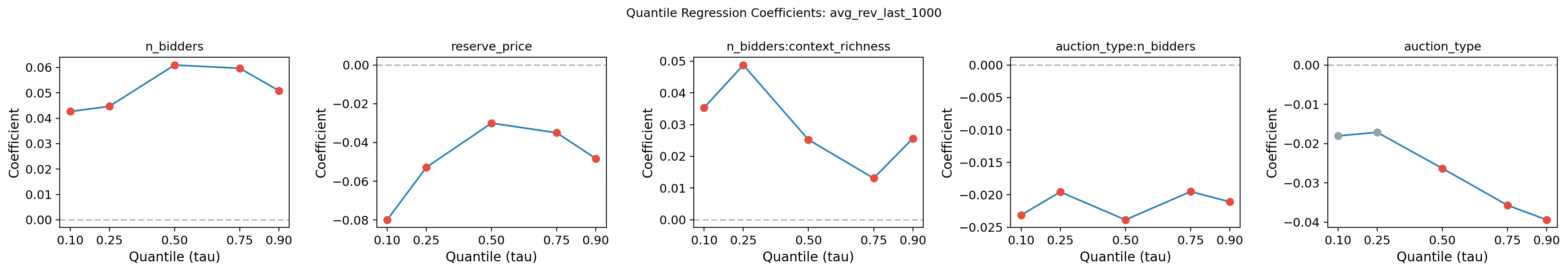}
  \caption{Experiment~2b: Quantile regression coefficients for average revenue (Thompson Sampling).}
  \label{fig:e2b_quantile_rev}
\end{figure}

Table~\ref{tab:exp2b_inference} shows that Holm--Bonferroni and Benjamini--Hochberg corrections retain the key effects. Quantile regression (Figure~\ref{fig:e2b_quantile_rev}) reveals that the auction format effect grows monotonically from $-0.018$ at the 10th percentile to $-0.039$ at the 90th, indicating that the first-price revenue penalty intensifies among the best-performing configurations.

\begin{longtable}{llrrr}
\caption{Experiment 2b: All significant effects ($p < 0.05$) across response variables, ranked by $|t|$.}
\label{tab:exp2b_sig}\\
\toprule
\textbf{Response} & \textbf{Effect} & \textbf{Coeff.} & \textbf{$|t|$} & \textbf{$p$-value} \\
\midrule
\endfirsthead
\multicolumn{5}{l}{\small\emph{Table~\ref{tab:exp2b_sig} continued}} \\
\toprule
\textbf{Response} & \textbf{Effect} & \textbf{Coeff.} & \textbf{$|t|$} & \textbf{$p$-value} \\
\midrule
\endhead
\midrule
\multicolumn{5}{r}{\small\emph{Continued on next page}} \\
\endfoot
\bottomrule
\endlastfoot
Average Revenue & Number of bidders & 0.0525 & 8.69 & < 0.0001 \\
Average Revenue & Reserve price & -0.0480 & 7.95 & < 0.0001 \\
Average Revenue & Number of bidders $\times$ Context richness & 0.0284 & 4.71 & < 0.0001 \\
Average Revenue & Auction format $\times$ Number of bidders & -0.0245 & 4.07 & < 0.0001 \\
Average Revenue & Auction format & -0.0231 & 3.83 & 0.0002 \\
Average Revenue & Exploration intensity & -0.0204 & 3.38 & 0.0009 \\
Average Revenue & Affiliation (linear) $\times$ Affiliation (quadratic) & -0.0117 & 3.18 & 0.0018 \\
Average Revenue & Affiliation (linear) & -0.0117 & 3.18 & 0.0018 \\
Average Revenue & Context richness & -0.0175 & 2.89 & 0.0043 \\
Average Revenue & Number of bidders $\times$ Reserve price & -0.0174 & 2.88 & 0.0045 \\
Average Revenue & Number of bidders $\times$ Affiliation (linear) & -0.0190 & 2.57 & 0.0111 \\
Lifetime Revenue & Number of bidders & 0.0525 & 9.44 & < 0.0001 \\
Lifetime Revenue & Reserve price & -0.0453 & 8.14 & < 0.0001 \\
Lifetime Revenue & Number of bidders $\times$ Context richness & 0.0260 & 4.68 & < 0.0001 \\
Lifetime Revenue & Exploration intensity & -0.0215 & 3.87 & 0.0002 \\
Lifetime Revenue & Auction format $\times$ Number of bidders & -0.0202 & 3.64 & 0.0004 \\
Lifetime Revenue & Auction format & -0.0195 & 3.51 & 0.0006 \\
Lifetime Revenue & Affiliation (linear) $\times$ Affiliation (quadratic) & -0.0118 & 3.47 & 0.0007 \\
Lifetime Revenue & Affiliation (linear) & -0.0118 & 3.47 & 0.0007 \\
Lifetime Revenue & Number of bidders $\times$ Reserve price & -0.0159 & 2.86 & 0.0048 \\
Lifetime Revenue & Context richness & -0.0158 & 2.84 & 0.0051 \\
Lifetime Revenue & Number of bidders $\times$ Affiliation (linear) & -0.0179 & 2.63 & 0.0093 \\
Lifetime Revenue & Exploration intensity $\times$ Context richness & 0.0132 & 2.37 & 0.0189 \\
Lifetime Revenue & Reserve price $\times$ Exploration intensity & 0.0123 & 2.20 & 0.0290 \\
Convergence Time & Auction format $\times$ Number of bidders & 6385.4167 & 4.10 & < 0.0001 \\
Convergence Time & Exploration intensity $\times$ Context richness & -5385.4167 & 3.46 & 0.0007 \\
Convergence Time & Number of bidders $\times$ Reserve price & 5020.8333 & 3.22 & 0.0015 \\
Convergence Time & Number of bidders & -4229.1667 & 2.72 & 0.0073 \\
Convergence Time & Auction format $\times$ Reserve price & 4229.1667 & 2.72 & 0.0073 \\
Convergence Time & Exploration intensity & 3895.8333 & 2.50 & 0.0133 \\
Convergence Time & Reserve price & 3843.7500 & 2.47 & 0.0146 \\
No-Sale Rate & Number of bidders & -0.0179 & 13.44 & < 0.0001 \\
No-Sale Rate & Reserve price & 0.0161 & 12.04 & < 0.0001 \\
No-Sale Rate & Number of bidders $\times$ Reserve price & -0.0145 & 10.87 & < 0.0001 \\
No-Sale Rate & Reserve price $\times$ Context richness & 0.0138 & 10.37 & < 0.0001 \\
No-Sale Rate & Context richness & 0.0137 & 10.24 & < 0.0001 \\
No-Sale Rate & Number of bidders $\times$ Context richness & -0.0123 & 9.21 & < 0.0001 \\
No-Sale Rate & Affiliation (linear) $\times$ Context richness & 0.0035 & 2.12 & 0.0357 \\
No-Sale Rate & Auction format $\times$ Context richness & 0.0028 & 2.09 & 0.0386 \\
Price Volatility & Auction format $\times$ Number of bidders & 0.0192 & 6.57 & < 0.0001 \\
Price Volatility & Number of bidders $\times$ Reserve price & 0.0180 & 6.16 & < 0.0001 \\
Price Volatility & Number of bidders & -0.0177 & 6.04 & < 0.0001 \\
Price Volatility & Number of bidders $\times$ Context richness & -0.0156 & 5.33 & < 0.0001 \\
Price Volatility & Auction format & -0.0124 & 4.24 & < 0.0001 \\
Price Volatility & Exploration intensity & 0.0119 & 4.06 & < 0.0001 \\
Price Volatility & Reserve price & 0.0106 & 3.62 & 0.0004 \\
Price Volatility & Auction format $\times$ Context richness & -0.0077 & 2.63 & 0.0095 \\
Price Volatility & Reserve price $\times$ Context richness & -0.0072 & 2.45 & 0.0153 \\
Price Volatility & Affiliation (linear) $\times$ Context richness & 0.0081 & 2.27 & 0.0248 \\
Price Volatility & Exploration intensity $\times$ Context richness & -0.0058 & 1.99 & 0.0482 \\
Winner Entropy & Number of bidders & 0.1924 & 11.41 & < 0.0001 \\
Winner Entropy & Reserve price & -0.1581 & 9.38 & < 0.0001 \\
Winner Entropy & Exploration intensity $\times$ Context richness & 0.0745 & 4.42 & < 0.0001 \\
Winner Entropy & Reserve price $\times$ Exploration intensity & 0.0650 & 3.85 & 0.0002 \\
Winner Entropy & Number of bidders $\times$ Exploration intensity & 0.0595 & 3.53 & 0.0005 \\
Winner Entropy & Exploration intensity & -0.0447 & 2.65 & 0.0088 \\
\end{longtable}

\subsection{Pacing Experiments}

\subsubsection{Experiment~3a}
\label{sec:exp3a_robustness}

\begin{table}[H]
\centering
\caption{Experiment 3a: OLS model fit summary across response variables.}
\label{tab:exp3a_fit}
\begin{tabular}{lrrrr}
\toprule
\textbf{Response} & \textbf{$R^2$} & \textbf{Adj.\,$R^2$} & \textbf{F-stat} & \textbf{F $p$-value} \\
\midrule
Platform Revenue & 0.8893 & 0.8845 & 187.357 & < 0.0001 \\
Liquid Welfare & 0.7970 & 0.7883 & 91.606 & < 0.0001 \\
Effective PoA (Greedy) & 0.6403 & 0.6249 & 41.540 & < 0.0001 \\
Budget Utilisation & 0.9569 & 0.9550 & 517.852 & < 0.0001 \\
Bid-to-Value Ratio & 0.1217 & 0.0840 & 3.233 & < 0.0001 \\
Allocative Efficiency & 0.8922 & 0.8876 & 193.096 & < 0.0001 \\
Dual Variable CV & 0.6210 & 0.6048 & 38.235 & < 0.0001 \\
No-Sale Rate & 0.7417 & 0.7306 & 67.000 & < 0.0001 \\
Winner Entropy & 0.8694 & 0.8638 & 155.316 & < 0.0001 \\
Warm-Start Benefit & 0.4752 & 0.4527 & 21.129 & < 0.0001 \\
Inter-Episode Volatility & 0.2818 & 0.2510 & 9.156 & < 0.0001 \\
Bid Suppression Ratio & 0.1228 & 0.0852 & 3.267 & < 0.0001 \\
Cross-Episode Drift & 0.1250 & 0.0875 & 3.333 & < 0.0001 \\
LP Offline Welfare & 0.7965 & 0.7878 & 91.344 & < 0.0001 \\
Effective PoA & 0.6399 & 0.6244 & 41.455 & < 0.0001 \\
Lifetime Revenue & 0.8893 & 0.8845 & 187.384 & < 0.0001 \\
\bottomrule
\end{tabular}
\end{table}

\begin{table}[H]
\centering
\caption{Experiment 3a: Model adequacy diagnostics.}
\label{tab:exp3a_adequacy}
\begin{tabular}{lrrrrr}
\toprule
\textbf{Response} & \textbf{$R^2$} & \textbf{Pred-$R^2$} & \textbf{Gap} & \textbf{LGBM $R^2$} & \textbf{LOF $p$} \\
\midrule
Platform Revenue & 0.8893 & 0.8791 & 0.0102 & 0.6067 & < 0.0001 \\
Effective PoA & 0.6399 & 0.6068 & 0.0331 & 0.4188 & < 0.0001 \\
Bid-to-Value Ratio & 0.1217 & 0.0411 & 0.0806 & -24.5214 & 0.0056 \\
\bottomrule
\end{tabular}
\par\smallskip\footnotesize Gap $= R^2 - \text{Pred-}R^2$. LGBM $R^2$: five-fold cross-validated LightGBM. LOF $p$: lack-of-fit $F$-test.
\end{table}

Table~\ref{tab:exp3a_adequacy} reports model adequacy for the three primary responses. Platform revenue ($R^2 = 0.89$) shows a small PRESS gap, indicating strong generalisability. Effective Price of Anarchy ($R^2 = 0.64$) shows moderate explanatory power. The bid-to-value ratio is poorly explained by the factorial model ($R^2 = 0.12$), reflecting seed-level noise in bidding behaviour.\footnote{The LightGBM cross-validated $R^2$ for bid-to-value is deeply negative ($-24.5$), indicating that the nonparametric model severely overfits on this noisy response.} LightGBM cross-validated $R^2$ is substantially lower than OLS $R^2$ for all responses, confirming that the $2^6$ factorial with a linear model is the correct specification.

\begin{table}[H]
\centering
\caption{Experiment 3a: Inference robustness under heteroskedasticity and multiple testing corrections.}
\label{tab:exp3a_inference}
\begin{tabular}{lrrrr}
\toprule
\textbf{Response} & \textbf{OLS Sig} & \textbf{HC3 Flipped} & \textbf{Holm Sig} & \textbf{BH Sig} \\
\midrule
Platform Revenue & 11/21 & 0/21 & 7/21 & 11/21 \\
Effective PoA & 8/21 & 0/21 & 6/21 & 7/21 \\
Bid-to-Value Ratio & 6/21 & 1/21 & 0/21 & 3/21 \\
\midrule
\textit{All responses} & 164/336 & 1/336 & 104/336 & 138/336 \\
\bottomrule
\end{tabular}
\par\smallskip\footnotesize OLS Sig: $p < 0.05$ under OLS standard errors. HC3 Flipped: effects changing significance under HC3 robust standard errors. Holm/BH Sig: effects surviving Holm--Bonferroni/Benjamini--Hochberg correction.
\end{table}

\begin{figure}[H]
  \centering
  \includegraphics[width=0.7\textwidth]{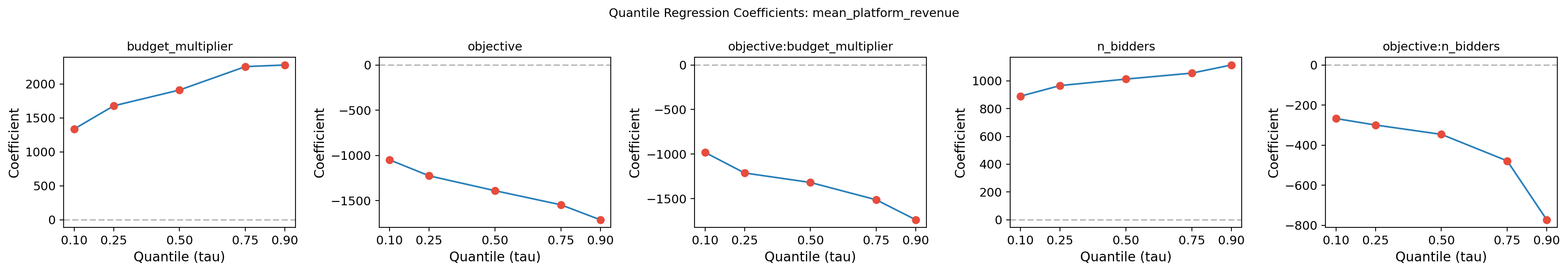}
  \caption{Experiment~3a: Quantile regression coefficients for platform revenue. The objective effect (utility-maximizer vs.\ value-maximizer) grows substantially toward the upper tail.}
  \label{fig:e3a_quantile_rev}
\end{figure}

Table~\ref{tab:exp3a_inference} reports inference robustness. Quantile regression reveals moderate heterogeneity for platform revenue (Figure~\ref{fig:e3a_quantile_rev}), where the objective coefficient grows from $-1{,}049$ at the 10th percentile to $-1{,}712$ at the 90th percentile, indicating that the revenue penalty from utility-maximizing objectives is concentrated among high-revenue configurations. The number of bidders effect remains stable across quantiles ($890$--$1{,}114$), confirming that market thickness benefits revenue uniformly.

\begin{longtable}{llrrr}
\caption{Experiment 3a: All significant effects ($p < 0.05$) across response variables, ranked by $|t|$.}
\label{tab:exp3a_sig}\\
\toprule
\textbf{Response} & \textbf{Effect} & \textbf{Coeff.} & \textbf{$|t|$} & \textbf{$p$-value} \\
\midrule
\endfirsthead
\multicolumn{5}{l}{\small\emph{Table~\ref{tab:exp3a_sig} continued}} \\
\toprule
\textbf{Response} & \textbf{Effect} & \textbf{Coeff.} & \textbf{$|t|$} & \textbf{$p$-value} \\
\midrule
\endhead
\midrule
\multicolumn{5}{r}{\small\emph{Continued on next page}} \\
\endfoot
\bottomrule
\endlastfoot
Platform Revenue & Budget multiplier & 2006.6019 & 39.44 & < 0.0001 \\
Platform Revenue & Bidder objective & -1428.0108 & 28.07 & < 0.0001 \\
Platform Revenue & Bidder objective $\times$ Budget multiplier & -1387.4815 & 27.27 & < 0.0001 \\
Platform Revenue & Number of bidders & 1158.2655 & 22.76 & < 0.0001 \\
Platform Revenue & Bidder objective $\times$ Number of bidders & -629.6958 & 12.38 & < 0.0001 \\
Platform Revenue & Number of bidders $\times$ Budget multiplier & 448.9495 & 8.82 & < 0.0001 \\
Platform Revenue & Value dispersion ($\sigma$) & 356.7849 & 7.01 & < 0.0001 \\
Platform Revenue & Auction format & 184.0336 & 3.62 & 0.0003 \\
Platform Revenue & Auction format $\times$ Bidder objective & 182.7252 & 3.59 & 0.0004 \\
Platform Revenue & Auction format $\times$ Budget multiplier & 140.8645 & 2.77 & 0.0058 \\
Platform Revenue & Number of bidders $\times$ Value dispersion ($\sigma$) & 124.7788 & 2.45 & 0.0145 \\
Liquid Welfare & Budget multiplier & 874.6949 & 33.28 & < 0.0001 \\
Liquid Welfare & Number of bidders & 467.7688 & 17.80 & < 0.0001 \\
Liquid Welfare & Value dispersion ($\sigma$) & 461.1103 & 17.54 & < 0.0001 \\
Liquid Welfare & Number of bidders $\times$ Budget multiplier & -228.5773 & 8.70 & < 0.0001 \\
Liquid Welfare & Budget multiplier $\times$ Value dispersion ($\sigma$) & 190.6957 & 7.25 & < 0.0001 \\
Liquid Welfare & Number of bidders $\times$ Value dispersion ($\sigma$) & 149.9667 & 5.71 & < 0.0001 \\
Liquid Welfare & Bidder objective $\times$ Budget multiplier & 92.8450 & 3.53 & 0.0005 \\
Liquid Welfare & Bidder objective & 71.3494 & 2.71 & 0.0069 \\
Effective PoA (Greedy) & Bidder objective $\times$ Budget multiplier & -0.0301 & 17.62 & < 0.0001 \\
Effective PoA (Greedy) & Bidder objective & -0.0205 & 12.02 & < 0.0001 \\
Effective PoA (Greedy) & Number of bidders & 0.0201 & 11.79 & < 0.0001 \\
Effective PoA (Greedy) & Value dispersion ($\sigma$) & -0.0181 & 10.62 & < 0.0001 \\
Effective PoA (Greedy) & Budget multiplier & 0.0148 & 8.68 & < 0.0001 \\
Effective PoA (Greedy) & Number of bidders $\times$ Value dispersion ($\sigma$) & -0.0142 & 8.30 & < 0.0001 \\
Effective PoA (Greedy) & Bidder objective $\times$ Number of bidders & -0.0042 & 2.47 & 0.0137 \\
Effective PoA (Greedy) & Bidder objective $\times$ Value dispersion ($\sigma$) & 0.0039 & 2.30 & 0.0219 \\
Budget Utilisation & Bidder objective & -0.1685 & 64.01 & < 0.0001 \\
Budget Utilisation & Bidder objective $\times$ Budget multiplier & -0.1472 & 55.91 & < 0.0001 \\
Budget Utilisation & Budget multiplier & -0.1452 & 55.17 & < 0.0001 \\
Budget Utilisation & Bidder objective $\times$ Number of bidders & -0.0351 & 13.34 & < 0.0001 \\
Budget Utilisation & Number of bidders & -0.0311 & 11.80 & < 0.0001 \\
Budget Utilisation & Auction format $\times$ Bidder objective & 0.0240 & 9.12 & < 0.0001 \\
Budget Utilisation & Auction format & 0.0221 & 8.41 & < 0.0001 \\
Budget Utilisation & Auction format $\times$ Budget multiplier & 0.0159 & 6.05 & < 0.0001 \\
Budget Utilisation & Value dispersion ($\sigma$) & 0.0147 & 5.58 & < 0.0001 \\
Budget Utilisation & Bidder objective $\times$ Value dispersion ($\sigma$) & 0.0128 & 4.85 & < 0.0001 \\
Budget Utilisation & Auction format $\times$ Value dispersion ($\sigma$) & 0.0099 & 3.78 & 0.0002 \\
Budget Utilisation & Number of bidders $\times$ Budget multiplier & -0.0099 & 3.75 & 0.0002 \\
Budget Utilisation & Number of bidders $\times$ Value dispersion ($\sigma$) & 0.0067 & 2.56 & 0.0109 \\
Budget Utilisation & Budget multiplier $\times$ Reserve price & -0.0053 & 2.02 & 0.0441 \\
Bid-to-Value Ratio & Budget multiplier & 1.2878 & 3.26 & 0.0012 \\
Bid-to-Value Ratio & Bidder objective & -1.1490 & 2.91 & 0.0038 \\
Bid-to-Value Ratio & Bidder objective $\times$ Budget multiplier & -1.1266 & 2.85 & 0.0045 \\
Bid-to-Value Ratio & Auction format & -0.8413 & 2.13 & 0.0337 \\
Bid-to-Value Ratio & Auction format $\times$ Bidder objective & 0.7996 & 2.02 & 0.0435 \\
Bid-to-Value Ratio & Auction format $\times$ Value dispersion ($\sigma$) & -0.7853 & 1.99 & 0.0474 \\
Allocative Efficiency & Bidder objective & 0.1045 & 35.19 & < 0.0001 \\
Allocative Efficiency & Value dispersion ($\sigma$) & 0.0881 & 29.65 & < 0.0001 \\
Allocative Efficiency & Bidder objective $\times$ Budget multiplier & 0.0800 & 26.94 & < 0.0001 \\
Allocative Efficiency & Number of bidders & -0.0610 & 20.53 & < 0.0001 \\
Allocative Efficiency & Budget multiplier & 0.0582 & 19.59 & < 0.0001 \\
Allocative Efficiency & Budget multiplier $\times$ Value dispersion ($\sigma$) & -0.0338 & 11.40 & < 0.0001 \\
Allocative Efficiency & Bidder objective $\times$ Number of bidders & 0.0304 & 10.23 & < 0.0001 \\
Allocative Efficiency & Bidder objective $\times$ Value dispersion ($\sigma$) & -0.0287 & 9.68 & < 0.0001 \\
Allocative Efficiency & Number of bidders $\times$ Value dispersion ($\sigma$) & 0.0190 & 6.40 & < 0.0001 \\
Allocative Efficiency & Number of bidders $\times$ Budget multiplier & 0.0117 & 3.93 & < 0.0001 \\
Allocative Efficiency & Auction format $\times$ Number of bidders & -0.0075 & 2.53 & 0.0117 \\
Allocative Efficiency & Auction format & -0.0062 & 2.09 & 0.0373 \\
Allocative Efficiency & Auction format $\times$ Budget multiplier & 0.0062 & 2.08 & 0.0382 \\
Dual Variable CV & Bidder objective $\times$ Budget multiplier & -0.0651 & 20.59 & < 0.0001 \\
Dual Variable CV & Budget multiplier & -0.0322 & 10.17 & < 0.0001 \\
Dual Variable CV & Value dispersion ($\sigma$) & 0.0295 & 9.31 & < 0.0001 \\
Dual Variable CV & Bidder objective $\times$ Number of bidders & 0.0184 & 5.81 & < 0.0001 \\
Dual Variable CV & Bidder objective & -0.0182 & 5.75 & < 0.0001 \\
Dual Variable CV & Bidder objective $\times$ Value dispersion ($\sigma$) & -0.0165 & 5.22 & < 0.0001 \\
Dual Variable CV & Budget multiplier $\times$ Value dispersion ($\sigma$) & 0.0153 & 4.82 & < 0.0001 \\
Dual Variable CV & Number of bidders $\times$ Value dispersion ($\sigma$) & -0.0150 & 4.76 & < 0.0001 \\
Dual Variable CV & Number of bidders & 0.0121 & 3.83 & 0.0001 \\
Dual Variable CV & Auction format $\times$ Number of bidders & 0.0096 & 3.03 & 0.0025 \\
Dual Variable CV & Auction format & -0.0096 & 3.02 & 0.0026 \\
Dual Variable CV & Auction format $\times$ Value dispersion ($\sigma$) & -0.0065 & 2.04 & 0.0416 \\
Dual Variable CV & Auction format $\times$ Budget multiplier & 0.0064 & 2.01 & 0.0452 \\
No-Sale Rate & Budget multiplier & -0.0052 & 20.21 & < 0.0001 \\
No-Sale Rate & Number of bidders & -0.0037 & 14.20 & < 0.0001 \\
No-Sale Rate & Number of bidders $\times$ Budget multiplier & 0.0033 & 12.98 & < 0.0001 \\
No-Sale Rate & Bidder objective & -0.0027 & 10.37 & < 0.0001 \\
No-Sale Rate & Reserve price & 0.0026 & 9.93 & < 0.0001 \\
No-Sale Rate & Budget multiplier $\times$ Reserve price & -0.0021 & 8.30 & < 0.0001 \\
No-Sale Rate & Bidder objective $\times$ Budget multiplier & 0.0020 & 7.82 & < 0.0001 \\
No-Sale Rate & Auction format $\times$ Reserve price & -0.0018 & 7.00 & < 0.0001 \\
No-Sale Rate & Number of bidders $\times$ Reserve price & -0.0017 & 6.69 & < 0.0001 \\
No-Sale Rate & Auction format $\times$ Budget multiplier & -0.0017 & 6.59 & < 0.0001 \\
No-Sale Rate & Auction format & 0.0014 & 5.61 & < 0.0001 \\
No-Sale Rate & Bidder objective $\times$ Number of bidders & 0.0014 & 5.45 & < 0.0001 \\
No-Sale Rate & Auction format $\times$ Bidder objective & -0.0014 & 5.38 & < 0.0001 \\
No-Sale Rate & Auction format $\times$ Value dispersion ($\sigma$) & 0.0011 & 4.18 & < 0.0001 \\
No-Sale Rate & Value dispersion ($\sigma$) & 0.0010 & 3.87 & 0.0001 \\
No-Sale Rate & Budget multiplier $\times$ Value dispersion ($\sigma$) & -0.0008 & 3.11 & 0.0020 \\
No-Sale Rate & Bidder objective $\times$ Reserve price & -0.0007 & 2.83 & 0.0049 \\
No-Sale Rate & Bidder objective $\times$ Value dispersion ($\sigma$) & 0.0007 & 2.61 & 0.0093 \\
No-Sale Rate & Auction format $\times$ Number of bidders & -0.0006 & 2.52 & 0.0121 \\
Winner Entropy & Number of bidders & 0.3106 & 48.11 & < 0.0001 \\
Winner Entropy & Bidder objective & -0.0977 & 15.14 & < 0.0001 \\
Winner Entropy & Budget multiplier & -0.0879 & 13.62 & < 0.0001 \\
Winner Entropy & Bidder objective $\times$ Budget multiplier & -0.0835 & 12.93 & < 0.0001 \\
Winner Entropy & Bidder objective $\times$ Value dispersion ($\sigma$) & 0.0711 & 11.02 & < 0.0001 \\
Winner Entropy & Value dispersion ($\sigma$) & 0.0669 & 10.36 & < 0.0001 \\
Winner Entropy & Budget multiplier $\times$ Value dispersion ($\sigma$) & 0.0590 & 9.13 & < 0.0001 \\
Winner Entropy & Bidder objective $\times$ Number of bidders & -0.0310 & 4.81 & < 0.0001 \\
Winner Entropy & Number of bidders $\times$ Value dispersion ($\sigma$) & 0.0242 & 3.75 & 0.0002 \\
Winner Entropy & Number of bidders $\times$ Budget multiplier & -0.0184 & 2.84 & 0.0047 \\
Warm-Start Benefit & Bidder objective $\times$ Budget multiplier & -28.6464 & 13.07 & < 0.0001 \\
Warm-Start Benefit & Number of bidders & 26.6488 & 12.16 & < 0.0001 \\
Warm-Start Benefit & Budget multiplier & 13.7893 & 6.29 & < 0.0001 \\
Warm-Start Benefit & Bidder objective & 11.7624 & 5.37 & < 0.0001 \\
Warm-Start Benefit & Auction format $\times$ Value dispersion ($\sigma$) & -9.5488 & 4.36 & < 0.0001 \\
Warm-Start Benefit & Auction format $\times$ Budget multiplier & 7.3303 & 3.35 & 0.0009 \\
Warm-Start Benefit & Auction format & -5.7391 & 2.62 & 0.0091 \\
Warm-Start Benefit & Auction format $\times$ Number of bidders & -5.4620 & 2.49 & 0.0130 \\
Warm-Start Benefit & Number of bidders $\times$ Budget multiplier & -5.2018 & 2.37 & 0.0180 \\
Inter-Episode Volatility & Budget multiplier $\times$ Value dispersion ($\sigma$) & 0.0022 & 6.03 & < 0.0001 \\
Inter-Episode Volatility & Value dispersion ($\sigma$) & 0.0020 & 5.39 & < 0.0001 \\
Inter-Episode Volatility & Budget multiplier & 0.0019 & 5.28 & < 0.0001 \\
Inter-Episode Volatility & Bidder objective & 0.0019 & 5.10 & < 0.0001 \\
Inter-Episode Volatility & Number of bidders $\times$ Budget multiplier & -0.0016 & 4.32 & < 0.0001 \\
Inter-Episode Volatility & Bidder objective $\times$ Number of bidders & 0.0014 & 3.79 & 0.0002 \\
Inter-Episode Volatility & Number of bidders $\times$ Value dispersion ($\sigma$) & -0.0012 & 3.19 & 0.0015 \\
Inter-Episode Volatility & Auction format & -0.0011 & 3.07 & 0.0023 \\
Inter-Episode Volatility & Auction format $\times$ Value dispersion ($\sigma$) & -0.0010 & 2.77 & 0.0058 \\
Bid Suppression Ratio & Budget multiplier & 1.4554 & 3.69 & 0.0003 \\
Bid Suppression Ratio & Bidder objective & -1.2361 & 3.13 & 0.0018 \\
Bid Suppression Ratio & Bidder objective $\times$ Budget multiplier & -1.2140 & 3.08 & 0.0022 \\
Bid Suppression Ratio & Auction format $\times$ Value dispersion ($\sigma$) & -0.8156 & 2.07 & 0.0393 \\
Cross-Episode Drift & Auction format & -0.0258 & 2.12 & 0.0343 \\
Cross-Episode Drift & Number of bidders & -0.0258 & 2.12 & 0.0343 \\
Cross-Episode Drift & Auction format $\times$ Bidder objective & 0.0258 & 2.12 & 0.0343 \\
Cross-Episode Drift & Bidder objective & -0.0258 & 2.12 & 0.0343 \\
Cross-Episode Drift & Value dispersion ($\sigma$) & 0.0258 & 2.12 & 0.0343 \\
Cross-Episode Drift & Budget multiplier & 0.0258 & 2.12 & 0.0343 \\
Cross-Episode Drift & Auction format $\times$ Number of bidders & 0.0258 & 2.12 & 0.0343 \\
Cross-Episode Drift & Auction format $\times$ Value dispersion ($\sigma$) & -0.0258 & 2.12 & 0.0343 \\
Cross-Episode Drift & Auction format $\times$ Budget multiplier & -0.0258 & 2.12 & 0.0343 \\
Cross-Episode Drift & Bidder objective $\times$ Budget multiplier & -0.0258 & 2.12 & 0.0344 \\
Cross-Episode Drift & Number of bidders $\times$ Budget multiplier & -0.0258 & 2.12 & 0.0344 \\
Cross-Episode Drift & Bidder objective $\times$ Number of bidders & 0.0258 & 2.12 & 0.0344 \\
Cross-Episode Drift & Number of bidders $\times$ Value dispersion ($\sigma$) & -0.0258 & 2.12 & 0.0344 \\
Cross-Episode Drift & Budget multiplier $\times$ Value dispersion ($\sigma$) & 0.0258 & 2.12 & 0.0344 \\
Cross-Episode Drift & Bidder objective $\times$ Value dispersion ($\sigma$) & -0.0258 & 2.12 & 0.0344 \\
LP Offline Welfare & Budget multiplier & 942.8678 & 34.18 & < 0.0001 \\
LP Offline Welfare & Number of bidders & 530.6935 & 19.24 & < 0.0001 \\
LP Offline Welfare & Value dispersion ($\sigma$) & 423.5321 & 15.36 & < 0.0001 \\
LP Offline Welfare & Number of bidders $\times$ Budget multiplier & -219.2078 & 7.95 & < 0.0001 \\
LP Offline Welfare & Budget multiplier $\times$ Value dispersion ($\sigma$) & 200.1427 & 7.26 & < 0.0001 \\
LP Offline Welfare & Number of bidders $\times$ Value dispersion ($\sigma$) & 118.6511 & 4.30 & < 0.0001 \\
Effective PoA & Bidder objective $\times$ Budget multiplier & -0.0301 & 17.59 & < 0.0001 \\
Effective PoA & Bidder objective & -0.0205 & 12.01 & < 0.0001 \\
Effective PoA & Number of bidders & 0.0201 & 11.76 & < 0.0001 \\
Effective PoA & Value dispersion ($\sigma$) & -0.0182 & 10.63 & < 0.0001 \\
Effective PoA & Budget multiplier & 0.0149 & 8.69 & < 0.0001 \\
Effective PoA & Number of bidders $\times$ Value dispersion ($\sigma$) & -0.0142 & 8.28 & < 0.0001 \\
Effective PoA & Bidder objective $\times$ Number of bidders & -0.0042 & 2.48 & 0.0136 \\
Effective PoA & Bidder objective $\times$ Value dispersion ($\sigma$) & 0.0039 & 2.29 & 0.0222 \\
Lifetime Revenue & Budget multiplier & 2006.8766 & 39.45 & < 0.0001 \\
Lifetime Revenue & Bidder objective & -1428.5804 & 28.08 & < 0.0001 \\
Lifetime Revenue & Bidder objective $\times$ Budget multiplier & -1387.3709 & 27.27 & < 0.0001 \\
Lifetime Revenue & Number of bidders & 1157.4546 & 22.75 & < 0.0001 \\
Lifetime Revenue & Bidder objective $\times$ Number of bidders & -629.4071 & 12.37 & < 0.0001 \\
Lifetime Revenue & Number of bidders $\times$ Budget multiplier & 448.7012 & 8.82 & < 0.0001 \\
Lifetime Revenue & Value dispersion ($\sigma$) & 357.2601 & 7.02 & < 0.0001 \\
Lifetime Revenue & Auction format & 183.6621 & 3.61 & 0.0003 \\
Lifetime Revenue & Auction format $\times$ Bidder objective & 182.9808 & 3.60 & 0.0004 \\
Lifetime Revenue & Auction format $\times$ Budget multiplier & 140.4831 & 2.76 & 0.0060 \\
Lifetime Revenue & Number of bidders $\times$ Value dispersion ($\sigma$) & 124.5254 & 2.45 & 0.0147 \\
\end{longtable}

\subsubsection{Experiment~3b}
\label{sec:exp3b_robustness}

\begin{table}[H]
\centering
\caption{Experiment 3b: OLS model fit summary across response variables.}
\label{tab:exp3b_fit}
\begin{tabular}{lrrrr}
\toprule
\textbf{Response} & \textbf{$R^2$} & \textbf{Adj.\,$R^2$} & \textbf{F-stat} & \textbf{F $p$-value} \\
\midrule
Platform Revenue & 0.7675 & 0.7576 & 77.046 & < 0.0001 \\
Liquid Welfare & 0.7316 & 0.7201 & 63.597 & < 0.0001 \\
Effective PoA (Greedy) & 0.3911 & 0.3650 & 14.987 & < 0.0001 \\
Budget Utilisation & 0.8225 & 0.8149 & 108.109 & < 0.0001 \\
Bid-to-Value Ratio & 0.8857 & 0.8808 & 180.891 & < 0.0001 \\
Allocative Efficiency & 0.8272 & 0.8198 & 111.699 & < 0.0001 \\
Dual Variable CV & 0.8823 & 0.8772 & 174.840 & < 0.0001 \\
No-Sale Rate & 0.6974 & 0.6844 & 53.776 & < 0.0001 \\
Winner Entropy & 0.7980 & 0.7894 & 92.186 & < 0.0001 \\
Warm-Start Benefit & 0.0298 & -0.0118 & 0.716 & 0.8177 \\
Inter-Episode Volatility & 0.6081 & 0.5914 & 36.212 & < 0.0001 \\
Bid Suppression Ratio & 0.8857 & 0.8808 & 180.891 & < 0.0001 \\
Cross-Episode Drift & 0.0541 & 0.0136 & 1.335 & 0.1462 \\
LP Offline Welfare & 0.7314 & 0.7199 & 63.545 & < 0.0001 \\
Effective PoA & 0.3909 & 0.3648 & 14.974 & < 0.0001 \\
Lifetime Revenue & 0.7676 & 0.7576 & 77.049 & < 0.0001 \\
\bottomrule
\end{tabular}
\end{table}

\begin{table}[H]
\centering
\caption{Experiment 3b: Model adequacy diagnostics.}
\label{tab:exp3b_adequacy}
\begin{tabular}{lrrrrr}
\toprule
\textbf{Response} & \textbf{$R^2$} & \textbf{Pred-$R^2$} & \textbf{Gap} & \textbf{LGBM $R^2$} & \textbf{LOF $p$} \\
\midrule
Platform Revenue & 0.7675 & 0.7462 & 0.0213 & 0.7089 & 0.0004 \\
Effective PoA & 0.3909 & 0.3350 & 0.0559 & 0.0195 & < 0.0001 \\
Bid-to-Value Ratio & 0.8857 & 0.8753 & 0.0105 & 0.8789 & 0.0698 \\
\bottomrule
\end{tabular}
\par\smallskip\footnotesize Gap $= R^2 - \text{Pred-}R^2$. LGBM $R^2$: five-fold cross-validated LightGBM. LOF $p$: lack-of-fit $F$-test.
\end{table}

Experiment~3b shares the $2^6$ factorial structure of Experiment~3a. Table~\ref{tab:exp3b_adequacy} reports model adequacy for the three primary responses. Platform revenue achieves $R^2 = 0.77$ with a PRESS gap of 0.021, comparable to Experiment~3a. The bid-to-value ratio is well-explained ($R^2 = 0.89$, gap 0.011), while effective Price of Anarchy has moderate explanatory power ($R^2 = 0.39$, gap 0.056). LightGBM cross-validated $R^2$ matches or falls below OLS for all responses, confirming that the linear factorial model captures the dominant variation.

\begin{table}[H]
\centering
\caption{Experiment 3b: Inference robustness under heteroskedasticity and multiple testing corrections.}
\label{tab:exp3b_inference}
\begin{tabular}{lrrrr}
\toprule
\textbf{Response} & \textbf{OLS Sig} & \textbf{HC3 Flipped} & \textbf{Holm Sig} & \textbf{BH Sig} \\
\midrule
Platform Revenue & 8/21 & 0/21 & 7/21 & 8/21 \\
Effective PoA & 10/21 & 0/21 & 4/21 & 6/21 \\
Bid-to-Value Ratio & 4/21 & 0/21 & 4/21 & 4/21 \\
\midrule
\textit{All responses} & 136/336 & 1/336 & 82/336 & 120/336 \\
\bottomrule
\end{tabular}
\par\smallskip\footnotesize OLS Sig: $p < 0.05$ under OLS standard errors. HC3 Flipped: effects changing significance under HC3 robust standard errors. Holm/BH Sig: effects surviving Holm--Bonferroni/Benjamini--Hochberg correction.
\end{table}

\begin{longtable}{llrrr}
\caption{Experiment 3b: All significant effects ($p < 0.05$) across response variables, ranked by $|t|$.}
\label{tab:exp3b_sig}\\
\toprule
\textbf{Response} & \textbf{Effect} & \textbf{Coeff.} & \textbf{$|t|$} & \textbf{$p$-value} \\
\midrule
\endfirsthead
\multicolumn{5}{l}{\small\emph{Table~\ref{tab:exp3b_sig} continued}} \\
\toprule
\textbf{Response} & \textbf{Effect} & \textbf{Coeff.} & \textbf{$|t|$} & \textbf{$p$-value} \\
\midrule
\endhead
\midrule
\multicolumn{5}{r}{\small\emph{Continued on next page}} \\
\endfoot
\bottomrule
\endlastfoot
Platform Revenue & Budget multiplier & 1316.1821 & 31.68 & < 0.0001 \\
Platform Revenue & Number of bidders & 667.7779 & 16.07 & < 0.0001 \\
Platform Revenue & Auction format & 458.3301 & 11.03 & < 0.0001 \\
Platform Revenue & Auction format $\times$ Budget multiplier & 393.6340 & 9.47 & < 0.0001 \\
Platform Revenue & Value dispersion ($\sigma$) & 360.6121 & 8.68 & < 0.0001 \\
Platform Revenue & Auction format $\times$ Value dispersion ($\sigma$) & 233.6535 & 5.62 & < 0.0001 \\
Platform Revenue & Budget multiplier $\times$ Value dispersion ($\sigma$) & 175.0143 & 4.21 & < 0.0001 \\
Platform Revenue & Number of bidders $\times$ Value dispersion ($\sigma$) & 146.9008 & 3.54 & 0.0004 \\
Liquid Welfare & Budget multiplier & 887.8179 & 28.55 & < 0.0001 \\
Liquid Welfare & Number of bidders & 493.8917 & 15.88 & < 0.0001 \\
Liquid Welfare & Value dispersion ($\sigma$) & 392.2036 & 12.61 & < 0.0001 \\
Liquid Welfare & Budget multiplier $\times$ Value dispersion ($\sigma$) & 202.6324 & 6.52 & < 0.0001 \\
Liquid Welfare & Number of bidders $\times$ Budget multiplier & -196.9716 & 6.33 & < 0.0001 \\
Liquid Welfare & Number of bidders $\times$ Value dispersion ($\sigma$) & 114.3764 & 3.68 & 0.0003 \\
Effective PoA (Greedy) & Auction format & 0.0141 & 10.71 & < 0.0001 \\
Effective PoA (Greedy) & Number of bidders $\times$ Budget multiplier & -0.0130 & 9.86 & < 0.0001 \\
Effective PoA (Greedy) & Auction format $\times$ Budget multiplier & 0.0073 & 5.57 & < 0.0001 \\
Effective PoA (Greedy) & Value dispersion ($\sigma$) & -0.0064 & 4.82 & < 0.0001 \\
Effective PoA (Greedy) & Number of bidders & 0.0039 & 2.95 & 0.0034 \\
Effective PoA (Greedy) & Auction format $\times$ Reserve price & 0.0036 & 2.74 & 0.0064 \\
Effective PoA (Greedy) & Number of bidders $\times$ Reserve price & 0.0031 & 2.36 & 0.0189 \\
Effective PoA (Greedy) & Number of bidders $\times$ Value dispersion ($\sigma$) & -0.0029 & 2.23 & 0.0263 \\
Effective PoA (Greedy) & Budget multiplier $\times$ Reserve price & -0.0029 & 2.19 & 0.0291 \\
Effective PoA (Greedy) & Reserve price $\times$ Value dispersion ($\sigma$) & 0.0029 & 2.17 & 0.0305 \\
Budget Utilisation & Budget multiplier & -0.2056 & 41.28 & < 0.0001 \\
Budget Utilisation & Auction format & 0.0605 & 12.14 & < 0.0001 \\
Budget Utilisation & Number of bidders $\times$ Budget multiplier & -0.0584 & 11.73 & < 0.0001 \\
Budget Utilisation & Number of bidders & -0.0542 & 10.87 & < 0.0001 \\
Budget Utilisation & Auction format $\times$ Budget multiplier & 0.0451 & 9.05 & < 0.0001 \\
Budget Utilisation & Auction format $\times$ Number of bidders & -0.0285 & 5.71 & < 0.0001 \\
Budget Utilisation & Value dispersion ($\sigma$) & 0.0192 & 3.85 & 0.0001 \\
Budget Utilisation & Budget multiplier $\times$ Value dispersion ($\sigma$) & 0.0174 & 3.49 & 0.0005 \\
Budget Utilisation & Auction format $\times$ Value dispersion ($\sigma$) & 0.0174 & 3.49 & 0.0005 \\
Budget Utilisation & Number of bidders $\times$ Reserve price & -0.0113 & 2.27 & 0.0237 \\
Bid-to-Value Ratio & Budget multiplier & 0.3873 & 54.37 & < 0.0001 \\
Bid-to-Value Ratio & Auction format & -0.1585 & 22.25 & < 0.0001 \\
Bid-to-Value Ratio & Auction format $\times$ Budget multiplier & 0.1233 & 17.31 & < 0.0001 \\
Bid-to-Value Ratio & Auction format $\times$ Number of bidders & 0.0410 & 5.76 & < 0.0001 \\
Allocative Efficiency & Budget multiplier & 0.2297 & 44.74 & < 0.0001 \\
Allocative Efficiency & Value dispersion ($\sigma$) & 0.0486 & 9.47 & < 0.0001 \\
Allocative Efficiency & Auction format & -0.0396 & 7.72 & < 0.0001 \\
Allocative Efficiency & Number of bidders $\times$ Budget multiplier & 0.0378 & 7.37 & < 0.0001 \\
Allocative Efficiency & Number of bidders & -0.0370 & 7.20 & < 0.0001 \\
Allocative Efficiency & Budget multiplier $\times$ Reserve price & 0.0254 & 4.94 & < 0.0001 \\
Allocative Efficiency & Number of bidders $\times$ Reserve price & 0.0203 & 3.96 & < 0.0001 \\
Allocative Efficiency & Reserve price & -0.0185 & 3.60 & 0.0004 \\
Allocative Efficiency & Auction format $\times$ Reserve price & -0.0182 & 3.55 & 0.0004 \\
Allocative Efficiency & Auction format $\times$ Budget multiplier & -0.0155 & 3.03 & 0.0026 \\
Allocative Efficiency & Reserve price $\times$ Value dispersion ($\sigma$) & -0.0103 & 2.01 & 0.0448 \\
Dual Variable CV & Budget multiplier & -0.4610 & 52.94 & < 0.0001 \\
Dual Variable CV & Auction format & 0.1925 & 22.11 & < 0.0001 \\
Dual Variable CV & Auction format $\times$ Budget multiplier & -0.1472 & 16.91 & < 0.0001 \\
Dual Variable CV & Auction format $\times$ Number of bidders & -0.0373 & 4.29 & < 0.0001 \\
Dual Variable CV & Aggressiveness & 0.0283 & 3.25 & 0.0012 \\
Dual Variable CV & Number of bidders $\times$ Budget multiplier & -0.0278 & 3.19 & 0.0015 \\
Dual Variable CV & Auction format $\times$ Value dispersion ($\sigma$) & 0.0278 & 3.19 & 0.0015 \\
Dual Variable CV & Number of bidders & 0.0260 & 2.98 & 0.0030 \\
Dual Variable CV & Auction format $\times$ Aggressiveness & 0.0247 & 2.84 & 0.0047 \\
Dual Variable CV & Budget multiplier $\times$ Value dispersion ($\sigma$) & -0.0230 & 2.64 & 0.0086 \\
Dual Variable CV & Number of bidders $\times$ Value dispersion ($\sigma$) & -0.0227 & 2.61 & 0.0094 \\
Dual Variable CV & Aggressiveness $\times$ Budget multiplier & -0.0225 & 2.58 & 0.0102 \\
Dual Variable CV & Number of bidders $\times$ Reserve price & -0.0201 & 2.30 & 0.0216 \\
No-Sale Rate & Budget multiplier & -0.0271 & 12.88 & < 0.0001 \\
No-Sale Rate & Auction format & 0.0267 & 12.68 & < 0.0001 \\
No-Sale Rate & Budget multiplier $\times$ Reserve price & -0.0267 & 12.68 & < 0.0001 \\
No-Sale Rate & Auction format $\times$ Budget multiplier & -0.0255 & 12.12 & < 0.0001 \\
No-Sale Rate & Reserve price & 0.0255 & 12.11 & < 0.0001 \\
No-Sale Rate & Auction format $\times$ Reserve price & 0.0243 & 11.55 & < 0.0001 \\
No-Sale Rate & Number of bidders $\times$ Budget multiplier & 0.0148 & 7.05 & < 0.0001 \\
No-Sale Rate & Number of bidders & -0.0136 & 6.48 & < 0.0001 \\
No-Sale Rate & Number of bidders $\times$ Reserve price & -0.0132 & 6.28 & < 0.0001 \\
No-Sale Rate & Auction format $\times$ Number of bidders & -0.0123 & 5.82 & < 0.0001 \\
No-Sale Rate & Budget multiplier $\times$ Value dispersion ($\sigma$) & -0.0073 & 3.45 & 0.0006 \\
No-Sale Rate & Value dispersion ($\sigma$) & 0.0061 & 2.88 & 0.0042 \\
No-Sale Rate & Reserve price $\times$ Value dispersion ($\sigma$) & 0.0056 & 2.68 & 0.0076 \\
No-Sale Rate & Auction format $\times$ Value dispersion ($\sigma$) & 0.0054 & 2.58 & 0.0102 \\
No-Sale Rate & Auction format $\times$ Aggressiveness & 0.0042 & 2.00 & 0.0465 \\
Winner Entropy & Number of bidders & 0.2765 & 36.40 & < 0.0001 \\
Winner Entropy & Budget multiplier & -0.1283 & 16.90 & < 0.0001 \\
Winner Entropy & Budget multiplier $\times$ Value dispersion ($\sigma$) & 0.0784 & 10.32 & < 0.0001 \\
Winner Entropy & Value dispersion ($\sigma$) & 0.0635 & 8.36 & < 0.0001 \\
Winner Entropy & Number of bidders $\times$ Budget multiplier & -0.0571 & 7.51 & < 0.0001 \\
Winner Entropy & Auction format & 0.0466 & 6.13 & < 0.0001 \\
Winner Entropy & Auction format $\times$ Budget multiplier & 0.0322 & 4.24 & < 0.0001 \\
Winner Entropy & Auction format $\times$ Value dispersion ($\sigma$) & -0.0236 & 3.11 & 0.0020 \\
Winner Entropy & Reserve price & -0.0208 & 2.74 & 0.0063 \\
Winner Entropy & Auction format $\times$ Number of bidders & -0.0198 & 2.60 & 0.0095 \\
Winner Entropy & Number of bidders $\times$ Value dispersion ($\sigma$) & 0.0195 & 2.57 & 0.0105 \\
Inter-Episode Volatility & Auction format & -0.0047 & 11.92 & < 0.0001 \\
Inter-Episode Volatility & Auction format $\times$ Budget multiplier & 0.0046 & 11.72 & < 0.0001 \\
Inter-Episode Volatility & Number of bidders & -0.0042 & 10.73 & < 0.0001 \\
Inter-Episode Volatility & Auction format $\times$ Number of bidders & 0.0042 & 10.63 & < 0.0001 \\
Inter-Episode Volatility & Number of bidders $\times$ Budget multiplier & 0.0038 & 9.51 & < 0.0001 \\
Inter-Episode Volatility & Budget multiplier $\times$ Value dispersion ($\sigma$) & 0.0030 & 7.73 & < 0.0001 \\
Inter-Episode Volatility & Budget multiplier & -0.0016 & 3.95 & < 0.0001 \\
Inter-Episode Volatility & Value dispersion ($\sigma$) & 0.0012 & 3.12 & 0.0019 \\
Inter-Episode Volatility & Aggressiveness $\times$ Budget multiplier & 0.0012 & 3.11 & 0.0020 \\
Inter-Episode Volatility & Aggressiveness & -0.0012 & 3.09 & 0.0021 \\
Inter-Episode Volatility & Auction format $\times$ Aggressiveness & 0.0011 & 2.77 & 0.0058 \\
Inter-Episode Volatility & Budget multiplier $\times$ Reserve price & 0.0010 & 2.62 & 0.0090 \\
Inter-Episode Volatility & Auction format $\times$ Value dispersion ($\sigma$) & 0.0010 & 2.62 & 0.0092 \\
Inter-Episode Volatility & Reserve price & -0.0010 & 2.53 & 0.0116 \\
Inter-Episode Volatility & Aggressiveness $\times$ Number of bidders & 0.0009 & 2.38 & 0.0176 \\
Inter-Episode Volatility & Auction format $\times$ Reserve price & 0.0009 & 2.32 & 0.0207 \\
Inter-Episode Volatility & Aggressiveness $\times$ Value dispersion ($\sigma$) & 0.0009 & 2.28 & 0.0232 \\
Inter-Episode Volatility & Number of bidders $\times$ Reserve price & 0.0009 & 2.26 & 0.0240 \\
Inter-Episode Volatility & Number of bidders $\times$ Value dispersion ($\sigma$) & 0.0009 & 2.20 & 0.0283 \\
Bid Suppression Ratio & Budget multiplier & 0.3873 & 54.37 & < 0.0001 \\
Bid Suppression Ratio & Auction format & -0.1585 & 22.25 & < 0.0001 \\
Bid Suppression Ratio & Auction format $\times$ Budget multiplier & 0.1233 & 17.31 & < 0.0001 \\
Bid Suppression Ratio & Auction format $\times$ Number of bidders & 0.0410 & 5.76 & < 0.0001 \\
Cross-Episode Drift & Number of bidders $\times$ Reserve price & 0.0000 & 3.24 & 0.0013 \\
LP Offline Welfare & Budget multiplier & 905.2377 & 28.68 & < 0.0001 \\
LP Offline Welfare & Number of bidders & 501.8241 & 15.90 & < 0.0001 \\
LP Offline Welfare & Value dispersion ($\sigma$) & 379.1168 & 12.01 & < 0.0001 \\
LP Offline Welfare & Number of bidders $\times$ Budget multiplier & -237.8532 & 7.54 & < 0.0001 \\
LP Offline Welfare & Budget multiplier $\times$ Value dispersion ($\sigma$) & 201.4717 & 6.38 & < 0.0001 \\
LP Offline Welfare & Number of bidders $\times$ Value dispersion ($\sigma$) & 103.1073 & 3.27 & 0.0012 \\
Effective PoA & Auction format & 0.0141 & 10.69 & < 0.0001 \\
Effective PoA & Number of bidders $\times$ Budget multiplier & -0.0130 & 9.87 & < 0.0001 \\
Effective PoA & Auction format $\times$ Budget multiplier & 0.0073 & 5.56 & < 0.0001 \\
Effective PoA & Value dispersion ($\sigma$) & -0.0064 & 4.84 & < 0.0001 \\
Effective PoA & Number of bidders & 0.0039 & 2.93 & 0.0036 \\
Effective PoA & Auction format $\times$ Reserve price & 0.0036 & 2.73 & 0.0065 \\
Effective PoA & Number of bidders $\times$ Reserve price & 0.0031 & 2.35 & 0.0192 \\
Effective PoA & Number of bidders $\times$ Value dispersion ($\sigma$) & -0.0029 & 2.21 & 0.0274 \\
Effective PoA & Budget multiplier $\times$ Reserve price & -0.0029 & 2.18 & 0.0296 \\
Effective PoA & Reserve price $\times$ Value dispersion ($\sigma$) & 0.0029 & 2.16 & 0.0311 \\
Lifetime Revenue & Budget multiplier & 1316.0047 & 31.68 & < 0.0001 \\
Lifetime Revenue & Number of bidders & 667.7049 & 16.07 & < 0.0001 \\
Lifetime Revenue & Auction format & 458.2580 & 11.03 & < 0.0001 \\
Lifetime Revenue & Auction format $\times$ Budget multiplier & 393.6021 & 9.47 & < 0.0001 \\
Lifetime Revenue & Value dispersion ($\sigma$) & 360.4600 & 8.68 & < 0.0001 \\
Lifetime Revenue & Auction format $\times$ Value dispersion ($\sigma$) & 233.6161 & 5.62 & < 0.0001 \\
Lifetime Revenue & Budget multiplier $\times$ Value dispersion ($\sigma$) & 174.8824 & 4.21 & < 0.0001 \\
Lifetime Revenue & Number of bidders $\times$ Value dispersion ($\sigma$) & 146.8410 & 3.53 & 0.0004 \\
\end{longtable}

\begin{figure}[H]
  \centering
  \includegraphics[width=0.7\textwidth]{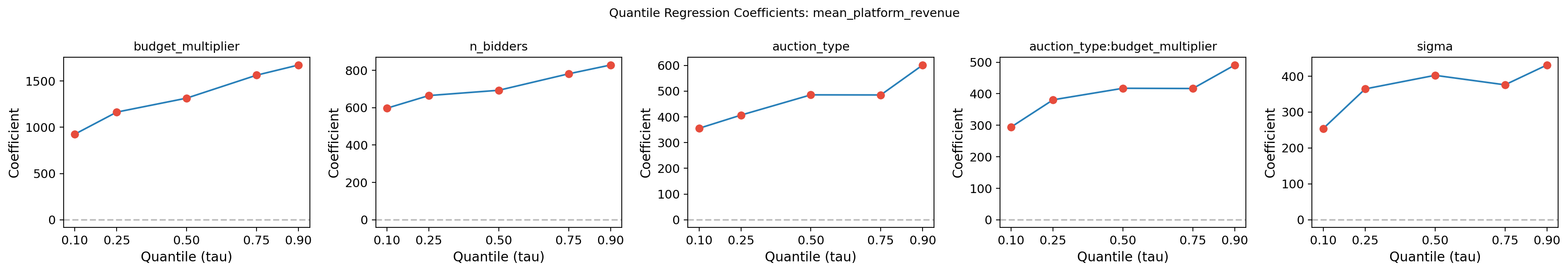}
  \caption{Experiment~3b: Quantile regression coefficients for platform revenue. The auction format effect grows monotonically toward the upper tail.}
  \label{fig:e3b_quantile_rev}
\end{figure}

Quantile regression reveals that the auction format effect on revenue grows monotonically across quantiles, from $+356$ at the 10th percentile to $+601$ at the 90th (Figure~\ref{fig:e3b_quantile_rev}). The first-price revenue advantage nearly doubles from the lowest to highest quantile, indicating that the best-performing configurations benefit most from first-price rules. The budget multiplier effect is stable across quantiles, confirming that budget tightness affects revenue uniformly.

\subsection{Sensitivity Analyses}

\subsubsection{Discretisation Sensitivity}
\label{sec:discretization_robust}

To verify that results are not driven by action space granularity, we re-run a representative subset of factorial cells at grid sizes of 6, 11, and 21 discrete bid levels for Experiments~1a--2b. Effect rankings are stable across grid sizes. The dominant factors (number of bidders, auction type) retain their relative ordering at all granularities. Mean response values shift modestly with grid size, consistent with the theoretical prediction that coarser grids introduce approximation error proportional to the grid spacing. Full results are available in \texttt{results/expN/robust/discretization/}.

\subsubsection{Budget Sensitivity (Experiment~3a)}
\label{sec:budget_robust}

We re-run Experiment~3a with budget multipliers $m \in \{0.25, 0.5, 1.0\}$, where $B_i = m \cdot \mathbb{E}[v_i] \cdot T$. The baseline ($m = 0.5$) produces moderately constrained budgets. Tighter budgets ($m = 0.25$) amplify the effect of bidder objective on revenue, while relaxed budgets ($m = 1.0$) reduce auction format effects as budget constraints become non-binding. The ranking of number of bidders as the dominant factor is preserved across all budget levels. Full results are reported in \texttt{results/exp3a/robust/budget/}.

\subsubsection{Detailed Sensitivity Analysis}
\label{sec:appendix_sensitivity_tables}

Tables~\ref{tab:sens_exp1a}--\ref{tab:sens_exp3b_secondary} report per-experiment Sobol' indices for all response variables. The cross-experiment synthesis appears in Section~\ref{sec:sens_synthesis}.

\begin{table}[H]
\centering
\caption{Experiment~1a: Analytical Sobol' indices for all five response variables. $S_1$ is the first-order (main effect) index; $S_T$ is the total-order index including all interactions.}
\label{tab:sens_exp1a}
\footnotesize
\resizebox{\textwidth}{!}{%
\begin{tabular}{l cc cc cc cc cc}
\toprule
& \multicolumn{2}{c}{Revenue} & \multicolumn{2}{c}{Conv.\ Time} & \multicolumn{2}{c}{No-Sale Rate} & \multicolumn{2}{c}{Volatility} & \multicolumn{2}{c}{Entropy} \\
\cmidrule(lr){2-3} \cmidrule(lr){4-5} \cmidrule(lr){6-7} \cmidrule(lr){8-9} \cmidrule(lr){10-11}
Factor & $S_1$ & $S_T$ & $S_1$ & $S_T$ & $S_1$ & $S_T$ & $S_1$ & $S_T$ & $S_1$ & $S_T$ \\
\midrule
Number of bidders     & 0.188 & 0.238 & 0.093 & 0.111 & 0.038 & 0.101 & 0.139 & 0.165 & 0.272 & 0.309 \\
Reserve price         & 0.015 & 0.043 & 0.009 & 0.022 & 0.152 & 0.266 & 0.098 & 0.138 & 0.001 & 0.014 \\
Update mode           & 0.033 & 0.057 & 0.022 & 0.037 & 0.009 & 0.086 & 0.039 & 0.060 & 0.060 & 0.127 \\
Discount factor ($\gamma$)  & 0.051 & 0.096 & 0.034 & 0.053 & 0.016 & 0.047 & 0.053 & 0.072 & 0.000 & 0.001 \\
Auction format        & 0.000 & 0.045 & 0.072 & 0.100 & 0.001 & 0.004 & 0.003 & 0.013 & 0.001 & 0.005 \\
Information feedback  & 0.008 & 0.036 & 0.000 & 0.008 & 0.003 & 0.008 & 0.000 & 0.029 & 0.003 & 0.065 \\
Exploration strategy  & 0.000 & 0.017 & 0.002 & 0.018 & 0.000 & 0.037 & 0.001 & 0.024 & 0.000 & 0.043 \\
Initialisation        & 0.002 & 0.006 & 0.000 & 0.010 & 0.009 & 0.026 & 0.011 & 0.025 & 0.000 & 0.006 \\
Decay type            & 0.001 & 0.006 & 0.010 & 0.023 & 0.007 & 0.020 & 0.005 & 0.013 & 0.000 & 0.011 \\
Learning rate ($\alpha$)    & 0.001 & 0.005 & 0.000 & 0.004 & 0.003 & 0.016 & 0.001 & 0.006 & 0.000 & 0.001 \\
\midrule
Residual              & \multicolumn{2}{c}{0.577} & \multicolumn{2}{c}{0.685} & \multicolumn{2}{c}{0.576} & \multicolumn{2}{c}{0.553} & \multicolumn{2}{c}{0.541} \\
\bottomrule
\end{tabular}}
\end{table}

\begin{table}[H]
\centering
\caption{Experiment~1b: Analytical Sobol' indices for market performance responses. $R^2$: revenue 0.40, convergence 0.18, volatility 0.36, entropy 0.29.}
\label{tab:sens_exp1b_a}
\small
\begin{tabular}{l cc cc cc cc}
\toprule
& \multicolumn{2}{c}{Revenue} & \multicolumn{2}{c}{Conv.\ Time} & \multicolumn{2}{c}{Volatility} & \multicolumn{2}{c}{Entropy} \\
\cmidrule(lr){2-3} \cmidrule(lr){4-5} \cmidrule(lr){6-7} \cmidrule(lr){8-9}
Factor & $S_1$ & $S_T$ & $S_1$ & $S_T$ & $S_1$ & $S_T$ & $S_1$ & $S_T$ \\
\midrule
Number of bidders     & 0.142 & 0.223 & 0.013 & 0.110 & 0.136 & 0.173 & 0.125 & 0.155 \\
State information     & 0.114 & 0.156 & 0.004 & 0.064 & 0.098 & 0.120 & 0.111 & 0.145 \\
Auction format        & 0.033 & 0.078 & 0.000 & 0.114 & 0.051 & 0.099 & 0.004 & 0.012 \\
Affiliation ($\eta$)  & 0.025 & 0.052 & 0.023 & 0.047 & 0.006 & 0.047 & 0.001 & 0.022 \\
\midrule
Residual              & \multicolumn{2}{c}{0.588} & \multicolumn{2}{c}{0.813} & \multicolumn{2}{c}{0.635} & \multicolumn{2}{c}{0.712} \\
\bottomrule
\end{tabular}
\end{table}

\begin{table}[H]
\centering
\caption{Experiment~1b: Analytical Sobol' indices for bidding behaviour responses. $R^2$: bid-to-value 0.49, winner's curse 0.55, bid dispersion 0.28, signal slope 0.41.}
\label{tab:sens_exp1b_b}
\small
\begin{tabular}{l cc cc cc cc}
\toprule
& \multicolumn{2}{c}{Bid-to-Value} & \multicolumn{2}{c}{Winner's Curse} & \multicolumn{2}{c}{Bid Disp.} & \multicolumn{2}{c}{Signal Slope} \\
\cmidrule(lr){2-3} \cmidrule(lr){4-5} \cmidrule(lr){6-7} \cmidrule(lr){8-9}
Factor & $S_1$ & $S_T$ & $S_1$ & $S_T$ & $S_1$ & $S_T$ & $S_1$ & $S_T$ \\
\midrule
Number of bidders     & 0.139 & 0.266 & 0.089 & 0.294 & 0.066 & 0.087 & 0.016 & 0.063 \\
State information     & 0.099 & 0.146 & 0.042 & 0.106 & 0.049 & 0.068 & 0.020 & 0.089 \\
Auction format        & 0.060 & 0.160 & 0.146 & 0.361 & 0.065 & 0.083 & 0.002 & 0.086 \\
Affiliation ($\eta$)  & 0.059 & 0.073 & 0.026 & 0.039 & 0.086 & 0.119 & 0.352 & 0.397 \\
\midrule
Residual              & \multicolumn{2}{c}{0.499} & \multicolumn{2}{c}{0.448} & \multicolumn{2}{c}{0.688} & \multicolumn{2}{c}{0.487} \\
\bottomrule
\end{tabular}
\end{table}

\begin{table}[H]
\centering
\caption{Experiment~2a: Analytical Sobol' indices for all five LinUCB response variables.}
\label{tab:sens_exp2a}
\footnotesize
\resizebox{\textwidth}{!}{%
\begin{tabular}{l cc cc cc cc cc}
\toprule
& \multicolumn{2}{c}{Revenue} & \multicolumn{2}{c}{Conv.\ Time} & \multicolumn{2}{c}{No-Sale Rate} & \multicolumn{2}{c}{Volatility} & \multicolumn{2}{c}{Entropy} \\
\cmidrule(lr){2-3} \cmidrule(lr){4-5} \cmidrule(lr){6-7} \cmidrule(lr){8-9} \cmidrule(lr){10-11}
Factor & $S_1$ & $S_T$ & $S_1$ & $S_T$ & $S_1$ & $S_T$ & $S_1$ & $S_T$ & $S_1$ & $S_T$ \\
\midrule
Number of bidders         & 0.479 & 0.551 & 0.158 & 0.210 & 0.131 & 0.330 & 0.080 & 0.236 & 0.746 & 0.752 \\
Reserve price             & 0.001 & 0.067 & 0.003 & 0.071 & 0.205 & 0.452 & 0.003 & 0.046 & 0.015 & 0.020 \\
Context richness          & 0.005 & 0.026 & 0.002 & 0.036 & 0.121 & 0.335 & 0.151 & 0.170 & 0.006 & 0.013 \\
Auction format            & 0.065 & 0.092 & 0.096 & 0.138 & 0.002 & 0.010 & 0.094 & 0.242 & 0.005 & 0.011 \\
Exploration intensity     & 0.008 & 0.020 & 0.020 & 0.036 & 0.000 & 0.003 & 0.024 & 0.040 & 0.010 & 0.015 \\
Affiliation ($\eta$)      & 0.027 & 0.037 & 0.000 & 0.015 & 0.003 & 0.007 & 0.022 & 0.032 & 0.000 & 0.002 \\
Regularisation ($\lambda$) & 0.002 & 0.008 & 0.003 & 0.014 & 0.002 & 0.006 & 0.029 & 0.037 & 0.003 & 0.016 \\
Memory decay ($\gamma_m$) & 0.002 & 0.006 & 0.013 & 0.032 & 0.000 & 0.003 & 0.015 & 0.017 & 0.000 & 0.002 \\
\midrule
Residual                  & \multicolumn{2}{c}{0.302} & \multicolumn{2}{c}{0.577} & \multicolumn{2}{c}{0.194} & \multicolumn{2}{c}{0.381} & \multicolumn{2}{c}{0.192} \\
\bottomrule
\end{tabular}}
\end{table}

\begin{table}[H]
\centering
\caption{Experiment~2b: Analytical Sobol' indices for all five Thompson Sampling response variables.}
\label{tab:sens_exp2b}
\footnotesize
\resizebox{\textwidth}{!}{%
\begin{tabular}{l cc cc cc cc cc}
\toprule
& \multicolumn{2}{c}{Revenue} & \multicolumn{2}{c}{Conv.\ Time} & \multicolumn{2}{c}{No-Sale Rate} & \multicolumn{2}{c}{Volatility} & \multicolumn{2}{c}{Entropy} \\
\cmidrule(lr){2-3} \cmidrule(lr){4-5} \cmidrule(lr){6-7} \cmidrule(lr){8-9} \cmidrule(lr){10-11}
Factor & $S_1$ & $S_T$ & $S_1$ & $S_T$ & $S_1$ & $S_T$ & $S_1$ & $S_T$ & $S_1$ & $S_T$ \\
\midrule
Number of bidders         & 0.176 & 0.306 & 0.030 & 0.156 & 0.193 & 0.412 & 0.093 & 0.374 & 0.292 & 0.324 \\
Reserve price             & 0.147 & 0.186 & 0.024 & 0.122 & 0.155 & 0.406 & 0.033 & 0.159 & 0.197 & 0.236 \\
Context richness          & 0.020 & 0.086 & 0.002 & 0.065 & 0.112 & 0.327 & 0.000 & 0.131 & 0.005 & 0.050 \\
Auction format            & 0.034 & 0.075 & 0.010 & 0.114 & 0.001 & 0.009 & 0.046 & 0.188 & 0.000 & 0.002 \\
Exploration intensity     & 0.027 & 0.050 & 0.025 & 0.108 & 0.000 & 0.007 & 0.042 & 0.065 & 0.016 & 0.123 \\
Affiliation ($\eta$)      & 0.048 & 0.081 & 0.005 & 0.023 & 0.007 & 0.019 & 0.004 & 0.027 & 0.005 & 0.015 \\
\midrule
Residual                  & \multicolumn{2}{c}{0.382} & \multicolumn{2}{c}{0.658} & \multicolumn{2}{c}{0.175} & \multicolumn{2}{c}{0.418} & \multicolumn{2}{c}{0.367} \\
\bottomrule
\end{tabular}}
\end{table}

\begin{table}[H]
\centering
\caption{Experiment~3a: Total-order Sobol' indices ($S_T$) for performance and efficiency responses. $R^2$: revenue 0.89, welfare 0.80, PoA 0.64, budget utilisation 0.96, allocative efficiency 0.89, no-sale rate 0.74, LP welfare 0.80.}
\label{tab:sens_exp3a_primary}
\small
\begin{tabular}{l ccccccc}
\toprule
Factor & Revenue & Welfare & Eff.\ PoA & Budget & Alloc.\ Eff. & No-Sale & LP Welfare \\
\midrule
Budget multiplier         & 0.540 & 0.519 & 0.285 & 0.548 & 0.277 & 0.401 & 0.535 \\
Bidder objective          & 0.384 & 0.010 & 0.345 & 0.661 & 0.477 & 0.128 & 0.001 \\
Number of bidders         & 0.171 & 0.177 & 0.159 & 0.030 & 0.130 & 0.238 & 0.188 \\
Value dispersion ($\sigma$) & 0.014 & 0.163 & 0.139 & 0.007 & 0.252 & 0.027 & 0.127 \\
Auction format            & 0.008 & 0.001 & 0.005 & 0.018 & 0.004 & 0.093 & 0.001 \\
Reserve price             & 0.001 & 0.003 & 0.001 & 0.001 & 0.001 & 0.142 & 0.003 \\
\midrule
Residual                  & 0.111 & 0.203 & 0.360 & 0.043 & 0.108 & 0.258 & 0.203 \\
\bottomrule
\end{tabular}
\end{table}

\begin{table}[H]
\centering
\caption{Experiment~3a: Total-order Sobol' indices ($S_T$) for stability and dynamics responses. $R^2$: pacing stability 0.62, entropy 0.87, warm start 0.48, inter-episode volatility 0.28, bid-to-value 0.12, bid suppression 0.12, cross-episode drift 0.12.}
\label{tab:sens_exp3a_secondary}
\small
\resizebox{\textwidth}{!}{%
\begin{tabular}{l ccccccc}
\toprule
Factor & Pacing Stab. & Entropy & Warm Start & Inter-Ep.\ Vol. & Bid-to-Value & Bid Supp. & Drift \\
\midrule
Budget multiplier         & 0.431 & 0.119 & 0.245 & 0.125 & 0.051 & 0.057 & 0.041 \\
Bidder objective          & 0.403 & 0.144 & 0.218 & 0.066 & 0.046 & 0.049 & 0.041 \\
Number of bidders         & 0.064 & 0.630 & 0.174 & 0.071 & 0.023 & 0.023 & 0.041 \\
Value dispersion ($\sigma$) & 0.127 & 0.087 & 0.025 & 0.124 & 0.030 & 0.030 & 0.041 \\
Auction format            & 0.022 & 0.001 & 0.047 & 0.035 & 0.036 & 0.025 & 0.041 \\
Reserve price             & 0.003 & 0.002 & 0.001 & 0.005 & 0.007 & 0.007 & 0.004 \\
\bottomrule
\end{tabular}}
\end{table}

\begin{table}[H]
\centering
\caption{Experiment~3b: Total-order Sobol' indices ($S_T$) for performance and efficiency responses. $R^2$: revenue 0.77, welfare 0.73, PoA 0.39, budget utilisation 0.82, allocative efficiency 0.83, no-sale rate 0.70, LP welfare 0.73.}
\label{tab:sens_exp3b_primary}
\small
\begin{tabular}{l ccccccc}
\toprule
Factor & Revenue & Welfare & Eff.\ PoA & Budget & Alloc.\ Eff. & No-Sale & LP Welfare \\
\midrule
Budget multiplier         & 0.529 & 0.495 & 0.170 & 0.702 & 0.738 & 0.332 & 0.506 \\
Auction format            & 0.116 & 0.004 & 0.191 & 0.099 & 0.030 & 0.300 & 0.001 \\
Number of bidders         & 0.131 & 0.168 & 0.147 & 0.107 & 0.044 & 0.104 & 0.176 \\
Value dispersion ($\sigma$) & 0.065 & 0.118 & 0.042 & 0.015 & 0.036 & 0.025 & 0.107 \\
Reserve price             & 0.001 & 0.004 & 0.032 & 0.002 & 0.025 & 0.302 & 0.003 \\
Aggressiveness            & 0.001 & 0.001 & 0.007 & 0.000 & 0.001 & 0.008 & 0.001 \\
\midrule
Residual                  & 0.232 & 0.268 & 0.609 & 0.178 & 0.173 & 0.303 & 0.269 \\
\bottomrule
\end{tabular}
\end{table}

\begin{table}[H]
\centering
\caption{Experiment~3b: Total-order Sobol' indices ($S_T$) for stability and dynamics responses. $R^2$: pacing stability 0.88, entropy 0.80, warm start 0.03, inter-episode volatility 0.61, bid-to-value 0.89, bid suppression 0.89, cross-episode drift 0.05.}
\label{tab:sens_exp3b_secondary}
\small
\resizebox{\textwidth}{!}{%
\begin{tabular}{l ccccccc}
\toprule
Factor & Pacing Stab. & Entropy & Warm Start & Inter-Ep.\ Vol. & Bid-to-Value & Bid Supp. & Drift \\
\midrule
Budget multiplier         & 0.748 & 0.192 & 0.014 & 0.256 & 0.759 & 0.759 & 0.010 \\
Auction format            & 0.195 & 0.031 & 0.003 & 0.330 & 0.194 & 0.194 & 0.012 \\
Number of bidders         & 0.012 & 0.576 & 0.005 & 0.267 & 0.010 & 0.010 & 0.026 \\
Value dispersion ($\sigma$) & 0.006 & 0.080 & 0.017 & 0.070 & 0.002 & 0.002 & 0.013 \\
Reserve price             & 0.001 & 0.005 & 0.011 & 0.022 & 0.001 & 0.001 & 0.034 \\
Aggressiveness            & 0.006 & 0.001 & 0.004 & 0.033 & 0.000 & 0.000 & 0.007 \\
\midrule
Residual                  & 0.118 & 0.202 & 0.970 & 0.392 & 0.114 & 0.114 & 0.946 \\
\bottomrule
\end{tabular}}
\end{table}

\end{document}